\documentclass[prl,twocolumn,twoside,superscriptaddress,longbibliography]{revtex4-2}
\usepackage[latin9]{inputenc}
\usepackage{color}
\usepackage{graphicx}
\usepackage{amsmath}
\usepackage{amssymb}
\usepackage[version=3]{mhchem}
\usepackage{float}
\usepackage{soul}
\usepackage{changes}
\usepackage{mathrsfs}
\usepackage{mathtools}
\usepackage{comment}
\usepackage{multirow}
\usepackage{xcolor}
\usepackage{physics}
\usepackage{soul,color}
\usepackage{ulem}

\DeclareMathAlphabet{\mathpzc}{OT1}{pzc}{m}{it}

\begin{document}
\title{First-principles open quantum dynamics for solids based on density-matrix formalism}
\setcounter{page}{1}   
\date{\today} 
\author{Jacopo Simoni}
\affiliation{Department of Materials Science and Engineering, University of Wisconsin-Madison, 53706, USA}
\author{Gabriele Riva}
\affiliation{Department of Materials Science and Engineering, University of Wisconsin-Madison, 53706, USA}
\author{Yuan Ping}
\email{yping3@wisc.edu} 
\affiliation{Department of Materials Science and Engineering, University of Wisconsin-Madison, 53706, USA}
\affiliation{Department of Physics, University of Wisconsin-Madison, 53706, USA}
\affiliation{Department of Chemistry, University of Wisconsin-Madison, 53706, USA}

\begin{abstract}
    \noindent
    The theoretical description of materials' properties driven out of equilibrium has important consequences in various fields such as semiconductor spintronics, nonlinear optics, continuous and discrete quantum information science and technology.
    The coupling of a quantum many-body system to an external bath can dramatically modify its dynamics compared to that of closed systems, new phenomena like relaxation and decoherence appear as a consequence of the non-unitary evolution of the quantum system. In addition, electron-electron correlations must be properly accounted for in order to go beyond a simple one-electron or mean-field description of the electronic system.
    Here we discuss a first-principles methodology based on the evolution of the electronic density matrix capable of treating electron-environment interactions and electron-electron correlations at the same level of description. The effect of the environment is separated into a coherent contribution, like the coupling to applied external electro-magnetic fields, and an incoherent contribution, like the interaction with lattice vibrations or the thermal background of radiation. Electron-electron interactions are included using the nonequilibrium Green's function plus generalized Kadanoff-Baym ansatz. The obtained non-Markovian coupled set of equations reduces to ordinary Lindblad quantum master equation form in the Markovian limit.
\end{abstract}

\maketitle
\section{Introduction - open quantum dynamics}
\begin{figure}[htbp]
    \centering
   \includegraphics[width=\columnwidth]{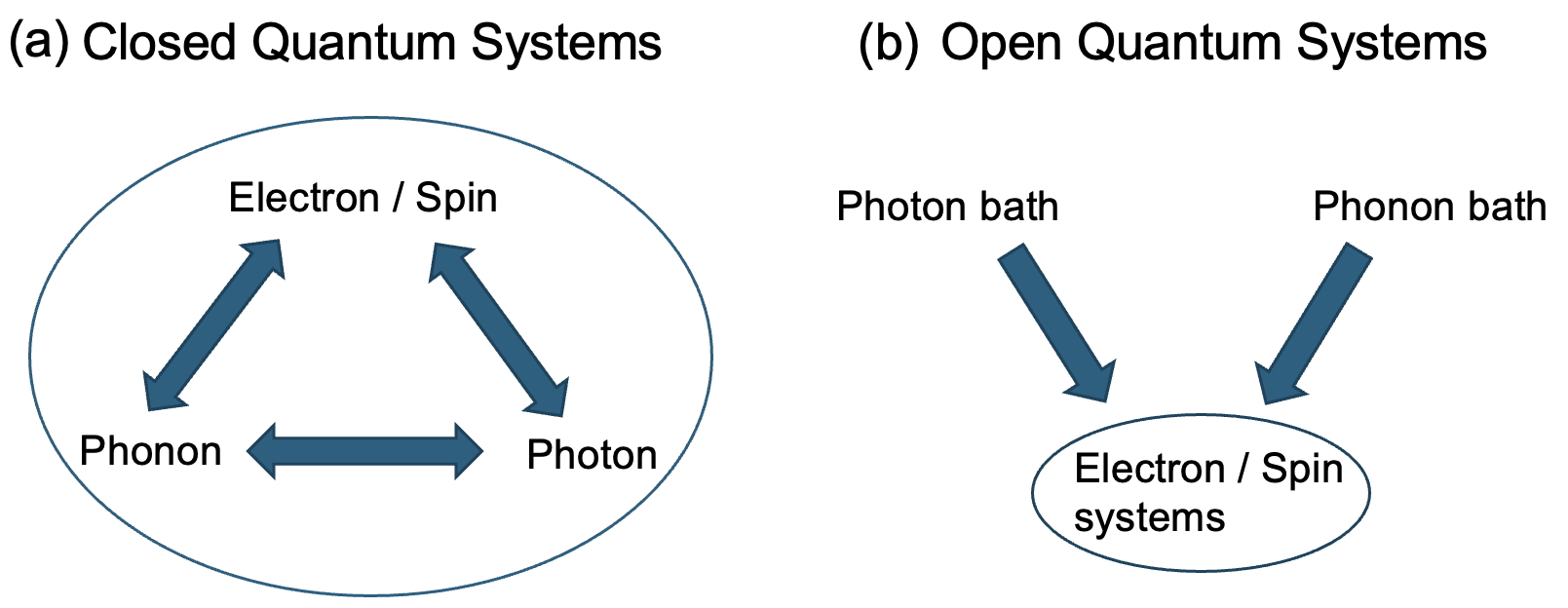}
    \caption{Schematic depiction between closed and open quantum systems. (a) The closed quantum systems treat interactions among different quasiparticles self-consistently at equal footing. (b) The open quantum systems partition the total to systems (e.g. electrons) and baths (e.g. photons and phonons), where the system-bath interactions are treated at various levels of approximations.}
    \label{fig:sys-env}
\end{figure}

\noindent
A closed quantum system is a system that does not interact with its surroundings. In Fig.~(\ref{fig:sys-env}a) the coupled system formed by electrons, phonons, and photons can be considered closed given that these three reservoirs do not interact with any additional degree of freedom.\\
\noindent
A closed system is clearly an idealization that is, most of the time, unnecessary.
Physically, one often picks certain subsystems of interest as the focus, such as electrons, then treats the rest (e.g. photons, phonons) as the environment where the electronic subsystem is immersed.  
Consequently, we call a quantum system {\rm S} open when it is coupled to an external system {\rm E} (the environment); in this case {\rm S} interacts with {\rm E}, as depicted in Fig.~(\ref{fig:sys-env}b) where the electron/spin system is the open system and the phonon and photon baths are the external environment.\\
\noindent
There are two forms of interaction (considering that the number of electrons must be conserved during the process) that we should take into account in the case of open systems, i.e. interactions between the system {\rm S} and the environment {\rm E}, and interactions within the system itself.
In Fig.(\ref{fig:sys-env}b) the interaction between the system and the environment, formed by the photonic and the phononic baths, is represented by arrows.
The combined supersystem ({\rm S}+{\rm E}) is a closed system given that all the interactions are internal to the supersystem, and the evolution of the density matrix $\rho$ is unitary; instead, the evolution of the reduced density matrix, $\rho_{\rm S}={\rm Tr}_{\rm E}\rho$ obtained by tracing out the degrees of freedom of the environment is not\cite{OQS_book1}.\\
%
%
When the evolution of the quantum system is not unitary, decoherence appears as the decay of the off-diagonal elements of the density matrix\cite{Shu2023}. In condensed matter physics, the most important forms of interaction with the environment are: {\bf (1)} interactions with phonons, or lattice vibrations; {\bf (2)} interaction with impurities, and {\bf (3)} interaction with photons. An accurate description of decoherence in solids and condensed matter systems requires a first-principles description of such interactions. 
A second important problem is that condensed matter systems are many-electron systems. Electron-electron interactions cannot be neglected, and this requires the development of a many-body open quantum systems formalism beyond the single-particle picture. There is no unified first-principles theory that accounts for all these different interactions at the same level of description. Non-equilibrium Green function methods are the standard approach for the study of many-body electronic systems driven out of equilibrium\cite{bonitz2015quantum}. The description of a quantum system that interacts with external phonon or photon baths is instead often based on 
Markovian approximations such as Lindbladian or Redfield equations, quite accurate for weak system-bath interactions, or taking additional approximations\cite{Petruccione07}. Here we will try to bridge the gap between these two approaches and develop a first-principles formalism based on the evolution of the one-particle density matrix of the many-body systems.\\
\noindent
The study of open quantum condensed matter systems has important applications in the description of highly correlated materials and their interactions with the environment. 
Quantum technologies such as quantum computing\cite{Bharti_2022,Georgescu_2014,Lloyd_1996}, quantum communications, and quantum sensing\cite{Bonato2016} crucially depend on the ability to preserve the coherence of quantum states for the time needed for reliable operation. For this purpose, various experiments have been designed to control decoherence in superconductors\cite{Chiorescu03,Martinis02,Yu02}, ion traps\cite{Myatt00,Turchette00}, quantum dots\cite{Kuhlmann13,Urbaszek13,Tighineanu18} and micromechanical resonators\cite{Fong12,Zhang14,Moser14}. The ability to predict decoherence times using first-principles theory is of fundamental importance for the development of these technologies.
Low power electronics based on spin or orbital angular momentum as information carriers \cite{RevModPhys.76.323,RevModPhys.80.1517,PhysRevLett.95.066601,Jo2024} also require long spin and OAM relaxation and decoherence times, as well as long spin or orbital diffusion lengths, for low-energy dissipation and stable operation.  
In the next section, we discuss different theoretical approaches for describing open quantum dynamics in solids and highly correlated materials.

\section{Theoretical approaches to open quantum dynamics}
\noindent
Historically, the theoretical methods used to describe the dynamics of open quantum systems fall into three main categories\cite{Weiss2008} as follows.\\
{\rm (i)} Approaches based on the modification of the quantization procedure to include additional dissipation terms in the Schr{\"o}dinger equation of the system {\rm S}\cite{Dekker77,Kostin72,Yasue78,Nelson66}. These methodologies were not successful and are no longer being used.\\
{\rm (ii)} Full Hilbert space treatment (of the system {\rm S}). Several approaches have been developed, which are briefly explained below.  {\rm (a)} Wave function Monte Carlo techniques based on an exact numerical treatment of the Hilbert space of the system {\rm S}\cite{PhysRevA.85.043620,PhysRevA.88.053627}; the stochastic Schr\"odinger equation method\cite{Gisin84,Gisin92} where the state vector behaves as a stochastic process; or the quantum jump method where additional non-Hermitian terms are present in the Hamiltonian and a quantum jump occurs when the norm of the state vector drops below a certain value\cite{Diosi88,Dalibard91}. {\rm (b)} Tensor network simulation techniques are limited to quantum states that are most relevant to the dynamical evolution\cite{Biamonte2017TensorNI}. {\rm (c)} Variational methods based on parameterization of the total density matrix, $\rho_{\rm S}$, of the system\cite{PhysRevLett.114.040402}.
These methods, although very successful, are hardly applicable to solid-state systems due to the high dimension of the Hilbert space in solids. We will not discuss these methods further, as they are explained in detail elsewhere\cite{RevModPhys.93.015008}.\\
{\rm (iii)} The reduced system + environment approach. In this case, the dynamics of a reduced quantum system is obtained from the dynamics of the complete system {\rm S+E}, by tracing out the degrees of freedom of the environment and part of the degrees of freedom of the quantum system {\rm S}. Within the Schr\"odinger picture, this formalism gives rise to quantum master equations for the reduced density operator\cite{Prigogine61,Petruccione07}; in the Heisenberg picture it leads to the quantum Langevin equations for the observables of the reduced system \cite{Mori65}.
The problem, in general, cannot be solved exactly due to the correlations that unavoidably arise between the reduced system and the degrees of freedom that have been traced out; as a consequence, the reduced system dynamics cannot be completely decoupled from the rest of the system or environmental degrees of freedom. All the methods most directly applicable to solid state systems fall within this category and will be analyzed in more detail below.\\
In the case of weak correlations with the environment (and close to equilibrium), relaxation times can often be reliably computed perturbatively using the Fermi Golden Rule (FGR)\cite{PhysRevLett.129.140402,PhysRevLett.59.1460}. A quantum FGR was derived that includes the off-diagonal contribution of the density matrix~\cite{Rossi2008}, which is meant to describe close-to-equilibrium decoherence processes. 
This scheme works more reliably in the limit of small perturbations and linear response, and is strongly limited in the description of out-of-equilibrium effects\cite{PhysRevLett.109.166604,PhysRevB.101.045202}.\\
The application of Keldysh functional integral techniques\cite{stefanucci2024kadanoff,THOMPSON2023169385} to the study of non-equilibrium evolution of many-body dissipative-driven quantum systems has been often limited to model Hamiltonians for bosons and fermions~\cite{Reeves2024,bonitz2020,PhysRevB.80.115107,PhysRevB.82.155108,PhysRevLett.103.176404,PhysRevB.93.054303,Reeves2024,Pav2024}.
Non-Equilibrium Green's Function (NEGF) methods based on Kadanoff-Baym equations (KBE) with the generalized Kadanoff-Baym ansatz (GKBA) are in active development for studying non-equilibrium many-body systems from first-principles \cite{Stef2022,Joost2020,Stef2021,Stef18,MARINI2022147189,refId0}. They have been applied to problems in nonlinear optics\cite{doi:10.1073/pnas.1906938118},  and ultrafast electron dynamics in low-dimensional materials\cite{PhysRevLett.128.016801,Perfett02023}.
In these studies, the system's coupling with the environment (photons and phonons) is described either semi-classically or evolving self-consistently electrons and phonons density matrices, which already represents a quite formidable task given the complexity of the problem\cite{Perfett02023}. However, given the current status of the field, a complete description of electrons, phonons and photons coupled dynamics remains a very difficult task for realistic systems\cite{PhysRevB.93.155102}. Another important application area for open quantum system dynamics based on NEGF is the study of non-equilibrium currents in molecular junctions, where the contribution of phonons to dynamics is more relevant\cite{di2008electrical,PhysRevB.82.085426,PhysRevLett.93.256601,PhysRevLett.100.226604}. Here, because of strong electron-phonon interactions, electrons and molecular vibrations are often treated using non adiabatic methods beyond the Born-Oppenheimer approximation. In general, the computational complexity of these methods makes their applicability still limited\cite{PhysRevB.100.235117}.\\
Time-dependent density functional theory (TDDFT)\cite{PhysRevLett.52.997,Marques2003} is the time-dependent extension of density functional theory (DFT) for simulating excited states and non-equilibrium phenomena\cite{Xu2024}. 
The ability of TDDFT to describe decoherence has previously been discussed\cite{10.1093/acprof:oso/9780199563029.001.0001}, and a generalization of TDDFT for open quantum systems has been proposed ~\cite{PhysRevLett.104.043001}.
However, a complete open-system Kohn-Sham TDDFT formulation requires the knowledge of an exchange-correlation functional for the interaction with the environment.
A possible alternative is to consistently simulate the full dynamics of the system composed of the electrons (the Kohn-Sham system) and the environment; the effect of the environment is then given by the direct interaction between the electronic system and the external degrees of freedom.
One problem here is that the interaction of electronic systems with phonon baths requires large supercells, leading to high computational costs. In addition, the combined dynamics of electrons with the atomic system has been mainly described using the surface hopping method\cite{10.1063/1.459170} and Ehrenfest dynamics\cite{Zhang_NL_2019,PhysRevB.102.184308,PhysRevB.94.184310}, where the exact description of detailed balance and decoherence remains challenging\cite{Parandekar2006,https://doi.org/10.1002/cphc.201200941}.\\
\noindent
Other methods based on the exact solution of the Schr{\"o}dinger equation for the single-particle system, like symmetry-adapted-cluster configuration-interaction (SAC-CI)\cite{Nakatsuji1979}, or multiconfiguration time-dependent Hartree-Fock (MCTDHF)\cite{Caillat2005} have been used to study the ultrafast dynamics of molecules under strong external fields\cite{10.1063/1.3553176,Zanghellini2003AnMA,PhysRevA.71.012712} or the dynamics of small magnetic clusters\cite{Barhoumi2023,Lefkidis24,PhysRevB.84.054415}. External fields are usually treated as classical potentials. These methods cannot be easily applied to solid-state systems. The description of environmental decoherence using such approaches is, in fact, computationally very expensive and difficult to achieve\cite{Baiardi2019}. 
The time-dependent Density Matrix Renormalization Group (t-DMRG) represents an attempt to overcome these difficulties and has been applied to the study of electronic dynamics in electron-phonon coupled systems\cite{10.1063/1.5125945}. Examples are given by the application of t-DMRG to the calculation of carrier mobilities in organic semiconductors\cite{tDMRG1} and to the study of exciton dynamics in a system with hundreds of atomic vibrations\cite{Borrelli2017}. However, the method becomes unfavorable in larger dimensions, limiting its practical applicability. Finally, some recent works have also explored the possibility of simulating open quantum dynamics based on the Lindblad master equations on quantum computers\cite{PhysRevLett.127.270503}.\\
\noindent
From considerations above, a clear need emerges, 
for a computationally-tractable first principles approach for open quantum dynamics, which is capable of simultaneously describing different interactions and sources of decoherence in nanostructures and solid-state materials.
Within the system plus the  environment framework the Von Neumann equation for the full density matrix $\hat{\rho}(t)$ in the interaction picture is written as
\begin{equation}\label{Eq:fullrho}
    \frac{d\hat{\rho}}{dt} = -i\big[\hat{H}_{\rm sys-env}(t), \hat{\rho}(t)\big]\,,
\end{equation}
$\hat{H}_{\rm sys-env}(t)$ is the interaction Hamiltonian between the system and the environment. Eq.~(\ref{Eq:fullrho}) cannot be solved exactly and in the next sections we will discuss different possible approximations.\\
In section (\ref{sec:DMformalism}) we describe the nature of the physical system's Hamiltonian, considering the different interaction terms contributing to $\hat{H}_{\rm sys-env}$, and writing the dynamical equations for the density matrix and the other correlation functions. In section (\ref{sec:markovlim}) we discuss the Markovian limits of the dynamical equations and the structure of the resulting quantum master equations. In section~(\ref{sec:OQSsolids}), we discuss possible applications of the formalism to open quantum dynamics in solids. In the next sections, to simplify the notation, we will use atomic units ($\hbar=1$) and refer to electronic states using integers, phonon states using the label ${\bf q}=({\bf Q},\lambda)$, which merges together the phonon momentum and modes, and photon states using the label ${\bf k}=({\bf K},\eta)$, which instead merges the photon momentum and modes. 
%
\noindent
\section{Density matrix based formalism}\label{sec:DMformalism}
\subsection{The Hamiltonian of system + environment}
\noindent
The total Hamiltonian including both the system and the environment degrees of freedom is given as
\begin{equation}
    \hat{H}(t) = \hat{H}_{\rm sys} + \hat{H}_{\rm env}(t) + \hat{H}_{\rm sys-env}(t)\,,
\end{equation}
where the last term incorporates all the possible interactions between the system and the environment.\\
The system corresponds to the gas of interacting electrons in the solid, and the environment is given by the solid's lattice and the external radiation field,
\begin{align}
    &\hat{H}_{\rm sys} = \hat{H}_{\rm e}^{0} + \hat{H}_{\rm ee}\nonumber\\
    &\hat{H}_{\rm env} (t) = \hat{H}_{\rm latt} + \hat{H}_{\rm rad}(t)\nonumber\\
    &\hat{H}_{\rm sys-env} (t)  = \hat{H}_{\rm e-ph} + \hat{H}_{\rm e-i} +\hat{H}_{\rm e-rad}(t)\,.\label{eq:sys-env}
\end{align}
In the previous expression $\hat{H}_{\rm e}^0$ is the single-particle electronic Hamiltonian, written in terms of the kinetic energy operator and of the unperturbed bare electron-ion potential $\hat{H}_{\rm e}^0=\hat{T}+\hat{V}_{\rm ei}^0$. $\hat{H}_{\rm ee}$ is the Coulomb electron-electron interaction. In the second quantization, the Hamiltonian of the electron gas in the solid is written as
\begin{equation}\label{Hsys}
    \hat{H}_{\rm sys} = \hat{H}_{\rm e}^{0} + \hat{H}_{\rm ee}=\sum_1 \varepsilon_1 \hat{c}_1^\dagger \hat{c}_1 + \frac{1}{2}\sum_{12;34}v_{12;34}\hat{c}_1^\dagger \hat{c}_2^\dagger \hat{c}_3 \hat{c}_4\,,
\end{equation}
where $\varepsilon_1$ is the energy of the single particle state in the crystal field potential $V_{\rm ei}^0$; $\hat{c}_1$ and $\hat{c}^\dagger_1$ are the electron annihilation and creation operators for the state $\ket{1}$. $v_{12;34}$ represents the single-particle matrix elements of the bare electron-electron potential using physics notation,
\begin{align}\label{Eq:potential}
    &v_{12;34} =\int d{\bf x}\!\!\int d{\bf y}\,\phi_1({\bf x})^*\phi_2({\bf y})^* v_{\rm c}({\bf x} - {\bf y})\phi_3({\bf y})\phi_4({\bf x}),
\end{align}
where $\phi_1({\bf x})$ are the eigenstates of the single particle Hamiltonian, and $v_{\rm c}({\bf x}-{\bf y})$ is the Coulomb potential.
The Hamiltonian of the environment is the sum of the atomic lattice and of the radiation field Hamiltonians
\begin{align} \label{Eq:Hrad}
    &\hat{H}_{\rm latt} = \bar{V}_{\rm nn} + \Delta\hat{V}_{\rm imp} + \sum_{\bf q}\epsilon_{\bf q}\Big(\hat{b}_{\bf q}^\dagger \hat{b}_{\bf q} +\frac{1}{2}\Big)\nonumber\\
    &\hat{H}_{\rm rad}(t) = E_{\rm rad}^{\rm ext}(t) + \sum_{\bf k}\omega_{\bf k} \Big(\hat{a}_{\bf k}^\dagger \hat{a}_{\bf k} + \frac{1}{2}\Big)\,.
\end{align}
The first two terms in $\hat{H}_{\rm latt}$ correspond to the energy of the static atomic lattice $\bar{V}_{\rm nn}$ and to the variation of lattice energy due to the presence of impurities ($\Delta\hat{V}_{\rm imp}$). The last term in $\hat{H}_{\rm latt}$ is the Hamiltonian of the free phonons $\hat{H}_{\rm ph}$, which describes the vibrations of the lattice. $\epsilon_{\bf q}$, $\hat{b}_{\bf q}$ and $\hat{b}_{\bf q}^\dagger$, are the energy, annihilation, and creation operators of the phonon mode. $\hat{H}_{\rm rad}(t)$ is the effective Hamiltonian of the radiation field acting on the system. To understand its structure we start from the full electro-magnetic field Hamiltonian ($\hat{H}_{\rm EM}(t)$), written as $\hat{H}_{\rm EM}(t)=\frac{1}{2}\int_V d{\bf r}(\frac{1}{4\pi}\abs*{\hat{{\bf E}}({\bf r},t)}^2 + \frac{c^2}{4\pi}\abs*{\hat{{\bf B}}({\bf r},t)}^2)$. In the presence of an external radiation source, we can separate the electric and magnetic field operators into their out-of-equilibrium 
and their thermal quantum fluctuations components. After some algebra given in appendix (\ref{app:em}), the effective Hamiltonian of the radiation field, $\hat{H}_{\rm rad}(t)$, is given as in Eq.~(\ref{Eq:Hrad}). There $E^{\rm ext}_{\rm rad}(t)=\frac{1}{2}\int_V d{\bf r}(\frac{1}{4\pi}\abs{\bar{{\bf E}}({\bf r},t)}^2 + \frac{c^2}{4\pi}\abs{\bar{\bf B}({\bf r},t)}^2)$ is the energy of the external electro-magnetic sources
\begin{align}
\bar{\bf E} &=\expval*{\hat{\bf E}}={\rm Tr}[\delta\hat{\rho}_{\rm rad}\hat{\bf E}] \,\\
\bar{\bf B} &=\expval*{\hat{\bf B}}={\rm Tr}[\delta\hat{\rho}_{\rm rad}\hat{\bf B}].
\end{align}
\noindent
The second term in $\hat{H}_{\rm rad}(t)$ from Eq.~(\ref{Eq:Hrad}) instead gives the energy of the thermal quantum fluctuations $\delta\hat{\bf E}({\bf r},t)=\hat{{\bf E}}({\bf r},t)-\bar{\bf E}({\bf r},t)$ and $\delta\hat{\bf B}({\bf r},t)=\hat{{\bf B}}({\bf r},t)-\bar{\bf B}({\bf r},t)$ with respect to the external macroscopic fields as shown in Eq.~(\ref{A11}).
$\omega_{\bf k}$ is the energy of the radiation mode {\bf k}; $\hat{a}_{\bf k}$ and $\hat{a}_{\bf k}^\dagger$ are the annihilation and creation operators of the eigenmode of the fluctuation field such that $\expval*{\hat{a}_{\bf k}}={\rm Tr}[\hat{\rho}_{\rm rad}\hat{a}_{\bf k}]=0$.
\subsection{The interaction between the electrons and the environment}
\noindent
We now analyze each interaction term, i.e. $\hat{H}_{\rm e-ph}$, $\hat{H}_{\rm e-i}$, $\hat{H}_{\rm e-rad}$, in $\hat{H}_{\rm sys-env}(t)$ in Eq.~(\ref{eq:sys-env}) respectively. At the first order expansion of $\hat{H}_{\rm e}^0$ over atomic positions, the electron-phonon Hamiltonian, $\hat{H}_{\rm e-ph}$, is linear in the phonon displacement field and can be written as
\begin{equation}\label{Eq:ephinter}
    \hat{H}_{\rm e-ph} = \sum_{\bf q}\sum_{1,2}\big(g_{12}^{{\bf q}-}\hat{c}_1^\dagger \hat{b}_{\bf q}\hat{c}_{2} + g_{12}^{{\bf q}+} \hat{c}_{2}^\dagger \hat{b}_{\bf q}^\dagger \hat{c}_1\big)\,,
\end{equation}
where $g_{12}^{{\bf q}-}$ is the electron-phonon coupling matrix, using the electronic states $\ket{1}=\ket{\mu{\bf K'}+{\bf Q}}$ and $\ket{2}=\ket{\nu{\bf K'}}$.
\begin{equation}\label{Eq:geph}
    g_{\mu\nu}^\lambda({\bf K'},{\bf Q})=\mel*{\mu{\bf K'}+{\bf Q}}{\Delta_{{\bf q}={\bf Q},\lambda}\hat{H}_{\rm sys}}{\nu{\bf K'}}\ ,
\end{equation}
that satisfies $g_{12}^{{\bf q}+}=g_{21}^{{\bf p}-}$, where ${\bf p}=(-{\bf Q},\lambda)$ and $\lambda$ is the phonon mode.
In practical calculations $\hat{H}_{\rm sys}$ in Eq.~(\ref{Eq:geph}) is replaced by an effective single-particle Hamiltonian, like the Kohn-Sham Hamiltonian of density functional theory, neglecting higher-order electronic correlations. However, we should keep in mind that $g_{12}^{{\bf q}-}$ defines the response of the full electronic system to the external phonon perturbation. A comprehensive review of first-principles electron-phonon calculations in solids is given in Ref.~\cite{RevModPhys.89.015003}.\\
Similarly, the interaction with impurities is given by the following term.
\begin{equation}
    \hat{H}_{\rm e-i} = \sum_{1,2}g_{12}^{\rm i}\hat{c}_1^\dagger\hat{c}_2\,,
\end{equation}
where $g_{12}^{\rm i}=\mel*{\mu{\bf K'}}{\Delta\hat{V}_{\rm imp}}{\nu{\bf K'}}$ is the matrix element of the elastic scattering between the impurity potential ($\Delta\hat{V}_{\rm imp}$) and the unperturbed electronic states.  $\hat{H}_{\rm e-rad}$ describes the coupling between the system and the electro-magnetic radiation. The interaction Hamiltonian of an electron within an external electromagnetic field is written as
\begin{equation}\label{Eq:Herad0}
    \hat{H}_{\rm e-rad}(t)\!\! = \frac{1}{2}\big(\hat{\bf p}\cdot\hat{\bf A}({\bf r},t) + \hat{\bf A}({\bf r},t)\cdot\hat{\bf p}\big) + \frac{\hat{\bf A}({\bf r},t)^2}{2} + \hat{\bf B}({\bf r},t)\cdot\hat{\bf S}\,,
\end{equation}
that corresponds to the Pauli Hamiltonian of a single electron interacting with the electro-magnetic field, which can be simplified with a choice of gauge for the vector potential. With the choice of Coulomb gauge, we have $-i\nabla\cdot\hat{\bf A}({\bf r},t)=0$. Following the procedure outlined in appendix (\ref{app:radmatinter}), we can express $\hat{\bf A}({\bf r},t)=\bar{\bf A}({\bf r},t)+\delta\hat{\bf A}({\bf r},t)$. In addition, by neglecting the magnetic field fluctuations and assuming that the field is spatially homogeneous, $\hat{\bf B}({\bf r},t)\simeq\bar{\bf B}(t)$, we can use the definition of the vector potential $\bar{\bf B}(t)=\nabla\times\hat{\bf A}({\bf r},t)$ and obtain $\hat{\bf A}({\bf r},t)=\frac{1}{2}\bar{\bf B}\times{\bf r}+\bar{\bf a}(t)+\delta\hat{\bf a}(t)$, such that ${\rm Tr}[\hat{\rho}_{\rm rad}(t)\delta\hat{\bf a}(t)]=0$.\\
If we neglect higher-order terms that give rise to multi-photon processes proportional to $\hat{\bf a}(t)^2$, we obtain the following expression\cite{sakurai1999advanced}
\begin{equation}\label{Eq:Herad}
    \hat{H}_{\rm e-rad}(t) = \big(\bar{\bf a}(t) + \delta\hat{\bf a}(t)\big)\cdot\hat{\bf p} + \frac{1}{2}\bar{\bf B}(t)\cdot\big(\hat{\bf L} + 2\hat{\bf S}\big)\,,
\end{equation}
where $\hat{\bf L}$ is the electron orbital angular momentum. In Eq.~(\ref{Eq:Herad}) we can distinguish between a coherent, macroscopic and single-particle, and an incoherent contribution to the electronic dynamics. The term $h^{\rm e-r}(t)=\bar{\bf a}(t)\cdot\hat{\bf p}+\frac{1}{2}\bar{\bf B}(t)\cdot(\hat{\bf L}+2\hat{\bf S})$ is a one-particle operator and can be fully expressed by means of a single-particle basis of the Hilbert space of the system {\rm S}. It is not affected when we trace out the environment degrees of freedom; as a consequence, it contributes only to the coherent or the single-electron part of the dynamics.
The term $\delta\hat{\bf a}(t)\cdot\hat{\bf p}$ is responsible for photon absorption and emission processes and it is described exactly only within the reduced system+environment Hilbert space. The trace out of the environment causes a loss of information that gives rise to an incoherent contribution to the electron system's density matrix dynamics. In second quantization we can then write as follows:
\begin{align} \label{Eq:Herad2ndq}
    &\hat{H}_{\rm e-rad}(t) =\nonumber\\
    &=\sum_{12}h^{\rm e-r}_{12}(t)\hat{c}_1^\dagger\hat{c}_{2} +\sum_{\bf k}\sum_{12}\big(\mathpzc{k}_{12}^{{\bf k}-}\hat{c}_1^\dagger\hat{a}_{\bf k}\hat{c}_{2} + \mathpzc{k}_{12}^{{\bf k}+}\hat{c}_{2}^\dagger\hat{a}_{\bf k}^\dagger\hat{c}_1\big)\,,
\end{align}
\noindent
where the second term on the right of Eq.~(\ref{Eq:Herad2ndq}) describes the incoherent photon absorption and emission processes, and $\mathpzc{k}_{12}^{{\bf k}-}$ are the matrix elements of the electric dipole between different electronic states. This is computed under the assumption that the photon momentum is much smaller than the electronic one, ${\bf K}\simeq 0$, and using $\ket{1}=\ket{\mu{\bf K'}}$ and $\ket{2}=\ket{\nu{\bf K'}}$, where $\mu,\nu$ are the electronic bands and ${\bf K'}$ the k-point.
\begin{equation}\label{Eq:dipole}
    \mathpzc{k}_{12}^{{\bf k}-} = {\sqrt{\frac{2\pi}{\omega_{\bf k}V}}}e_{\bf k}\cdot\mel*{\mu{\bf K'}}{\hat{\bf p}}{\nu{\bf K'}}
\end{equation}
$V$ is the volume of the system, and $e_{\bf k}$ is the wave polarization vector. 

\noindent
In the next section, we will see how starting from Eq.~(\ref{Eq:fullrho}) we can obtain a quantum master equation for the single-particle density matrix time evolution. Table~(\ref{Tab:1}) provides a summary of the symbols associated with the different quantum correlation functions that will be used in the next sections.\\
\begin{table}[ht]
\begin{tabular}{ |p{3.2cm}|p{3cm}|  }
\hline
\multicolumn{2}{|c|}{Variables reference} \\
\hline
Quantum correlation function& Definition\\
\hline
$\rho_{12}^{(1)}=\expval*{\hat{c}_2^\dagger\hat{c}_1}$ & one-particle density matrix\\
$K_{1234}=\expval*{\hat{c}_4^\dagger\hat{c}_3^\dagger\hat{c}_2\hat{c}_1}$ & two-particle density matrix\\
$\rho_{12}^{\bf q}=\expval*{\hat{c}_2^\dagger\hat{b}_{\bf q}\hat{c}_1}$ & electron-phonon density matrix\\
$\rho_{12}^{\bf k}=\expval*{\hat{c}_2^\dagger\hat{a}_{\bf k}\hat{c}_1}$ & electron-photon density matrix\\
\hline
\end{tabular}
 \caption{Summary of different quantum correlation functions.}\label{Tab:1}
\end{table}
\subsection{The one-particle density-matrix master equation}
\noindent
The dynamics of the electronic system is obtained, in principle, by solving Eq.~(\ref{Eq:fullrho}) and computing $\rho_{\rm S}$. This requires the knowledge of the eigenvalues and eigenstates of the Hamiltonian of the electronic system $\hat{H}_{\rm sys}=\hat{H}_{\rm e}^0+\hat{H}_{\rm ee}$. This is a formidable task that requires the solution of the many-body Schr{\"o}dinger equation of a solid-state system. We can instead write the equation for the single-particle density matrix, $\rho_{12}^{(1)}={\rm Tr}_{\rm S}\big\{\hat{\rho}_{\rm S}\hat{c}_2^\dagger\hat{c}_1\big\}=\expval*{\hat{c}_2^\dagger\hat{c}_1}$, where we take $<\ldots>$ as a short hand notation for ${\rm Tr}\big\{\hat{\rho}\ldots\big\}$. The trace is computed over the system (electronic) degrees of freedom. The traces over the system and the environment are defined respectively as ${\rm Tr}_{\rm S}(\ldots)=\sum_{s_i}\mel*{s_i}{\ldots}{s_i}$, where $\ket*{s_i}$ are many-body eigenstates of the system Hamiltonian $\hat{H}_{\rm sys}$. Analogously, for the environment we have ${\rm Tr}_{\rm E}(\ldots)=\sum_{e_i}\mel*{e_i}{\ldots}{e_i}$, where $\ket*{e_i}$ are eigenstates of the static $\hat{H}_{\rm env}$ without external fields.\\
The time derivative of the one-particle density matrix has two contributions,
\begin{equation}\label{Eq:rho1A}
    \frac{d\rho_{12}^{(1)}}{dt} = {\rm Tr}_{\rm S}\big\{\dot{\hat{\rho}}_{\rm S}\hat{c}_2^\dagger\hat{c}_1\big\} + {\rm Tr}_{\rm S}\bigg\{\hat{\rho}_{\rm S}\frac{d}{dt}\big(\hat{c}_2^\dagger\hat{c}_1\big)\bigg\}\,.
\end{equation}
By using Eq.~(\ref{Eq:fullrho}) for the time evolution of density matrices of the system in the interaction picture, we obtain the following.
\begin{equation}
    \dot{\hat{\rho}}_{\rm S} = -i{\rm Tr}_{\rm E}\big\{\big[\hat{H}_{\rm sys-env}(t), \hat{\rho}(t)\big]\big\}
\end{equation}
$\hat{c}_2^\dagger\hat{c}_1$ is the one-particle operator and evolves, in the interaction picture, according to the Heisenberg equation, $\frac{d}{dt}\big(\hat{c}_2^\dagger\hat{c}_1\big)=i\big[\hat{H}_{\rm sys},\hat{c}_2^\dagger\hat{c}_1\big]$. Eq.~(\ref{Eq:rho1A}) should be then written as
\begin{align}\label{Eq:rho1B}
    \frac{d\rho_{12}^{(1)}}{dt} &= i{\rm Tr}_{\rm S}{\rm Tr}_{\rm E}\big\{\hat{\rho}(t)\big[\hat{H}_{\rm sys-env}(t), \hat{c}_2^\dagger\hat{c}_1\big]\big\} +\nonumber\\
    &+i{\rm Tr}_{\rm S}\big\{\hat{\rho}_{\rm S}(t)\big[\hat{H}_{\rm sys}, \hat{c}_2^\dagger\hat{c}_1\big]\big\}.
\end{align}
We emphasize that Eq.~(\ref{Eq:rho1B}) is still exact and has two contributions, i.e. the first term coming from the interactions with the environment and the second from the dynamics of the isolated quantum system itself. The first term requires knowledge of the complete density matrix (system + environment), $\hat{\rho}(t)$, while the second depends on the system alone via $\hat{\rho}_{\rm S}(t)$. However, we notice that the two terms are not independent. $\hat{\rho}_{\rm S}$ is influenced by $\hat{H}_{\rm sys-env}(t)$ via Eq.~(\ref{Eq:fullrho}).\\
We refer to the first term in Eq.~(\ref{Eq:rho1B}) as $\dot{\rho}_{12}^{(1)}|_{\rm e-env}$, and to the second term as $\dot{\rho}_{12}^{(1)}|_{\rm e-e}$. The structure of $\hat{H}_{\rm sys-env}$ allows us to separate
the electron-environment term into three contributions coming from phonons (e-ph), radiation (e-rad), and impurities (e-i) as follows, 
\begin{equation}\label{Eq:rho1pNM}
    \dot{\rho}_{12}^{(1)} = \dot{\rho}_{12}^{(1)}|_{\rm e-e} + \dot{\rho}_{12}^{(1)}|_{\rm e-ph} + \dot{\rho}_{12}^{(1)}|_{\rm e-rad} + \dot{\rho}_{12}^{(1)}|_{\rm e-i}.
\end{equation}
We will discuss each term separately in the next sections.
%
\subsubsection {Electron-electron contribution to the dynamics}
\noindent
In this section, we examine the contribution to the dynamics of the density matrix originating from $\dot{\rho}_{12}^{(1)}|_{\rm e-e}$.
Taking the commutator with $\hat{H}_{\rm sys}$ in the second term on the right hand side of Eq.~(\ref{Eq:rho1B}) we obtain 
\begin{align}\label{Eq:rhoee}
  &\dot{\rho}_{12}^{(1)}|_{\rm e-e}=\nonumber\\
  &=-i\big(\varepsilon_{1} - \varepsilon_{2}\big)\rho_{12}^{(1)} + i\sum_{345}\big(v_{34;52}K_{1543} - v_{13;54}K_{4532}\big)\,,
\end{align}
where the first term is derived from $\hat{H}_{\rm e}^0$ and the second term is from $\hat{H}_{\rm ee}$ of electronic Hamiltonian $\hat{H}_{\rm sys}$ (Eq.~(\ref{Hsys})). 
$K$ is the two-particle correlation function, i.e. $K_{1234}=\expval*{\hat{c}_4^\dagger\hat{c}_{3}^\dagger\hat{c}_{2}\hat{c}_{1}}$.
%
%
The solution of Eq.~(\ref{Eq:rhoee}) requires the knowledge of $K_{1234}$ as a function of time\cite{scholl2011theory}. The two-particle correlation function can be decomposed into a Hartree-Fock (HF) and a quantum fluctuation contribution,
\begin{equation}\label{Eq:K2}
    K_{1234} = \underbrace{\rho_{14}^{(1)}\rho_{23}^{(1)} - \rho_{13}^{(1)}\rho_{24}^{(1)}}_{\text{HF}} + \underbrace{\delta K_{1234}}_{\text{Quantum Fluctuations}}.
\end{equation}
\noindent
Using this separation within the density matrix equation Eq.(\ref{Eq:rhoee}), we obtain the following
\begin{align}\label{Eq:rhoEOM}
    \dot{\rho}_{12}^{(1)}|_{\rm e-e} &= -i\sum_3\big(\xi_{13}\rho_{32}^{(1)} - \rho_{13}^{(1)}\xi_{32}\big) +\nonumber\\
    &+i\sum_{345}\big(v_{34;52}\delta K_{1543} - v_{13;54}\delta K_{4532}\big)
\end{align}
\noindent
and the HF energies are given by
\begin{equation}\label{Eq:xi}
    \xi_{12} = \varepsilon_1\delta_{12} + \sum_{34}\big(v_{13;42}\rho_{43}^{(1)} - v_{31;42}\rho_{43}^{(1)}\big)
\end{equation}
where the last two terms correspond to the direct (Hartree) and the exchange contributions respectively. The second term in Eq.~(\ref{Eq:rhoEOM}) contains all the many-body corrections beyond the mean-field HF approximation and requires the knowledge of the two-particle correlation function. The equation for $\delta K$ is considerably more complex\cite{scholl2011theory}. 
\begin{align}\label{Eq:fulldeltaK}
    \frac{d}{dt}\delta K_{1234} &= \underbrace{i{\rm Tr}_{\rm S}{\rm Tr}_{\rm E}\big\{\hat{\rho}(t)\big[\hat{H}_{\rm sys-env}(t), \hat{c}_4^\dagger\hat{c}_3^\dagger\hat{c}_2\hat{c}_1\big]\big\}}_{\text{exciton-phonon//polaritons}} +\nonumber\\
    &+i{\rm Tr}_{\rm S}\big\{\hat{\rho}_{\rm S}(t)\big[\hat{H}_{\rm sys},\hat{c}_4^\dagger\hat{c}_3^\dagger\hat{c}_2\hat{c}_1\big]\big\} -\dot{K}^0_{1234}.
\end{align}
We assume here that the dynamics of $\delta K$ is only governed by $\hat{H}_{\rm sys}$ and we can neglect the effect of the environment, which gives higher-order corrections to the excitation dynamics. This is equivalent to neglecting exciton-phonon and exciton-photon couplings. This approximation breaks down when polaritonic effects are important due to the strong coupling between electrons and light. In such a case, this scheme is not sufficient and an additional quantum correlation, $\expval*{\hat{c}_4^\dagger\hat{c}_3^\dagger\hat{a}_{\bf k}\hat{c}_2\hat{c}_1}$, should be consistently included by explicitly solving Eq.~(\ref{Eq:fulldeltaK}) with the effect of the interaction with radiation. Similar considerations are valid for exciton dynamics in the presence of electron-phonon interactions, which also demands the introduction of an additional correlation function, $\expval*{\hat{c}_4^\dagger\hat{c}_3^\dagger\hat{b}_{\bf q}\hat{c}_2\hat{c}_1}$.\\
If we neglect such additional many-body effects, equation (\ref{Eq:fulldeltaK}) can then be written in the following form
%
%
\begin{align}\label{Eq:bornapprox}
    \frac{d}{dt}\delta K_{1234} &= i{\rm Tr}_{\rm S}\big\{\hat{\rho}_{\rm S}(t)\big[\hat{H}_{\rm sys}, \hat{c}_4^\dagger\hat{c}_3^\dagger\hat{c}_2\hat{c}_1\big]\big\} - \dot{K}_{1234}^0\nonumber\\
    &= -i\big(\varepsilon_1 + \varepsilon_2 - \varepsilon_3 - \varepsilon_4\big)\delta K_{1234} + \nonumber\\
    &+i\sum_{567}\Big[  v_{56;74}(\delta_{73}-\rho_{73}^{(1)}) + v_{56;73}\rho_{74}^{(1)}\Big]K^0_{1265} +\nonumber\\
    &+i\sum_{567}\Big[  v_{51;76}(\delta_{25}-\rho_{25}^{(1)}) + v_{52;76}\rho_{15}^{(1)}\Big]K^0_{6734}  +\nonumber\\
    &+\delta S_{12;34}
\end{align}
where $K^0_{1234}$ is the zero-order component of the two-particle correlation function in Eq.~(\ref{Eq:K2}) and $\delta S_{12;34}$ is written as: 
\begin{align}\label{Eq:threescatt}
    &\delta S_{12;34} =\nonumber\\
    &=\!\! -i\sum_{567}\big[v_{56;74}\delta\expval*{\hat{c}_5^\dagger\hat{c}_6^\dagger\hat{c}_3^\dagger\hat{c}_7\hat{c}_2\hat{c}_1}\!\!-\!\! v_{56;73}\delta\expval*{\hat{c}_5^\dagger\hat{c}_6^\dagger\hat{c}_4^\dagger\hat{c}_7\hat{c}_2\hat{c}_1}\big]+\nonumber\\
    &+i\sum_{567}\big[v_{52;76}\delta\expval*{\hat{c}_4^\dagger\hat{c}_3^\dagger\hat{c}_5^\dagger\hat{c}_1\hat{c}_7\hat{c}_6}-v_{51;76}\delta\expval*{\hat{c}_4^\dagger\hat{c}_3^\dagger\hat{c}_5^\dagger\hat{c}_2\hat{c}_7\hat{c}_6}\big]+\nonumber\\
    &+i\sum_{567}\Big[ v_{56;74}\delta_{73}\delta K_{1265} + v_{51;76}\delta_{25}\delta K_{6734}\Big]
\end{align}
where each $\delta \expval*{\hat{c}_5^\dagger\hat{c}_6^\dagger\hat{c}_3^\dagger\hat{c}_7\hat{c}_2\hat{c}_1}$ is composed by the five connected combination of $\rho^{(1)}\delta K$ plus the three-particle correlations.
Eq. (\ref{Eq:threescatt}) cannot be solved analytically in an exact form and makes the treatment of screening effects complicated. We need, in fact, a complete closed-form solution for $\delta K$ and $\rho^{(1)}$ that could be used in Eq.~(\ref{Eq:rhoEOM}).\\
This procedure is analogous to the BBGKY hierarchical expansion in plasma physics\cite{mcquarrie76a}. At this point different approximations can be taken. If we set $\delta S$ to zero in Eq.~(\ref{Eq:bornapprox})\cite{scholl2011theory} we obtain the Born approximation.
In such a case, we leave the effects of electronic screening out of the density matrix dynamics, which makes such a solution hardly applicable to solids.
If we instead set the higher-order quantum fluctuations (i.e. the three-particle correlation) to zero but keep the second order, 
the results can produce electronic screening, exact T-matrix, and vertex corrections\cite{Wyld1963,Iotti2005,Rossi2002}.
%
%
\noindent
An alternative approach to derive many-body effects of e-e interaction consists of using the non-equilibrium Green functions (NEGF)~\cite{Stef2013,Joost2020,Stef2021}.
Within the NEGF formalism, Eq.~\eqref{Eq:rhoEOM} can be translated to the equation of motion (EOM) for the lesser component of the non-equilibrium (NE) one-particle Green function evaluated at equal time $t=t'$. Eq.~\eqref{Eq:bornapprox} is the equivalent of the EOM for the NE two-particle Green function. This equation can be solved at different levels of approximation (2-Born, GW + exchange (X), particle-particle (pp) or particle-hole (ph) T-matrix + (X))~\cite{Stef2022}.\\
\noindent
In this paper, we focus on an alternative way to write Eq.~\eqref{Eq:rhoEOM}. 
Using the adiabatic assumption~\cite{Stef2013}, Eq.~\eqref{Eq:rhoEOM} is written in a different form,
\begin{align}\label{Eq:rhoEOM_NEGF}
    &\dot{\rho}_{12}^{(1)}|_{\rm e-e} = -i\sum_{3}\big(\bar{\xi}_{13}\rho_{32}^{(1)} - \rho_{13}^{(1)}\bar{\xi}_{32}\big) - I_{12}.
\end{align}
This form introduces the collision integral, $I_{12}$, in terms of the non-equilibrium self-energy (e.g. at GW approximation~\cite{C3CS00007A}) to include the screening effects directly in the density matrix dynamics in Eq.~(\ref{Eq:rhoEOM_NEGF})~\cite{haug2007,Stef18,Marini2013}.
\begin{align}\label{Eq:collision}
    I_{12} = \sum_{3} (\Sigma^>_{13} * G^<_{32} - \Sigma^<_{13} * G^>_{32}) + \text{H.c.}\,,
\end{align}
\begin{figure}
    \centering
    \includegraphics[width=0.9\linewidth]{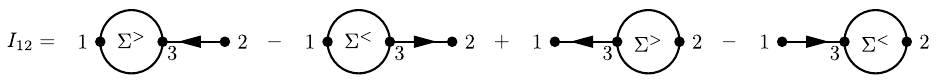}
    \caption{Schematic representation of the collision integral for e-e contributions to the dynamics. }
    \label{fig:collision integral}
\end{figure}
\noindent
where we have used the notation $\Sigma^{\lessgtr} * G^{\gtrless} = \int_{-\infty}^t d\tau \Sigma^{\lessgtr}(t,\tau) G^{\gtrless}(\tau,t)$. Eq.~\eqref{Eq:collision} is schematically represented in Fig.~(\ref{fig:collision integral}). The collision integral contains only the nonlocal-time part of the non-equilibrium self-energy. Its local-time contribution enters $\bar{\xi}$ instead. This is written in general as $\bar{\xi}_{12}=\varepsilon_1\delta_{12}+V_{12}^{\rm H}[\rho]+\Sigma_{12}^{\rm s}[\rho]$, where $V^{\rm H}_{12}$ is the Hartree potential and $\Sigma_{12}^{\rm s}$ is the static self-energy. In practice, $\Sigma_{12}^{\rm s}$ requires some approximations. Within the Fock approximation for $\Sigma^{\rm s}$, $\bar{\xi}$ reduces to the matrix in Eq.~\eqref{Eq:xi}; a more complex formulation for static self-energy is given by the so-called Coulomb-Hole + Screened Exchange (COHSEX) self-energy \cite{RevModPhys.74.601,C3CS00007A,PhysRevB.85.045116,Govoni2015}. The collision integral, $I_{12}$, is then evaluated using out-of-equilibrium GW self-energy defined as
\begin{equation}
    \Sigma^\lessgtr_{12}(t,t')=-iG^\lessgtr_{12}(t,t')W^\lessgtr_{12}(t,t').
\end{equation}
We consider the screening at the RPA level
\begin{align} \label{Eq:RPA_screening}
    W^\lessgtr_{12}(t,t')&=-i{\sum_{34}}\int_{-\infty}^tdt'' \int_{\infty}^{t'}  dt''' \nonumber \\ &\times W^R_{13}(t,t'') G^\lessgtr_{34}(t'',t''')G_{43}^\gtrless(t''',t'') W^A_{42}(t''',t').
\end{align}
The two-time dependence of the screening is important for studying femtosecond experiments with pulses shorter than or equal to the inverse plasma frequency. We focus our attention on a longer timescale $t\omega_{pl}\gg1$, i.e., we consider the retarded and advanced screened interaction in the semi-static approximation $W^R(t,t')=W^A(t,t')=W\delta(t-t')$~\cite{haug2007,Marini2013}.
To close the EOM we use the generalized Kadanoff-Baym Ansatz GKBA~\cite{Lipa1986},
\begin{align}
    &G^\lessgtr_{12}(t,t')=-{\sum_3}[G^R_{13}(t,t')\rho_{32}^\lessgtr(t')-\rho^\lessgtr_{13}(t)G^A_{32}(t,t')], \\
    &\rho^<_{12}(t)=\rho^{(1)}_{12}(t),  \;\;\;\;\;\;\;\;\;\;\;\;\;\;\;\;\;\;\; \rho^>_{12}(t)=\delta_{12}-\rho_{12}^{(1)}(t),
\end{align}
%
additionally, treating advanced and retarded GF at the quasi-particle (QP) level, i.e.~\cite{Per2015,Per2015_2}
\begin{align}\label{Eq:QA_approx}
G^{R}_{12}(t,t')&=[G^A_{12}(t',t)]^*\nonumber\\
&\approx -i\delta_{12}\theta(t-t')e^{-\frac{(t-t')^2}{2\tau_{\rm QP}^2}}e^{-i(t-t')\bar{\varepsilon}_{1}}\ .
\end{align}
Here, the GF has been written on the basis of the eigenstates of the $\bar{\xi}_{12}$ matrix. The energies $\bar{\varepsilon}$ then correspond to the eigenvalues of such states and can be identified, based on the approximation used for $\Sigma^{\rm s}$, either with the HF eigenenergy or its COHSEX quasiparticle energies. In general, we have $\bar{\varepsilon}_n={\rm Re}\mel*{n}{\bar{\xi}}{n}$. $\tau_{\rm QP}$ is the finite lifetime of the quasiparticle. 
The collision integral then becomes~\cite{Rivaprep}
%
\begin{align}\label{Eq:eescatter}
    &I_{12}(t)=\nonumber\\
    &\sum_{\substack{3456\\789}} W_{15;89} W_{67;34}\!\!\int_{0}^{t} dt'{\frac{e^{-\frac{(t-t')^2}{2\bar{\tau}_{\rm QP}^2}}}{\big(\sqrt{2\pi}\bar{\tau}_{\rm QP}\big)^{1/2}}}e^{i(t-t')(\bar{\varepsilon}_{2}+\bar{\varepsilon}_{5}-\bar{\varepsilon}_{8}-\bar{\varepsilon}_{9})}   \nonumber \\& \times\Big[ -\big(\delta_{32}-\rho_{32}^{(1)}(t')\big) \rho_{86}^{(1)}(t')   \rho_{97}^{(1)}(t')\big(\delta_{45}-\rho_{45}^{(1)}(t') \big)  + \nonumber \\&+ \big(\delta_{86}-\rho_{86}^{(1)}(t')\big)\rho_{32}^{(1)}(t') \rho_{45}^{(1)}(t') \big(\delta_{97} -\rho_{97}^{(1)}(t')\big)\Big] +\nonumber \\
    &+\text{H.c.} \ ,
\end{align}
\begin{figure}
    \centering
    \includegraphics[width=0.9\linewidth]{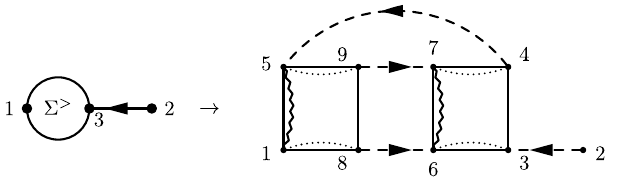}
    \caption{Diagrammatic representation of one of the four terms in Eq.~\eqref{Eq:eescatter}. The dashed lines represent the density matrix, the squares with wiggled lines represent the screened interaction, and the dotted lines link indices related to the same position argument, accordingly to Eq.~\eqref{Eq:potential}.}
    \label{fig:scatteringGW}
\end{figure}
\noindent
where $\bar{\tau}_{\rm QP}$ is the effective lifetime of the quasi-particles and the exponential is multiplied by an additional renormalization factor.
Equation~(\ref{Eq:eescatter}) can be considered as the direct term in incoherent dynamics, and it is diagrammatically represented in Fig.~(\ref{fig:scatteringGW}).  Vertex corrections of the non-equilibrium self-energy give rise to an additional exchange term~\cite{Rossi2002,PhysRevB.104.184418}. $W$ is the statically screened electron-electron interaction, at the random phase approximation (RPA) level,  $W({\bf Q})=\epsilon^{-1}({\bf Q},\omega=0)v({\bf Q})$, where {\bf Q} is the electronic momentum transfer.
%
We finally note that by combining Eq.~(\ref{Eq:rhoEOM}) with Eq.~(\ref{Eq:bornapprox}) within the Born approximation, after we neglect the exchange contribution to the two-particle correlation $K_0$, it is possible to obtain a similar structure as Eq.~(\ref{Eq:eescatter}), with $W$ replaced by the bare Coulomb potential and $\bar{\varepsilon}$ by the Hartree-Fock energies.
\subsubsection{Electron-phonon interaction}
\noindent
Now we consider the contribution to the electron dynamics coming from the different interactions with the environment in $\hat{H}_{\rm sys-env}$. The electron-phonon interaction is intrinsic, independent from external potentials, thus will be considered first. Under the Born approximation, which allows one to separate electronic and environmental degrees of freedom, we have $\hat{\rho}(t)=\hat{\rho}_{\rm S}(t)\otimes\hat{\rho}_{\rm E}(t)$. This is valid in the limit of weak coupling and when the correlations between system and environment decay fast compared to the dynamics of the system. We emphasize that Eq.~(\ref{Eq:rho1B}) is written in the interaction picture; the operators in the system subspace evolve in time under the QP energy spectrum introduced in the previous section $\{\bar{\varepsilon}_n\}$. The operators in the phonon subspace are instead driven by the $\hat{H}_{\rm ph}$ spectrum. Using Eq.~(\ref{Eq:rho1B}) and tracing out the remaining environmental degrees of freedom, we can write
\begin{equation}
    \dot{\rho}_{12}^{(1)}|_{\rm e-ph} =i{\rm Tr}_{\rm S}{\rm Tr}_{\rm ph}\big\{\hat{\rho}_{\rm S}(t)\otimes\hat{\rho}_{\rm ph}(t)\big[\hat{H}_{\rm e-ph},\hat{c}_2^\dagger\hat{c}_1\big]\big\}.
\end{equation}
Here $\hat{\rho}_{\rm ph}$ is obtained after tracing out all the non-phononic degrees of freedom.
If we separate the phonon distribution into a thermal one, $\hat{\rho}_{\rm ph}^0$, and a fluctuation term $\delta\hat{\rho}_{\rm ph}$, the phonon field, $\hat{b}_{\bf q}$, can be written as a combination of a coherent term, $B_{\bf q}$, that corresponds to a finite atomic displacement in the lattice and a fluctuation, $\delta\hat{b}_{\bf q}$, such that
\begin{equation} \label{Eq:bq}
    \hat{b}_{\bf q} = B_{\bf q} + \delta\hat{b}_{\bf q}\,,
\end{equation}
where the thermal expectation value of the fluctuation field is ${\rm Tr}_{\rm ph}\{\hat{\rho}_{\rm ph}^0\delta\hat{b}_{\bf q}\}=\expval*{\delta\hat{b}_{\bf q}}=0$ and $B_{\bf q}={\rm Tr}_{\rm ph}\{\delta\hat{\rho}_{\rm ph}\hat{b}_{\bf q}\}$.\\
The electron-phonon density matrix is instead defined as 
\begin{equation}
    \rho_{12}^{\bf q}={\rm Tr}_{\rm S}{\rm Tr}_{\rm ph}\{\hat{\rho}_{\rm S}(t)\otimes\hat{\rho}_{\rm ph}(t)\hat{c}^\dagger_{2}\hat{b}_{\bf q}\hat{c}_{1}\}=\expval*{\hat{c}^\dagger_{2}\hat{b}_{\bf q}\hat{c}_{1}}.
\end{equation}
Using Eq.~(\ref{Eq:bq}) this can be separated into $\rho^{\bf q}_{12}=B_{\bf q}\rho_{12}^{(1)} + \delta\rho^{\bf q}_{12}$.
It is often assumed that the coherent contribution of phonons to electron dynamics is inherently small and could be neglected, $B_{\bf q} \simeq 0$. However, this is not true in general, the approximation is valid only close to a fully thermalized equilibrium phonon distribution. It can be shown, in fact, that $B_{\bf q}$ contributes with a polaronic correction to the coherent part of the Hamiltonian, this corresponds to the so-called Ehrenfest self-energy\cite{PhysRevB.106.075119,stefanucci2024kadanoff,blommel2025unifiedsimulationframeworkcorrelated}, but we will not discuss these effects further here.
The electron-phonon contribution to density matrix dynamics is written as
\begin{align}\label{Eq:eph}
    \dot{\rho}_{12}^{(1)}|_{\rm e-ph} &= -i\sum_{{\bf q}}\sum_3\big(g_{31}^{{\bf q}+}\delta\rho_{23}^{{\bf q}\,*} +g_{13}^{{\bf q}-}\delta\rho_{32}^{\bf q} \big) + \text{H.c.}\,.
\end{align}
In order to solve the equation above, we need the equation of motion for the fluctuation field $\delta\rho_{12}^{\bf q}$. We can write its dynamics as follows.
\begin{align}\label{Eq:eph0}
    &\frac{d}{dt}\delta\rho_{12}^{\bf q} =\big[-\dot{B}_{\bf q}\rho_{12}^{(1)} - B_{\bf q}\dot{\rho}_{12}^{(1)}|_{\rm e-ph+ee} + i{\rm Tr}_{\rm S}{\rm Tr}_{\rm ph}\big\{\hat{\rho}_{\rm S}(t)\otimes\nonumber\\
    &\otimes\hat{\rho}_{\rm ph}(t)\big[\hat{H}_{\rm e}+\hat{H}_{\rm ph}+\hat{H}_{\rm e-ph}, \hat{c}_2^\dagger\hat{b}_{\bf q}\hat{c}_1\big]\big\}\big]_{B_{\bf q}=0},
\end{align}
where $\dot{\rho}_{12}^{(1)}|_{\rm e-ph+ee}$ is a combination of $\dot{\rho}_{12}^{(1)}|_{\rm e-ph}$ and the contribution of $\dot{\rho}_{12}^{(1)}|_{\rm e-e}$ approximated by the effective Hamiltonian $\hat{H}_{\rm e}=\sum_1\bar{\varepsilon}_1\hat{c}_1^\dagger\hat{c}_1$.\\
After a lengthy calculation, it is possible to obtain\cite{Iotti_2017}
%
\begin{equation}\label{Eq:eph2}
    \frac{d}{dt}\delta\rho_{12}^{\bf q}=-i(\bar{\varepsilon}_{1}-\bar{\varepsilon}_{2}+\epsilon_{\bf q})\delta\rho_{12}^{\bf q}+iF_{12}^{\bf q}(t)\,,
\end{equation}
where
\begin{align} \label{Eq:F12q}
    F_{12}^{\bf q}(t) &= \sum_{34}g_{43}^{{\bf q}+}n_{\bf q}(t)\rho_{13}^{(1)}(t)(\delta_{42}-\rho_{42}^{(1)}(t)) -\nonumber\\
    &-\sum_{34}g_{43}^{{\bf q}+}(1+n_{\bf q}(t))\rho_{42}^{(1)}(t)(\delta_{13}-\rho_{13}^{(1)}(t))\,.
\end{align}
%
In Eq.~(\ref{Eq:F12q}) we have used $\expval*{\hat{b}_{\bf q}^\dagger\hat{b}_{\bf q'}}\simeq\delta_{\bf qq'}n_{\bf q}(t)$. 
This requires an additional dynamical equation for $n_{\bf q}(t)$ instead of the full phonon density matrix $\rho_{\bf qq'}={\rm Tr}_{\rm ph}\{\hat{\rho}_{\rm ph}\hat{b}_{\bf q'}^\dagger\hat{b}_{\bf q}\}$. The off-diagonal terms in {\bf q}-space are important whenever the phonon distribution is not a simple superposition of phonon occupation eigenstates. Such cases, however, will not be discussed further here and we will not consider phonon dynamics either.\\
The combination of Eq.~(\ref{Eq:eph}) for the electronic density matrix and Eq.~(\ref{Eq:eph2}) for the electron-phonon density matrix produces the lowest order contribution to the non-Markovian electron-phonon dynamics. In order to solve Eq.~\eqref{Eq:eph2} we assume that the electron-phonon interaction, $\hat H_{\rm e-ph}$, is switched on at time $t=0$. This means that the electron-phonon density matrix is zero at the beginning of the dynamics, $\delta\rho_{12}^{\bf q}(t=0)=0$. The differential equation~\eqref{Eq:eph2} can be integrated and gives
\begin{equation}
    \delta\rho_{12}^{\bf q}(t) = i\int_0^t d\tau e^{-i(\bar{\varepsilon}_{12}+\epsilon_{\bf q})(t-\tau)}F_{12}^{\bf q}(\tau)\,,
\end{equation}
where we have used $\bar{\varepsilon}_{12}=\bar{\varepsilon}_{1}-\bar{\varepsilon}_{2}$.
Using the previous expression inside Eq.~\eqref{Eq:eph}, after some work, we obtain the following,
\begin{align}\label{Eq:e-ph_non_M}
     &\dot{\rho}_{12}^{(1)}|_{\rm e-ph}\!\!= \!\!\sum_{{\bf q},s=\pm}  \sum_{345}   g^{{\bf q}+}_{23}g^{{\bf q}-}_{54} 
     \int_{0}^{t}\,d t'\left(n_{\bf q}(t')+\frac{s}{2}+\frac{1}{2}\right)\times \nonumber \\ &\times (\delta_{15}-\rho_{15}^{(1)}(t')) \rho_{43}^{(1)}(t')  e^{-i(\bar{\varepsilon}_{13}+s\epsilon_{\bf q})(t-t')}  - \nonumber \\
     &-  \sum_{{\bf q},s=\pm}   \sum_{345}   g^{{\bf q}-}_{13} g^{{\bf q}+}_{54}
     \int_{0}^{t}\,d t'\left(n_{\bf q}(t')+\frac{s}{2}+\frac{1}{2}\right)\times\nonumber\\ &\times (\delta_{34}-\rho_{34}^{(1)}(t')) \rho_{52}^{(1)} (t')   e^{-i(\bar{\varepsilon}_{32}+s\epsilon_{\bf q})(t-t')} + \text{H.c.}.
\end{align}
This equation is analogous to Eq.~\eqref{Eq:eescatter} but for the electron-phonon interaction, and it can also be obtained from the collision integral in Eq.~\eqref{Eq:rhoEOM_NEGF} by using the non-equilibrium Fan-Migdal self-energy with the GKBA for both the electron and phonon propagators~\cite{Kar2021}. It can describe both the phonon absorption and emission processes ($s=-1$ and $s=1$, respectively). It also accounts for electronic energy renormalization effects that come from the real part of the Fan-Migdal self-energy, while the imaginary part gives rise in the equilibrium limit to the finite linewidth of the electronic levels.
%
%
\subsubsection{Electron-radiation field interaction}
\noindent
The electron-radiation contribution to the dynamics of the one-particle density matrix $\rho^{(1)}$ can be derived in the same way under the Born approximation, from the electron-radiation contribution to Eq.~(\ref{Eq:rho1B}). Removing the remaining environmental degrees of freedom, we obtain as follows.
\begin{align}
    \dot{\rho}_{12}^{(1)}|_{\rm e-rad} =i{\rm Tr}_{\rm S}{\rm Tr}_{\rm rad}\big\{\hat{\rho}_{\rm S}(t)\otimes\hat{\rho}_{\rm rad}(t)\big[\hat{H}_{\rm e-rad},\hat{c}_2^\dagger\hat{c}_1\big]\big\},
\end{align}
which produces two contributions based on the expression of $\hat{H}_{\rm e-rad}$ in Eq.~(\ref{Eq:Herad2ndq}). The coherent contribution, $\dot{\rho}_{12}^{(1)}|_{\rm e-rad}^{\rm coher}$, comes from the time-dependent electromagnetic fields in $h^{\rm e-r}(t)$. This term, due to its single-particle nature, can be added directly to $\dot{\rho}_{12}^{(1)}|_{\rm e-e}$ in Eq.~(\ref{Eq:rhoEOM_NEGF}) and produces an expression of the following form,
%
\begin{align}\label{Eq:rhoEOM_NEGF+EMF}
    &\dot{\rho}_{12}^{(1)}|_{\rm e-e}\!\!+\!\!\dot{\rho}_{12}^{(1)}|_{\rm e-rad}^{\rm coher} =\!\! -i\sum_{3}\big(h_{13}(t)\rho_{32}^{(1)} - \rho_{13}^{(1)}h_{32}(t)\big) -\!\! I_{12}, 
\end{align}
where we have introduced an effective one-particle Hamiltonian,
\begin{equation}\label{Eq:tdMFham}
    h_{12}(t)= \varepsilon_1\delta_{12}+V^{\rm H}_{12}+\Sigma^{\rm s}_{12}+h_{12}^{\rm e-r}(t).
\end{equation}
By approximating $\Sigma^{\rm s}\simeq\Sigma^{\rm Fock}$ and neglecting the scattering term $I_{12}$, Eq.~(\ref{Eq:rhoEOM_NEGF+EMF}) is equivalent to the time-dependent Hartree-Fock equations. 
Under $\Sigma^{\rm s}\simeq\Sigma^{\rm COHSEX}$ we obtain instead the time-dependent COHSEX equation\cite{doi:10.1073/pnas.2301957120,doi:10.1073/pnas.1906938118,PhysRevLett.128.016801}, which at the limit of small external perturbations 
is equivalent to the statically-screened Bethe-Salpeter equation~\cite{RevModPhys.74.601} and with $\Sigma^{\rm s}\simeq V^{\rm XC}$ we recover the equations of adiabatic TDDFT\cite{Runge_1984}, where $V^{\rm XC}$ is the exchange-correlation potential.
The incoherent contribution of the electron-radiation field interaction comes instead from the second term in Eq.~(\ref{Eq:Herad2ndq}), and behaves in the same way as the electron-phonon interaction term in Eq.~(\ref{Eq:eph}) discussed in the previous section, 
\begin{align}\label{Eq:rho_erad}
    \dot{\rho}_{12}^{(1)}|_{\rm e-rad}^{\rm incoher} &= -i\sum_{\bf k}\sum_3\big( \mathpzc{k}_{31}^{{\bf k}+}\delta\rho^{{\bf k}\,*}_{23}+\mathpzc{k}_{13}^{{\bf k}-}\delta\rho_{32}^{\bf k}\big) + \text{H.c.}\,.
\end{align}
\noindent
$\mathpzc{k}$ is the light-matter interaction matrix element (e.g. electronic transition dipole) expressed in Eq.~(\ref{Eq:dipole}) and we have introduced the so-called electron-photon density matrix.
\begin{equation}
    \rho^{\bf k}_{12}={\rm Tr}_{\rm S}{\rm Tr}_{\rm rad}\big\{\hat{\rho}_{\rm S}(t)\otimes\hat{\rho}_{\rm rad}(t)\hat{c}_2^\dagger\hat{a}_{\bf k}\hat{c}_1\big\} = \expval*{\hat{c}_2^\dagger\hat{a}_{\bf k}\hat{c}_1}.
\end{equation}
We then need a dynamical equation for the fluctuation of the photon density matrix $\delta\rho^{\bf k}_{12}=\rho^{\bf k}_{12}-\expval*{\hat{a}_{\bf k}}\rho_{12}^{(1)}=\rho_{12}^{\bf k}$, given that the expectation value $\expval*{\hat{a}_{\bf k}}={\rm Tr}\big\{\hat{\rho}_{\rm rad}(t)\hat{a}_{\bf k}\big\}=0$.
%
The equation of motion for the fluctuation field is given by
\begin{align}\label{Eq:erad0}
    \frac{d}{dt}&\delta\rho_{12}^{\bf k} =\Big[-\frac{d}{dt}\big[\expval*{\hat{a}_{\bf k}}\rho_{12}^{(1)}\big] + i{\rm Tr}_{\rm S}{\rm Tr}_{\rm rad}\big\{\hat{\rho}_{\rm S}(t)\otimes\nonumber\\
    &\otimes\hat{\rho}_{\rm rad}(t)\big[\hat{H}_{\rm e}+\hat{H}_{\rm rad}+\hat{H}_{\rm e-rad}, \hat{c}_2^\dagger\hat{a}_{\bf k}\hat{c}_1\big]\big\}\Big]_{\expval*{\hat{a}_{\bf k}}=0},
\end{align}
where $\hat{H}_{\rm rad}$ is the Hamiltonian of the radiation field in Eq.~(\ref{Eq:Hrad}), $\hat{H}_{\rm e-rad}$ is the interaction Hamiltonian between the radiation field and the electron system, and $\hat{H}_{\rm e}$ is the effective electronic Hamiltonian that we have introduced in the previous section.
Following the same procedure as in the electron-phonon case, integrating Eq.~(\ref{Eq:erad0}) and using it inside Eq.~(\ref{Eq:rho_erad}), we obtain the following result.
\begin{align}\label{Eq:e-rad_incoher}
     &\dot{\rho}_{12}^{(1)}|_{\rm e-rad}^{\rm incoher} =\nonumber\\
     &=\sum_{{\bf k},s=\pm}  \sum_{345} \mathpzc{k}^{{\bf k}+}_{23}\mathpzc{k}^{{\bf k}-}_{54} 
     \int_{0}^{t}\,d t'\left(n_{\bf k}(t')+\frac{s}{2}+\frac{1}{2}\right)\times \nonumber \\ &\times (\delta_{15}-\rho_{15}^{(1)}(t')) \rho_{43}^{(1)}(t')  e^{-i(\bar{\varepsilon}_{13}+s\omega_{\bf k})(t-t')}  - \nonumber \\
     &-  \sum_{{\bf k},s=\pm}   \sum_{345}   \mathpzc{k}^{{\bf k}-}_{13} \mathpzc{k}^{{\bf k}+}_{54}
     \int_{0}^{t}\,d{t'}\left(n_{\bf k}(t')+\frac{s}{2}+\frac{1}{2}\right)\times\nonumber\\ &\times (\delta_{34}-\rho_{34}^{(1)}(t')) \rho_{52}^{(1)} (t')   e^{-i(\bar{\varepsilon}_{32}+s\omega_{\bf k})(t-t')} + \text{H.c.}.
\end{align}
%
Eq.~(\ref{Eq:e-rad_incoher}) includes photon absorption and emission processes through the sum over the index $s=\pm$. 
The coherent excitation of the electronic system under external electromagnetic pulses or waves is controlled by the interaction in $h^{\rm e-r}(t)$ in Eq.~(\ref{Eq:tdMFham}), but the description of radiative relaxation and decoherence is given entirely by Eq.~(\ref{Eq:e-rad_incoher}).\\
%
We finally note that we are not evolving the external radiation field coherently with the electronic system here, which may be necessary under a very strong laser field or cavity confinement. This would require a complete quantum electrodynamics framework and a set of equations for ${\bf a}(t)$ and ${\bf B}(t)$ that is beyond the scope of this work.
\subsubsection{Real-time density matrix dynamics}
%
%
\noindent
If we combine all the contributions discussed earlier in this section, Eq.~(\ref{Eq:rho1pNM}) reduces to
\begin{align}\label{Eq:rho00}
    \dot{\rho}_{12}^{(1)} &= -i\sum_3\big(h_{13}(t)\rho_{32}^{(1)} - \rho_{13}^{(1)}h_{32}(t)\big) + \dot{\rho}_{12}^{(1)}|_{\rm incoher},
\end{align}
where $\dot{\rho}_{12}^{(1)}|_{\rm incoher}$ contains all the incoherent contributions defined in Eqs.~(\ref{Eq:eescatter}), (\ref{Eq:e-ph_non_M}), and (\ref{Eq:e-rad_incoher}). $h_{12}(t)$ is the time-dependent Hamiltonian of Eq.~(\ref{Eq:tdMFham}), responsible for non-dissipative coherent dynamics of one-particle density matrices. 
The expression can be written more conveniently on the basis of the ground-state Hamiltonian $\bar{\xi}_{12}$ ($\bar{\xi}_{12}=\varepsilon_1\delta_{12}+V^{\rm H}_{12}[\rho^{(1)}_{GS}]+\Sigma_{12}^{\rm s}[\rho^{(1)}_{GS}]$). 
\begin{equation}\label{eq:unperturbed-e}
    h_{12}(t) = \delta_{12}\bar{\varepsilon}_1 +\Delta V^{\rm H}_{12} + \Delta\Sigma_{12}^{\rm s} + h_{12}^{\rm e-r}(t),
\end{equation}
where $\bar{\varepsilon}_n={\rm Re}\mel*{n}{\bar{\xi}}{n}$, $\Delta V^{\rm H}_{12}=V^{\rm H}_{12}[\rho^{(1)}]-V^{\rm H}_{12}[\rho^{(1)}_{GS}]$, $\Delta\Sigma_{12}^{\rm s}=\Sigma_{12}^{\rm s}[\rho^{(1)}]-\Sigma_{12}^{\rm s}[\rho^{(1)}_{\rm GS}]$ are the dynamical variation of the Hartree potential and of the static self-energy compared to their value in the ground state, and $\rho^{(1)}_{\rm GS}$ is the one-particle density matrix of the ground state.\\
In contrast to the coherent part of the dynamics in Eq.~(\ref{Eq:rho00}), that is local in time, the incoherent part given by $\dot{\rho}_{12}^{(1)}|_{\rm incoher}$ is non-local. Such a non-locality, combined with the large number of electronic states and the dense {\bf k} and {\bf q} grids required for accurate description of electronic and phonon properties in solids, makes the full non-Markovian dynamical evolution of the density matrix in most cases prohibitive. Memory effects in Eqs.~(\ref{Eq:eescatter}), (\ref{Eq:e-ph_non_M}) and (\ref{Eq:e-rad_incoher}) enter through the double-time integral and require knowledge of the density matrix at earlier times. We may ask whether a complete description of memory effects in Eqs.~(\ref{Eq:eescatter}), (\ref{Eq:e-ph_non_M}) and (\ref{Eq:e-rad_incoher}) is always necessary or, in some cases, it is possible to take some simplified assumptions.\\
In the next sections, we discuss the Markovian limits of the different incoherent terms. This approximation allows us to drastically simplify the set of equations for the different correlation functions. We discuss in which limit this approximation is valid and when instead it breaks down.
%
\section{Markovian limit of the density matrix equations}\label{sec:markovlim}
\noindent
A Markov process is a stochastic process independent of the history of the system. In the weak system-environment coupling regime and, generally, when the environment correlation functions decay on a time scale that is much shorter compared to the relaxation time of the system, we have the following condition,
\begin{equation} \label{Eq:env_correl}
    \expval*{\hat{E}(t)\hat{E}(0)}\sim \expval*{\hat{E}(0)^2}e^{-\frac{t^2}{2\tau_{\rm E}^2}} << e^{-\frac{t^2}{2\tau_{\rm S}^2}},
\end{equation}
where $\tau_{\rm S} >> \tau_{\rm E}$ are the relaxation times of the system and the environment, respectively. This relation is valid for an environment that is infinitely large and characterized by a continuum of frequencies, e.g. a phonon bath.\\
The equation of motion that we have obtained for the single-particle density matrix, Eq.~(\ref{Eq:rho1pNM}), contains several incoherent contributions to the dynamics, Eq.~(\ref{Eq:eescatter}) for e-e, Eq.~(\ref{Eq:e-ph_non_M}) for e-ph and Eq.~(\ref{Eq:e-rad_incoher}) for e-rad. It is easy to observe, for instance in Eq.~(\ref{Eq:e-ph_non_M}), that under the aforementioned conditions, the integral $\mathcal{I}(t)=\int_{0}^{t}\,d t'\,(n_{\bf q}(t')+s/2+1/2)(\delta_{15}-\rho_{15}^{(1)}(t')) \rho_{43}^{(1)}(t')  e^{-i(\bar{\varepsilon}_{13}+s\epsilon_{\bf q})(t-t')}$ can be rewritten as follows,
\begin{align}
    \mathcal{I}(t) &\sim\Big(n_{\bf q}(t)+\frac{s}{2}+\frac{1}{2}\Big)(\delta_{15}-\rho_{15}^{(1)}(t)) \rho_{43}^{(1)}(t)\times\nonumber\\
    &\times\int_{0}^\infty d{\tau}e^{-\frac{\tau^2}{2\tau_{\rm E}^2}} e^{-i(\bar{\varepsilon}_{13}+s\epsilon_{\bf q})\tau}\nonumber\\
    &\sim\Big(n_{\bf q}(t)+\frac{s}{2}+\frac{1}{2}\Big)(\delta_{15}-\rho_{15}^{(1)}(t)) \rho_{43}^{(1)}(t)\times\nonumber\\
    &\times\exp\Big\{-\frac{\big(\bar{\varepsilon}_{13}+s\epsilon_{\bf q}\big)^2}{2\epsilon^2}\Big\}\,
\end{align}
where $\epsilon=1/\tau_{\rm E}$. In the first step, we introduced an exponential factor that expresses the fast decay of the environment correlations. This allows us to take the product of density operators out of the integral, whose dynamics is much slower and could be effectively approximated with the value at time $t$; this procedure is usually known as \emph{first Markov approximation}. The fast decay of the integrand allows us to extend the integration limit to $\infty$ thus obtain the Gaussian factor, which is known as \emph{second Markov approximation}\cite{RevModPhys.89.015001}. This is the conventional procedure to perform the Markov limit. When it is applied to the incoherent contributions to the density matrix dynamics, it produces a set of equations that are local in time. These quantum master equations have been applied to the study of electronic relaxation and Coulomb blockade in semiconductors and nanostructures\cite{PhysRevB.38.3342,PhysRevB.56.13177,PhysRevB.59.10748,PhysRevB.70.205334}.\\
However, the way to perform the Markovian limit is not unique. In Appendix (\ref{app:a}) we show how different ways of performing this limit gives rise to different forms of scattering operators and potentially influence the incoherent part of the dynamics. In order to clarify this point, we observe that the system density matrix dynamics, after the environment has been traced out, can always be written in the following form\cite{RevModPhys.89.015001,OQS_book1}:
\begin{align}
    \rho_{\rm S}(t) &= {\rm Tr}_{\rm B}\{\hat{\mathcal{U}}(t)\hat{\rho}_{\rm B}\otimes\hat{\rho}_{\rm S}(t=0)\hat{\mathcal{U}}^\dagger(t)\}\nonumber\\
    &=\sum_{l}\hat{\mathcal{K}}_{l}(t)\hat{\rho}_{\rm S}(t=0)\hat{\mathcal{K}}_{l}^\dagger(t) = \mathcal{M}(t)[\hat{\rho}_{\rm S}(t=0)],
\end{align}
where the Kraus decomposition is expressed in terms of a universal dynamical map $\mathcal{M}$ that preserves the complete positivity of the reduced density matrix. The Markovian limit does not necessarily guarantee the preservation of such complete positivity. For example, it is known that the Redfield-type quantum master equations\cite{REDFIELD19651} do not preserve the complete positivity of the density matrix\cite{Davies1974}.
Only a Lindblad-type dynamical equation preserves this exact property\cite{RevModPhys.89.015001}.  A new strategy developed to preserve such exact property has been applied to the case of electron-phonon and electron-electron interactions~\cite{taj2009microscopic,PhysRevB.90.125140}. These density matrix equations produce the correct semiclassical limit by taking the diagonal approximation, reducing to the so-called Pauli master equation based on semiclassical scattering superoperators\cite{PhysRevB.59.4901,PhysRevB.77.125301}, then further recovering the generalized Fermi's golden rule in the perturbative limit. Note that after the diagonal approximation is taken, it was found that both Markovian limits give the same semiclassical equation. 
%
In the next sections, we apply this ansatz to the different incoherent non-Markovian expressions and obtain the Lindblad form of the Markovian limit. More details on this procedure can be found in Ref.~\cite{PhysRevB.90.125140}.
%
\subsection{Markovian limit of electron-phonon (Eq.~(\ref{Eq:e-ph_non_M})) and electron-photon interaction (Eq.~(\ref{Eq:e-rad_incoher})) in the Lindblad form}
\noindent
As explained in Ref.~\cite{taj2009microscopic}, the Lindblad limit of Eqs.~\eqref{Eq:e-ph_non_M} and~\eqref{Eq:e-rad_incoher}, can be  obtained as follows.
We first compute the integral in Eq.~(\ref{Eq:e-ph_non_M}) under a change of variables, $\rho_{12}^{(1)}=\tilde{\rho}^{(1)}_{12}e^{-i\bar{\varepsilon}_{12}t}$ and introduce the macroscopic $T=(t+t')/2$ and microscopic $\tau=t-t'$ time variables. To simplify the expressions, we introduce $\mathcal{K}_{23541}^{{\bf q}s}=g_{23}^{{\bf q}+}g_{54}^{{\bf q}-}\big(n_{\bf q}+\frac{s}{2}+\frac{1}{2}\big)\rho_{43}^{(1)}(\delta_{15}-\rho_{15}^{(1)})$ and ${\mathcal{K}'}_{54132}^{{\bf q}s}=g_{54}^{{\bf q}+}g_{13}^{{\bf q}-}\big(n_{\bf q}+\frac{s}{2}+\frac{1}{2}\big)\rho_{52}^{(1)}(\delta_{34}-\rho_{34}^{(1)})$.
\begin{align}
    &\Delta\tilde{\rho}_{12}^{(1)}(t)|_{\rm e-ph} = \sum_{{\bf q},s=\pm}\sum_{345}\int_0^{t}d{T}\int_0^{g(t,T)}d{\tau}\Big\{e^{-i\bar{\varepsilon}_{23}\big(T+\frac{\tau}{2}\big)}\nonumber\\
    &\times e^{i\bar{\varepsilon}_{54}\big(T-\frac{\tau}{2}\big)}e^{-is\epsilon_{\bf q}\tau}\tilde{\mathcal{K}}_{23541}^{{\bf q}s}\Big(T-\frac{\tau}{2}\Big) - e^{-i\bar{\varepsilon}_{31}\big(T+\frac{\tau}{2}\big)}\times\nonumber\\
    &\times e^{-i\bar{\varepsilon}_{54}\big(T-\frac{\tau}{2}\big)}e^{-is\epsilon_{\bf q}\tau}\tilde{\mathcal{K}'}_{54132}^{{\bf q}s}\Big(T-\frac{\tau}{2}\Big)\Big\} +\text{H.c.}
\end{align}
Here $\Delta \tilde{\rho}_{12}^{(1)}(t)|_{\rm e-ph}=\tilde{\rho}_{12}^{(1)}(t)|_{\rm e-ph}-{\rho}_{12}^{(1)}(t=0)$.
We can now apply Eq.~(\ref{Eq:env_correl}) to the previous integral in $\tau$. This allows to write $\int_0^{g(t,T)}d{\tau}\stackrel{(1)}{\simeq}\int_0^\infty d{\tau}e^{-\tau^2/(2\tau_{\rm E}^2)}/\big(\sqrt{2\pi}\tau_{\rm E}\big)^{1/2}$. As already discussed, this approximation is valid when the phonon auto-correlation functions decay fast in time. This requires a large number of phonon states potentially contributing to electronic transitions, which is quite typical in the case of solid state systems with large phonon bandwidths. After we take the time derivative, we perform an additional coarse-grain integration, integrating over a new variable $\tau'$ with weight $\exp(-{\tau'}^2/(2\tau_{\rm E}^2))/(\sqrt{2\pi}\tau_{\rm E})^{1/2}$, and obtain the following.
\begin{align} \label{Eq:coarsegr}
    &\dot{\tilde{\rho}}_{12}^{(1)}|_{\rm e-ph}\stackrel{(2)}{\simeq}\sum_{{\bf q},s=\pm}\sum_{345}\int_0^\infty d{\tau}\frac{e^{-\frac{\tau^2}{2\tau_{\rm E}^2}}}{\big(\sqrt{2\pi}\tau_{\rm E}\big)^{1/2}}\times\nonumber\\
    &\times\int_{-\infty}^\infty d{\tau'}\frac{e^{-\frac{{\tau'}^2}{2\tau_{\rm E}^2}}}{\big(\sqrt{2\pi}\tau_{\rm E}\big)^{1/2}}\Big\{\tilde{\mathcal{K}}_{23541}^{{\bf q}s}\Big(t+\frac{\tau'-\tau}{2}\Big)\times\nonumber\\
    &\times\,e^{-i(\bar{\varepsilon}_{23}+s\epsilon_{\bf q})\big(t+\frac{\tau+\tau'}{2}\big)}e^{i(\bar{\varepsilon}_{54}+s\epsilon_{\bf q})\big(t+\frac{\tau'-\tau}{2}\big)}-\nonumber\\
    &-\tilde{\mathcal{K}'}_{54132}^{{\bf q}s}\Big(t+\frac{\tau'-\tau}{2}\Big)e^{-i(\bar{\varepsilon}_{31}+s\epsilon_{\bf q})\big(t+\frac{\tau+\tau'}{2}\big)}\times\nonumber\\
    &\times e^{-i(\bar{\varepsilon}_{54}-s\epsilon_{\bf q})\big(t+\frac{\tau'-\tau}{2}\big)}\Big\}+\text{H.c.}\ .
\end{align}
By observing that the dynamics of $\tilde{\mathcal{K}}$ and $\tilde{\mathcal{K}'}$ is much slower compared to the fast oscillations in the integrand, we can set $\tilde{\mathcal{K}}^{{\bf q}s}\big(t+\frac{\tau'-\tau}{2}\big)\simeq\tilde{\mathcal{K}}^{{\bf q}s}(t)$, and do the same for $\tilde{\mathcal{K}'}$.
\begin{align}\label{Eq:ephmark1}
    &\dot{\rho}_{12}^{(1)}|_{\rm e-ph}\nonumber\\
    &\stackrel{(3)}{\simeq}\frac{1}{4}\sum_{{\bf q},s=\pm}\sum_{345}\Big\{\mathcal{K}_{23541}^{{\bf q}s}(t)\int_{-\infty}^\infty d{t_1}\frac{e^{-\frac{t_1^2}{4\tau_{\rm E}^2}}}{{\big(\sqrt{2\pi}\tau_{\rm E}\big)^{1/2}}}\times\nonumber\\
    &\times\,e^{-i(\bar{\varepsilon}_{23}+s\epsilon_{\bf q})\frac{t_1}{2}}\int_{-\infty}^\infty d{t_2}\frac{e^{-\frac{t_2^2}{4\tau_{\rm E}^2}}}{{\big(\sqrt{2\pi}\tau_{\rm E}\big)^{1/2}}}e^{i(\bar{\varepsilon}_{54}+s\epsilon_{\bf q})\frac{t_2}{2}} -\nonumber\\
    &-{\mathcal{K}'}^{{\bf q}s}_{54132}(t)\int_{-\infty}^\infty\!\!d{t_1}\frac{e^{-\frac{t_1^2}{4\tau_{\rm E}^2}}}{{\big(\sqrt{2\pi}\tau_{\rm E}\big)^{1/2}}}e^{-i(\bar{\varepsilon}_{31}+s\epsilon_{\bf q})\frac{t_1}{2}}\times\nonumber\\
    &\times\int_{-\infty}^\infty d{t_2}\frac{e^{-\frac{t_2^2}{4\tau_{\rm E}^2}}}{{\big(\sqrt{2\pi}\tau_{\rm E}\big)^{1/2}}}e^{-i(\bar{\varepsilon}_{54}-s\epsilon_{\bf q})\frac{t_2}{2}}\Big\}+\text{H.c.}
\end{align}
Eq.~(\ref{Eq:ephmark1}) can now be easily integrated and produces the following form for the electron-phonon dynamics~\cite{Iotti_2017}.
\noindent
\begin{align}\label{Eq:e-ph_general}
    \dot{\rho}_{12}^{(1)}|_{\rm e-ph}\stackrel{(3)}{\simeq}&\frac{1}{2}\sum_{345}\Big[\big(\delta_{15}-\rho_{15}^{(1)}\big)\mathcal{P}_{52;43}^{\rm e-ph}\rho_{43}^{(1)} -\nonumber\\
    &-\big(\delta_{34} - \rho_{34}^{(1)}\big){\mathcal{P}_{34;15}^{\rm e-ph}}^*\rho_{52}^{(1)}\Big] + \text{H.c}
\end{align}
The matrix elements for the electron-phonon scattering operator are given below.
\begin{equation}\label{Eq:Peph}
    \mathcal{P}_{52;43}^{\rm e-ph} = \sum_{\bf q}\sum_{s=\pm}A_{54}^{\bf q}(s){A_{23}^{{\bf q}}(s)}^*
\end{equation}
The scattering amplitudes are in turn given by
\begin{align}\label{Eq:e-env-scatt}
    &A_{12}^{\bf q}(s) =\nonumber\\
    &=\sqrt{2\pi}\frac{\exp{-\Big(\frac{\bar{\varepsilon}_{12}+s\epsilon_{\bf q}}{2\epsilon}\Big)^2}}{(2\pi\epsilon^2)^{1/4}}g_{12}^{{\bf q}-}\Big(n_{\bf q}^0+\frac{1}{2}+\frac{s}{2}\Big)^{1/2}
\end{align}

\begin{table}[ht]
\begin{tabular}{ |p{1.25cm}|p{2.5cm}|p{1.25cm}|  }
\hline
\multicolumn{3}{|c|}{Electron-phonon coupled dynamics approximations}\\
\hline
\!\!\!\!\!\!Approximation && Validity\\
\hline
Born approximation & $\hat{\rho}(t)\simeq\hat{\rho}_{\rm S}(t)\otimes\hat{\rho}_{\rm ph}(t)$&\multirow{2}{0.17\columnwidth}{weak electron-phonon interaction}\\
\cline{1-2}
No polaron formation & $\expval*{\hat{b}_{\bf q}}=B_{\bf q}\simeq 0$&\\
\hline
First Markov approximation & $\begin{aligned}&\rho_{\bf qq'}(t-\tau)\\
&\simeq\rho_{\bf qq'}(t)e^{-\frac{t^2}{2{\tau_{\rm E}}^2}}\end{aligned}$& \multirow{2}{0.17\columnwidth}{fast decay of $\rho_{\bf qq'}$ memory. $\tau_{\rm E}\ll\tau_{\rm S}$ large phonon bandwidth $\Delta_{\bf q}$}\\
\cline{1-2}
Coarse grain time integration & $\begin{aligned}\\\dot{\hat{\boldsymbol{\rho}}}^{(1)}&\simeq\hat{\boldsymbol{\rho}}^{(1)}(\hat{\boldsymbol{I}}-\hat{\boldsymbol{\rho}}^{(1)})\\
&\times\mathcal{P}_{\rm e-ph}\end{aligned}$&\\
\hline
Thermal phonon distribution & $\rho_{\bf qq'}(t)\simeq\delta_{\bf qq'}n_{\bf q}^0$ & near equilibrium\\
\hline
\end{tabular}
 \caption{Summary of approximations used for electron-phonon coupled dynamics.}\label{Tab:2}
\end{table}

where, again, we have replaced $\tau_{\rm E}$ with $\epsilon=1/\tau_{\rm E}$. It is easy to observe from Eq.~(\ref{Eq:Peph}) that the electron-phonon scattering has the same form of the scattering operator in the Lindblad Markovian limit, Eq.~(\ref{Eq:scatt_mark2}) in Appendix (\ref{app:a}), and preserves the positive definiteness of the density matrix.\\
Eq.~(\ref{Eq:e-ph_general}) is based on the assumption that the phonons are stationary in their equilibrium thermal distribution. In such cases $n_{\bf q}^0$, the thermal phonon distribution, replaces the distribution $n_{\bf q}(t)$ at an arbitrary time $t$. Such an assumption is questionable in the presence of external perturbations which can induce non-equilibrium distributions in the phonon population.\\
Eq.~(\ref{Eq:e-ph_general}), Eq.~(\ref{Eq:Peph}) and Eq.~(\ref{Eq:e-env-scatt}) are based on a number of approximations. These are listed in Table~(\ref{Tab:2}): (i) The Born approximation, allows us to evaluate the second-order electron-phonon correlations (in the number of phonon processes), by  approximating it with its mean-field value and obtaining Eq.~(\ref{Eq:F12q}). This approximation is valid in the limit of weak electron-phonon coupling. When the electron-phonon interaction is strong, it is not possible to neglect correlations between electrons and phonons and an additional equation for these second-order correlations (in the phonon processes) is required.\\
(ii) Second, we have neglected the coherent contribution to the dynamics of phonons, the term $B_{\bf q}$ in Eq.~(\ref{Eq:bq}). This term describes coherent atomic motion with net non-zero atomic displacement and is zero only when the system is a simple eigenstate of the phonon occupation number. 
\\
(iii) Third, the first Markov approximation (Eq.~(\ref{Eq:env_correl})), allows us to neglect the memory effects in the phonon correlation function, $\rho_{\bf qq'}$ or $n_{\bf q}$. If the phonon correlation decays faster compared to the electronic one, this can be considered a good approximation. Additionally, its validity is not guaranteed in the presence of small phonon baths.\\
(iv) The additional coarse grain time integration, Eq.~(\ref{Eq:coarsegr}) is required to recover the Lindbladian dynamics of Eq.~(\ref{Eq:e-ph_general}). This approximation is considered very effective for large phonon baths, but it is limited in the case of small phonon bandwidths, $\Delta_{\bf q}$, when non-Markovian effects become important\cite{Iotti_2017}. In such cases, a complete evolution of the electron-phonon density matrix fluctuations $\delta\rho_{12}^{\bf q}$ is required. This situation has been analyzed in both semiconducting quantum dots\cite{PhysRevLett.88.146803,PhysRevB.76.241304} and solid state qubits\cite{PhysRevA.101.022110}.\\
(v) Finally, we replace the phonon distribution $\rho_{\bf qq'}(t)$ with the thermal distribution $n_{\bf q}^0$. In a more general case, an additional dynamical variable $\rho_{\bf qq'}={\rm Tr}_{\rm ph}\{\hat{\rho}_{\rm ph}\hat{b}_{\bf q'}^\dagger\hat{b}_{\bf q}\}$ should enter the set of equations consistently with $B_{\bf q}$, the electronic density matrix and $\delta\rho^{\bf q}$, in particular in presence of strong electron-phonon interactions.\\
In the case of the electron-photon interaction~\cite{XuPing2024}, with an analogous procedure we obtain an expression equivalent to Eq. (\ref{Eq:e-ph_general}) where $\mathcal{P}_{52;43}^{\rm e-ph}$ is replaced by
\begin{equation}\label{eq:Pe-radI}
    \mathcal{P}_{52;43}^{\rm e-rad}=\sum_{\bf k}\sum_{s=\pm}A_{54}^{\bf k}(s){A_{23}^{\bf k}(s)}^*
\end{equation}
\begin{align}\label{eq:Pe-radII}
    &A_{12}^{\bf k}(s) =\nonumber\\
    &=\sqrt{2\pi}\frac{\exp{-\Big(\frac{\bar{\varepsilon}_{12}+s\omega_{\bf k}}{2\epsilon}\Big)^2}}{(2\pi\epsilon^2)^{1/4}}\mathpzc{k}^{{\bf k}-}_{12}\Big(n^0_{\bf k}+\frac{1}{2}+\frac{s}{2}\Big)^{1/2},
\end{align}
where $s=+/-1$ provides photon absorption and emission processes. 
Eqs.~(\ref{eq:Pe-radI}) and~(\ref{eq:Pe-radII}) can describe the electronic decoherence induced by photon emission and absorption, as already mentioned. For the interaction with radiation, we make similar considerations as in the phononic case. The background radiation defines a large environmental bath characterized by a continuum of modes. In such a situation, the Markovian approximation and the thermal photon occupation approximations, $n_{\bf k}(t)\simeq n_{\bf k}^0$, is justified. However, there are also cases like in cavity quantum electrodynamics\cite{Walther_2006} in which light interacts with atoms and molecules in small cavities and the condition of a large radiation bath breaks down. In such cases, the quantum nature of photons is significant and the Markovian approximation could be not as accurate.
\subsection{Markovian limit of electron-electron interaction in Eq.~(\ref{Eq:eescatter}) in the Lindblad form}
\noindent
In the case of electron-electron interaction, we can take the Markovian limit with the same procedure, this time starting from Eq.~(\ref{Eq:eescatter}).
\noindent
To simplify the notation, we define $\mathcal{K}^{18349}_{6257}=W_{15;89}W_{67;34}[(\delta_{86}-\rho^{(1)}_{86})\rho_{32}^{(1)}\rho_{45}^{(1)}(\delta_{97}-\rho_{97}^{(1)})-(\delta_{32}-\rho_{32}^{(1)})\rho_{97}^{(1)}\rho_{86}^{(1)}(\delta_{45}-\rho_{45}^{(1)})]$. After a change in variables, $\rho_{12}^{(1)}=\tilde{\rho}_{12}^{(1)}e^{-i\bar{\varepsilon}_{12}t}$, where $\bar{\varepsilon}_{12}=\bar{\varepsilon}_1-\bar{\varepsilon}_2$, the integral of Eq.~(\ref{Eq:eescatter}) gives the following contribution to the density matrix.
\begin{align}
    &\Delta\tilde{\rho}_{12}^{(1)}(t)|_{\rm e-e}^{\rm incoher} =\nonumber\\
    &=-\sum_{\substack{3456\\789}}\int_0^t d{t_1}\frac{e^{-\frac{(t_1-t_2)^2}{2\bar{\tau}_{\rm QP}^2}}}{\big(\sqrt{2\pi}\bar{\tau}_{\rm QP}\big)^{1/2}}e^{i(\bar{\varepsilon}_1+\bar{\varepsilon}_5-\bar{\varepsilon}_8-\bar{\varepsilon}_9)t_1}\times\nonumber\\
    &\times\int_{0}^{t_1}d{t_2}e^{-i(\bar{\varepsilon}_3+\bar{\varepsilon}_4-\bar{\varepsilon}_7-\bar{\varepsilon}_6)t_2}\tilde{\mathcal{K}}^{18349}_{6257}(t_2) + \text{H.c.}
\end{align}
Under a change of variables, $T=(t_1+t_2)/2$, $\tau=t_1-t_2$, the exponential decay allows us to extend the integral to $\infty$.
\begin{align}
    \dot{\tilde{\rho}}_{12}^{(1)}&|_{\rm e-e}^{\rm incoher} \stackrel{(1)}{\simeq} -\sum_{\substack{3456\\789}}\int_0^{\infty}d{\tau}\frac{e^{-\frac{\tau^2}{2\bar{\tau}_{\rm QP}^2}}}{(\sqrt{2\pi}\bar{\tau}_{\rm QP})^{1/2}}e^{i\bar{\varepsilon}_{15;89}(t+\tau/2)}\times\nonumber\\
    &\times e^{-i\bar{\varepsilon}_{34;76}(t-\tau/2)}\tilde{\mathcal{K}}^{18349}_{6257}(t-\tau/2) + \text{H.c.}
\end{align}
Then perform an additional coarse grain integration
\begin{align}
    &\dot{\tilde{\rho}}^{(1)}_{12}|_{\rm e-e}^{\rm incoher} \stackrel{(2)}{\simeq}\nonumber\\
    &\stackrel{(2)}{\simeq}-\sum_{\substack{3456\\789}}\int_0^{\infty}\!\!d{\tau}\frac{e^{-\frac{\tau^2}{2\bar{\tau}_{\rm QP}^2}}}{(\sqrt{2\pi}\bar{\tau}_{\rm QP})^{1/2}}\int_{-\infty}^{\infty}\!\!d{\tau'}\frac{e^{-\frac{{\tau'}^2}{2\bar{\tau}_{\rm QP}^2}}}{(\sqrt{2\pi}\bar{\tau}_{\rm QP})^{1/2}}\times\nonumber\\
    &\times e^{i\bar{\varepsilon}_{15;89}\big(t+\frac{\tau+\tau'}{2}\big)}e^{-i\bar{\varepsilon}_{34;76}\big(t+\frac{\tau'-\tau}{2}\big)}\tilde{\mathcal{K}}^{18349}_{6257}\Big(t+\frac{\tau'-\tau}{2}\Big) +\nonumber\\ 
    &+\text{H.c.}
\end{align}
Under a new change of variables, $t_1 = \tau+\tau'$, $t_2 = \tau' - \tau$ and considering that the dynamics of $\tilde{\mathcal{K}}$ are much slower than the oscillations of the exponential factors, $\tilde{\mathcal{K}}$ can be extracted from the integral.
\begin{align}
    &\dot{\rho}_{12}^{(1)}|_{\rm e-e}^{\rm incoher} \stackrel{(3)}{\simeq}\nonumber\\
    &\stackrel{(3)}{\simeq}-\frac{1}{4}\sum_{\substack{3456\\789}}\mathcal{K}^{18349}_{6257}(t)\int_{-\infty}^\infty\!\! d{t_1}\frac{e^{-\frac{t_1^2}{4\bar{\tau}_{\rm QP}^2}}}{(\sqrt{2\pi}\bar{\tau}_{\rm QP})^{1/2}}e^{i\bar{\varepsilon}_{15;89}\frac{t_1}{2}}\times\nonumber\\
    &\times\int_{-\infty}^\infty d{t_2}\frac{e^{-\frac{t_2^2}{4\bar{\tau}_{\rm QP}^2}}}{(\sqrt{2\pi}\bar{\tau}_{\rm QP})^{1/2}}e^{i\bar{\varepsilon}_{76;34}\frac{t_2}{2}} + \text{H.c.}
\end{align}
%
After some manipulations, the following expression is obtained.
\begin{align}\label{Eq:rho12_ee}
    &\dot{\rho}_{12}^{(1)}|_{\rm e-e}^{\rm incoher}\stackrel{(3)}{\simeq}\nonumber\\
    &\frac{1}{2}\sum_{345}\big[(\delta_{13}-\rho_{13}^{(1)}) \mathcal{P}_{32;45}^{\rm e-e}\rho_{45}^{(1)} -\rho_{32}^{(1)}{\mathcal{P}_{54;13}^{\rm e-e}}^*(\delta_{54} - \rho_{54}^{(1)})\big] +\nonumber\\
    &+\text{H.c.}
\end{align}
Here, the scattering matrix is defined as
\begin{equation}\label{Eq:e-e-scattP}
       \mathcal{P}_{52;43}^{\rm e-e} = \sum_{6789} \rho_{89}^{(1)} A_{56;48}A^*_{27;39} (\delta_{76}-\rho_{76}^{(1)})
\end{equation}
\begin{align}\label{Eq:e-e-scatt}
    A_{56;48}=\sqrt{2\pi}\frac{\exp{-\Big(\frac{\bar{\varepsilon}_{5}+\bar{\varepsilon}_{6}-\bar{\varepsilon}_{4}-\bar{\varepsilon}_{8}}{2\epsilon}\Big)^2}}{(2\pi\epsilon^2)^{1/4}}W_{56;48}
\end{align}
\begin{table}[!ht]
\begin{tabular}{ |p{1.2cm}|p{2.5cm}|p{1.2cm}|  }
\hline
\multicolumn{3}{|c|}{Electron-electron interaction approximations} \\
\hline
\!\!\!\!\!\!Approximation& &Validity\\
\hline
COHSEX approximation in the coherent dynamics & $\Delta\Sigma^{\rm s}\simeq\Delta\Sigma^{\rm s}_{\rm COHSEX}$ &wide gap semiconductors and insulators\\
\hline
RPA limit & $\begin{aligned}W^\lessgtr\simeq W^{\rm R}G^\lessgtr G^\gtrless W^{\rm A}\end{aligned}$ & long range correlations\\
\hline
Static screening approximation & $\begin{aligned}\\W^{\rm R/A}(t,t')\simeq&\\
\simeq\,W\delta(t,t')\end{aligned}$ & fast decay of electronic screening memory $\tau_{\rm scr}\!\!\ll\!\!\tau_{\rm S}$\\
\hline
Quasi-particle approximation & $\begin{aligned}\\G^{\rm R}(t,t')\simeq -i\theta(t-t')&\\
\times e^{-\frac{(t-t')^2}{2\tau_{\rm QP}^2}}e^{-i(t-t')\bar{\varepsilon}}\end{aligned}$ & weakly interacting QPs\\
\hline
Removal of exchange in the scattering integral & $\begin{aligned}I&\simeq I^{\rm d}\\
I^{\rm ex}&\simeq 0\end{aligned}$ & scattering between delocalized states
\\
\hline
First Markov approximation & $\begin{aligned}\int_0^{g(t,T)}d{\tau}e^{-\frac{\tau^2}{2\bar{\tau}_{\rm QP}^2}}&\simeq\\
\int_0^{\infty}d{\tau}e^{-\frac{\tau^2}{2\bar{\tau}_{\rm QP}^2}}\end{aligned}$ & \multirow{2}{0.17\columnwidth}{$\bar{\tau}_{\rm QP}\ll\tau_{\rm S}$ large electronic reservoir}\\
\cline{1-2}
Coarse grain time integration & $\begin{aligned}\\\mathcal{I}_{\rm s}(t)&\simeq\hat{\boldsymbol{\rho}}^{(1)}(\hat{\boldsymbol{I}}-\hat{\boldsymbol{\rho}}^{(1)})\\
&\times\mathcal{P}_{\rm ee}\end{aligned}$&\\
\hline
\end{tabular}
 \caption{Summary of approximations used to derive Eq.~(\ref{Eq:rho12_ee}).
 }\label{Tab:3}
\end{table}
That is again in the Lindbladian form (see Eq.~(\ref{Eq:scatt_mark2})). This approach is not unique, other ways of studying the Lindblad equations from the NEGF formalism have recently been developed~\cite{Sieberer_2016,Hans_20220,stefanucci2024kadanoff,THOMPSON2023169385}. By considering a non-interacting system in the stationary case, their dynamical equation reduces to the Lyapunov equation~\cite{stefanucci2024kadanoff}. This equation is currently being investigated to study different properties, such as topological phases, exceptional points, and bulk-edge correspondence~\cite{Lieu2020,Alt2021,He2022,Woj2022,Shen2018,Yao2018,Yok2019}. We notice that although in the case of the phonons and radiation fields the Markovian condition is more commonly satisfied, the electronic system does not necessarily behave as an external bath. The condition of a quasi-continuum of states is more suitable for hot metals, or even warm dense systems, but in solid-state systems it could break down easily, especially in the case of localized defect states, molecular states, and low-dimensional strongly correlated systems, the non-Markovian approach should be preferred\cite{Rivaprep}. In Table~(\ref{Tab:3}) we summarize the approximations required to derive the Lindblad equation (\ref{Eq:rho12_ee}):\\
(i) We applied the COHSEX approximation to the variation of the static self-energy $\Delta \Sigma^{\rm s}$. This approximation is known to overestimate band gaps\cite{PhysRevB.82.195108,PhysRevB.98.115125}. Solving only the coherent dynamics with this self-energy is formally equivalent to solve the BSE~\cite{attaccalite2011}.
\\
(ii) We used the RPA form in Eq.~(\ref{Eq:RPA_screening}) to approximate $W^\lessgtr$. This clearly works better for weakly correlated systems; i.e. RPA is limited in describing localized electronic states with no account of exchange\cite{PhysRevB.104.045134}.\\
(iii) The static approximation for the electronic screening replaces $W^{\rm R/A}(t,t')$ with $W\delta(t,t')$ where $W$ is the static RPA-screened Coulomb interaction. This is valid in the case of extremely fast screening relaxation $\tau_{\rm scr}\ll\tau_{\rm S}$ compared to the decay time of the electronic system correlations. This is typically a good approximation unless the electronic system is excited on a time scale of the order of the inverse of the plasma frequency\cite{haug2007}.\\
(iv) The quasi-particle approximation (Eq.~(\ref{Eq:QA_approx})) for $G^{\rm R/A}$ is also valid in the case of moderately correlated systems (or Fermi liquid theory still holds).\\
(v) The removal of the exchange term in the scattering operator is analogous to the RPA limit for the screening, by considering only direct long-range scattering we neglect short-range contributions more relevant for localized electronic states or spin-exchange interactions.\\
(vi) The first Markov approximation is a good approximation in the limit of short quasi-particle lifetimes $\bar{\tau}_{\rm QP}\ll\tau_{\rm S}$ compared to the lifetime of electronic correlations. This works best in systems with relatively large density of states close to the Fermi level facilitating scattering events. \\
(vii) Finally, we have coarse grained time integration, this is tightly connected to approximation (vi) and works best in the case of large electronic baths at finite temperature with a dense populated manifold of states around the Fermi level
~\cite{PhysRevB.111.075139}. Among these (iv), (vi) and (vii) are crucial to obtain the Lindblad limit form of electron-electron scattering.

\section{Computational approach to first-principles open quantum dynamics}\label{sec:FPDMD}
\begin{figure}[htbp]
    \centering
   \includegraphics[width=0.8\columnwidth]{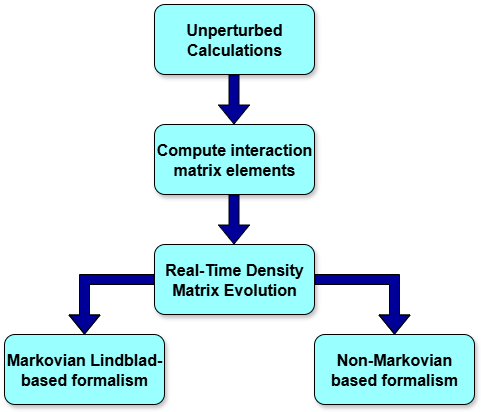}
    \caption{Workflow of the First-Principles density matrix dynamics (FPDMD) formalism. This is divided in three major steps: {\bf (1)} First we compute the unperturbed electronic states together with the phonon states; {\bf (2)} all the necessary matrix elements are computed from the unperturbed electronic structure and the environmental degrees of freedom; {\bf (3)} we perform a complete density matrix evolution using a Markovian or non-Markovian set of equations.
    }
    \label{fig:workflow}
\end{figure}

After detailed discussions of theory and approximations earlier, we will discuss the first-principles implementation of this open quantum dynamics framework. We have reviewed and summarized the detailed first-principles calculation procedure and codes for a specific application of spin relaxation in Ref.~\cite{Xu2024JCTC}. The general formalism workflow is schematically depicted in Fig.(\ref{fig:workflow}).\\
The First-Principles density matrix approach is characterized by three major steps as explained in detail below.
\subsubsection{First-Principles Calculations of Unperturbed Electron and Phonon States}

In the first phase we perform the ground state calculation of the electronic system to obtain quasiparticle wavefunctions, $\{\ket*{\mu{\bf K'}}\}$ and eigenvalues $\{\varepsilon_\mu({\bf K'})\}$. Depending on the level of theory used, these are unperturbed electronic eigenstates and eigenenergies of the matrix $\bar{\xi}$ in Eq.~(\ref{eq:unperturbed-e}). Depending on the level of theory employed for the unperturbed electronic structure calculation, $\bar{\xi}$ will correspond to the Hartree-Fock or Kohn-Sham Hamiltonian. From many-body perturbation theory it is possible to include higher-order correlation effects, for example by adding GW corrections. 
In the case where electron-phonon coupled dynamics is performed, a calculation of the phonon band structure is also required to obtain phonon eigenvectors $\{{\bf u}_{\bf q}\}$ and eigenenergies $\{\epsilon_{\bf q}\}$. Additionally, in presence of spin-orbit interaction or external magnetic fields the electronic states $\ket*{\mu{\bf K'}}$ are two-dimensional spinors.
\subsubsection{Matrix Elements Calculations}
Here depending on the interactions we account for in the simulation, additional calculations are required. We start by considering the electron-electron interaction given in Eq.~(\ref{Eq:potential}).\\
\paragraph{Electron-electron interaction}
The matrix elements of the screened Coulomb interaction are usually computed in reciprocal space by replacing $v_{\rm C}$ with the screened potential $W$. At the level of the RPA approximation and neglecting local field corrections\cite{giuliani2005quantum}, the electron-electron susceptibility is written as
\begin{align}
    &\chi_{\rm S}[\rho^{(1)}]({\bf Q},\omega)=\nonumber\\
    &=\!\!\sum_{{\bf K}\in{\rm BZ}}\sum_{\mu\nu}\frac{f_\mu({\bf K'})-f_\nu({\bf K'}+{\bf Q})}{\varepsilon_\mu({\bf K'})-\varepsilon_\nu({\bf K'}+{\bf Q})+\omega+i\eta}\abs{M_{\mu\nu}({\bf K'},{\bf Q})}^2\ ,
\end{align}
where $M_{\mu\nu}({\bf K'},{\bf Q})=\mel*{\mu{\bf K'}}{e^{-i{\bf Q}\cdot{\bf r}}}{\nu{\bf K'}+{\bf Q}}$ and $f_\mu({\bf K'})$ are the occupations of the states that can be computed in the ground state or updated at every step in the simulation. This can be used to compute the static screening $W({\bf Q})=v({\bf Q})/\big(\epsilon_{\rm s}\big[1 - v({\bf Q})\chi_{\rm S}({\bf Q},0)\big]\big)$ with the bare Coulomb potential given by $v({\bf Q})=e^2/(V\epsilon_0\abs{\bf Q}^2)$ and $\epsilon_0$, $\epsilon_{\rm s}$ being the vacuum permittivity and static dielectric function respectively\cite{PhysRevB.72.035105}; $V$ is the volume of the system. The electron-electron matrix elements are then computed from the screened interaction using the electronic wave functions\cite{Xu2024JCTC}. This can be also extended including local field corrections.
%
\paragraph{Electron-phonon coupling}
The fundamental quantity required to perform electron-phonon coupled dynamics is the $g_{12}^{{\bf q}-}$ coefficient of Eq.~(\ref{Eq:geph}). In general, the operator $\Delta_{\bf q}\hat{H}_{\rm sys}$ is extremely hard to compute and must be approximated. In density functional perturbation theory\cite{RevModPhys.73.515}, it is replaced with $\Delta_{\bf q}v_{\rm KS}$, where $v_{\rm KS}$ is the Kohn-Sham potential of the electronic system. This provides a reasonable account of screening. It is, however, problematic in the case of highly correlated systems, in such cases a more accurate account of many-body effects is required\cite{PhysRevB.106.125403}.\\
\paragraph{Electron-radiation coupling}
In order to describe the coupling between electrons and the radiation field, at the long wavelength limit, we need to compute the electric-dipole matrix elements (Eq.~(\ref{Eq:dipole})). These can be conveniently obtained from the knowledge of the electronic wave functions, but care needs to be taken to include the contribution of non-local part of poseudopotential at the velocity gauge or using the Berry phase for a proper description of position operator at the length gauge.\\
\newline
All these matrix elements are computed using density functional perturbation theory on a coarse ${\bf K'}$ and {\bf Q} mesh. However, a correct representation of the scattering operators used in the real-time dynamics requires a much finer mesh. This is a huge computational bottleneck for our method and can be solved by numerical interpolation techniques, for example Wannier interpolation\cite{RevModPhys.84.1419}. Specifically, one transforms the matrix elements from k-space Bloch basis to real space maximally-localized Wannier function basis\cite{PhysRevB.56.12847}, and then interpolates to a much finer ${\bf K'}$ and {\bf Q} mesh when transforming back to the k space\cite{https://doi.org/10.1002/adom.201600914,PhysRevB.94.075120,SUNDARARAMAN2017278}.
\subsubsection{Real-time Density Matrix Evolution}
The real-time evolution of the density matrix can be performed either at the Markovian or the non-Markovian level as discussed earlier.\\
\paragraph{Markovian evolution}
The Markovian evolution is based on the solution of Eq.~(\ref{Eq:rho00}), where the incoherent part is given by Eq.~(\ref{Eq:rho12_ee}) for the electron-electron interaction and Eq.~(\ref{Eq:e-ph_general}) for the electron-phonon coupling. The electron-phonon scattering $\mathcal{P}_{\rm e-ph}$ are pre-computed and do not need to be updated in time. The electron-electron scattering, $\mathcal{P}_{\rm e-e}$ must be, instead, updated over time. These coefficients are in fact explicitly dependent on the density matrices in Eq.~(\ref{Eq:e-e-scattP}). The electron-phonon scattering contribution\footnote{the same considerations are valid for the incoherent electron-radiation scattering term} has algorithm complexity $O(N^4N_{\bf K})$, where $N$ is the number of electronic bands and $N_{\bf K}$ the number of {\bf K}-points used. The inclusion of the electron-electron scattering makes the temporal update of $\dot{\rho}^{(1)}|_{\rm e-e}^{\rm incoher}$ the major bottleneck of the real-time algorithm. Its complexity is $O(N^7N_{\bf K})$ making this prohibitive for large electronic systems. The use of optimized matrix multiplication algorithms can reduce the overall complexity of the matrix operations; however, it is clear that the calculation of $\dot{\rho}^{(1)}|_{\rm e-e}^{\rm incoher}$ represents the more time-consuming part of the real-time Lindblad evolution.\\
\paragraph{Non-Markovian evolution}
Here with non-Markovian evolution we refer to the integration of Eq.~(\ref{Eq:eescatter}) for the electron-electron scattering, and to Eq.~(\ref{Eq:e-ph_non_M}) and Eq.~(\ref{Eq:e-rad_incoher}) for electron-phonon and electron-radiation scatterings respectively. Within the static screening approximation the electron-electron contribution does not require the knowledge of any additional correlation function, only of the density matrix. However, contrary to the Lindblad Markovian case, now we have to integrate over its full history (i.e. with two time-integrations). One possible approximation is based on the assumption that contributions at $t' = t-\Delta{t}$ will become less important as $\Delta{t}$ increases. This is well justified in the case of metals and not-too-strong-correlated electronic systems.\\
The non-Markovian electron-phonon dynamics requires the time evolution of the phonon distribution $n_{\bf q}(t)$ in combination with the electronic density operator. This increases the algorithm complexity by a factor $N_{\bf q}$ given that we have $N_{\bf q}$ correlation functions to update at each step. For strong electron-phonon interaction $B_{\bf q}(t)$ must also be updated over time. The algorithm complexity could be potentially reduced by considering only a few modes {\bf q} as major contributors to the dynamics and neglect or include the others via the Lindblad operator. The validity of these approximations, however, is strongly dependent on the nature of the physical system studied.
\section{Applications to open quantum dynamics in solids}\label{sec:OQSsolids}
\noindent
In this section we discuss the potential application of the real-time density matrix formalism to the simulation of the quantum dynamics in real materials. For each application, we discuss the range of validity of the Markovian-Lindblad approximation and the condition that nonMarkovian effects become more relevant.
\paragraph*{{\bf Semiconductor Spintronics applications}}
\noindent
Spintronic devices must exhibit long spin relaxation times and diffusion lengths for stable operations. The discovery of ultralong spin relaxation time and diffusion lengths in graphene at room temperature\cite{Drogeler2016} has fueled interest in 2D materials and their heterojunctions, considered promising candidates for spin transport and hosting spin qubits. Different spin decoherence mechanisms coexist in semiconductors\cite{WU201061}. The spin-orbit interaction is responsible for the two most efficient spin relaxation mechanisms in nonmagnetic semiconductors, the Elliott-Yafet (EY) spin relaxation\cite{Elli1954,Yafet1963} and the D{\'{y}}akonov-Pere{\'{l}} (DP) mechanism\cite{Dyakonov71}. The former often appears in centrosymmetric systems or systems with strong spin-momentum locking; the latter is more common in broken inversion-symmetry systems with strong scatterings~\cite{Xu2024,WU201061}. 
Additional relaxation processes are the s(p)-d exchange, where s(p) electrons interact with localized spins, and the Bir-Aranov-Pikus process due to electron-hole exchange in heavily p-doped semiconductors\cite{Pikus1971}, as well as hyperfine interactions with nuclear spins and paramagnetic impurities. The first-principles density matrix dynamics (FPDMD) approach based on the Markovian-Lindblad approximation has been successfully applied to the study of spin dynamics in diverse semiconducting systems where disparate symmetry and spin-orbit couplings are treated on an equal footing, e.g. III-V semiconductors~\cite{XuJ2020,Xu2021}, 
halide perovskites\cite{PhysRevB.104.184418,JXu2024,KLi2024},  two-dimensional Dirac materials\cite{XuJ2021}, as well as  transition metals dichalcogenides\cite{XuJ2020}. A recent study addresses the effect of $g$-factor fluctuations on $T_2$ and $T_2^*$ time in solids through FPDMD~\cite{JXu2024,PhysRevB.111.115113}. In all these cases, the Lindblad formalism has demonstrated a high level of accuracy in predicting spin decoherence and relaxation mechanisms in semiconductors.\\
Orbital AM relaxation and decoherence, which is of fundamental importance in the fields of orbitronics and spin-orbitronics\cite{Jo2024,Gu2024}, can be calculated in the same way as spin relaxation within the FPDMD framework.

\paragraph*{{\bf Nonlinear optics and photocurrents applications}}
Another interesting phenomenon for open quantum dynamics applications is the simulation of quantum kinetic processes in nonlinear optics and the calculation of induced photocurrents. 
The photogalvanic effect (PGE) that generates DC spin, charge, and orbital currents in homogeneous solids under linear and circularly-polarized light, without external fields or p-n junctions, is extremely relevant for bulk photovoltaic and spintronics applications. It is also important for characterizing symmetry\cite{PhysRevX.11.011001}, non-trivial topologies\cite{deJuan2017,PhysRevLett.116.237402}, and spin-orbit properties in non-centrosymmetric systems\cite{Huang2021,Mu2021,Xu2021}. 
The description of transient and steady-state PGE requires taking into account kinetic processes of excitations, scatterings, and recombinations, as discussed in Refs.~\cite{Sturman2020-mu,PhysRevB.110.115108}.
%
%
Current theories are mostly based on perturbative approaches\cite{PhysRevB.110.075159}, or TDDFT and TDGW calculations\cite{doi:10.1073/pnas.1906938118}, in which the description of decoherence effects is mostly based on a single relaxation time. In order to go beyond the existing theories, a more general formalism that can describe all the different kinetic contributions in real time on an equal footing is needed. The FPDMD provides a promising pathway to do this, by including different scattering sources in the incoherent light-driven dynamics.
\noindent
\paragraph*{{\bf Angular momentum flow in chiral solids}}
Interest in chiral solids has grown considerably in the last decade\cite{ma15175812}. Chirality is a geometric property associated with the lack of mirror and inversion symmetry, and it is shown to have fundamental connections with the generation and transfer of Angular Momentum (AM) in solids. In this context, a very interesting effect is the generation of spin polarization via the injection of charge current in chiral nonmagnetic materials. This process is known as \emph{chiral-induced spin selectivity} (CISS)\cite{CISS} and it is observed in weak spin-orbit systems. The details of the mechanisms governing the flux of angular momentum between spins, electrons, and phonons in these systems are still under debate\cite{Dalum2019}, e.g. in molecules the spin-orbit interaction is often considered not strong enough to explain the magnitude of observed spin polarization\cite{10.1063/1.3167404}. Spin-orbit coupling from interfaces in proximity with chiral molecules has been also suggested as a potential source of spin polarization~\cite{10.1063/1.4795319,Liu2021}. 
On the other hand, chiral materials possess chiral phonons, whose out-of-equilibrium distribution produces net AM, which then possibly converts to orbital AM and spin polarization\cite{Kim2023,Tauchert2022}.   Understanding such phenomena demands a non-adiabatic, and potentially dynamical description of electron-phonon interactions which is inherently non-Markovian given that equilibrium phonon distributions cannot produce net non-zero angular momentum even in chiral systems.
\noindent
\paragraph*{{\bf Ultrafast magnetism in solids}} Magnetically ordered systems represent an important area of applications for open quantum dynamics, for both metallic and semiconducting systems. Spin-phonon interactions are known to dominate the decoherence dynamics on a time scale of hundreds of picoseconds\cite{Illg2011}. Ultrafast dynamics, under intense femtosecond pulses\cite{PhysRevLett.76.4250} and below $100$ fs, are more complex due to the presence of different competing mechanisms. These range from electron-electron scattering\cite{PhysRevB.80.180407}, relativistic electron-light interaction\cite{HINSCHBERGER2012813}, out-of-equilibrium exchange mechanism\cite{PhysRevB.95.024412}, electron-magnon interactions\cite{PhysRevB.78.174422} and superdiffusive spin currents\cite{PhysRevLett.105.027203}. In these regimes, non-Markovian effects are very important and should not be neglected in order to obtain a quantitatively correct dynamics\cite{IMAMURA2022169209}. The FPDMD approach, with its ability to coherently describe all these different processes beyond the Markovian approximation, has the potential to provide quantitatively accurate predictions for the ultrafast demagnetization observed in solids. The short timescales involved, usually of the order of few hundreds of femtoseconds, require an accurate description of electron-electron scattering processes. The static screening approximation is in fact not fully justified in these regimes.
\noindent
\paragraph*{\bf Quantum Magnonics applications}
Quantum magnonics is a rapidly rising field studying the quantum states of magnons and the entanglement of magnons with other quasiparticles\cite{YUAN20221}. The integration of magnons with other quantum systems like cavity photons\cite{PhysRevLett.104.077202}, superconducting qubits, defect centers, and mechanical oscillators demands, first of all, the hybrid system to be controllable. One of the most important requirements is the ability to maintain the entanglement between magnons and other quantum states. It has been shown, for instance, that the entanglement between magnons and qubits enables the detection of single magnon states\cite{https://doi.org/10.1126/science.aaz9236}. The Lindblad-based FPDMD approach is applicable to the study of the relaxation and decoherence of quantum magnon states, including magnon-magnon, magnon-photon, and magnon-phonon couplings~\cite{PhysRevB.111.104431}.However, the description of the entanglement between different quasiparticles requires the inclusion of effects beyond the Born-Markovian approximation into the dynamics.  
\noindent
\paragraph*{{\bf Spin qubits and quantum information science}}
Another critical area for applications of Born-Markovian and non-Markovian open quantum dynamics in solids is the control and manipulation of spin qubits in semiconductors\cite{RevModPhys.95.025003,Hu_foundQM}. This is relevant for different quantum information technologies, including computation, communication, and sensing\cite{Awschalom2018}. These qubits are particularly susceptible to decoherence due to the strong interaction with phonons and require accurate modeling of spin-phonon interactions in order to predict their decoherence times\cite{Ghosh2021}. In the so-called \emph{central spin problem} the single spin is coupled to the environment represented by the surrounding nuclear spins via the hyperfine interaction. Another example is given by the electronic spin in the NV$^-$ centers coupled to the spin environment of substitutional nitrogen defects\cite{doi:10.1126/science.1155400}, or superconducting qubits interacting with other qubits\cite{RevModPhys.85.623}. Non-Markovian effects arising from the interaction with the phononic bath are also expected to be important, in particular, when the spin system is coupled only to few localized vibrational modes\cite{PhysRevA.101.022110}.
Non-Markovian effects may also arise in the dynamics of Bose-Einstein condensates (BEC) in a trap that are coupled to an atomic state outside the trap\cite{PhysRevA.55.R2531,PhysRevA.61.023603}. This behavior also appears in quantum dots coupled to a superfluid via laser transitions\cite{JAKSCH200552}, or to a BEC in a double-well potential\cite{PhysRevA.79.032106}, or in the case of atoms trapped in an optical lattice and coupled to an untrapped quantum level\cite{Navarrete2011}.
\noindent
\noindent
\paragraph*{\bf Excitonics applications}
\noindent
The study of the dynamics of quantum excitations in low-dimensional systems is a topic of great interest\cite{Spat2005}. The dynamics of excitons in two-dimensional materials has been explored using the NEGF scheme\cite{PhysRevLett.128.016801}, time-dependent Bethe-Salpeter equation\cite{attaccalite2011} and simulations based on a fully non Markovian scheme for electrons and phonons have been performed in the case of the \ce{MoS2} monolayer\cite{Perfett02023}. The formalism based on the density matrix evolution can potentially account for both non-equilibrium electron-electron and exciton-phonon interactions\cite{PhysRevB.92.235208,Malakhov2024,ant2022,PhysRevB.106.125403} in highly correlated systems, where the inclusion of non-Markovian effects is of fundamental importance. Similar to magnon dynamics, the coupling with both photons and phonons could be included within the FPDMD framework to extract exciton decoherence and relaxation times\cite{PhysRevLett.132.126902,amit2025abinitiodensitymatrixapproachexciton,guo2025phononassistedradiativelifetimesexciton}.
\noindent
\paragraph*{{\bf Applications beyond condensed matter}}
Other possible applications for the FPDMD formalism is the dynamics of matter under extreme conditions, hot solids, and warm dense systems, where the Born-Oppenheimer approximation could potentially break down and nonadiabatic effects become important\cite{PhysRevResearch.2.043139}. A better understanding of the relaxation mechanisms in these systems has potential applications in fusion energy studies\cite{10.1063/5.0138955} and X-ray free electron laser experiments\cite{a90936bc1d9c41fe8f767ace2bf4e491}. The theory of warm dense systems is particularly challenging due to the complex interplay of electron-electron correlations and quantum degeneracy\cite{MOLDABEKOV2025104144}. In such conditions, thermal excitations cannot be ignored. The Markovian limit is then considered a good approximation for both electron-electron and electron-phonon scattering, and the FPDMD formalism is potentially applicable in these regimes.
Other possible applications are given by vibrational relaxation of molecules in liquids\cite{velsko80}, solid impurities\cite{BERKOWITZ1977260} or metal surfaces\cite{Tully1995,Rudge2024}.
%
\section{Outlook} \label{sec:concl}
\noindent
In summary, the first-principles open quantum dynamics formalism that we are discussing here has the potential to describe a variety of nonequilibrium phenomena in real time, by treating electron-electron scattering and the scattering between electrons and environmental sources like phonons and photons accurately from weak to strong correlated regimes. The developed formalism and numerical implementation can account for both coherent and incoherent quantum effects for realistic materials from first-principles, opening the pathway for a complete \textit{ab-initio} description of quantum kinetic processes in solids and nanostructures, critical in a wide range of applications, from spintronics, quantum information science, to warm dense matter. 

\appendix
\section{Electro-magnetic field Hamiltonian}\label{app:em}
\noindent
The Hamiltonian of the electromagnetic field is written as
\begin{equation}\label{Eq:HEM}
\hat{H}_{\rm EM}(t)=\frac{1}{2}\int_V d{\bf r}\Big(\frac{1}{4\pi}\abs*{\hat{{\bf E}}({\bf r},t)}^2 + \frac{c^2}{4\pi}\abs*{\hat{{\bf B}}({\bf r},t)}^2\Big)
\end{equation}
Where the electric and magnetic fields operators  can be expressed as
\begin{align}
    \hat{\bf E}({\bf r},t) &= \bar{\bf E}({\bf r},t) + \delta\hat{\bf E}({\bf r},t) \, \label{Eq:E}\\
    \hat{\bf B}({\bf r},t) &= \bar{\bf B}({\bf r},t) + \delta\hat{\bf B}({\bf r},t)\label{Eq:B}
    \end{align}
Here we have assumed that the electric and magnetic fields can be separated into a macroscopic classical field and a quantum fluctuation component. This is quite accurate, for instance, when the radiation bath is an out of equilibrium configuration under the application of an external laser pulse. In such a case the density matrix of the radiation bath can be expressed as
\begin{equation}
    \hat{\rho}_{\rm rad}(t) = \hat{\rho}_{\rm rad}^0 + \delta\hat{\rho}_{\rm rad}(t) \,
\end{equation}
If we measure, in such conditions, the electric field acting on the open quantum system, we obtain
\begin{align}
    {\rm Tr}\big[\hat{\rho}_{\rm rad}\hat{\bf E}({\bf r},t)\big] &= {\rm Tr}\big[\hat{\rho}_{\rm rad}^0\hat{\bf E}] + {\rm Tr}\big[\delta\hat{\rho}_{\rm rad}(t)\hat{\bf E}\big]\nonumber\\
    &={\rm Tr}\big[\delta\hat{\rho}_{\rm rad}(t)\hat{\bf E}\big]
\end{align}
This is due to the fact that the trace of the electric field on the equilibrium radiation distribution is zero. In this way, using Eq.~(\ref{Eq:E}), the macroscopic measured electric field is
\begin{equation}
    \expval*{\hat{\bf E}({\bf r},t)} = \bar{\bf E}({\bf r},t) + \expval*{\delta\hat{\bf E}({\bf r},t)} = \bar{\bf E}({\bf r},t)
\end{equation}
We can now expand Eq.~(\ref{Eq:HEM}) using Eq.~(\ref{Eq:E}) and (\ref{Eq:B}).
If we compute the energy of the electro-magnetic radiation field, the cross contributions $\bar{\bf E}\cdot\delta\hat{\bf E}$ do not contribute to the total energy of the field and (after we neglect the higher order contribution ${\rm Tr}[\delta\hat{\rho}_{\rm rad}(t)\delta\hat{\bf E}({\bf r},t)\delta\hat{\bf E}({\bf r},t)^\dagger]\simeq 0$) we obtain
\begin{align}
    &E_{\rm EM}(t) \simeq \int_V d{\bf r}\Big\{\frac{\varepsilon_0}{2}\abs{\bar{\bf E}({\bf r},t)}^2 + \frac{1}{2\mu_0}\abs{\bar{\bf B}({\bf r},t)}^2\Big\} +\nonumber\\
    &+\int_V d{\bf r}\Big\{\frac{\varepsilon_0}{2}\expval*{\delta\hat{\bf E}({\bf r},t)\cdot\delta\hat{\bf E}({\bf r},t)^\dagger}_0 +\nonumber\\
    &+\frac{1}{2\mu_0}\expval*{\delta\hat{\bf B}({\bf r},t)\cdot\delta\hat{\bf B}({\bf r},t)^\dagger}_0\Big\}
\end{align}
In this way we can introduce an effective quasi-equilibrium radiation Hamiltonian, $\hat{H}_{\rm rad}(t)$, such that $\expval*{\hat{H}_{\rm EM}(t)}=E_{\rm EM}(t)\simeq\expval*{\hat{H}_{\rm rad}(t)}_0$.
\begin{align}
    &\hat{H}_{\rm rad}(t) = \int_V d{\bf r}\Big\{\frac{\varepsilon_0}{2}\abs{\bar{\bf E}({\bf r},t)}^2 + \frac{1}{2\mu_0}\abs{\bar{\bf B}({\bf r},t)}^2\Big\} +\nonumber\\
    &+\int_V d{\bf r}\Big\{\frac{\varepsilon_0}{2}\delta\hat{\bf E}({\bf r},t)\cdot\delta\hat{\bf E}({\bf r},t)^\dagger + \frac{1}{2\mu_0}\delta\hat{\bf B}({\bf r},t)\cdot\delta\hat{\bf B}({\bf r},t)^\dagger\Big\}
\end{align}
The first term is the energy of the external electro-magnetic field and the second contribution is the Hamiltonian of the background radiation field. By introducing the vector potentials
\begin{equation}\label{Eq:At}
    \hat{\bf A}({\bf r},t) = \bar{\bf A}({\bf r},t) + \delta\hat{\bf A}({\bf r},t),
\end{equation}
we have
\begin{align}
    \delta\hat{\bf E}({\bf r},t) &= \partial_t\delta\hat{\bf A}({\bf r},t)\nonumber\\
    \delta\hat{\bf B}({\bf r},t) &= \nabla_{\bf r}\times\delta\hat{\bf A}({\bf r},t)\nonumber\\
    \delta\hat{\bf A}({\bf r},t) &= \sum_{\bf k}\Big\{\delta\hat{A}_{{\bf k}}(t)\cdot\frac{e^{i{\bf K}\cdot{\bf r}}}{\sqrt{V}} + \delta\hat{A}_{{\bf k}}(t)^\dagger\cdot\frac{e^{-i{\bf K}\cdot{\bf r}}}{\sqrt{V}}\Big\}\, .
\end{align}
After some algebra and neglecting the terms that do not conserve the number of photons\footnote{these would not survive after we take the average and compute the energy} the background radiation contribution can be rewritten as
\begin{align}
    &\delta\hat{H}_{\rm rad}(t)\!\!=\!\!\frac{\varepsilon_0}{2}\sum_{{\bf k}}\omega_{\bf k}^2\Big\{\delta\hat{A}_{{\bf k}}(t)\delta\hat{A}_{{\bf k}}(t)^\dagger +\delta\hat{A}_{{\bf k}}(t)^\dagger\delta\hat{A}_{{\bf k}}(t)\Big\}\nonumber\\
    &+\frac{1}{2\mu_0}\sum_{{\bf k}}\abs{\bf K}^2\Big\{\delta\hat{A}_{{\bf k}}(t)\delta\hat{A}_{{\bf k}}(t)^\dagger + \delta\hat{A}_{{\bf k}}(t)^\dagger\delta\hat{A}_{{\bf k}}(t)\Big\}\nonumber\\
    &=\varepsilon_0\sum_{{\bf k}}\omega_{\bf k}^2\Big\{\delta\hat{A}_{{\bf k}}(t)\delta\hat{A}_{{\bf k}}(t)^\dagger + \delta\hat{A}_{{\bf k}}(t)^\dagger\delta\hat{A}_{{\bf k}}(t)\Big\}\, .
\end{align}
Where we used $1/(\mu_0\varepsilon_0)=c^2$ and $\omega_{\bf k}^2=c^2\abs{\bf K}^2$. If we then replace $\delta\hat{A}_{{\bf k}}(t)=\sqrt{\hbar / (2\varepsilon_0\omega_{\bf k})}\hat{a}_{{\bf k}}(t)$, the radiation Hamiltonian becomes
\begin{align}\label{A11}
    &\delta\hat{H}_{\rm rad}(t) = \frac{\hbar}{2}\sum_{{\bf k}}\omega_{\bf k}\big\{\hat{a}_{{\bf k}}(t)\hat{a}_{{\bf k}}(t)^\dagger + \hat{a}_{{\bf k}}(t)^\dagger\hat{a}_{{\bf k}}(t)\big\}\nonumber\\
    &=\hbar\sum_{{\bf k}}\omega_{\bf k}\Big(\hat{a}_{{\bf k}}(t)^\dagger\hat{a}_{{\bf k}}(t) + \frac{1}{2}\Big)
\end{align}
We can then write the quasi-equilibrium Hamiltonian as
\begin{align}
    &\hat{H}_{\rm rad}(t) = \int_V d{\bf r}\Big\{\frac{\varepsilon_0}{2}\abs{\bar{\bf E}({\bf r},t)}^2 + \frac{1}{2\mu_0}\abs{\bar{\bf B}({\bf r},t)}^2\Big\} + \nonumber\\
    &+\hbar\sum_{{\bf k}}\omega_{\bf k}\Big(\hat{a}_{{\bf k}}(t)^\dagger\hat{a}_{{\bf k}}(t) + \frac{1}{2}\Big)
\end{align}
That corresponds to Eq.~(\ref{Eq:Hrad}) in the main text.
\section{Interaction between matter and radiation}\label{app:radmatinter}
We can use (\ref{Eq:At}) inside the $\hat{H}_{\rm e-rad}(t)$ given in Eq.~(\ref{Eq:Herad0}).
\begin{align}
    \hat{H}_{\rm e-rad}(t) &= \frac{1}{2}[\hat{\bf p}, \hat{\bf A}({\bf r},t)] + \hat{\bf A}({\bf r},t)\cdot\hat{\bf p} + \frac{\hat{\bf A}({\bf r},t)^2}{2} +\nonumber\\
    &+\hat{\bf B}({\bf r},t)\cdot\hat{\bf S}
\end{align}
In Coulomb gauge we have $[\hat{\bf p}, \hat{\bf A}({\bf r},t)]=-i\hbar\nabla_{\bf r}\cdot\hat{\bf A}({\bf r},t)=0$. In addition, we neglect the $\hat{\bf A}({\bf r},t)^2$ term given that we are not considering two-photons scattering processes.
\begin{align}
    \hat{H}_{\rm e-rad}(t) &\simeq \hat{\bf A}({\bf r},t)\cdot\hat{\bf p} + \hat{\bf B}({\bf r},t)\cdot\hat{\bf S}
\end{align}
We then use $\hat{\bf B}({\bf r},t)=\nabla_{\bf r}\times\hat{\bf A}({\bf r},t)$, and assuming the magnetic field is spatially homogeneous and we can neglect the quantum field fluctuations
\begin{align}
    \bar{\bf A}({\bf r},t) &= \frac{1}{2}\bar{\bf B}\times{\bf r} + \bar{\bf a}(t)\\
    \delta\hat{\bf A}({\bf r},t) &= \delta\hat{\bf a}(t)
\end{align}
Such that we have $\nabla_{\bf r}\times\bar{\bf A}({\bf r},t) = \bar{\bf B}$ and $\nabla_{\bf r}\times\delta\hat{\bf A}({\bf r},t)=0$. $\hat{H}_{\rm e-rad}(t)$ then becomes
\begin{align}
    \hat{H}_{\rm e-rad}(t) &= \bar{\bf a}(t)\cdot\hat{\bf p} + \delta\hat{\bf a}(t)\cdot\hat{\bf p} + \frac{1}{2}\bar{\bf B}\cdot(\hat{\bf L} + 2\hat{\bf S})\ .
\end{align}
That is Eq.~(\ref{Eq:Herad}) in the main text.
\section{Markovian limits}\label{app:a}
Starting from Eq.~(\ref{Eq:fullrho}) and writing the Hamiltonian in the interaction picture as $\mathcal{H}(t)$, the equation of motion for the density operator is
\begin{equation}
  \frac{d\rho}{dt} = -i\big[\mathcal{H}(t),\rho(t_0)\big] - \big[\mathcal{H}(t), \big[\int_{t_0}^t dt'\mathcal{H}(t'), \rho(t')\big]\big]
\end{equation}
The contribution to the dynamics comes from the second term on the right-hand side. We now examine how to take the conventional and Lindblad Markovian limit of the previous expression.
\subsection{Conventional Markovian limit}
\noindent
Here, we introduce the operator
\begin{equation}
  \mathcal{K}(t) = 2\int_{t_0}^t dt'\,\mathcal{H}(t')\,,
\end{equation}
and we consider the time variation of $\rho(t')$ slow compared to that of the Hamiltonian.
The equation of motion is then written as
\begin{align}
  &\frac{d\rho}{dt} = -\frac{1}{2}\big[\mathcal{H}(t), \big[\mathcal{K}(t), \rho(t)\big]\big]\nonumber\\
  &=-\frac{1}{2}\big[\mathcal{H}(t)\mathcal{K}(t)\rho(t) - \mathcal{H}(t)\rho(t)\mathcal{K}(t) -\mathcal{K}(t)\rho(t)\mathcal{H}(t) +\nonumber\\ &+\rho(t)\mathcal{K}(t)\mathcal{H}(t)\big)
\end{align}
We can expand the previous expression over a complete basis set of the Hilbert space of the system.
\begin{align}
  \frac{d\rho_{12}}{dt} &= \frac{1}{2}\sum_{34}\big(\mathcal{P}_{12;34}\,\rho_{34} - \mathcal{P}^*_{33;14}\,\rho_{42}\big) + \text{H.c.}
\end{align}
The matrix elements of the scattering matrix are written as follows.
\begin{equation}\label{Eq:scatt_mark1}
  \mathcal{P}_{12;34} = \mathcal{H}_{13}\mathcal{K}^*_{24}
\end{equation}
\subsection{Lindblad-type Markovian limit}
The integrated equation for the density matrix in interaction picture is
\begin{align}
  &\rho(t) =\rho(t_0) -\frac{1}{2}\int_{t_0}^t dt_1\int_{t_0}^{t_1}dt_2\big[\mathcal{H}(t_1), \big[\mathcal{H}(t_2), \rho(t_2)\big]\big]
\end{align}
\noindent
We do here a change of variables that introduces the macroscopic time $T$ as follows
$$
\tau = t_1 - t_2\qquad T = \frac{t_1 + t_2}{2}
$$
\begin{align}
  &\rho(t)=\rho(t_0) - \frac{1}{2}\int_{t_0}^t d{T} \times\nonumber\\
  &\times\int_{t_0}^{g(t,T)}\!\!d{\tau}\bigg[\mathcal{H}\bigg(T+\frac{\tau}{2}\bigg), \bigg[\mathcal{H}\bigg(T-\frac{\tau}{2}\bigg), \rho\bigg(T-\frac{\tau}{2}\bigg)\bigg]\bigg]
\end{align}
The second integral can be approximated using a cutoff function. This is often justified in case the environment correlations decay faster in time compared to the electronic ones.
\begin{align}
  \int_{t_0}^{g(t,T)}d\tau f(\tau) &= \int_{-\infty}^\infty d\tau e^{-\frac{\tau^2}{2\bar{t}^2}}f(\tau)
\end{align}
where we let $t_0\rightarrow -\infty$. This produces the following expression.
\begin{align}
  &\frac{d\rho}{dt} =\nonumber\\
  &=\!\!-\frac{1}{2}\int_{-\infty}^\infty\!\! d{\tau}e^{-\frac{\tau^2}{2\bar{t}^2}}\bigg[\mathcal{H}\bigg(t+\frac{\tau}{2}\bigg),\bigg[\mathcal{H}\bigg(t-\frac{\tau}{2}\bigg), \rho\bigg(t-\frac{\tau}{2}\bigg)\bigg]\bigg]
\end{align}
by averaging over time the previous expression and approximating the density with its value at $t$.
\begin{align}
  &\frac{d\rho}{dt}=-\frac{1}{2}\int_{-\infty}^\infty d{\tau'}\frac{e^{-\frac{{\tau'}^2}{2\bar{t}^2}}}{\sqrt{2\pi\bar{t}^2}}\times\nonumber\\
  &\times\int_{-\infty}^\infty d\tau e^{-\frac{\tau^2}{2\bar{t}^2}}\bigg[\mathcal{H}\bigg(t-\frac{\tau'-\tau}{2}\bigg), \bigg[\mathcal{H}\bigg(t-\frac{\tau'+\tau}{2}\bigg), \rho(t)\bigg]\bigg]
\end{align}
we do again a change of variables
$$
\tau' = \tau_1 - \tau_2\qquad\tau = \tau_1 + \tau_2
$$
The determinant of the Jacobian of the transformation is
$$
||\mathcal{J}|| =
\begin{vmatrix}
1 & 1\\
1 &-1
\end{vmatrix}
= 2
$$
\begin{align}
  &\frac{d\rho}{dt}=-\frac{1}{2}\sqrt{\frac{2}{\pi\bar{t}^2}}\int_{-\infty}^\infty d\tau_1\,e^{-\frac{\tau_1^2}{\bar{t}^2}}\int_{-\infty}^\infty d\tau_2 e^{-\frac{\tau_2^2}{\bar{t}^2}}\times\nonumber\\
  &\times\big[\mathcal{H}(t+\tau_2), \big[\mathcal{H}(t-\tau_1), \rho(t)\big]\big]
\end{align}
We can define the operator
\begin{equation}
  \mathcal{L}(t) = \bigg(\frac{2}{\sqrt{2\pi}\bar{t}}\bigg)^{1/2}\int_{-\infty}^\infty d\tau e^{-\frac{\tau^2}{\bar{t}^2}}\mathcal{H}(t+\tau)
\end{equation}
and write the equation of motion in a more compact form.
\begin{align} \label{Eq:linbl}
  \frac{d\rho}{dt} &= -\frac{1}{2}\big[\mathcal{L}(t), \big[\mathcal{L}(t), \rho(t)\big]\big]\nonumber\\
  &=-\frac{1}{2}\big(\mathcal{L}^\dagger\mathcal{L}\rho - \mathcal{L}^\dagger\rho\mathcal{L}\big) + \text{H.c.}\nonumber\\
  &=-\frac{1}{2}\big\{\mathcal{L}^\dagger\mathcal{L}, \rho\big\} + \mathcal{L}^\dagger\rho\mathcal{L}
\end{align}
Eq.~(\ref{Eq:linbl}) corresponds to a Lindblad dynamics for the density matrix $\rho$. It can be rewritten in terms of the scattering operators.
\begin{align}
  \frac{d\rho_{12}}{dt} &= \frac{1}{2}\sum_{34}\big(\mathcal{P}_{12;34}\rho_{34} - \mathcal{P}_{33;14}^*\rho_{42}\big)
\end{align}
Where the scattering matrix acquires the following form.
\begin{equation}\label{Eq:scatt_mark2}
  \mathcal{P}_{12;34} = \mathcal{L}_{13}\mathcal{L}_{24}^*
\end{equation}
That is fundamentally different from Eq.~(\ref{Eq:scatt_mark1}) in the conventional Markovian limit, although the structure of the two scattering equations is the same.

\section*{acknowledgement}
J.S. acknowledges the support of the DOE Computational Chemical Science program within the Office of Science at DOE under grant No. DE-SC0023301. G. R. and Y. P.
acknowledge the support from the Air Force Office of Scientific Research under Award No. FA9550-21-1-0087.

\nocite{*}
\bibliography{ref}

\begin{thebibliography}{292}%
\makeatletter
\providecommand \@ifxundefined [1]{%
 \@ifx{#1\undefined}
}%
\providecommand \@ifnum [1]{%
 \ifnum #1\expandafter \@firstoftwo
 \else \expandafter \@secondoftwo
 \fi
}%
\providecommand \@ifx [1]{%
 \ifx #1\expandafter \@firstoftwo
 \else \expandafter \@secondoftwo
 \fi
}%
\providecommand \natexlab [1]{#1}%
\providecommand \enquote  [1]{``#1''}%
\providecommand \bibnamefont  [1]{#1}%
\providecommand \bibfnamefont [1]{#1}%
\providecommand \citenamefont [1]{#1}%
\providecommand \href@noop [0]{\@secondoftwo}%
\providecommand \href [0]{\begingroup \@sanitize@url \@href}%
\providecommand \@href[1]{\@@startlink{#1}\@@href}%
\providecommand \@@href[1]{\endgroup#1\@@endlink}%
\providecommand \@sanitize@url [0]{\catcode `\\12\catcode `\$12\catcode `\&12\catcode `\#12\catcode `\^12\catcode `\_12\catcode `\%12\relax}%
\providecommand \@@startlink[1]{}%
\providecommand \@@endlink[0]{}%
\providecommand \url  [0]{\begingroup\@sanitize@url \@url }%
\providecommand \@url [1]{\endgroup\@href {#1}{\urlprefix }}%
\providecommand \urlprefix  [0]{URL }%
\providecommand \Eprint [0]{\href }%
\providecommand \doibase [0]{http://dx.doi.org/}%
\providecommand \selectlanguage [0]{\@gobble}%
\providecommand \bibinfo  [0]{\@secondoftwo}%
\providecommand \bibfield  [0]{\@secondoftwo}%
\providecommand \translation [1]{[#1]}%
\providecommand \BibitemOpen [0]{}%
\providecommand \bibitemStop [0]{}%
\providecommand \bibitemNoStop [0]{.\EOS\space}%
\providecommand \EOS [0]{\spacefactor3000\relax}%
\providecommand \BibitemShut  [1]{\csname bibitem#1\endcsname}%
\let\auto@bib@innerbib\@empty
\bibitem [{\citenamefont {Rivas}\ and\ \citenamefont {Huelga}(2011)}]{OQS_book1}%
  \BibitemOpen
  \bibfield  {author} {\bibinfo {author} {\bibfnamefont {A.}~\bibnamefont {Rivas}}\ and\ \bibinfo {author} {\bibfnamefont {S.~F.}\ \bibnamefont {Huelga}},\ }\href {\doibase https://doi.org/10.1007/978-3-642-23354-8} {\emph {\bibinfo {title} {Open Quantum Systems}}}\ (\bibinfo  {publisher} {Springer Berlin, Heidelberg},\ \bibinfo {year} {2011})\BibitemShut {NoStop}%
\bibitem [{\citenamefont {Shu}\ and\ \citenamefont {Truhlar}(2023)}]{Shu2023}%
  \BibitemOpen
  \bibfield  {author} {\bibinfo {author} {\bibfnamefont {Y.}~\bibnamefont {Shu}}\ and\ \bibinfo {author} {\bibfnamefont {D.~G.}\ \bibnamefont {Truhlar}},\ }\href {\doibase 10.1021/acs.jctc.2c00988} {\bibfield  {journal} {\bibinfo  {journal} {J. Chem. Theory Comput.}\ }\textbf {\bibinfo {volume} {19}},\ \bibinfo {pages} {380} (\bibinfo {year} {2023})}\BibitemShut {NoStop}%
\bibitem [{\citenamefont {Bonitz}(2015)}]{bonitz2015quantum}%
  \BibitemOpen
  \bibfield  {author} {\bibinfo {author} {\bibfnamefont {M.}~\bibnamefont {Bonitz}},\ }\href {https://books.google.com/books?id=wW7_CgAAQBAJ} {\emph {\bibinfo {title} {Quantum Kinetic Theory}}}\ (\bibinfo  {publisher} {Springer International Publishing},\ \bibinfo {year} {2015})\BibitemShut {NoStop}%
\bibitem [{\citenamefont {Breuer}\ and\ \citenamefont {Petruccione}(2007)}]{Petruccione07}%
  \BibitemOpen
  \bibfield  {author} {\bibinfo {author} {\bibfnamefont {H.-P.}\ \bibnamefont {Breuer}}\ and\ \bibinfo {author} {\bibfnamefont {F.}~\bibnamefont {Petruccione}},\ }\href {\doibase https://doi.org/10.1093/acprof:oso/9780199213900.001.0001} {\emph {\bibinfo {title} {The Theory of Open Quantum Systems}}}\ (\bibinfo  {publisher} {Oxford University Press},\ \bibinfo {year} {2007})\BibitemShut {NoStop}%
\bibitem [{\citenamefont {Bharti}\ \emph {et~al.}(2022)\citenamefont {Bharti}, \citenamefont {Cervera-Lierta}, \citenamefont {Kyaw}, \citenamefont {Haug}, \citenamefont {Alperin-Lea}, \citenamefont {Anand}, \citenamefont {Degroote}, \citenamefont {Heimonen}, \citenamefont {Kottmann}, \citenamefont {Menke}, \citenamefont {Mok}, \citenamefont {Sim}, \citenamefont {Kwek},\ and\ \citenamefont {Aspuru-Guzik}}]{Bharti_2022}%
  \BibitemOpen
  \bibfield  {author} {\bibinfo {author} {\bibfnamefont {K.}~\bibnamefont {Bharti}}, \bibinfo {author} {\bibfnamefont {A.}~\bibnamefont {Cervera-Lierta}}, \bibinfo {author} {\bibfnamefont {T.~H.}\ \bibnamefont {Kyaw}}, \bibinfo {author} {\bibfnamefont {T.}~\bibnamefont {Haug}}, \bibinfo {author} {\bibfnamefont {S.}~\bibnamefont {Alperin-Lea}}, \bibinfo {author} {\bibfnamefont {A.}~\bibnamefont {Anand}}, \bibinfo {author} {\bibfnamefont {M.}~\bibnamefont {Degroote}}, \bibinfo {author} {\bibfnamefont {H.}~\bibnamefont {Heimonen}}, \bibinfo {author} {\bibfnamefont {J.~S.}\ \bibnamefont {Kottmann}}, \bibinfo {author} {\bibfnamefont {T.}~\bibnamefont {Menke}}, \bibinfo {author} {\bibfnamefont {W.-K.}\ \bibnamefont {Mok}}, \bibinfo {author} {\bibfnamefont {S.}~\bibnamefont {Sim}}, \bibinfo {author} {\bibfnamefont {L.-C.}\ \bibnamefont {Kwek}}, \ and\ \bibinfo {author} {\bibfnamefont {A.}~\bibnamefont {Aspuru-Guzik}},\ }\href {\doibase https://doi.org/10.1103/RevModPhys.94.015004} {\bibfield  {journal} {\bibinfo
  {journal} {Rev. Mod. Phys.}\ }\textbf {\bibinfo {volume} {94}},\ \bibinfo {pages} {015004} (\bibinfo {year} {2022})}\BibitemShut {NoStop}%
\bibitem [{\citenamefont {Georgescu}\ \emph {et~al.}(2014)\citenamefont {Georgescu}, \citenamefont {Ashhab},\ and\ \citenamefont {Nori}}]{Georgescu_2014}%
  \BibitemOpen
  \bibfield  {author} {\bibinfo {author} {\bibfnamefont {I.}~\bibnamefont {Georgescu}}, \bibinfo {author} {\bibfnamefont {S.}~\bibnamefont {Ashhab}}, \ and\ \bibinfo {author} {\bibfnamefont {F.}~\bibnamefont {Nori}},\ }\href {\doibase https://doi.org/10.1103/RevModPhys.86.153} {\bibfield  {journal} {\bibinfo  {journal} {Rev. Mod. Phys.}\ }\textbf {\bibinfo {volume} {86}},\ \bibinfo {pages} {153} (\bibinfo {year} {2014})}\BibitemShut {NoStop}%
\bibitem [{\citenamefont {Lloyd}(1996)}]{Lloyd_1996}%
  \BibitemOpen
  \bibfield  {author} {\bibinfo {author} {\bibfnamefont {S.}~\bibnamefont {Lloyd}},\ }\href {\doibase 10.1126/science.273.5278.1073} {\bibfield  {journal} {\bibinfo  {journal} {Science}\ }\textbf {\bibinfo {volume} {273}},\ \bibinfo {pages} {1073} (\bibinfo {year} {1996})}\BibitemShut {NoStop}%
\bibitem [{\citenamefont {Bonato}\ \emph {et~al.}(2016)\citenamefont {Bonato}, \citenamefont {Blok}, \citenamefont {Dinani}, \citenamefont {Berry}, \citenamefont {Markham}, \citenamefont {Twitchen},\ and\ \citenamefont {Hanson}}]{Bonato2016}%
  \BibitemOpen
  \bibfield  {author} {\bibinfo {author} {\bibfnamefont {C.}~\bibnamefont {Bonato}}, \bibinfo {author} {\bibfnamefont {M.}~\bibnamefont {Blok}}, \bibinfo {author} {\bibfnamefont {H.}~\bibnamefont {Dinani}}, \bibinfo {author} {\bibfnamefont {D.}~\bibnamefont {Berry}}, \bibinfo {author} {\bibfnamefont {M.}~\bibnamefont {Markham}}, \bibinfo {author} {\bibfnamefont {D.}~\bibnamefont {Twitchen}}, \ and\ \bibinfo {author} {\bibfnamefont {R.}~\bibnamefont {Hanson}},\ }\href {\doibase https://doi.org/10.1038/nnano.2015.261} {\bibfield  {journal} {\bibinfo  {journal} {Nat. Nanotechnol.}\ }\textbf {\bibinfo {volume} {11}},\ \bibinfo {pages} {247} (\bibinfo {year} {2016})}\BibitemShut {NoStop}%
\bibitem [{\citenamefont {Chiorescu}\ \emph {et~al.}(2003)\citenamefont {Chiorescu}, \citenamefont {Nakamura}, \citenamefont {Harmans},\ and\ \citenamefont {Mooij}}]{Chiorescu03}%
  \BibitemOpen
  \bibfield  {author} {\bibinfo {author} {\bibfnamefont {I.}~\bibnamefont {Chiorescu}}, \bibinfo {author} {\bibfnamefont {Y.}~\bibnamefont {Nakamura}}, \bibinfo {author} {\bibfnamefont {C.}~\bibnamefont {Harmans}}, \ and\ \bibinfo {author} {\bibfnamefont {J.}~\bibnamefont {Mooij}},\ }\href {\doibase https://doi.org/10.1126/science.1081045} {\bibfield  {journal} {\bibinfo  {journal} {Science}\ }\textbf {\bibinfo {volume} {299}},\ \bibinfo {pages} {1869} (\bibinfo {year} {2003})}\BibitemShut {NoStop}%
\bibitem [{\citenamefont {Martinis}\ \emph {et~al.}(2002)\citenamefont {Martinis}, \citenamefont {Nam}, \citenamefont {Aumentado},\ and\ \citenamefont {Urbina}}]{Martinis02}%
  \BibitemOpen
  \bibfield  {author} {\bibinfo {author} {\bibfnamefont {J.}~\bibnamefont {Martinis}}, \bibinfo {author} {\bibfnamefont {S.}~\bibnamefont {Nam}}, \bibinfo {author} {\bibfnamefont {J.}~\bibnamefont {Aumentado}}, \ and\ \bibinfo {author} {\bibfnamefont {C.}~\bibnamefont {Urbina}},\ }\href {\doibase https://doi.org/10.1103/PhysRevLett.89.117901} {\bibfield  {journal} {\bibinfo  {journal} {Phys. Rev. Lett.}\ }\textbf {\bibinfo {volume} {89}},\ \bibinfo {pages} {117901} (\bibinfo {year} {2002})}\BibitemShut {NoStop}%
\bibitem [{\citenamefont {Yu}\ \emph {et~al.}(2002)\citenamefont {Yu}, \citenamefont {Han}, \citenamefont {Chu}, \citenamefont {Chu},\ and\ \citenamefont {Wang}}]{Yu02}%
  \BibitemOpen
  \bibfield  {author} {\bibinfo {author} {\bibfnamefont {Y.}~\bibnamefont {Yu}}, \bibinfo {author} {\bibfnamefont {S.}~\bibnamefont {Han}}, \bibinfo {author} {\bibfnamefont {X.}~\bibnamefont {Chu}}, \bibinfo {author} {\bibfnamefont {S.-I.}\ \bibnamefont {Chu}}, \ and\ \bibinfo {author} {\bibfnamefont {Z.}~\bibnamefont {Wang}},\ }\href {\doibase https://doi.org/10.1126/science.1069452} {\bibfield  {journal} {\bibinfo  {journal} {Science}\ }\textbf {\bibinfo {volume} {296}},\ \bibinfo {pages} {889} (\bibinfo {year} {2002})}\BibitemShut {NoStop}%
\bibitem [{\citenamefont {Myatt}\ \emph {et~al.}(2000)\citenamefont {Myatt}, \citenamefont {King}, \citenamefont {Turchette}, \citenamefont {Sackett}, \citenamefont {Kielpinski}, \citenamefont {Itano}, \citenamefont {Monroe},\ and\ \citenamefont {Wineland}}]{Myatt00}%
  \BibitemOpen
  \bibfield  {author} {\bibinfo {author} {\bibfnamefont {C.}~\bibnamefont {Myatt}}, \bibinfo {author} {\bibfnamefont {B.}~\bibnamefont {King}}, \bibinfo {author} {\bibfnamefont {Q.}~\bibnamefont {Turchette}}, \bibinfo {author} {\bibfnamefont {C.}~\bibnamefont {Sackett}}, \bibinfo {author} {\bibfnamefont {D.}~\bibnamefont {Kielpinski}}, \bibinfo {author} {\bibfnamefont {W.}~\bibnamefont {Itano}}, \bibinfo {author} {\bibfnamefont {C.}~\bibnamefont {Monroe}}, \ and\ \bibinfo {author} {\bibfnamefont {D.}~\bibnamefont {Wineland}},\ }\href {\doibase https://doi.org/10.1038/35002001} {\bibfield  {journal} {\bibinfo  {journal} {Nature}\ }\textbf {\bibinfo {volume} {403}},\ \bibinfo {pages} {269} (\bibinfo {year} {2000})}\BibitemShut {NoStop}%
\bibitem [{\citenamefont {Turchette}\ \emph {et~al.}(2000)\citenamefont {Turchette}, \citenamefont {Myatt}, \citenamefont {King}, \citenamefont {Sackett}, \citenamefont {Kielpinski}, \citenamefont {Itano}, \citenamefont {Monroe},\ and\ \citenamefont {Wineland}}]{Turchette00}%
  \BibitemOpen
  \bibfield  {author} {\bibinfo {author} {\bibfnamefont {Q.}~\bibnamefont {Turchette}}, \bibinfo {author} {\bibfnamefont {C.}~\bibnamefont {Myatt}}, \bibinfo {author} {\bibfnamefont {B.}~\bibnamefont {King}}, \bibinfo {author} {\bibfnamefont {C.}~\bibnamefont {Sackett}}, \bibinfo {author} {\bibfnamefont {D.}~\bibnamefont {Kielpinski}}, \bibinfo {author} {\bibfnamefont {W.}~\bibnamefont {Itano}}, \bibinfo {author} {\bibfnamefont {C.}~\bibnamefont {Monroe}}, \ and\ \bibinfo {author} {\bibfnamefont {D.}~\bibnamefont {Wineland}},\ }\href {\doibase https://doi.org/10.1103/PhysRevA.62.053807} {\bibfield  {journal} {\bibinfo  {journal} {Phys. Rev. A}\ }\textbf {\bibinfo {volume} {62}},\ \bibinfo {pages} {053807} (\bibinfo {year} {2000})}\BibitemShut {NoStop}%
\bibitem [{\citenamefont {Kuhlmann}\ \emph {et~al.}(2013)\citenamefont {Kuhlmann}, \citenamefont {Houel}, \citenamefont {Ludwig}, \citenamefont {Greuter}, \citenamefont {Reuter}, \citenamefont {Wieck}, \citenamefont {Poggio},\ and\ \citenamefont {Warburton}}]{Kuhlmann13}%
  \BibitemOpen
  \bibfield  {author} {\bibinfo {author} {\bibfnamefont {A.}~\bibnamefont {Kuhlmann}}, \bibinfo {author} {\bibfnamefont {J.}~\bibnamefont {Houel}}, \bibinfo {author} {\bibfnamefont {A.}~\bibnamefont {Ludwig}}, \bibinfo {author} {\bibfnamefont {L.}~\bibnamefont {Greuter}}, \bibinfo {author} {\bibfnamefont {D.}~\bibnamefont {Reuter}}, \bibinfo {author} {\bibfnamefont {A.}~\bibnamefont {Wieck}}, \bibinfo {author} {\bibfnamefont {M.}~\bibnamefont {Poggio}}, \ and\ \bibinfo {author} {\bibfnamefont {R.}~\bibnamefont {Warburton}},\ }\href {\doibase https://doi.org/10.1038/nphys2688} {\bibfield  {journal} {\bibinfo  {journal} {Nature Physics}\ }\textbf {\bibinfo {volume} {9}},\ \bibinfo {pages} {570} (\bibinfo {year} {2013})}\BibitemShut {NoStop}%
\bibitem [{\citenamefont {Urbaszek}\ \emph {et~al.}(2013)\citenamefont {Urbaszek}, \citenamefont {Marie}, \citenamefont {Amand}, \citenamefont {Krebs}, \citenamefont {Voisin}, \citenamefont {Maletinsky}, \citenamefont {H\"ogele},\ and\ \citenamefont {Imamoglu}}]{Urbaszek13}%
  \BibitemOpen
  \bibfield  {author} {\bibinfo {author} {\bibfnamefont {B.}~\bibnamefont {Urbaszek}}, \bibinfo {author} {\bibfnamefont {X.}~\bibnamefont {Marie}}, \bibinfo {author} {\bibfnamefont {T.}~\bibnamefont {Amand}}, \bibinfo {author} {\bibfnamefont {O.}~\bibnamefont {Krebs}}, \bibinfo {author} {\bibfnamefont {P.}~\bibnamefont {Voisin}}, \bibinfo {author} {\bibfnamefont {P.}~\bibnamefont {Maletinsky}}, \bibinfo {author} {\bibfnamefont {A.}~\bibnamefont {H\"ogele}}, \ and\ \bibinfo {author} {\bibfnamefont {A.}~\bibnamefont {Imamoglu}},\ }\href {\doibase https://doi.org/10.1103/RevModPhys.85.79} {\bibfield  {journal} {\bibinfo  {journal} {Rev. Mod. Phys.}\ }\textbf {\bibinfo {volume} {85}},\ \bibinfo {pages} {79} (\bibinfo {year} {2013})}\BibitemShut {NoStop}%
\bibitem [{\citenamefont {Tighineanu}\ \emph {et~al.}(2018)\citenamefont {Tighineanu}, \citenamefont {Dree{\ss}en}, \citenamefont {Flindt}, \citenamefont {Lodahl},\ and\ \citenamefont {S\o{}rensen}}]{Tighineanu18}%
  \BibitemOpen
  \bibfield  {author} {\bibinfo {author} {\bibfnamefont {P.}~\bibnamefont {Tighineanu}}, \bibinfo {author} {\bibfnamefont {C.}~\bibnamefont {Dree{\ss}en}}, \bibinfo {author} {\bibfnamefont {C.}~\bibnamefont {Flindt}}, \bibinfo {author} {\bibfnamefont {P.}~\bibnamefont {Lodahl}}, \ and\ \bibinfo {author} {\bibfnamefont {A.}~\bibnamefont {S\o{}rensen}},\ }\href {\doibase https://doi.org/10.1103/PhysRevLett.120.257401} {\bibfield  {journal} {\bibinfo  {journal} {Phys. Rev. Lett.}\ }\textbf {\bibinfo {volume} {120}},\ \bibinfo {pages} {257401} (\bibinfo {year} {2018})}\BibitemShut {NoStop}%
\bibitem [{\citenamefont {Fong}\ \emph {et~al.}(2012)\citenamefont {Fong}, \citenamefont {Pernice},\ and\ \citenamefont {Tang}}]{Fong12}%
  \BibitemOpen
  \bibfield  {author} {\bibinfo {author} {\bibfnamefont {K.}~\bibnamefont {Fong}}, \bibinfo {author} {\bibfnamefont {W.}~\bibnamefont {Pernice}}, \ and\ \bibinfo {author} {\bibfnamefont {H.}~\bibnamefont {Tang}},\ }\href {\doibase https://doi.org/10.1103/PhysRevB.85.161410} {\bibfield  {journal} {\bibinfo  {journal} {Phys. Rev. B}\ }\textbf {\bibinfo {volume} {85}},\ \bibinfo {pages} {161410(R)} (\bibinfo {year} {2012})}\BibitemShut {NoStop}%
\bibitem [{\citenamefont {Zhang}\ \emph {et~al.}(2014)\citenamefont {Zhang}, \citenamefont {Moser}, \citenamefont {G\"uttinger}, \citenamefont {Bachtold},\ and\ \citenamefont {Dykman}}]{Zhang14}%
  \BibitemOpen
  \bibfield  {author} {\bibinfo {author} {\bibfnamefont {Y.}~\bibnamefont {Zhang}}, \bibinfo {author} {\bibfnamefont {J.}~\bibnamefont {Moser}}, \bibinfo {author} {\bibfnamefont {J.}~\bibnamefont {G\"uttinger}}, \bibinfo {author} {\bibfnamefont {A.}~\bibnamefont {Bachtold}}, \ and\ \bibinfo {author} {\bibfnamefont {M.}~\bibnamefont {Dykman}},\ }\href {\doibase https://doi.org/10.1103/PhysRevLett.113.255502} {\bibfield  {journal} {\bibinfo  {journal} {Phys. Rev. Lett.}\ }\textbf {\bibinfo {volume} {113}},\ \bibinfo {pages} {255502} (\bibinfo {year} {2014})}\BibitemShut {NoStop}%
\bibitem [{\citenamefont {Moser}\ \emph {et~al.}(2014)\citenamefont {Moser}, \citenamefont {Eichler}, \citenamefont {G\"uttinger}, \citenamefont {Dykman},\ and\ \citenamefont {Bachtold}}]{Moser14}%
  \BibitemOpen
  \bibfield  {author} {\bibinfo {author} {\bibfnamefont {J.}~\bibnamefont {Moser}}, \bibinfo {author} {\bibfnamefont {A.}~\bibnamefont {Eichler}}, \bibinfo {author} {\bibfnamefont {J.}~\bibnamefont {G\"uttinger}}, \bibinfo {author} {\bibfnamefont {M.}~\bibnamefont {Dykman}}, \ and\ \bibinfo {author} {\bibfnamefont {A.}~\bibnamefont {Bachtold}},\ }\href {\doibase https://doi.org/10.1038/nnano.2014.234} {\bibfield  {journal} {\bibinfo  {journal} {Nature Nanotech.}\ }\textbf {\bibinfo {volume} {9}},\ \bibinfo {pages} {1007} (\bibinfo {year} {2014})}\BibitemShut {NoStop}%
\bibitem [{\citenamefont {\ifmmode \check{Z}\else \v{Z}\fi{}uti\ifmmode~\acute{c}\else \'{c}\fi{}}\ \emph {et~al.}(2004)\citenamefont {\ifmmode \check{Z}\else \v{Z}\fi{}uti\ifmmode~\acute{c}\else \'{c}\fi{}}, \citenamefont {Fabian},\ and\ \citenamefont {Das~Sarma}}]{RevModPhys.76.323}%
  \BibitemOpen
  \bibfield  {author} {\bibinfo {author} {\bibfnamefont {I.}~\bibnamefont {\ifmmode \check{Z}\else \v{Z}\fi{}uti\ifmmode~\acute{c}\else \'{c}\fi{}}}, \bibinfo {author} {\bibfnamefont {J.}~\bibnamefont {Fabian}}, \ and\ \bibinfo {author} {\bibfnamefont {S.}~\bibnamefont {Das~Sarma}},\ }\href {\doibase 10.1103/RevModPhys.76.323} {\bibfield  {journal} {\bibinfo  {journal} {Rev. Mod. Phys.}\ }\textbf {\bibinfo {volume} {76}},\ \bibinfo {pages} {323} (\bibinfo {year} {2004})}\BibitemShut {NoStop}%
\bibitem [{\citenamefont {Fert}(2008)}]{RevModPhys.80.1517}%
  \BibitemOpen
  \bibfield  {author} {\bibinfo {author} {\bibfnamefont {A.}~\bibnamefont {Fert}},\ }\href {\doibase 10.1103/RevModPhys.80.1517} {\bibfield  {journal} {\bibinfo  {journal} {Rev. Mod. Phys.}\ }\textbf {\bibinfo {volume} {80}},\ \bibinfo {pages} {1517} (\bibinfo {year} {2008})}\BibitemShut {NoStop}%
\bibitem [{\citenamefont {Bernevig}\ \emph {et~al.}(2005)\citenamefont {Bernevig}, \citenamefont {Hughes},\ and\ \citenamefont {Zhang}}]{PhysRevLett.95.066601}%
  \BibitemOpen
  \bibfield  {author} {\bibinfo {author} {\bibfnamefont {B.~A.}\ \bibnamefont {Bernevig}}, \bibinfo {author} {\bibfnamefont {T.~L.}\ \bibnamefont {Hughes}}, \ and\ \bibinfo {author} {\bibfnamefont {S.-C.}\ \bibnamefont {Zhang}},\ }\href {\doibase 10.1103/PhysRevLett.95.066601} {\bibfield  {journal} {\bibinfo  {journal} {Phys. Rev. Lett.}\ }\textbf {\bibinfo {volume} {95}},\ \bibinfo {pages} {066601} (\bibinfo {year} {2005})}\BibitemShut {NoStop}%
\bibitem [{\citenamefont {Jo}\ \emph {et~al.}(2024)\citenamefont {Jo}, \citenamefont {Go}, \citenamefont {Choi},\ and\ \citenamefont {Lee}}]{Jo2024}%
  \BibitemOpen
  \bibfield  {author} {\bibinfo {author} {\bibfnamefont {D.}~\bibnamefont {Jo}}, \bibinfo {author} {\bibfnamefont {D.}~\bibnamefont {Go}}, \bibinfo {author} {\bibfnamefont {G.-M.}\ \bibnamefont {Choi}}, \ and\ \bibinfo {author} {\bibfnamefont {H.-W.}\ \bibnamefont {Lee}},\ }\href {\doibase 10.1038/s44306-024-00023-6} {\bibfield  {journal} {\bibinfo  {journal} {npj Spintronics}\ }\textbf {\bibinfo {volume} {2}},\ \bibinfo {pages} {19} (\bibinfo {year} {2024})}\BibitemShut {NoStop}%
\bibitem [{\citenamefont {Weiss}(2008)}]{Weiss2008}%
  \BibitemOpen
  \bibfield  {author} {\bibinfo {author} {\bibfnamefont {U.}~\bibnamefont {Weiss}},\ }\href@noop {} {\emph {\bibinfo {title} {Quantum Dissipative Systems}}}\ (\bibinfo  {publisher} {World Scientific Publishing Co. Pvt. Ltd., Singapore},\ \bibinfo {year} {2008})\BibitemShut {NoStop}%
\bibitem [{\citenamefont {Dekker}(1977)}]{Dekker77}%
  \BibitemOpen
  \bibfield  {author} {\bibinfo {author} {\bibfnamefont {H.}~\bibnamefont {Dekker}},\ }\href@noop {} {\bibfield  {journal} {\bibinfo  {journal} {Phys. Rev. A}\ }\textbf {\bibinfo {volume} {16}},\ \bibinfo {pages} {2126} (\bibinfo {year} {1977})}\BibitemShut {NoStop}%
\bibitem [{\citenamefont {Kostin}(1972)}]{Kostin72}%
  \BibitemOpen
  \bibfield  {author} {\bibinfo {author} {\bibfnamefont {M.}~\bibnamefont {Kostin}},\ }\href {\doibase https://doi.org/10.1063/1.1678812} {\bibfield  {journal} {\bibinfo  {journal} {J. Chem. Phys.}\ }\textbf {\bibinfo {volume} {57}},\ \bibinfo {pages} {3589} (\bibinfo {year} {1972})}\BibitemShut {NoStop}%
\bibitem [{\citenamefont {Yasue}(1978)}]{Yasue78}%
  \BibitemOpen
  \bibfield  {author} {\bibinfo {author} {\bibfnamefont {K.}~\bibnamefont {Yasue}},\ }\href {\doibase https://doi.org/10.1016/0003-4916(78)90279-8} {\bibfield  {journal} {\bibinfo  {journal} {Annals of Physics}\ }\textbf {\bibinfo {volume} {114}},\ \bibinfo {pages} {479} (\bibinfo {year} {1978})}\BibitemShut {NoStop}%
\bibitem [{\citenamefont {Nelson}(1966)}]{Nelson66}%
  \BibitemOpen
  \bibfield  {author} {\bibinfo {author} {\bibfnamefont {E.}~\bibnamefont {Nelson}},\ }\href {\doibase https://doi.org/10.1103/PhysRev.150.1079} {\bibfield  {journal} {\bibinfo  {journal} {Phys. Rev.}\ }\textbf {\bibinfo {volume} {150}},\ \bibinfo {pages} {1079} (\bibinfo {year} {1966})}\BibitemShut {NoStop}%
\bibitem [{\citenamefont {Ates}\ \emph {et~al.}(2012)\citenamefont {Ates}, \citenamefont {Olmos}, \citenamefont {Garrahan},\ and\ \citenamefont {Lesanovsky}}]{PhysRevA.85.043620}%
  \BibitemOpen
  \bibfield  {author} {\bibinfo {author} {\bibfnamefont {C.}~\bibnamefont {Ates}}, \bibinfo {author} {\bibfnamefont {B.}~\bibnamefont {Olmos}}, \bibinfo {author} {\bibfnamefont {J.~P.}\ \bibnamefont {Garrahan}}, \ and\ \bibinfo {author} {\bibfnamefont {I.}~\bibnamefont {Lesanovsky}},\ }\href {\doibase 10.1103/PhysRevA.85.043620} {\bibfield  {journal} {\bibinfo  {journal} {Phys. Rev. A}\ }\textbf {\bibinfo {volume} {85}},\ \bibinfo {pages} {043620} (\bibinfo {year} {2012})}\BibitemShut {NoStop}%
\bibitem [{\citenamefont {Hu}\ \emph {et~al.}(2013)\citenamefont {Hu}, \citenamefont {Lee},\ and\ \citenamefont {Clark}}]{PhysRevA.88.053627}%
  \BibitemOpen
  \bibfield  {author} {\bibinfo {author} {\bibfnamefont {A.}~\bibnamefont {Hu}}, \bibinfo {author} {\bibfnamefont {T.~E.}\ \bibnamefont {Lee}}, \ and\ \bibinfo {author} {\bibfnamefont {C.~W.}\ \bibnamefont {Clark}},\ }\href {\doibase 10.1103/PhysRevA.88.053627} {\bibfield  {journal} {\bibinfo  {journal} {Phys. Rev. A}\ }\textbf {\bibinfo {volume} {88}},\ \bibinfo {pages} {053627} (\bibinfo {year} {2013})}\BibitemShut {NoStop}%
\bibitem [{\citenamefont {Gisin}(1984)}]{Gisin84}%
  \BibitemOpen
  \bibfield  {author} {\bibinfo {author} {\bibfnamefont {N.}~\bibnamefont {Gisin}},\ }\href {\doibase https://doi.org/10.1103/PhysRevLett.52.1657} {\bibfield  {journal} {\bibinfo  {journal} {Phys. Rev. Lett.}\ }\textbf {\bibinfo {volume} {52}},\ \bibinfo {pages} {1657} (\bibinfo {year} {1984})}\BibitemShut {NoStop}%
\bibitem [{\citenamefont {Gisin}\ and\ \citenamefont {Percival}(1992)}]{Gisin92}%
  \BibitemOpen
  \bibfield  {author} {\bibinfo {author} {\bibfnamefont {N.}~\bibnamefont {Gisin}}\ and\ \bibinfo {author} {\bibfnamefont {I.}~\bibnamefont {Percival}},\ }\href {\doibase 10.1088/0305-4470/25/21/023} {\bibfield  {journal} {\bibinfo  {journal} {J. Phys. A: Math. Gen.}\ }\textbf {\bibinfo {volume} {25}},\ \bibinfo {pages} {5677} (\bibinfo {year} {1992})}\BibitemShut {NoStop}%
\bibitem [{\citenamefont {Diosi}(1988)}]{Diosi88}%
  \BibitemOpen
  \bibfield  {author} {\bibinfo {author} {\bibfnamefont {L.}~\bibnamefont {Diosi}},\ }\href {\doibase 10.1088/0305-4470/21/13/013} {\bibfield  {journal} {\bibinfo  {journal} {J. Phys. A: Math. Gen.}\ }\textbf {\bibinfo {volume} {21}},\ \bibinfo {pages} {2885} (\bibinfo {year} {1988})}\BibitemShut {NoStop}%
\bibitem [{\citenamefont {Dalibard}\ \emph {et~al.}(1992)\citenamefont {Dalibard}, \citenamefont {Castin},\ and\ \citenamefont {M{\o}lmer}}]{Dalibard91}%
  \BibitemOpen
  \bibfield  {author} {\bibinfo {author} {\bibfnamefont {J.}~\bibnamefont {Dalibard}}, \bibinfo {author} {\bibfnamefont {Y.}~\bibnamefont {Castin}}, \ and\ \bibinfo {author} {\bibfnamefont {K.}~\bibnamefont {M{\o}lmer}},\ }\href {\doibase https://doi.org/10.1103/PhysRevLett.68.580} {\bibfield  {journal} {\bibinfo  {journal} {Phys. Rev. Lett.}\ }\textbf {\bibinfo {volume} {68}},\ \bibinfo {pages} {580} (\bibinfo {year} {1992})}\BibitemShut {NoStop}%
\bibitem [{\citenamefont {Biamonte}\ and\ \citenamefont {Bergholm}(2017)}]{Biamonte2017TensorNI}%
  \BibitemOpen
  \bibfield  {author} {\bibinfo {author} {\bibfnamefont {J.~D.}\ \bibnamefont {Biamonte}}\ and\ \bibinfo {author} {\bibfnamefont {V.}~\bibnamefont {Bergholm}},\ }\href {https://api.semanticscholar.org/CorpusID:119679246} {\bibfield  {journal} {\bibinfo  {journal} {arXiv: Quantum Physics}\ } (\bibinfo {year} {2017})}\BibitemShut {NoStop}%
\bibitem [{\citenamefont {Weimer}(2015)}]{PhysRevLett.114.040402}%
  \BibitemOpen
  \bibfield  {author} {\bibinfo {author} {\bibfnamefont {H.}~\bibnamefont {Weimer}},\ }\href {\doibase 10.1103/PhysRevLett.114.040402} {\bibfield  {journal} {\bibinfo  {journal} {Phys. Rev. Lett.}\ }\textbf {\bibinfo {volume} {114}},\ \bibinfo {pages} {040402} (\bibinfo {year} {2015})}\BibitemShut {NoStop}%
\bibitem [{\citenamefont {Weimer}\ \emph {et~al.}(2021)\citenamefont {Weimer}, \citenamefont {Kshetrimayum},\ and\ \citenamefont {Or\'us}}]{RevModPhys.93.015008}%
  \BibitemOpen
  \bibfield  {author} {\bibinfo {author} {\bibfnamefont {H.}~\bibnamefont {Weimer}}, \bibinfo {author} {\bibfnamefont {A.}~\bibnamefont {Kshetrimayum}}, \ and\ \bibinfo {author} {\bibfnamefont {R.}~\bibnamefont {Or\'us}},\ }\href {\doibase 10.1103/RevModPhys.93.015008} {\bibfield  {journal} {\bibinfo  {journal} {Rev. Mod. Phys.}\ }\textbf {\bibinfo {volume} {93}},\ \bibinfo {pages} {015008} (\bibinfo {year} {2021})}\BibitemShut {NoStop}%
\bibitem [{\citenamefont {Prigogine}\ and\ \citenamefont {R\'esibois}(1961)}]{Prigogine61}%
  \BibitemOpen
  \bibfield  {author} {\bibinfo {author} {\bibfnamefont {I.}~\bibnamefont {Prigogine}}\ and\ \bibinfo {author} {\bibfnamefont {P.}~\bibnamefont {R\'esibois}},\ }\href {\doibase https://doi.org/10.1016/0031-8914(61)90008-8} {\bibfield  {journal} {\bibinfo  {journal} {Physica}\ }\textbf {\bibinfo {volume} {27}},\ \bibinfo {pages} {629} (\bibinfo {year} {1961})}\BibitemShut {NoStop}%
\bibitem [{\citenamefont {Mori}(1965)}]{Mori65}%
  \BibitemOpen
  \bibfield  {author} {\bibinfo {author} {\bibfnamefont {H.}~\bibnamefont {Mori}},\ }\href {\doibase https://doi.org/10.1143/PTP.33.423} {\bibfield  {journal} {\bibinfo  {journal} {Progr. Theor. Phys.}\ }\textbf {\bibinfo {volume} {33}},\ \bibinfo {pages} {423} (\bibinfo {year} {1965})}\BibitemShut {NoStop}%
\bibitem [{\citenamefont {Micklitz}\ \emph {et~al.}(2022)\citenamefont {Micklitz}, \citenamefont {Morningstar}, \citenamefont {Altland},\ and\ \citenamefont {Huse}}]{PhysRevLett.129.140402}%
  \BibitemOpen
  \bibfield  {author} {\bibinfo {author} {\bibfnamefont {T.}~\bibnamefont {Micklitz}}, \bibinfo {author} {\bibfnamefont {A.}~\bibnamefont {Morningstar}}, \bibinfo {author} {\bibfnamefont {A.}~\bibnamefont {Altland}}, \ and\ \bibinfo {author} {\bibfnamefont {D.~A.}\ \bibnamefont {Huse}},\ }\href {\doibase 10.1103/PhysRevLett.129.140402} {\bibfield  {journal} {\bibinfo  {journal} {Phys. Rev. Lett.}\ }\textbf {\bibinfo {volume} {129}},\ \bibinfo {pages} {140402} (\bibinfo {year} {2022})}\BibitemShut {NoStop}%
\bibitem [{\citenamefont {Allen}(1987)}]{PhysRevLett.59.1460}%
  \BibitemOpen
  \bibfield  {author} {\bibinfo {author} {\bibfnamefont {P.~B.}\ \bibnamefont {Allen}},\ }\href {\doibase 10.1103/PhysRevLett.59.1460} {\bibfield  {journal} {\bibinfo  {journal} {Phys. Rev. Lett.}\ }\textbf {\bibinfo {volume} {59}},\ \bibinfo {pages} {1460} (\bibinfo {year} {1987})}\BibitemShut {NoStop}%
\bibitem [{\citenamefont {Rossi}(2008)}]{Rossi2008}%
  \BibitemOpen
  \bibfield  {author} {\bibinfo {author} {\bibfnamefont {F.}~\bibnamefont {Rossi}},\ }\href {\doibase https://doi.org/10.1002/pssc.200776506} {\bibfield  {journal} {\bibinfo  {journal} {physica status solidi c}\ }\textbf {\bibinfo {volume} {5}},\ \bibinfo {pages} {35} (\bibinfo {year} {2008})}\BibitemShut {NoStop}%
\bibitem [{\citenamefont {Restrepo}\ and\ \citenamefont {Windl}(2012)}]{PhysRevLett.109.166604}%
  \BibitemOpen
  \bibfield  {author} {\bibinfo {author} {\bibfnamefont {O.~D.}\ \bibnamefont {Restrepo}}\ and\ \bibinfo {author} {\bibfnamefont {W.}~\bibnamefont {Windl}},\ }\href {\doibase 10.1103/PhysRevLett.109.166604} {\bibfield  {journal} {\bibinfo  {journal} {Phys. Rev. Lett.}\ }\textbf {\bibinfo {volume} {109}},\ \bibinfo {pages} {166604} (\bibinfo {year} {2012})}\BibitemShut {NoStop}%
\bibitem [{\citenamefont {Park}\ \emph {et~al.}(2020)\citenamefont {Park}, \citenamefont {Zhou},\ and\ \citenamefont {Bernardi}}]{PhysRevB.101.045202}%
  \BibitemOpen
  \bibfield  {author} {\bibinfo {author} {\bibfnamefont {J.}~\bibnamefont {Park}}, \bibinfo {author} {\bibfnamefont {J.-J.}\ \bibnamefont {Zhou}}, \ and\ \bibinfo {author} {\bibfnamefont {M.}~\bibnamefont {Bernardi}},\ }\href {\doibase 10.1103/PhysRevB.101.045202} {\bibfield  {journal} {\bibinfo  {journal} {Phys. Rev. B}\ }\textbf {\bibinfo {volume} {101}},\ \bibinfo {pages} {045202} (\bibinfo {year} {2020})}\BibitemShut {NoStop}%
\bibitem [{\citenamefont {Stefanucci}(2024)}]{stefanucci2024kadanoff}%
  \BibitemOpen
  \bibfield  {author} {\bibinfo {author} {\bibfnamefont {G.}~\bibnamefont {Stefanucci}},\ }\href@noop {} {\bibfield  {journal} {\bibinfo  {journal} {Physical Review Letters}\ }\textbf {\bibinfo {volume} {133}},\ \bibinfo {pages} {066901} (\bibinfo {year} {2024})}\BibitemShut {NoStop}%
\bibitem [{\citenamefont {Thompson}\ and\ \citenamefont {Kamenev}(2023)}]{THOMPSON2023169385}%
  \BibitemOpen
  \bibfield  {author} {\bibinfo {author} {\bibfnamefont {F.}~\bibnamefont {Thompson}}\ and\ \bibinfo {author} {\bibfnamefont {A.}~\bibnamefont {Kamenev}},\ }\href {\doibase https://doi.org/10.1016/j.aop.2023.169385} {\bibfield  {journal} {\bibinfo  {journal} {Annals of Physics}\ }\textbf {\bibinfo {volume} {455}},\ \bibinfo {pages} {169385} (\bibinfo {year} {2023})}\BibitemShut {NoStop}%
\bibitem [{\citenamefont {Reeves}\ and\ \citenamefont {Vl\ifmmode~\check{c}\else \v{c}\fi{}ek}(2024)}]{Reeves2024}%
  \BibitemOpen
  \bibfield  {author} {\bibinfo {author} {\bibfnamefont {C.~C.}\ \bibnamefont {Reeves}}\ and\ \bibinfo {author} {\bibfnamefont {V.~c.~v.}\ \bibnamefont {Vl\ifmmode~\check{c}\else \v{c}\fi{}ek}},\ }\href {\doibase 10.1103/PhysRevLett.133.226902} {\bibfield  {journal} {\bibinfo  {journal} {Phys. Rev. Lett.}\ }\textbf {\bibinfo {volume} {133}},\ \bibinfo {pages} {226902} (\bibinfo {year} {2024})}\BibitemShut {NoStop}%
\bibitem [{\citenamefont {Schl\"unzen}\ \emph {et~al.}(2020)\citenamefont {Schl\"unzen}, \citenamefont {Joost},\ and\ \citenamefont {Bonitz}}]{bonitz2020}%
  \BibitemOpen
  \bibfield  {author} {\bibinfo {author} {\bibfnamefont {N.}~\bibnamefont {Schl\"unzen}}, \bibinfo {author} {\bibfnamefont {J.-P.}\ \bibnamefont {Joost}}, \ and\ \bibinfo {author} {\bibfnamefont {M.}~\bibnamefont {Bonitz}},\ }\href {\doibase 10.1103/PhysRevLett.124.076601} {\bibfield  {journal} {\bibinfo  {journal} {Phys. Rev. Lett.}\ }\textbf {\bibinfo {volume} {124}},\ \bibinfo {pages} {076601} (\bibinfo {year} {2020})}\BibitemShut {NoStop}%
\bibitem [{\citenamefont {My\"oh\"anen}\ \emph {et~al.}(2009)\citenamefont {My\"oh\"anen}, \citenamefont {Stan}, \citenamefont {Stefanucci},\ and\ \citenamefont {van Leeuwen}}]{PhysRevB.80.115107}%
  \BibitemOpen
  \bibfield  {author} {\bibinfo {author} {\bibfnamefont {P.}~\bibnamefont {My\"oh\"anen}}, \bibinfo {author} {\bibfnamefont {A.}~\bibnamefont {Stan}}, \bibinfo {author} {\bibfnamefont {G.}~\bibnamefont {Stefanucci}}, \ and\ \bibinfo {author} {\bibfnamefont {R.}~\bibnamefont {van Leeuwen}},\ }\href {\doibase 10.1103/PhysRevB.80.115107} {\bibfield  {journal} {\bibinfo  {journal} {Phys. Rev. B}\ }\textbf {\bibinfo {volume} {80}},\ \bibinfo {pages} {115107} (\bibinfo {year} {2009})}\BibitemShut {NoStop}%
\bibitem [{\citenamefont {Puig~von Friesen}\ \emph {et~al.}(2010)\citenamefont {Puig~von Friesen}, \citenamefont {Verdozzi},\ and\ \citenamefont {Almbladh}}]{PhysRevB.82.155108}%
  \BibitemOpen
  \bibfield  {author} {\bibinfo {author} {\bibfnamefont {M.}~\bibnamefont {Puig~von Friesen}}, \bibinfo {author} {\bibfnamefont {C.}~\bibnamefont {Verdozzi}}, \ and\ \bibinfo {author} {\bibfnamefont {C.-O.}\ \bibnamefont {Almbladh}},\ }\href {\doibase 10.1103/PhysRevB.82.155108} {\bibfield  {journal} {\bibinfo  {journal} {Phys. Rev. B}\ }\textbf {\bibinfo {volume} {82}},\ \bibinfo {pages} {155108} (\bibinfo {year} {2010})}\BibitemShut {NoStop}%
\bibitem [{\citenamefont {von Friesen}\ \emph {et~al.}(2009)\citenamefont {von Friesen}, \citenamefont {Verdozzi},\ and\ \citenamefont {Almbladh}}]{PhysRevLett.103.176404}%
  \BibitemOpen
  \bibfield  {author} {\bibinfo {author} {\bibfnamefont {M.~P.}\ \bibnamefont {von Friesen}}, \bibinfo {author} {\bibfnamefont {C.}~\bibnamefont {Verdozzi}}, \ and\ \bibinfo {author} {\bibfnamefont {C.-O.}\ \bibnamefont {Almbladh}},\ }\href {\doibase 10.1103/PhysRevLett.103.176404} {\bibfield  {journal} {\bibinfo  {journal} {Phys. Rev. Lett.}\ }\textbf {\bibinfo {volume} {103}},\ \bibinfo {pages} {176404} (\bibinfo {year} {2009})}\BibitemShut {NoStop}%
\bibitem [{\citenamefont {Sch\"uler}\ \emph {et~al.}(2016)\citenamefont {Sch\"uler}, \citenamefont {Berakdar},\ and\ \citenamefont {Pavlyukh}}]{PhysRevB.93.054303}%
  \BibitemOpen
  \bibfield  {author} {\bibinfo {author} {\bibfnamefont {M.}~\bibnamefont {Sch\"uler}}, \bibinfo {author} {\bibfnamefont {J.}~\bibnamefont {Berakdar}}, \ and\ \bibinfo {author} {\bibfnamefont {Y.}~\bibnamefont {Pavlyukh}},\ }\href {\doibase 10.1103/PhysRevB.93.054303} {\bibfield  {journal} {\bibinfo  {journal} {Phys. Rev. B}\ }\textbf {\bibinfo {volume} {93}},\ \bibinfo {pages} {054303} (\bibinfo {year} {2016})}\BibitemShut {NoStop}%
\bibitem [{\citenamefont {Pavlyukh}(2024)}]{Pav2024}%
  \BibitemOpen
  \bibfield  {author} {\bibinfo {author} {\bibfnamefont {Y.}~\bibnamefont {Pavlyukh}},\ }\href {\doibase https://doi.org/10.1002/pssb.202300510} {\bibfield  {journal} {\bibinfo  {journal} {physica status solidi (b)}\ }\textbf {\bibinfo {volume} {261}},\ \bibinfo {pages} {2300510} (\bibinfo {year} {2024})}\BibitemShut {NoStop}%
\bibitem [{\citenamefont {Pavlyukh}\ \emph {et~al.}(2022)\citenamefont {Pavlyukh}, \citenamefont {Perfetto}, \citenamefont {Karlsson}, \citenamefont {van Leeuwen},\ and\ \citenamefont {Stefanucci}}]{Stef2022}%
  \BibitemOpen
  \bibfield  {author} {\bibinfo {author} {\bibfnamefont {Y.}~\bibnamefont {Pavlyukh}}, \bibinfo {author} {\bibfnamefont {E.}~\bibnamefont {Perfetto}}, \bibinfo {author} {\bibfnamefont {D.}~\bibnamefont {Karlsson}}, \bibinfo {author} {\bibfnamefont {R.}~\bibnamefont {van Leeuwen}}, \ and\ \bibinfo {author} {\bibfnamefont {G.}~\bibnamefont {Stefanucci}},\ }\href {\doibase 10.1103/PhysRevB.105.125134} {\bibfield  {journal} {\bibinfo  {journal} {Phys. Rev. B}\ }\textbf {\bibinfo {volume} {105}},\ \bibinfo {pages} {125134} (\bibinfo {year} {2022})}\BibitemShut {NoStop}%
\bibitem [{\citenamefont {Joost}\ \emph {et~al.}(2020{\natexlab{a}})\citenamefont {Joost}, \citenamefont {Schl\"unzen},\ and\ \citenamefont {Bonitz}}]{Joost2020}%
  \BibitemOpen
  \bibfield  {author} {\bibinfo {author} {\bibfnamefont {J.-P.}\ \bibnamefont {Joost}}, \bibinfo {author} {\bibfnamefont {N.}~\bibnamefont {Schl\"unzen}}, \ and\ \bibinfo {author} {\bibfnamefont {M.}~\bibnamefont {Bonitz}},\ }\href {\doibase 10.1103/PhysRevB.101.245101} {\bibfield  {journal} {\bibinfo  {journal} {Phys. Rev. B}\ }\textbf {\bibinfo {volume} {101}},\ \bibinfo {pages} {245101} (\bibinfo {year} {2020}{\natexlab{a}})}\BibitemShut {NoStop}%
\bibitem [{\citenamefont {Pavlyukh}\ \emph {et~al.}(2021)\citenamefont {Pavlyukh}, \citenamefont {Perfetto},\ and\ \citenamefont {Stefanucci}}]{Stef2021}%
  \BibitemOpen
  \bibfield  {author} {\bibinfo {author} {\bibfnamefont {Y.}~\bibnamefont {Pavlyukh}}, \bibinfo {author} {\bibfnamefont {E.}~\bibnamefont {Perfetto}}, \ and\ \bibinfo {author} {\bibfnamefont {G.}~\bibnamefont {Stefanucci}},\ }\href {\doibase 10.1103/PhysRevB.104.035124} {\bibfield  {journal} {\bibinfo  {journal} {Phys. Rev. B}\ }\textbf {\bibinfo {volume} {104}},\ \bibinfo {pages} {035124} (\bibinfo {year} {2021})}\BibitemShut {NoStop}%
\bibitem [{\citenamefont {Karlsson}\ \emph {et~al.}(2018)\citenamefont {Karlsson}, \citenamefont {van Leeuwen}, \citenamefont {Perfetto},\ and\ \citenamefont {Stefanucci}}]{Stef18}%
  \BibitemOpen
  \bibfield  {author} {\bibinfo {author} {\bibfnamefont {D.}~\bibnamefont {Karlsson}}, \bibinfo {author} {\bibfnamefont {R.}~\bibnamefont {van Leeuwen}}, \bibinfo {author} {\bibfnamefont {E.}~\bibnamefont {Perfetto}}, \ and\ \bibinfo {author} {\bibfnamefont {G.}~\bibnamefont {Stefanucci}},\ }\href {\doibase 10.1103/PhysRevB.98.115148} {\bibfield  {journal} {\bibinfo  {journal} {Phys. Rev. B}\ }\textbf {\bibinfo {volume} {98}},\ \bibinfo {pages} {115148} (\bibinfo {year} {2018})}\BibitemShut {NoStop}%
\bibitem [{\citenamefont {Marini}\ \emph {et~al.}(2022)\citenamefont {Marini}, \citenamefont {Perfetto},\ and\ \citenamefont {Stefanucci}}]{MARINI2022147189}%
  \BibitemOpen
  \bibfield  {author} {\bibinfo {author} {\bibfnamefont {A.}~\bibnamefont {Marini}}, \bibinfo {author} {\bibfnamefont {E.}~\bibnamefont {Perfetto}}, \ and\ \bibinfo {author} {\bibfnamefont {G.}~\bibnamefont {Stefanucci}},\ }\href {\doibase https://doi.org/10.1016/j.elspec.2022.147189} {\bibfield  {journal} {\bibinfo  {journal} {Journal of Electron Spectroscopy and Related Phenomena}\ }\textbf {\bibinfo {volume} {257}},\ \bibinfo {pages} {147189} (\bibinfo {year} {2022})}\BibitemShut {NoStop}%
\bibitem [{\citenamefont {Sangalli}\ \emph {et~al.}(2018)\citenamefont {Sangalli}, \citenamefont {Perfetto}, \citenamefont {Stefanucci},\ and\ \citenamefont {Marini}}]{refId0}%
  \BibitemOpen
  \bibfield  {author} {\bibinfo {author} {\bibfnamefont {D.}~\bibnamefont {Sangalli}}, \bibinfo {author} {\bibfnamefont {E.}~\bibnamefont {Perfetto}}, \bibinfo {author} {\bibfnamefont {G.}~\bibnamefont {Stefanucci}}, \ and\ \bibinfo {author} {\bibfnamefont {A.}~\bibnamefont {Marini}},\ }\href {\doibase 10.1140/epjb/e2018-90126-5} {\bibfield  {journal} {\bibinfo  {journal} {Eur. Phys. J. B}\ }\textbf {\bibinfo {volume} {91}},\ \bibinfo {pages} {171} (\bibinfo {year} {2018})}\BibitemShut {NoStop}%
\bibitem [{\citenamefont {Chan}\ \emph {et~al.}(2021)\citenamefont {Chan}, \citenamefont {Qiu}, \citenamefont {da~Jornada},\ and\ \citenamefont {Louie}}]{doi:10.1073/pnas.1906938118}%
  \BibitemOpen
  \bibfield  {author} {\bibinfo {author} {\bibfnamefont {Y.-H.}\ \bibnamefont {Chan}}, \bibinfo {author} {\bibfnamefont {D.~Y.}\ \bibnamefont {Qiu}}, \bibinfo {author} {\bibfnamefont {F.~H.}\ \bibnamefont {da~Jornada}}, \ and\ \bibinfo {author} {\bibfnamefont {S.~G.}\ \bibnamefont {Louie}},\ }\href {\doibase 10.1073/pnas.1906938118} {\bibfield  {journal} {\bibinfo  {journal} {Proceedings of the National Academy of Sciences}\ }\textbf {\bibinfo {volume} {118}},\ \bibinfo {pages} {e1906938118} (\bibinfo {year} {2021})}\BibitemShut {NoStop}%
\bibitem [{\citenamefont {Perfetto}\ \emph {et~al.}(2022)\citenamefont {Perfetto}, \citenamefont {Pavlyukh},\ and\ \citenamefont {Stefanucci}}]{PhysRevLett.128.016801}%
  \BibitemOpen
  \bibfield  {author} {\bibinfo {author} {\bibfnamefont {E.}~\bibnamefont {Perfetto}}, \bibinfo {author} {\bibfnamefont {Y.}~\bibnamefont {Pavlyukh}}, \ and\ \bibinfo {author} {\bibfnamefont {G.}~\bibnamefont {Stefanucci}},\ }\href {\doibase 10.1103/PhysRevLett.128.016801} {\bibfield  {journal} {\bibinfo  {journal} {Phys. Rev. Lett.}\ }\textbf {\bibinfo {volume} {128}},\ \bibinfo {pages} {016801} (\bibinfo {year} {2022})}\BibitemShut {NoStop}%
\bibitem [{\citenamefont {Perfetto}\ and\ \citenamefont {Stefanucci}(2023)}]{Perfett02023}%
  \BibitemOpen
  \bibfield  {author} {\bibinfo {author} {\bibfnamefont {E.}~\bibnamefont {Perfetto}}\ and\ \bibinfo {author} {\bibfnamefont {G.}~\bibnamefont {Stefanucci}},\ }\href {\doibase 10.1021/acs.nanolett.3c01772} {\bibfield  {journal} {\bibinfo  {journal} {Nano Letters}\ }\textbf {\bibinfo {volume} {23}},\ \bibinfo {pages} {7029} (\bibinfo {year} {2023})},\ \bibinfo {note} {pMID: 37493350},\ \Eprint {http://arxiv.org/abs/https://doi.org/10.1021/acs.nanolett.3c01772} {https://doi.org/10.1021/acs.nanolett.3c01772} \BibitemShut {NoStop}%
\bibitem [{\citenamefont {de~Melo}\ and\ \citenamefont {Marini}(2016)}]{PhysRevB.93.155102}%
  \BibitemOpen
  \bibfield  {author} {\bibinfo {author} {\bibfnamefont {P.~M. M.~C.}\ \bibnamefont {de~Melo}}\ and\ \bibinfo {author} {\bibfnamefont {A.}~\bibnamefont {Marini}},\ }\href {\doibase 10.1103/PhysRevB.93.155102} {\bibfield  {journal} {\bibinfo  {journal} {Phys. Rev. B}\ }\textbf {\bibinfo {volume} {93}},\ \bibinfo {pages} {155102} (\bibinfo {year} {2016})}\BibitemShut {NoStop}%
\bibitem [{\citenamefont {Di~Ventra}(2008)}]{di2008electrical}%
  \BibitemOpen
  \bibfield  {author} {\bibinfo {author} {\bibfnamefont {M.}~\bibnamefont {Di~Ventra}},\ }\href {https://books.google.com/books?id=hLyryP7zZmsC} {\emph {\bibinfo {title} {Electrical Transport in Nanoscale Systems}}}\ (\bibinfo  {publisher} {Cambridge University Press},\ \bibinfo {year} {2008})\BibitemShut {NoStop}%
\bibitem [{\citenamefont {Ness}\ \emph {et~al.}(2010)\citenamefont {Ness}, \citenamefont {Dash},\ and\ \citenamefont {Godby}}]{PhysRevB.82.085426}%
  \BibitemOpen
  \bibfield  {author} {\bibinfo {author} {\bibfnamefont {H.}~\bibnamefont {Ness}}, \bibinfo {author} {\bibfnamefont {L.~K.}\ \bibnamefont {Dash}}, \ and\ \bibinfo {author} {\bibfnamefont {R.~W.}\ \bibnamefont {Godby}},\ }\href {\doibase 10.1103/PhysRevB.82.085426} {\bibfield  {journal} {\bibinfo  {journal} {Phys. Rev. B}\ }\textbf {\bibinfo {volume} {82}},\ \bibinfo {pages} {085426} (\bibinfo {year} {2010})}\BibitemShut {NoStop}%
\bibitem [{\citenamefont {Frederiksen}\ \emph {et~al.}(2004)\citenamefont {Frederiksen}, \citenamefont {Brandbyge}, \citenamefont {Lorente},\ and\ \citenamefont {Jauho}}]{PhysRevLett.93.256601}%
  \BibitemOpen
  \bibfield  {author} {\bibinfo {author} {\bibfnamefont {T.}~\bibnamefont {Frederiksen}}, \bibinfo {author} {\bibfnamefont {M.}~\bibnamefont {Brandbyge}}, \bibinfo {author} {\bibfnamefont {N.}~\bibnamefont {Lorente}}, \ and\ \bibinfo {author} {\bibfnamefont {A.-P.}\ \bibnamefont {Jauho}},\ }\href {\doibase 10.1103/PhysRevLett.93.256601} {\bibfield  {journal} {\bibinfo  {journal} {Phys. Rev. Lett.}\ }\textbf {\bibinfo {volume} {93}},\ \bibinfo {pages} {256601} (\bibinfo {year} {2004})}\BibitemShut {NoStop}%
\bibitem [{\citenamefont {Paulsson}\ \emph {et~al.}(2008)\citenamefont {Paulsson}, \citenamefont {Frederiksen}, \citenamefont {Ueba}, \citenamefont {Lorente},\ and\ \citenamefont {Brandbyge}}]{PhysRevLett.100.226604}%
  \BibitemOpen
  \bibfield  {author} {\bibinfo {author} {\bibfnamefont {M.}~\bibnamefont {Paulsson}}, \bibinfo {author} {\bibfnamefont {T.}~\bibnamefont {Frederiksen}}, \bibinfo {author} {\bibfnamefont {H.}~\bibnamefont {Ueba}}, \bibinfo {author} {\bibfnamefont {N.}~\bibnamefont {Lorente}}, \ and\ \bibinfo {author} {\bibfnamefont {M.}~\bibnamefont {Brandbyge}},\ }\href {\doibase 10.1103/PhysRevLett.100.226604} {\bibfield  {journal} {\bibinfo  {journal} {Phys. Rev. Lett.}\ }\textbf {\bibinfo {volume} {100}},\ \bibinfo {pages} {226604} (\bibinfo {year} {2008})}\BibitemShut {NoStop}%
\bibitem [{\citenamefont {Gole\ifmmode~\check{z}\else \v{z}\fi{}}\ \emph {et~al.}(2019)\citenamefont {Gole\ifmmode~\check{z}\else \v{z}\fi{}}, \citenamefont {Eckstein},\ and\ \citenamefont {Werner}}]{PhysRevB.100.235117}%
  \BibitemOpen
  \bibfield  {author} {\bibinfo {author} {\bibfnamefont {D.}~\bibnamefont {Gole\ifmmode~\check{z}\else \v{z}\fi{}}}, \bibinfo {author} {\bibfnamefont {M.}~\bibnamefont {Eckstein}}, \ and\ \bibinfo {author} {\bibfnamefont {P.}~\bibnamefont {Werner}},\ }\href {\doibase 10.1103/PhysRevB.100.235117} {\bibfield  {journal} {\bibinfo  {journal} {Phys. Rev. B}\ }\textbf {\bibinfo {volume} {100}},\ \bibinfo {pages} {235117} (\bibinfo {year} {2019})}\BibitemShut {NoStop}%
\bibitem [{\citenamefont {Runge}\ and\ \citenamefont {Gross}(1984{\natexlab{a}})}]{PhysRevLett.52.997}%
  \BibitemOpen
  \bibfield  {author} {\bibinfo {author} {\bibfnamefont {E.}~\bibnamefont {Runge}}\ and\ \bibinfo {author} {\bibfnamefont {E.~K.~U.}\ \bibnamefont {Gross}},\ }\href {\doibase 10.1103/PhysRevLett.52.997} {\bibfield  {journal} {\bibinfo  {journal} {Phys. Rev. Lett.}\ }\textbf {\bibinfo {volume} {52}},\ \bibinfo {pages} {997} (\bibinfo {year} {1984}{\natexlab{a}})}\BibitemShut {NoStop}%
\bibitem [{\citenamefont {Marques}\ and\ \citenamefont {Gross}(2003)}]{Marques2003}%
  \BibitemOpen
  \bibfield  {author} {\bibinfo {author} {\bibfnamefont {M.~A.~L.}\ \bibnamefont {Marques}}\ and\ \bibinfo {author} {\bibfnamefont {E.~K.~U.}\ \bibnamefont {Gross}},\ }\enquote {\bibinfo {title} {Time-dependent density functional theory},}\ in\ \href {\doibase 10.1007/3-540-37072-2_4} {\emph {\bibinfo {booktitle} {A Primer in Density Functional Theory}}},\ \bibinfo {editor} {edited by\ \bibinfo {editor} {\bibfnamefont {C.}~\bibnamefont {Fiolhais}}, \bibinfo {editor} {\bibfnamefont {F.}~\bibnamefont {Nogueira}}, \ and\ \bibinfo {editor} {\bibfnamefont {M.~A.~L.}\ \bibnamefont {Marques}}}\ (\bibinfo  {publisher} {Springer Berlin Heidelberg},\ \bibinfo {address} {Berlin, Heidelberg},\ \bibinfo {year} {2003})\ pp.\ \bibinfo {pages} {144--184}\BibitemShut {NoStop}%
\bibitem [{\citenamefont {Xu}\ \emph {et~al.}(2024{\natexlab{a}})\citenamefont {Xu}, \citenamefont {Carney}, \citenamefont {Zhou}, \citenamefont {Shepard},\ and\ \citenamefont {Kanai}}]{Xu2024}%
  \BibitemOpen
  \bibfield  {author} {\bibinfo {author} {\bibfnamefont {J.}~\bibnamefont {Xu}}, \bibinfo {author} {\bibfnamefont {T.}~\bibnamefont {Carney}}, \bibinfo {author} {\bibfnamefont {R.}~\bibnamefont {Zhou}}, \bibinfo {author} {\bibfnamefont {C.}~\bibnamefont {Shepard}}, \ and\ \bibinfo {author} {\bibfnamefont {Y.}~\bibnamefont {Kanai}},\ }\href {\doibase 10.1021/jacs.3c08226} {\bibfield  {journal} {\bibinfo  {journal} {J. Am. Chem. Soc.}\ }\textbf {\bibinfo {volume} {146}},\ \bibinfo {pages} {5011} (\bibinfo {year} {2024}{\natexlab{a}})}\BibitemShut {NoStop}%
\bibitem [{\citenamefont {Ullrich}(2011)}]{10.1093/acprof:oso/9780199563029.001.0001}%
  \BibitemOpen
  \bibfield  {author} {\bibinfo {author} {\bibfnamefont {C.~A.}\ \bibnamefont {Ullrich}},\ }\href {\doibase 10.1093/acprof:oso/9780199563029.001.0001} {\emph {\bibinfo {title} {Time-Dependent Density-Functional Theory: Concepts and Applications}}}\ (\bibinfo  {publisher} {Oxford University Press},\ \bibinfo {year} {2011})\BibitemShut {NoStop}%
\bibitem [{\citenamefont {Yuen-Zhou}\ \emph {et~al.}(2010)\citenamefont {Yuen-Zhou}, \citenamefont {Tempel}, \citenamefont {Rodr\'{\i}guez-Rosario},\ and\ \citenamefont {Aspuru-Guzik}}]{PhysRevLett.104.043001}%
  \BibitemOpen
  \bibfield  {author} {\bibinfo {author} {\bibfnamefont {J.}~\bibnamefont {Yuen-Zhou}}, \bibinfo {author} {\bibfnamefont {D.~G.}\ \bibnamefont {Tempel}}, \bibinfo {author} {\bibfnamefont {C.~A.}\ \bibnamefont {Rodr\'{\i}guez-Rosario}}, \ and\ \bibinfo {author} {\bibfnamefont {A.}~\bibnamefont {Aspuru-Guzik}},\ }\href {\doibase 10.1103/PhysRevLett.104.043001} {\bibfield  {journal} {\bibinfo  {journal} {Phys. Rev. Lett.}\ }\textbf {\bibinfo {volume} {104}},\ \bibinfo {pages} {043001} (\bibinfo {year} {2010})}\BibitemShut {NoStop}%
\bibitem [{\citenamefont {Tully}(1990)}]{10.1063/1.459170}%
  \BibitemOpen
  \bibfield  {author} {\bibinfo {author} {\bibfnamefont {J.~C.}\ \bibnamefont {Tully}},\ }\href {\doibase 10.1063/1.459170} {\bibfield  {journal} {\bibinfo  {journal} {The Journal of Chemical Physics}\ }\textbf {\bibinfo {volume} {93}},\ \bibinfo {pages} {1061} (\bibinfo {year} {1990})}\BibitemShut {NoStop}%
\bibitem [{\citenamefont {Zhang}\ \emph {et~al.}(2019)\citenamefont {Zhang}, \citenamefont {Lian}, \citenamefont {Guan}, \citenamefont {Ma}, \citenamefont {Fu}, \citenamefont {Guo},\ and\ \citenamefont {Meng}}]{Zhang_NL_2019}%
  \BibitemOpen
  \bibfield  {author} {\bibinfo {author} {\bibfnamefont {J.}~\bibnamefont {Zhang}}, \bibinfo {author} {\bibfnamefont {C.}~\bibnamefont {Lian}}, \bibinfo {author} {\bibfnamefont {M.}~\bibnamefont {Guan}}, \bibinfo {author} {\bibfnamefont {W.}~\bibnamefont {Ma}}, \bibinfo {author} {\bibfnamefont {H.}~\bibnamefont {Fu}}, \bibinfo {author} {\bibfnamefont {H.}~\bibnamefont {Guo}}, \ and\ \bibinfo {author} {\bibfnamefont {S.}~\bibnamefont {Meng}},\ }\href {\doibase 10.1021/acs.nanolett.9b01865} {\bibfield  {journal} {\bibinfo  {journal} {Nano Lett.}\ }\textbf {\bibinfo {volume} {19}},\ \bibinfo {pages} {6027} (\bibinfo {year} {2019})}\BibitemShut {NoStop}%
\bibitem [{\citenamefont {Liu}\ \emph {et~al.}(2020)\citenamefont {Liu}, \citenamefont {Luo}, \citenamefont {Li},\ and\ \citenamefont {Wang}}]{PhysRevB.102.184308}%
  \BibitemOpen
  \bibfield  {author} {\bibinfo {author} {\bibfnamefont {W.-H.}\ \bibnamefont {Liu}}, \bibinfo {author} {\bibfnamefont {J.-W.}\ \bibnamefont {Luo}}, \bibinfo {author} {\bibfnamefont {S.-S.}\ \bibnamefont {Li}}, \ and\ \bibinfo {author} {\bibfnamefont {L.-W.}\ \bibnamefont {Wang}},\ }\href {\doibase 10.1103/PhysRevB.102.184308} {\bibfield  {journal} {\bibinfo  {journal} {Phys. Rev. B}\ }\textbf {\bibinfo {volume} {102}},\ \bibinfo {pages} {184308} (\bibinfo {year} {2020})}\BibitemShut {NoStop}%
\bibitem [{\citenamefont {Lian}\ \emph {et~al.}(2016)\citenamefont {Lian}, \citenamefont {Zhang},\ and\ \citenamefont {Meng}}]{PhysRevB.94.184310}%
  \BibitemOpen
  \bibfield  {author} {\bibinfo {author} {\bibfnamefont {C.}~\bibnamefont {Lian}}, \bibinfo {author} {\bibfnamefont {S.~B.}\ \bibnamefont {Zhang}}, \ and\ \bibinfo {author} {\bibfnamefont {S.}~\bibnamefont {Meng}},\ }\href {\doibase 10.1103/PhysRevB.94.184310} {\bibfield  {journal} {\bibinfo  {journal} {Phys. Rev. B}\ }\textbf {\bibinfo {volume} {94}},\ \bibinfo {pages} {184310} (\bibinfo {year} {2016})}\BibitemShut {NoStop}%
\bibitem [{\citenamefont {Parandekar}(2006)}]{Parandekar2006}%
  \BibitemOpen
  \bibfield  {author} {\bibinfo {author} {\bibfnamefont {P.}~\bibnamefont {Parandekar}},\ }\href {\doibase doi: 10.1021/ct050213k} {\bibfield  {journal} {\bibinfo  {journal} {J. Chem. Theory Comput.}\ }\textbf {\bibinfo {volume} {2}},\ \bibinfo {pages} {229} (\bibinfo {year} {2006})}\BibitemShut {NoStop}%
\bibitem [{\citenamefont {Curchod}\ \emph {et~al.}(2013)\citenamefont {Curchod}, \citenamefont {Rothlisberger},\ and\ \citenamefont {Tavernelli}}]{https://doi.org/10.1002/cphc.201200941}%
  \BibitemOpen
  \bibfield  {author} {\bibinfo {author} {\bibfnamefont {B.~F.~E.}\ \bibnamefont {Curchod}}, \bibinfo {author} {\bibfnamefont {U.}~\bibnamefont {Rothlisberger}}, \ and\ \bibinfo {author} {\bibfnamefont {I.}~\bibnamefont {Tavernelli}},\ }\href {\doibase https://doi.org/10.1002/cphc.201200941} {\bibfield  {journal} {\bibinfo  {journal} {ChemPhysChem}\ }\textbf {\bibinfo {volume} {14}},\ \bibinfo {pages} {1314} (\bibinfo {year} {2013})}\BibitemShut {NoStop}%
\bibitem [{\citenamefont {Nakatsuji}(1979)}]{Nakatsuji1979}%
  \BibitemOpen
  \bibfield  {author} {\bibinfo {author} {\bibfnamefont {H.}~\bibnamefont {Nakatsuji}},\ }\href {\doibase https://doi.org/10.1016/0009-2614(79)85172-6} {\bibfield  {journal} {\bibinfo  {journal} {Chem. Phys. Lett.}\ }\textbf {\bibinfo {volume} {67}},\ \bibinfo {pages} {329} (\bibinfo {year} {1979})}\BibitemShut {NoStop}%
\bibitem [{\citenamefont {Caillat}\ \emph {et~al.}(2005{\natexlab{a}})\citenamefont {Caillat}, \citenamefont {Zanghellini}, \citenamefont {Kitzler}, \citenamefont {Koch}, \citenamefont {Kreuzer},\ and\ \citenamefont {Scrinzi}}]{Caillat2005}%
  \BibitemOpen
  \bibfield  {author} {\bibinfo {author} {\bibfnamefont {J.}~\bibnamefont {Caillat}}, \bibinfo {author} {\bibfnamefont {J.}~\bibnamefont {Zanghellini}}, \bibinfo {author} {\bibfnamefont {M.}~\bibnamefont {Kitzler}}, \bibinfo {author} {\bibfnamefont {O.}~\bibnamefont {Koch}}, \bibinfo {author} {\bibfnamefont {W.}~\bibnamefont {Kreuzer}}, \ and\ \bibinfo {author} {\bibfnamefont {A.}~\bibnamefont {Scrinzi}},\ }\href {\doibase https://doi.org/10.1103/PhysRevA.71.012712} {\bibfield  {journal} {\bibinfo  {journal} {Phys. Rev. A}\ }\textbf {\bibinfo {volume} {71}},\ \bibinfo {pages} {012712} (\bibinfo {year} {2005}{\natexlab{a}})}\BibitemShut {NoStop}%
\bibitem [{\citenamefont {Hochstuhl}\ and\ \citenamefont {Bonitz}(2011)}]{10.1063/1.3553176}%
  \BibitemOpen
  \bibfield  {author} {\bibinfo {author} {\bibfnamefont {D.}~\bibnamefont {Hochstuhl}}\ and\ \bibinfo {author} {\bibfnamefont {M.}~\bibnamefont {Bonitz}},\ }\href {\doibase 10.1063/1.3553176} {\bibfield  {journal} {\bibinfo  {journal} {The Journal of Chemical Physics}\ }\textbf {\bibinfo {volume} {134}},\ \bibinfo {pages} {084106} (\bibinfo {year} {2011})},\ \Eprint {http://arxiv.org/abs/https://pubs.aip.org/aip/jcp/article-pdf/doi/10.1063/1.3553176/14905108/084106\_1\_online.pdf} {https://pubs.aip.org/aip/jcp/article-pdf/doi/10.1063/1.3553176/14905108/084106\_1\_online.pdf} \BibitemShut {NoStop}%
\bibitem [{\citenamefont {Zanghellini}\ \emph {et~al.}(2003)\citenamefont {Zanghellini}, \citenamefont {Kitzler}, \citenamefont {Fabian}, \citenamefont {Brabec},\ and\ \citenamefont {Scrinzi}}]{Zanghellini2003AnMA}%
  \BibitemOpen
  \bibfield  {author} {\bibinfo {author} {\bibfnamefont {J.}~\bibnamefont {Zanghellini}}, \bibinfo {author} {\bibfnamefont {M.}~\bibnamefont {Kitzler}}, \bibinfo {author} {\bibfnamefont {C.}~\bibnamefont {Fabian}}, \bibinfo {author} {\bibfnamefont {T.}~\bibnamefont {Brabec}}, \ and\ \bibinfo {author} {\bibfnamefont {A.}~\bibnamefont {Scrinzi}},\ }\href {https://api.semanticscholar.org/CorpusID:124426050} {\bibfield  {journal} {\bibinfo  {journal} {Laser Physics}\ }\textbf {\bibinfo {volume} {13}},\ \bibinfo {pages} {1064} (\bibinfo {year} {2003})}\BibitemShut {NoStop}%
\bibitem [{\citenamefont {Caillat}\ \emph {et~al.}(2005{\natexlab{b}})\citenamefont {Caillat}, \citenamefont {Zanghellini}, \citenamefont {Kitzler}, \citenamefont {Koch}, \citenamefont {Kreuzer},\ and\ \citenamefont {Scrinzi}}]{PhysRevA.71.012712}%
  \BibitemOpen
  \bibfield  {author} {\bibinfo {author} {\bibfnamefont {J.}~\bibnamefont {Caillat}}, \bibinfo {author} {\bibfnamefont {J.}~\bibnamefont {Zanghellini}}, \bibinfo {author} {\bibfnamefont {M.}~\bibnamefont {Kitzler}}, \bibinfo {author} {\bibfnamefont {O.}~\bibnamefont {Koch}}, \bibinfo {author} {\bibfnamefont {W.}~\bibnamefont {Kreuzer}}, \ and\ \bibinfo {author} {\bibfnamefont {A.}~\bibnamefont {Scrinzi}},\ }\href {\doibase 10.1103/PhysRevA.71.012712} {\bibfield  {journal} {\bibinfo  {journal} {Phys. Rev. A}\ }\textbf {\bibinfo {volume} {71}},\ \bibinfo {pages} {012712} (\bibinfo {year} {2005}{\natexlab{b}})}\BibitemShut {NoStop}%
\bibitem [{\citenamefont {Barhoumi}\ \emph {et~al.}(2023)\citenamefont {Barhoumi}, \citenamefont {Liu}, \citenamefont {Lefkidis},\ and\ \citenamefont {H\"ubner}}]{Barhoumi2023}%
  \BibitemOpen
  \bibfield  {author} {\bibinfo {author} {\bibfnamefont {M.}~\bibnamefont {Barhoumi}}, \bibinfo {author} {\bibfnamefont {J.}~\bibnamefont {Liu}}, \bibinfo {author} {\bibfnamefont {G.}~\bibnamefont {Lefkidis}}, \ and\ \bibinfo {author} {\bibfnamefont {W.}~\bibnamefont {H\"ubner}},\ }\href {\doibase 10.1063/5.0158160} {\bibfield  {journal} {\bibinfo  {journal} {J. Chem. Phys.}\ }\textbf {\bibinfo {volume} {159}},\ \bibinfo {pages} {084304} (\bibinfo {year} {2023})}\BibitemShut {NoStop}%
\bibitem [{\citenamefont {Lefkidis}\ \emph {et~al.}(2024)\citenamefont {Lefkidis}, \citenamefont {Chaudhuri}, \citenamefont {Jin}, \citenamefont {Li}, \citenamefont {Dutta},\ and\ \citenamefont {H\"ubner}}]{Lefkidis24}%
  \BibitemOpen
  \bibfield  {author} {\bibinfo {author} {\bibfnamefont {G.}~\bibnamefont {Lefkidis}}, \bibinfo {author} {\bibfnamefont {D.}~\bibnamefont {Chaudhuri}}, \bibinfo {author} {\bibfnamefont {W.}~\bibnamefont {Jin}}, \bibinfo {author} {\bibfnamefont {C.}~\bibnamefont {Li}}, \bibinfo {author} {\bibfnamefont {D.}~\bibnamefont {Dutta}}, \ and\ \bibinfo {author} {\bibfnamefont {W.}~\bibnamefont {H\"ubner}},\ }\href {\doibase 10.1088/1402-4896/ad2140} {\bibfield  {journal} {\bibinfo  {journal} {Phys. Scr.}\ }\textbf {\bibinfo {volume} {99}},\ \bibinfo {pages} {035909} (\bibinfo {year} {2024})}\BibitemShut {NoStop}%
\bibitem [{\citenamefont {Li}\ \emph {et~al.}(2011)\citenamefont {Li}, \citenamefont {Jin}, \citenamefont {Xiang}, \citenamefont {Lefkidis},\ and\ \citenamefont {H\"ubner}}]{PhysRevB.84.054415}%
  \BibitemOpen
  \bibfield  {author} {\bibinfo {author} {\bibfnamefont {C.}~\bibnamefont {Li}}, \bibinfo {author} {\bibfnamefont {W.}~\bibnamefont {Jin}}, \bibinfo {author} {\bibfnamefont {H.}~\bibnamefont {Xiang}}, \bibinfo {author} {\bibfnamefont {G.}~\bibnamefont {Lefkidis}}, \ and\ \bibinfo {author} {\bibfnamefont {W.}~\bibnamefont {H\"ubner}},\ }\href {\doibase 10.1103/PhysRevB.84.054415} {\bibfield  {journal} {\bibinfo  {journal} {Phys. Rev. B}\ }\textbf {\bibinfo {volume} {84}},\ \bibinfo {pages} {054415} (\bibinfo {year} {2011})}\BibitemShut {NoStop}%
\bibitem [{\citenamefont {Baiardi}\ and\ \citenamefont {Reiher}(2019)}]{Baiardi2019}%
  \BibitemOpen
  \bibfield  {author} {\bibinfo {author} {\bibfnamefont {A.}~\bibnamefont {Baiardi}}\ and\ \bibinfo {author} {\bibfnamefont {M.}~\bibnamefont {Reiher}},\ }\href {https://doi.org/10.1021/acs.jctc.9b00301} {\bibfield  {journal} {\bibinfo  {journal} {J. Chem. Theory Comput.}\ }\textbf {\bibinfo {volume} {15}},\ \bibinfo {pages} {3481} (\bibinfo {year} {2019})}\BibitemShut {NoStop}%
\bibitem [{\citenamefont {Xie}\ \emph {et~al.}(2019)\citenamefont {Xie}, \citenamefont {Liu}, \citenamefont {Yao}, \citenamefont {Schollw{\"o}ck}, \citenamefont {Liu},\ and\ \citenamefont {Ma}}]{10.1063/1.5125945}%
  \BibitemOpen
  \bibfield  {author} {\bibinfo {author} {\bibfnamefont {X.}~\bibnamefont {Xie}}, \bibinfo {author} {\bibfnamefont {Y.}~\bibnamefont {Liu}}, \bibinfo {author} {\bibfnamefont {Y.}~\bibnamefont {Yao}}, \bibinfo {author} {\bibfnamefont {U.}~\bibnamefont {Schollw{\"o}ck}}, \bibinfo {author} {\bibfnamefont {C.}~\bibnamefont {Liu}}, \ and\ \bibinfo {author} {\bibfnamefont {H.}~\bibnamefont {Ma}},\ }\href {\doibase 10.1063/1.5125945} {\bibfield  {journal} {\bibinfo  {journal} {The Journal of Chemical Physics}\ }\textbf {\bibinfo {volume} {151}},\ \bibinfo {pages} {224101} (\bibinfo {year} {2019})},\ \Eprint {http://arxiv.org/abs/https://pubs.aip.org/aip/jcp/article-pdf/doi/10.1063/1.5125945/13331146/224101\_1\_online.pdf} {https://pubs.aip.org/aip/jcp/article-pdf/doi/10.1063/1.5125945/13331146/224101\_1\_online.pdf} \BibitemShut {NoStop}%
\bibitem [{\citenamefont {Li}\ \emph {et~al.}(2020)\citenamefont {Li}, \citenamefont {Ren},\ and\ \citenamefont {Shuai}}]{tDMRG1}%
  \BibitemOpen
  \bibfield  {author} {\bibinfo {author} {\bibfnamefont {W.}~\bibnamefont {Li}}, \bibinfo {author} {\bibfnamefont {J.}~\bibnamefont {Ren}}, \ and\ \bibinfo {author} {\bibfnamefont {Z.}~\bibnamefont {Shuai}},\ }\href {\doibase 10.1021/acs.jpclett.0c01072} {\bibfield  {journal} {\bibinfo  {journal} {J. Phys. Chem. Lett.}\ }\textbf {\bibinfo {volume} {11}},\ \bibinfo {pages} {4930} (\bibinfo {year} {2020})}\BibitemShut {NoStop}%
\bibitem [{\citenamefont {Borrelli}\ and\ \citenamefont {Gelin}(2017)}]{Borrelli2017}%
  \BibitemOpen
  \bibfield  {author} {\bibinfo {author} {\bibfnamefont {R.}~\bibnamefont {Borrelli}}\ and\ \bibinfo {author} {\bibfnamefont {M.}~\bibnamefont {Gelin}},\ }\href {\doibase 10.1038/s41598-017-08901-2} {\bibfield  {journal} {\bibinfo  {journal} {Scientific Reports}\ }\textbf {\bibinfo {volume} {7}},\ \bibinfo {pages} {9127} (\bibinfo {year} {2017})}\BibitemShut {NoStop}%
\bibitem [{\citenamefont {Schlimgen}\ \emph {et~al.}(2021)\citenamefont {Schlimgen}, \citenamefont {Head-Marsden}, \citenamefont {Sager}, \citenamefont {Narang},\ and\ \citenamefont {Mazziotti}}]{PhysRevLett.127.270503}%
  \BibitemOpen
  \bibfield  {author} {\bibinfo {author} {\bibfnamefont {A.~W.}\ \bibnamefont {Schlimgen}}, \bibinfo {author} {\bibfnamefont {K.}~\bibnamefont {Head-Marsden}}, \bibinfo {author} {\bibfnamefont {L.~M.}\ \bibnamefont {Sager}}, \bibinfo {author} {\bibfnamefont {P.}~\bibnamefont {Narang}}, \ and\ \bibinfo {author} {\bibfnamefont {D.~A.}\ \bibnamefont {Mazziotti}},\ }\href {\doibase 10.1103/PhysRevLett.127.270503} {\bibfield  {journal} {\bibinfo  {journal} {Phys. Rev. Lett.}\ }\textbf {\bibinfo {volume} {127}},\ \bibinfo {pages} {270503} (\bibinfo {year} {2021})}\BibitemShut {NoStop}%
\bibitem [{\citenamefont {Giustino}(2017)}]{RevModPhys.89.015003}%
  \BibitemOpen
  \bibfield  {author} {\bibinfo {author} {\bibfnamefont {F.}~\bibnamefont {Giustino}},\ }\href {\doibase 10.1103/RevModPhys.89.015003} {\bibfield  {journal} {\bibinfo  {journal} {Rev. Mod. Phys.}\ }\textbf {\bibinfo {volume} {89}},\ \bibinfo {pages} {015003} (\bibinfo {year} {2017})}\BibitemShut {NoStop}%
\bibitem [{\citenamefont {Sakurai}(1999)}]{sakurai1999advanced}%
  \BibitemOpen
  \bibfield  {author} {\bibinfo {author} {\bibfnamefont {J.}~\bibnamefont {Sakurai}},\ }\href {https://books.google.com/books?id=BKtoOgAACAAJ} {\emph {\bibinfo {title} {Advanced Quantum Mechanics}}},\ Addison-Wesley series in advanced physics\ (\bibinfo  {publisher} {Addison-Wesley Longman},\ \bibinfo {year} {1999})\BibitemShut {NoStop}%
\bibitem [{\citenamefont {Sch{\"o}ll}(2011)}]{scholl2011theory}%
  \BibitemOpen
  \bibfield  {author} {\bibinfo {author} {\bibfnamefont {E.}~\bibnamefont {Sch{\"o}ll}},\ }\href {https://books.google.com/books?id=E7UYswEACAAJ} {\emph {\bibinfo {title} {Theory of Transport Properties of Semiconductor Nanostructures}}},\ Electronic Materials Series\ (\bibinfo  {publisher} {Springer US},\ \bibinfo {year} {2011})\BibitemShut {NoStop}%
\bibitem [{\citenamefont {McQuarrie}(1976)}]{mcquarrie76a}%
  \BibitemOpen
  \bibfield  {author} {\bibinfo {author} {\bibfnamefont {D.~A.}\ \bibnamefont {McQuarrie}},\ }\href@noop {} {\emph {\bibinfo {title} {Statistical Mechanics}}},\ Harper's chemistry series\ (\bibinfo  {publisher} {Harper Collins},\ \bibinfo {address} {New York},\ \bibinfo {year} {1976})\BibitemShut {NoStop}%
\bibitem [{\citenamefont {Wyld~Jr.}\ and\ \citenamefont {Fried}(1963)}]{Wyld1963}%
  \BibitemOpen
  \bibfield  {author} {\bibinfo {author} {\bibfnamefont {H.}~\bibnamefont {Wyld~Jr.}}\ and\ \bibinfo {author} {\bibfnamefont {B.}~\bibnamefont {Fried}},\ }\href {\doibase https://doi.org/10.1016/0003-4916(63)90260-4} {\bibfield  {journal} {\bibinfo  {journal} {Annals of Physics}\ }\textbf {\bibinfo {volume} {23}},\ \bibinfo {pages} {374} (\bibinfo {year} {1963})}\BibitemShut {NoStop}%
\bibitem [{\citenamefont {Iotti}\ \emph {et~al.}(2005)\citenamefont {Iotti}, \citenamefont {Ciancio},\ and\ \citenamefont {Rossi}}]{Iotti2005}%
  \BibitemOpen
  \bibfield  {author} {\bibinfo {author} {\bibfnamefont {R.}~\bibnamefont {Iotti}}, \bibinfo {author} {\bibfnamefont {E.}~\bibnamefont {Ciancio}}, \ and\ \bibinfo {author} {\bibfnamefont {F.}~\bibnamefont {Rossi}},\ }\href {\doibase https://doi.org/10.1103/PhysRevB.72.125347} {\bibfield  {journal} {\bibinfo  {journal} {Phys. Rev. B}\ }\textbf {\bibinfo {volume} {72}},\ \bibinfo {pages} {125347} (\bibinfo {year} {2005})}\BibitemShut {NoStop}%
\bibitem [{\citenamefont {Rossi}\ and\ \citenamefont {Kuhn}(2002)}]{Rossi2002}%
  \BibitemOpen
  \bibfield  {author} {\bibinfo {author} {\bibfnamefont {F.}~\bibnamefont {Rossi}}\ and\ \bibinfo {author} {\bibfnamefont {T.}~\bibnamefont {Kuhn}},\ }\href {\doibase 10.1103/RevModPhys.74.895} {\bibfield  {journal} {\bibinfo  {journal} {Rev. Mod. Phys.}\ }\textbf {\bibinfo {volume} {74}},\ \bibinfo {pages} {895} (\bibinfo {year} {2002})}\BibitemShut {NoStop}%
\bibitem [{\citenamefont {Stefanucci}\ and\ \citenamefont {van Leeuwen}(2013)}]{Stef2013}%
  \BibitemOpen
  \bibfield  {author} {\bibinfo {author} {\bibfnamefont {G.}~\bibnamefont {Stefanucci}}\ and\ \bibinfo {author} {\bibfnamefont {R.}~\bibnamefont {van Leeuwen}},\ }\href@noop {} {\emph {\bibinfo {title} {Nonequilibrium Many-Body Theory of Quantum Systems: A Modern Introduction}}}\ (\bibinfo  {publisher} {Cambridge University Press},\ \bibinfo {year} {2013})\BibitemShut {NoStop}%
\bibitem [{\citenamefont {Ping}\ \emph {et~al.}(2013)\citenamefont {Ping}, \citenamefont {Rocca},\ and\ \citenamefont {Galli}}]{C3CS00007A}%
  \BibitemOpen
  \bibfield  {author} {\bibinfo {author} {\bibfnamefont {Y.}~\bibnamefont {Ping}}, \bibinfo {author} {\bibfnamefont {D.}~\bibnamefont {Rocca}}, \ and\ \bibinfo {author} {\bibfnamefont {G.}~\bibnamefont {Galli}},\ }\href {\doibase 10.1039/C3CS00007A} {\bibfield  {journal} {\bibinfo  {journal} {Chem. Soc. Rev.}\ }\textbf {\bibinfo {volume} {42}},\ \bibinfo {pages} {2437} (\bibinfo {year} {2013})}\BibitemShut {NoStop}%
\bibitem [{\citenamefont {Haug}\ and\ \citenamefont {Jauho}(2007)}]{haug2007}%
  \BibitemOpen
  \bibfield  {author} {\bibinfo {author} {\bibfnamefont {H.}~\bibnamefont {Haug}}\ and\ \bibinfo {author} {\bibfnamefont {A.}~\bibnamefont {Jauho}},\ }\href {https://books.google.com/books?id=w1am24ZE9jQC} {\emph {\bibinfo {title} {Quantum Kinetics in Transport and Optics of Semiconductors}}},\ Springer Series in Solid-State Sciences\ (\bibinfo  {publisher} {Springer Berlin Heidelberg},\ \bibinfo {year} {2007})\BibitemShut {NoStop}%
\bibitem [{\citenamefont {Marini}(2013)}]{Marini2013}%
  \BibitemOpen
  \bibfield  {author} {\bibinfo {author} {\bibfnamefont {A.}~\bibnamefont {Marini}},\ }\href {\doibase 10.1088/1742-6596/427/1/012003} {\bibfield  {journal} {\bibinfo  {journal} {Journal of Physics: Conference Series}\ }\textbf {\bibinfo {volume} {427}},\ \bibinfo {pages} {012003} (\bibinfo {year} {2013})}\BibitemShut {NoStop}%
\bibitem [{\citenamefont {Onida}\ \emph {et~al.}(2002)\citenamefont {Onida}, \citenamefont {Reining},\ and\ \citenamefont {Rubio}}]{RevModPhys.74.601}%
  \BibitemOpen
  \bibfield  {author} {\bibinfo {author} {\bibfnamefont {G.}~\bibnamefont {Onida}}, \bibinfo {author} {\bibfnamefont {L.}~\bibnamefont {Reining}}, \ and\ \bibinfo {author} {\bibfnamefont {A.}~\bibnamefont {Rubio}},\ }\href {\doibase 10.1103/RevModPhys.74.601} {\bibfield  {journal} {\bibinfo  {journal} {Rev. Mod. Phys.}\ }\textbf {\bibinfo {volume} {74}},\ \bibinfo {pages} {601} (\bibinfo {year} {2002})}\BibitemShut {NoStop}%
\bibitem [{\citenamefont {Rocca}\ \emph {et~al.}(2012)\citenamefont {Rocca}, \citenamefont {Ping}, \citenamefont {Gebauer},\ and\ \citenamefont {Galli}}]{PhysRevB.85.045116}%
  \BibitemOpen
  \bibfield  {author} {\bibinfo {author} {\bibfnamefont {D.}~\bibnamefont {Rocca}}, \bibinfo {author} {\bibfnamefont {Y.}~\bibnamefont {Ping}}, \bibinfo {author} {\bibfnamefont {R.}~\bibnamefont {Gebauer}}, \ and\ \bibinfo {author} {\bibfnamefont {G.}~\bibnamefont {Galli}},\ }\href {\doibase 10.1103/PhysRevB.85.045116} {\bibfield  {journal} {\bibinfo  {journal} {Phys. Rev. B}\ }\textbf {\bibinfo {volume} {85}},\ \bibinfo {pages} {045116} (\bibinfo {year} {2012})}\BibitemShut {NoStop}%
\bibitem [{\citenamefont {Govoni}\ and\ \citenamefont {Galli}(2015)}]{Govoni2015}%
  \BibitemOpen
  \bibfield  {author} {\bibinfo {author} {\bibfnamefont {M.}~\bibnamefont {Govoni}}\ and\ \bibinfo {author} {\bibfnamefont {G.}~\bibnamefont {Galli}},\ }\href {\doibase 10.1021/ct500958p} {\bibfield  {journal} {\bibinfo  {journal} {Journal of Chemical Theory and Computation}\ }\textbf {\bibinfo {volume} {11}},\ \bibinfo {pages} {2680} (\bibinfo {year} {2015})}\BibitemShut {NoStop}%
\bibitem [{\citenamefont {Lipavsk\'y}\ \emph {et~al.}(1986)\citenamefont {Lipavsk\'y}, \citenamefont {\ifmmode \check{S}\else \v{S}\fi{}pi\ifmmode~\check{c}\else \v{c}\fi{}ka},\ and\ \citenamefont {Velick\'y}}]{Lipa1986}%
  \BibitemOpen
  \bibfield  {author} {\bibinfo {author} {\bibfnamefont {P.}~\bibnamefont {Lipavsk\'y}}, \bibinfo {author} {\bibfnamefont {V.}~\bibnamefont {\ifmmode \check{S}\else \v{S}\fi{}pi\ifmmode~\check{c}\else \v{c}\fi{}ka}}, \ and\ \bibinfo {author} {\bibfnamefont {B.}~\bibnamefont {Velick\'y}},\ }\href {\doibase 10.1103/PhysRevB.34.6933} {\bibfield  {journal} {\bibinfo  {journal} {Phys. Rev. B}\ }\textbf {\bibinfo {volume} {34}},\ \bibinfo {pages} {6933} (\bibinfo {year} {1986})}\BibitemShut {NoStop}%
\bibitem [{\citenamefont {Perfetto}\ \emph {et~al.}(2015{\natexlab{a}})\citenamefont {Perfetto}, \citenamefont {Uimonen}, \citenamefont {van Leeuwen},\ and\ \citenamefont {Stefanucci}}]{Per2015}%
  \BibitemOpen
  \bibfield  {author} {\bibinfo {author} {\bibfnamefont {E.}~\bibnamefont {Perfetto}}, \bibinfo {author} {\bibfnamefont {A.-M.}\ \bibnamefont {Uimonen}}, \bibinfo {author} {\bibfnamefont {R.}~\bibnamefont {van Leeuwen}}, \ and\ \bibinfo {author} {\bibfnamefont {G.}~\bibnamefont {Stefanucci}},\ }\href {\doibase 10.1103/PhysRevA.92.033419} {\bibfield  {journal} {\bibinfo  {journal} {Phys. Rev. A}\ }\textbf {\bibinfo {volume} {92}},\ \bibinfo {pages} {033419} (\bibinfo {year} {2015}{\natexlab{a}})}\BibitemShut {NoStop}%
\bibitem [{\citenamefont {Perfetto}\ \emph {et~al.}(2015{\natexlab{b}})\citenamefont {Perfetto}, \citenamefont {Sangalli}, \citenamefont {Marini},\ and\ \citenamefont {Stefanucci}}]{Per2015_2}%
  \BibitemOpen
  \bibfield  {author} {\bibinfo {author} {\bibfnamefont {E.}~\bibnamefont {Perfetto}}, \bibinfo {author} {\bibfnamefont {D.}~\bibnamefont {Sangalli}}, \bibinfo {author} {\bibfnamefont {A.}~\bibnamefont {Marini}}, \ and\ \bibinfo {author} {\bibfnamefont {G.}~\bibnamefont {Stefanucci}},\ }\href {\doibase 10.1103/PhysRevB.92.205304} {\bibfield  {journal} {\bibinfo  {journal} {Phys. Rev. B}\ }\textbf {\bibinfo {volume} {92}},\ \bibinfo {pages} {205304} (\bibinfo {year} {2015}{\natexlab{b}})}\BibitemShut {NoStop}%
\bibitem [{\citenamefont {Riva}\ \emph {et~al.}(2025)\citenamefont {Riva}, \citenamefont {Simoni},\ and\ \citenamefont {Ping}}]{Rivaprep}%
  \BibitemOpen
  \bibfield  {author} {\bibinfo {author} {\bibfnamefont {G.}~\bibnamefont {Riva}}, \bibinfo {author} {\bibfnamefont {J.}~\bibnamefont {Simoni}}, \ and\ \bibinfo {author} {\bibfnamefont {Y.}~\bibnamefont {Ping}},\ }\href@noop {} {\bibfield  {journal} {\bibinfo  {journal} {Manuscript in preparation}\ } (\bibinfo {year} {2025})}\BibitemShut {NoStop}%
\bibitem [{\citenamefont {Xu}\ \emph {et~al.}(2021{\natexlab{a}})\citenamefont {Xu}, \citenamefont {Habib}, \citenamefont {Sundararaman},\ and\ \citenamefont {Ping}}]{PhysRevB.104.184418}%
  \BibitemOpen
  \bibfield  {author} {\bibinfo {author} {\bibfnamefont {J.}~\bibnamefont {Xu}}, \bibinfo {author} {\bibfnamefont {A.}~\bibnamefont {Habib}}, \bibinfo {author} {\bibfnamefont {R.}~\bibnamefont {Sundararaman}}, \ and\ \bibinfo {author} {\bibfnamefont {Y.}~\bibnamefont {Ping}},\ }\href {\doibase 10.1103/PhysRevB.104.184418} {\bibfield  {journal} {\bibinfo  {journal} {Phys. Rev. B}\ }\textbf {\bibinfo {volume} {104}},\ \bibinfo {pages} {184418} (\bibinfo {year} {2021}{\natexlab{a}})}\BibitemShut {NoStop}%
\bibitem [{\citenamefont {Lafuente-Bartolome}\ \emph {et~al.}(2022)\citenamefont {Lafuente-Bartolome}, \citenamefont {Lian}, \citenamefont {Sio}, \citenamefont {Gurtubay}, \citenamefont {Eiguren},\ and\ \citenamefont {Giustino}}]{PhysRevB.106.075119}%
  \BibitemOpen
  \bibfield  {author} {\bibinfo {author} {\bibfnamefont {J.}~\bibnamefont {Lafuente-Bartolome}}, \bibinfo {author} {\bibfnamefont {C.}~\bibnamefont {Lian}}, \bibinfo {author} {\bibfnamefont {W.~H.}\ \bibnamefont {Sio}}, \bibinfo {author} {\bibfnamefont {I.~G.}\ \bibnamefont {Gurtubay}}, \bibinfo {author} {\bibfnamefont {A.}~\bibnamefont {Eiguren}}, \ and\ \bibinfo {author} {\bibfnamefont {F.}~\bibnamefont {Giustino}},\ }\href {\doibase 10.1103/PhysRevB.106.075119} {\bibfield  {journal} {\bibinfo  {journal} {Phys. Rev. B}\ }\textbf {\bibinfo {volume} {106}},\ \bibinfo {pages} {075119} (\bibinfo {year} {2022})}\BibitemShut {NoStop}%
\bibitem [{\citenamefont {Blommel}\ \emph {et~al.}(2025)\citenamefont {Blommel}, \citenamefont {Perfetto}, \citenamefont {Stefanucci},\ and\ \citenamefont {Vl\v{c}ek}}]{blommel2025unifiedsimulationframeworkcorrelated}%
  \BibitemOpen
  \bibfield  {author} {\bibinfo {author} {\bibfnamefont {T.}~\bibnamefont {Blommel}}, \bibinfo {author} {\bibfnamefont {E.}~\bibnamefont {Perfetto}}, \bibinfo {author} {\bibfnamefont {G.}~\bibnamefont {Stefanucci}}, \ and\ \bibinfo {author} {\bibfnamefont {V.}~\bibnamefont {Vl\v{c}ek}},\ }\href {https://arxiv.org/abs/2505.01541} {\enquote {\bibinfo {title} {A unified simulation framework for correlated driven-dissipative quantum dynamics},}\ } (\bibinfo {year} {2025}),\ \Eprint {http://arxiv.org/abs/2505.01541} {arXiv:2505.01541 [cond-mat.mtrl-sci]} \BibitemShut {NoStop}%
\bibitem [{\citenamefont {Iotti}\ and\ \citenamefont {Rossi}(2017)}]{Iotti_2017}%
  \BibitemOpen
  \bibfield  {author} {\bibinfo {author} {\bibfnamefont {R.~C.}\ \bibnamefont {Iotti}}\ and\ \bibinfo {author} {\bibfnamefont {F.}~\bibnamefont {Rossi}},\ }\href {\doibase 10.1140/epjb/e2017-80462-3} {\bibfield  {journal} {\bibinfo  {journal} {Eur. Phys. J. B}\ }\textbf {\bibinfo {volume} {90}},\ \bibinfo {pages} {250} (\bibinfo {year} {2017})}\BibitemShut {NoStop}%
\bibitem [{\citenamefont {Karlsson}\ \emph {et~al.}(2021)\citenamefont {Karlsson}, \citenamefont {van Leeuwen}, \citenamefont {Pavlyukh}, \citenamefont {Perfetto},\ and\ \citenamefont {Stefanucci}}]{Kar2021}%
  \BibitemOpen
  \bibfield  {author} {\bibinfo {author} {\bibfnamefont {D.}~\bibnamefont {Karlsson}}, \bibinfo {author} {\bibfnamefont {R.}~\bibnamefont {van Leeuwen}}, \bibinfo {author} {\bibfnamefont {Y.}~\bibnamefont {Pavlyukh}}, \bibinfo {author} {\bibfnamefont {E.}~\bibnamefont {Perfetto}}, \ and\ \bibinfo {author} {\bibfnamefont {G.}~\bibnamefont {Stefanucci}},\ }\href {\doibase 10.1103/PhysRevLett.127.036402} {\bibfield  {journal} {\bibinfo  {journal} {Phys. Rev. Lett.}\ }\textbf {\bibinfo {volume} {127}},\ \bibinfo {pages} {036402} (\bibinfo {year} {2021})}\BibitemShut {NoStop}%
\bibitem [{\citenamefont {Chan}\ \emph {et~al.}(2023)\citenamefont {Chan}, \citenamefont {Qiu}, \citenamefont {da~Jornada},\ and\ \citenamefont {Louie}}]{doi:10.1073/pnas.2301957120}%
  \BibitemOpen
  \bibfield  {author} {\bibinfo {author} {\bibfnamefont {Y.-H.}\ \bibnamefont {Chan}}, \bibinfo {author} {\bibfnamefont {D.~Y.}\ \bibnamefont {Qiu}}, \bibinfo {author} {\bibfnamefont {F.~H.}\ \bibnamefont {da~Jornada}}, \ and\ \bibinfo {author} {\bibfnamefont {S.~G.}\ \bibnamefont {Louie}},\ }\href {\doibase 10.1073/pnas.2301957120} {\bibfield  {journal} {\bibinfo  {journal} {Proceedings of the National Academy of Sciences}\ }\textbf {\bibinfo {volume} {120}},\ \bibinfo {pages} {e2301957120} (\bibinfo {year} {2023})},\ \Eprint {http://arxiv.org/abs/https://www.pnas.org/doi/pdf/10.1073/pnas.2301957120} {https://www.pnas.org/doi/pdf/10.1073/pnas.2301957120} \BibitemShut {NoStop}%
\bibitem [{\citenamefont {Runge}\ and\ \citenamefont {Gross}(1984{\natexlab{b}})}]{Runge_1984}%
  \BibitemOpen
  \bibfield  {author} {\bibinfo {author} {\bibfnamefont {E.}~\bibnamefont {Runge}}\ and\ \bibinfo {author} {\bibfnamefont {E.~K.~U.}\ \bibnamefont {Gross}},\ }\href {\doibase https://doi.org/10.1103/PhysRevLett.52.997} {\bibfield  {journal} {\bibinfo  {journal} {Phys. Rev. Lett.}\ }\textbf {\bibinfo {volume} {52}},\ \bibinfo {pages} {997} (\bibinfo {year} {1984}{\natexlab{b}})}\BibitemShut {NoStop}%
\bibitem [{\citenamefont {de~Vega}\ and\ \citenamefont {Alonso}(2017)}]{RevModPhys.89.015001}%
  \BibitemOpen
  \bibfield  {author} {\bibinfo {author} {\bibfnamefont {I.}~\bibnamefont {de~Vega}}\ and\ \bibinfo {author} {\bibfnamefont {D.}~\bibnamefont {Alonso}},\ }\href {\doibase 10.1103/RevModPhys.89.015001} {\bibfield  {journal} {\bibinfo  {journal} {Rev. Mod. Phys.}\ }\textbf {\bibinfo {volume} {89}},\ \bibinfo {pages} {015001} (\bibinfo {year} {2017})}\BibitemShut {NoStop}%
\bibitem [{\citenamefont {Lindberg}\ and\ \citenamefont {Koch}(1988)}]{PhysRevB.38.3342}%
  \BibitemOpen
  \bibfield  {author} {\bibinfo {author} {\bibfnamefont {M.}~\bibnamefont {Lindberg}}\ and\ \bibinfo {author} {\bibfnamefont {S.~W.}\ \bibnamefont {Koch}},\ }\href {\doibase 10.1103/PhysRevB.38.3342} {\bibfield  {journal} {\bibinfo  {journal} {Phys. Rev. B}\ }\textbf {\bibinfo {volume} {38}},\ \bibinfo {pages} {3342} (\bibinfo {year} {1988})}\BibitemShut {NoStop}%
\bibitem [{\citenamefont {Hohenester}\ and\ \citenamefont {P\"otz}(1997)}]{PhysRevB.56.13177}%
  \BibitemOpen
  \bibfield  {author} {\bibinfo {author} {\bibfnamefont {U.}~\bibnamefont {Hohenester}}\ and\ \bibinfo {author} {\bibfnamefont {W.}~\bibnamefont {P\"otz}},\ }\href {\doibase 10.1103/PhysRevB.56.13177} {\bibfield  {journal} {\bibinfo  {journal} {Phys. Rev. B}\ }\textbf {\bibinfo {volume} {56}},\ \bibinfo {pages} {13177} (\bibinfo {year} {1997})}\BibitemShut {NoStop}%
\bibitem [{\citenamefont {Bi~Sun}\ and\ \citenamefont {Milburn}(1999)}]{PhysRevB.59.10748}%
  \BibitemOpen
  \bibfield  {author} {\bibinfo {author} {\bibfnamefont {H.}~\bibnamefont {Bi~Sun}}\ and\ \bibinfo {author} {\bibfnamefont {G.~J.}\ \bibnamefont {Milburn}},\ }\href {\doibase 10.1103/PhysRevB.59.10748} {\bibfield  {journal} {\bibinfo  {journal} {Phys. Rev. B}\ }\textbf {\bibinfo {volume} {59}},\ \bibinfo {pages} {10748} (\bibinfo {year} {1999})}\BibitemShut {NoStop}%
\bibitem [{\citenamefont {Flindt}\ \emph {et~al.}(2004)\citenamefont {Flindt}, \citenamefont {Novotn\'y},\ and\ \citenamefont {Jauho}}]{PhysRevB.70.205334}%
  \BibitemOpen
  \bibfield  {author} {\bibinfo {author} {\bibfnamefont {C.}~\bibnamefont {Flindt}}, \bibinfo {author} {\bibfnamefont {T.~c.~v.}\ \bibnamefont {Novotn\'y}}, \ and\ \bibinfo {author} {\bibfnamefont {A.-P.}\ \bibnamefont {Jauho}},\ }\href {\doibase 10.1103/PhysRevB.70.205334} {\bibfield  {journal} {\bibinfo  {journal} {Phys. Rev. B}\ }\textbf {\bibinfo {volume} {70}},\ \bibinfo {pages} {205334} (\bibinfo {year} {2004})}\BibitemShut {NoStop}%
\bibitem [{\citenamefont {REDFIELD}(1965)}]{REDFIELD19651}%
  \BibitemOpen
  \bibfield  {author} {\bibinfo {author} {\bibfnamefont {A.}~\bibnamefont {REDFIELD}},\ }in\ \href {\doibase https://doi.org/10.1016/B978-1-4832-3114-3.50007-6} {\emph {\bibinfo {booktitle} {Advances in Magnetic Resonance}}},\ \bibinfo {series} {Advances in Magnetic and Optical Resonance}, Vol.~\bibinfo {volume} {1},\ \bibinfo {editor} {edited by\ \bibinfo {editor} {\bibfnamefont {J.~S.}\ \bibnamefont {Waugh}}}\ (\bibinfo  {publisher} {Academic Press},\ \bibinfo {year} {1965})\ pp.\ \bibinfo {pages} {1--32}\BibitemShut {NoStop}%
\bibitem [{\citenamefont {Davies}(1974)}]{Davies1974}%
  \BibitemOpen
  \bibfield  {author} {\bibinfo {author} {\bibfnamefont {E.~B.}\ \bibnamefont {Davies}},\ }\href@noop {} {\bibfield  {journal} {\bibinfo  {journal} {Communications in Mathematical Physics}\ }\textbf {\bibinfo {volume} {39}},\ \bibinfo {pages} {91 } (\bibinfo {year} {1974})}\BibitemShut {NoStop}%
\bibitem [{\citenamefont {Taj}\ \emph {et~al.}(2009)\citenamefont {Taj}, \citenamefont {Iotti},\ and\ \citenamefont {Rossi}}]{taj2009microscopic}%
  \BibitemOpen
  \bibfield  {author} {\bibinfo {author} {\bibfnamefont {D.}~\bibnamefont {Taj}}, \bibinfo {author} {\bibfnamefont {R.~C.}\ \bibnamefont {Iotti}}, \ and\ \bibinfo {author} {\bibfnamefont {F.}~\bibnamefont {Rossi}},\ }\href@noop {} {\bibfield  {journal} {\bibinfo  {journal} {The European Physical Journal B}\ }\textbf {\bibinfo {volume} {72}},\ \bibinfo {pages} {305} (\bibinfo {year} {2009})}\BibitemShut {NoStop}%
\bibitem [{\citenamefont {Rosati}\ \emph {et~al.}(2014)\citenamefont {Rosati}, \citenamefont {Iotti}, \citenamefont {Dolcini},\ and\ \citenamefont {Rossi}}]{PhysRevB.90.125140}%
  \BibitemOpen
  \bibfield  {author} {\bibinfo {author} {\bibfnamefont {R.}~\bibnamefont {Rosati}}, \bibinfo {author} {\bibfnamefont {R.~C.}\ \bibnamefont {Iotti}}, \bibinfo {author} {\bibfnamefont {F.}~\bibnamefont {Dolcini}}, \ and\ \bibinfo {author} {\bibfnamefont {F.}~\bibnamefont {Rossi}},\ }\href {\doibase 10.1103/PhysRevB.90.125140} {\bibfield  {journal} {\bibinfo  {journal} {Phys. Rev. B}\ }\textbf {\bibinfo {volume} {90}},\ \bibinfo {pages} {125140} (\bibinfo {year} {2014})}\BibitemShut {NoStop}%
\bibitem [{\citenamefont {Fischetti}(1999)}]{PhysRevB.59.4901}%
  \BibitemOpen
  \bibfield  {author} {\bibinfo {author} {\bibfnamefont {M.~V.}\ \bibnamefont {Fischetti}},\ }\href {\doibase 10.1103/PhysRevB.59.4901} {\bibfield  {journal} {\bibinfo  {journal} {Phys. Rev. B}\ }\textbf {\bibinfo {volume} {59}},\ \bibinfo {pages} {4901} (\bibinfo {year} {1999})}\BibitemShut {NoStop}%
\bibitem [{\citenamefont {Knezevic}(2008)}]{PhysRevB.77.125301}%
  \BibitemOpen
  \bibfield  {author} {\bibinfo {author} {\bibfnamefont {I.}~\bibnamefont {Knezevic}},\ }\href {\doibase 10.1103/PhysRevB.77.125301} {\bibfield  {journal} {\bibinfo  {journal} {Phys. Rev. B}\ }\textbf {\bibinfo {volume} {77}},\ \bibinfo {pages} {125301} (\bibinfo {year} {2008})}\BibitemShut {NoStop}%
\bibitem [{\citenamefont {Verzelen}\ \emph {et~al.}(2002)\citenamefont {Verzelen}, \citenamefont {Ferreira},\ and\ \citenamefont {Bastard}}]{PhysRevLett.88.146803}%
  \BibitemOpen
  \bibfield  {author} {\bibinfo {author} {\bibfnamefont {O.}~\bibnamefont {Verzelen}}, \bibinfo {author} {\bibfnamefont {R.}~\bibnamefont {Ferreira}}, \ and\ \bibinfo {author} {\bibfnamefont {G.}~\bibnamefont {Bastard}},\ }\href {\doibase 10.1103/PhysRevLett.88.146803} {\bibfield  {journal} {\bibinfo  {journal} {Phys. Rev. Lett.}\ }\textbf {\bibinfo {volume} {88}},\ \bibinfo {pages} {146803} (\bibinfo {year} {2002})}\BibitemShut {NoStop}%
\bibitem [{\citenamefont {Grange}\ \emph {et~al.}(2007)\citenamefont {Grange}, \citenamefont {Ferreira},\ and\ \citenamefont {Bastard}}]{PhysRevB.76.241304}%
  \BibitemOpen
  \bibfield  {author} {\bibinfo {author} {\bibfnamefont {T.}~\bibnamefont {Grange}}, \bibinfo {author} {\bibfnamefont {R.}~\bibnamefont {Ferreira}}, \ and\ \bibinfo {author} {\bibfnamefont {G.}~\bibnamefont {Bastard}},\ }\href {\doibase 10.1103/PhysRevB.76.241304} {\bibfield  {journal} {\bibinfo  {journal} {Phys. Rev. B}\ }\textbf {\bibinfo {volume} {76}},\ \bibinfo {pages} {241304} (\bibinfo {year} {2007})}\BibitemShut {NoStop}%
\bibitem [{\citenamefont {Norambuena}\ \emph {et~al.}(2020)\citenamefont {Norambuena}, \citenamefont {Maze}, \citenamefont {Rabl},\ and\ \citenamefont {Coto}}]{PhysRevA.101.022110}%
  \BibitemOpen
  \bibfield  {author} {\bibinfo {author} {\bibfnamefont {A.}~\bibnamefont {Norambuena}}, \bibinfo {author} {\bibfnamefont {J.~R.}\ \bibnamefont {Maze}}, \bibinfo {author} {\bibfnamefont {P.}~\bibnamefont {Rabl}}, \ and\ \bibinfo {author} {\bibfnamefont {R.}~\bibnamefont {Coto}},\ }\href {\doibase 10.1103/PhysRevA.101.022110} {\bibfield  {journal} {\bibinfo  {journal} {Phys. Rev. A}\ }\textbf {\bibinfo {volume} {101}},\ \bibinfo {pages} {022110} (\bibinfo {year} {2020})}\BibitemShut {NoStop}%
\bibitem [{\citenamefont {Xu}\ and\ \citenamefont {Ping}(2024{\natexlab{a}})}]{XuPing2024}%
  \BibitemOpen
  \bibfield  {author} {\bibinfo {author} {\bibfnamefont {J.}~\bibnamefont {Xu}}\ and\ \bibinfo {author} {\bibfnamefont {Y.}~\bibnamefont {Ping}},\ }\href {\doibase 10.1021/acs.jctc.3c00598} {\bibfield  {journal} {\bibinfo  {journal} {J. Chem. Theory Comput.}\ }\textbf {\bibinfo {volume} {20}},\ \bibinfo {pages} {492} (\bibinfo {year} {2024}{\natexlab{a}})}\BibitemShut {NoStop}%
\bibitem [{\citenamefont {Walther}\ \emph {et~al.}(2006)\citenamefont {Walther}, \citenamefont {Varcoe}, \citenamefont {Englert},\ and\ \citenamefont {Becker}}]{Walther_2006}%
  \BibitemOpen
  \bibfield  {author} {\bibinfo {author} {\bibfnamefont {H.}~\bibnamefont {Walther}}, \bibinfo {author} {\bibfnamefont {B.~T.~H.}\ \bibnamefont {Varcoe}}, \bibinfo {author} {\bibfnamefont {B.-G.}\ \bibnamefont {Englert}}, \ and\ \bibinfo {author} {\bibfnamefont {T.}~\bibnamefont {Becker}},\ }\href {\doibase 10.1088/0034-4885/69/5/R02} {\bibfield  {journal} {\bibinfo  {journal} {Reports on Progress in Physics}\ }\textbf {\bibinfo {volume} {69}},\ \bibinfo {pages} {1325} (\bibinfo {year} {2006})}\BibitemShut {NoStop}%
\bibitem [{\citenamefont {Sieberer}\ \emph {et~al.}(2016)\citenamefont {Sieberer}, \citenamefont {Buchhold},\ and\ \citenamefont {Diehl}}]{Sieberer_2016}%
  \BibitemOpen
  \bibfield  {author} {\bibinfo {author} {\bibfnamefont {L.~M.}\ \bibnamefont {Sieberer}}, \bibinfo {author} {\bibfnamefont {M.}~\bibnamefont {Buchhold}}, \ and\ \bibinfo {author} {\bibfnamefont {S.}~\bibnamefont {Diehl}},\ }\href {\doibase 10.1088/0034-4885/79/9/096001} {\bibfield  {journal} {\bibinfo  {journal} {Reports on Progress in Physics}\ }\textbf {\bibinfo {volume} {79}},\ \bibinfo {pages} {096001} (\bibinfo {year} {2016})}\BibitemShut {NoStop}%
\bibitem [{\citenamefont {Fogedby}(2022)}]{Hans_20220}%
  \BibitemOpen
  \bibfield  {author} {\bibinfo {author} {\bibfnamefont {H.~C.}\ \bibnamefont {Fogedby}},\ }\href {\doibase 10.1103/PhysRevA.106.022205} {\bibfield  {journal} {\bibinfo  {journal} {Phys. Rev. A}\ }\textbf {\bibinfo {volume} {106}},\ \bibinfo {pages} {022205} (\bibinfo {year} {2022})}\BibitemShut {NoStop}%
\bibitem [{\citenamefont {Lieu}\ \emph {et~al.}(2020)\citenamefont {Lieu}, \citenamefont {McGinley},\ and\ \citenamefont {Cooper}}]{Lieu2020}%
  \BibitemOpen
  \bibfield  {author} {\bibinfo {author} {\bibfnamefont {S.}~\bibnamefont {Lieu}}, \bibinfo {author} {\bibfnamefont {M.}~\bibnamefont {McGinley}}, \ and\ \bibinfo {author} {\bibfnamefont {N.~R.}\ \bibnamefont {Cooper}},\ }\href {\doibase 10.1103/PhysRevLett.124.040401} {\bibfield  {journal} {\bibinfo  {journal} {Phys. Rev. Lett.}\ }\textbf {\bibinfo {volume} {124}},\ \bibinfo {pages} {040401} (\bibinfo {year} {2020})}\BibitemShut {NoStop}%
\bibitem [{\citenamefont {Altland}\ \emph {et~al.}(2021)\citenamefont {Altland}, \citenamefont {Fleischhauer},\ and\ \citenamefont {Diehl}}]{Alt2021}%
  \BibitemOpen
  \bibfield  {author} {\bibinfo {author} {\bibfnamefont {A.}~\bibnamefont {Altland}}, \bibinfo {author} {\bibfnamefont {M.}~\bibnamefont {Fleischhauer}}, \ and\ \bibinfo {author} {\bibfnamefont {S.}~\bibnamefont {Diehl}},\ }\href {\doibase 10.1103/PhysRevX.11.021037} {\bibfield  {journal} {\bibinfo  {journal} {Phys. Rev. X}\ }\textbf {\bibinfo {volume} {11}},\ \bibinfo {pages} {021037} (\bibinfo {year} {2021})}\BibitemShut {NoStop}%
\bibitem [{\citenamefont {He}\ and\ \citenamefont {Chien}(2022)}]{He2022}%
  \BibitemOpen
  \bibfield  {author} {\bibinfo {author} {\bibfnamefont {Y.}~\bibnamefont {He}}\ and\ \bibinfo {author} {\bibfnamefont {C.-C.}\ \bibnamefont {Chien}},\ }\href {\doibase 10.1088/1361-648X/ac53da} {\bibfield  {journal} {\bibinfo  {journal} {Journal of Physics: Condensed Matter}\ }\textbf {\bibinfo {volume} {34}},\ \bibinfo {pages} {175403} (\bibinfo {year} {2022})}\BibitemShut {NoStop}%
\bibitem [{\citenamefont {Wojcik}\ \emph {et~al.}(2022)\citenamefont {Wojcik}, \citenamefont {Wang}, \citenamefont {Dutt}, \citenamefont {Zhong},\ and\ \citenamefont {Fan}}]{Woj2022}%
  \BibitemOpen
  \bibfield  {author} {\bibinfo {author} {\bibfnamefont {C.~C.}\ \bibnamefont {Wojcik}}, \bibinfo {author} {\bibfnamefont {K.}~\bibnamefont {Wang}}, \bibinfo {author} {\bibfnamefont {A.}~\bibnamefont {Dutt}}, \bibinfo {author} {\bibfnamefont {J.}~\bibnamefont {Zhong}}, \ and\ \bibinfo {author} {\bibfnamefont {S.}~\bibnamefont {Fan}},\ }\href {\doibase 10.1103/PhysRevB.106.L161401} {\bibfield  {journal} {\bibinfo  {journal} {Phys. Rev. B}\ }\textbf {\bibinfo {volume} {106}},\ \bibinfo {pages} {L161401} (\bibinfo {year} {2022})}\BibitemShut {NoStop}%
\bibitem [{\citenamefont {Shen}\ \emph {et~al.}(2018)\citenamefont {Shen}, \citenamefont {Zhen},\ and\ \citenamefont {Fu}}]{Shen2018}%
  \BibitemOpen
  \bibfield  {author} {\bibinfo {author} {\bibfnamefont {H.}~\bibnamefont {Shen}}, \bibinfo {author} {\bibfnamefont {B.}~\bibnamefont {Zhen}}, \ and\ \bibinfo {author} {\bibfnamefont {L.}~\bibnamefont {Fu}},\ }\href {\doibase 10.1103/PhysRevLett.120.146402} {\bibfield  {journal} {\bibinfo  {journal} {Phys. Rev. Lett.}\ }\textbf {\bibinfo {volume} {120}},\ \bibinfo {pages} {146402} (\bibinfo {year} {2018})}\BibitemShut {NoStop}%
\bibitem [{\citenamefont {Yao}\ and\ \citenamefont {Wang}(2018)}]{Yao2018}%
  \BibitemOpen
  \bibfield  {author} {\bibinfo {author} {\bibfnamefont {S.}~\bibnamefont {Yao}}\ and\ \bibinfo {author} {\bibfnamefont {Z.}~\bibnamefont {Wang}},\ }\href {\doibase 10.1103/PhysRevLett.121.086803} {\bibfield  {journal} {\bibinfo  {journal} {Phys. Rev. Lett.}\ }\textbf {\bibinfo {volume} {121}},\ \bibinfo {pages} {086803} (\bibinfo {year} {2018})}\BibitemShut {NoStop}%
\bibitem [{\citenamefont {Yokomizo}\ and\ \citenamefont {Murakami}(2019)}]{Yok2019}%
  \BibitemOpen
  \bibfield  {author} {\bibinfo {author} {\bibfnamefont {K.}~\bibnamefont {Yokomizo}}\ and\ \bibinfo {author} {\bibfnamefont {S.}~\bibnamefont {Murakami}},\ }\href {\doibase 10.1103/PhysRevLett.123.066404} {\bibfield  {journal} {\bibinfo  {journal} {Phys. Rev. Lett.}\ }\textbf {\bibinfo {volume} {123}},\ \bibinfo {pages} {066404} (\bibinfo {year} {2019})}\BibitemShut {NoStop}%
\bibitem [{\citenamefont {Kang}\ and\ \citenamefont {Hybertsen}(2010)}]{PhysRevB.82.195108}%
  \BibitemOpen
  \bibfield  {author} {\bibinfo {author} {\bibfnamefont {W.}~\bibnamefont {Kang}}\ and\ \bibinfo {author} {\bibfnamefont {M.~S.}\ \bibnamefont {Hybertsen}},\ }\href {\doibase 10.1103/PhysRevB.82.195108} {\bibfield  {journal} {\bibinfo  {journal} {Phys. Rev. B}\ }\textbf {\bibinfo {volume} {82}},\ \bibinfo {pages} {195108} (\bibinfo {year} {2010})}\BibitemShut {NoStop}%
\bibitem [{\citenamefont {Tan}\ \emph {et~al.}(2018)\citenamefont {Tan}, \citenamefont {Kas},\ and\ \citenamefont {Rehr}}]{PhysRevB.98.115125}%
  \BibitemOpen
  \bibfield  {author} {\bibinfo {author} {\bibfnamefont {T.~S.}\ \bibnamefont {Tan}}, \bibinfo {author} {\bibfnamefont {J.~J.}\ \bibnamefont {Kas}}, \ and\ \bibinfo {author} {\bibfnamefont {J.~J.}\ \bibnamefont {Rehr}},\ }\href {\doibase 10.1103/PhysRevB.98.115125} {\bibfield  {journal} {\bibinfo  {journal} {Phys. Rev. B}\ }\textbf {\bibinfo {volume} {98}},\ \bibinfo {pages} {115125} (\bibinfo {year} {2018})}\BibitemShut {NoStop}%
\bibitem [{\citenamefont {Attaccalite}\ \emph {et~al.}(2011)\citenamefont {Attaccalite}, \citenamefont {Gr\"uning},\ and\ \citenamefont {Marini}}]{attaccalite2011}%
  \BibitemOpen
  \bibfield  {author} {\bibinfo {author} {\bibfnamefont {C.}~\bibnamefont {Attaccalite}}, \bibinfo {author} {\bibfnamefont {M.}~\bibnamefont {Gr\"uning}}, \ and\ \bibinfo {author} {\bibfnamefont {A.}~\bibnamefont {Marini}},\ }\href {\doibase 10.1103/PhysRevB.84.245110} {\bibfield  {journal} {\bibinfo  {journal} {Phys. Rev. B}\ }\textbf {\bibinfo {volume} {84}},\ \bibinfo {pages} {245110} (\bibinfo {year} {2011})}\BibitemShut {NoStop}%
\bibitem [{\citenamefont {van Loon}\ \emph {et~al.}(2021)\citenamefont {van Loon}, \citenamefont {R\"osner}, \citenamefont {Katsnelson},\ and\ \citenamefont {Wehling}}]{PhysRevB.104.045134}%
  \BibitemOpen
  \bibfield  {author} {\bibinfo {author} {\bibfnamefont {E.~G. C.~P.}\ \bibnamefont {van Loon}}, \bibinfo {author} {\bibfnamefont {M.}~\bibnamefont {R\"osner}}, \bibinfo {author} {\bibfnamefont {M.~I.}\ \bibnamefont {Katsnelson}}, \ and\ \bibinfo {author} {\bibfnamefont {T.~O.}\ \bibnamefont {Wehling}},\ }\href {\doibase 10.1103/PhysRevB.104.045134} {\bibfield  {journal} {\bibinfo  {journal} {Phys. Rev. B}\ }\textbf {\bibinfo {volume} {104}},\ \bibinfo {pages} {045134} (\bibinfo {year} {2021})}\BibitemShut {NoStop}%
\bibitem [{\citenamefont {Aryasetiawan}\ and\ \citenamefont {Karlsson}(2025)}]{PhysRevB.111.075139}%
  \BibitemOpen
  \bibfield  {author} {\bibinfo {author} {\bibfnamefont {F.}~\bibnamefont {Aryasetiawan}}\ and\ \bibinfo {author} {\bibfnamefont {K.}~\bibnamefont {Karlsson}},\ }\href {\doibase 10.1103/PhysRevB.111.075139} {\bibfield  {journal} {\bibinfo  {journal} {Phys. Rev. B}\ }\textbf {\bibinfo {volume} {111}},\ \bibinfo {pages} {075139} (\bibinfo {year} {2025})}\BibitemShut {NoStop}%
\bibitem [{\citenamefont {Xu}\ and\ \citenamefont {Ping}(2024{\natexlab{b}})}]{Xu2024JCTC}%
  \BibitemOpen
  \bibfield  {author} {\bibinfo {author} {\bibfnamefont {J.}~\bibnamefont {Xu}}\ and\ \bibinfo {author} {\bibfnamefont {Y.}~\bibnamefont {Ping}},\ }\href {\doibase 10.1021/acs.jctc.3c00598} {\bibfield  {journal} {\bibinfo  {journal} {Journal of Chemical Theory and Computation}\ }\textbf {\bibinfo {volume} {20}},\ \bibinfo {pages} {492} (\bibinfo {year} {2024}{\natexlab{b}})},\ \bibinfo {note} {pMID: 38157422},\ \Eprint {http://arxiv.org/abs/https://doi.org/10.1021/acs.jctc.3c00598} {https://doi.org/10.1021/acs.jctc.3c00598} \BibitemShut {NoStop}%
\bibitem [{\citenamefont {Giuliani}\ and\ \citenamefont {Vignale}(2005)}]{giuliani2005quantum}%
  \BibitemOpen
  \bibfield  {author} {\bibinfo {author} {\bibfnamefont {G.}~\bibnamefont {Giuliani}}\ and\ \bibinfo {author} {\bibfnamefont {G.}~\bibnamefont {Vignale}},\ }\href {https://books.google.com/books?id=kFkIKRfgUpsC} {\emph {\bibinfo {title} {Quantum Theory of the Electron Liquid}}},\ Masters Series in Physics and Astronomy\ (\bibinfo  {publisher} {Cambridge University Press},\ \bibinfo {year} {2005})\BibitemShut {NoStop}%
\bibitem [{\citenamefont {Wu}\ \emph {et~al.}(2005)\citenamefont {Wu}, \citenamefont {Vanderbilt},\ and\ \citenamefont {Hamann}}]{PhysRevB.72.035105}%
  \BibitemOpen
  \bibfield  {author} {\bibinfo {author} {\bibfnamefont {X.}~\bibnamefont {Wu}}, \bibinfo {author} {\bibfnamefont {D.}~\bibnamefont {Vanderbilt}}, \ and\ \bibinfo {author} {\bibfnamefont {D.~R.}\ \bibnamefont {Hamann}},\ }\href {\doibase 10.1103/PhysRevB.72.035105} {\bibfield  {journal} {\bibinfo  {journal} {Phys. Rev. B}\ }\textbf {\bibinfo {volume} {72}},\ \bibinfo {pages} {035105} (\bibinfo {year} {2005})}\BibitemShut {NoStop}%
\bibitem [{\citenamefont {Baroni}\ \emph {et~al.}(2001)\citenamefont {Baroni}, \citenamefont {de~Gironcoli}, \citenamefont {Dal~Corso},\ and\ \citenamefont {Giannozzi}}]{RevModPhys.73.515}%
  \BibitemOpen
  \bibfield  {author} {\bibinfo {author} {\bibfnamefont {S.}~\bibnamefont {Baroni}}, \bibinfo {author} {\bibfnamefont {S.}~\bibnamefont {de~Gironcoli}}, \bibinfo {author} {\bibfnamefont {A.}~\bibnamefont {Dal~Corso}}, \ and\ \bibinfo {author} {\bibfnamefont {P.}~\bibnamefont {Giannozzi}},\ }\href {\doibase 10.1103/RevModPhys.73.515} {\bibfield  {journal} {\bibinfo  {journal} {Rev. Mod. Phys.}\ }\textbf {\bibinfo {volume} {73}},\ \bibinfo {pages} {515} (\bibinfo {year} {2001})}\BibitemShut {NoStop}%
\bibitem [{\citenamefont {Paleari}\ and\ \citenamefont {Marini}(2022)}]{PhysRevB.106.125403}%
  \BibitemOpen
  \bibfield  {author} {\bibinfo {author} {\bibfnamefont {F.}~\bibnamefont {Paleari}}\ and\ \bibinfo {author} {\bibfnamefont {A.}~\bibnamefont {Marini}},\ }\href {\doibase 10.1103/PhysRevB.106.125403} {\bibfield  {journal} {\bibinfo  {journal} {Phys. Rev. B}\ }\textbf {\bibinfo {volume} {106}},\ \bibinfo {pages} {125403} (\bibinfo {year} {2022})}\BibitemShut {NoStop}%
\bibitem [{\citenamefont {Marzari}\ \emph {et~al.}(2012)\citenamefont {Marzari}, \citenamefont {Mostofi}, \citenamefont {Yates}, \citenamefont {Souza},\ and\ \citenamefont {Vanderbilt}}]{RevModPhys.84.1419}%
  \BibitemOpen
  \bibfield  {author} {\bibinfo {author} {\bibfnamefont {N.}~\bibnamefont {Marzari}}, \bibinfo {author} {\bibfnamefont {A.~A.}\ \bibnamefont {Mostofi}}, \bibinfo {author} {\bibfnamefont {J.~R.}\ \bibnamefont {Yates}}, \bibinfo {author} {\bibfnamefont {I.}~\bibnamefont {Souza}}, \ and\ \bibinfo {author} {\bibfnamefont {D.}~\bibnamefont {Vanderbilt}},\ }\href {\doibase 10.1103/RevModPhys.84.1419} {\bibfield  {journal} {\bibinfo  {journal} {Rev. Mod. Phys.}\ }\textbf {\bibinfo {volume} {84}},\ \bibinfo {pages} {1419} (\bibinfo {year} {2012})}\BibitemShut {NoStop}%
\bibitem [{\citenamefont {Marzari}\ and\ \citenamefont {Vanderbilt}(1997)}]{PhysRevB.56.12847}%
  \BibitemOpen
  \bibfield  {author} {\bibinfo {author} {\bibfnamefont {N.}~\bibnamefont {Marzari}}\ and\ \bibinfo {author} {\bibfnamefont {D.}~\bibnamefont {Vanderbilt}},\ }\href {\doibase 10.1103/PhysRevB.56.12847} {\bibfield  {journal} {\bibinfo  {journal} {Phys. Rev. B}\ }\textbf {\bibinfo {volume} {56}},\ \bibinfo {pages} {12847} (\bibinfo {year} {1997})}\BibitemShut {NoStop}%
\bibitem [{\citenamefont {Narang}\ \emph {et~al.}(2017)\citenamefont {Narang}, \citenamefont {Zhao}, \citenamefont {Claybrook},\ and\ \citenamefont {Sundararaman}}]{https://doi.org/10.1002/adom.201600914}%
  \BibitemOpen
  \bibfield  {author} {\bibinfo {author} {\bibfnamefont {P.}~\bibnamefont {Narang}}, \bibinfo {author} {\bibfnamefont {L.}~\bibnamefont {Zhao}}, \bibinfo {author} {\bibfnamefont {S.}~\bibnamefont {Claybrook}}, \ and\ \bibinfo {author} {\bibfnamefont {R.}~\bibnamefont {Sundararaman}},\ }\href {\doibase https://doi.org/10.1002/adom.201600914} {\bibfield  {journal} {\bibinfo  {journal} {Advanced Optical Materials}\ }\textbf {\bibinfo {volume} {5}},\ \bibinfo {pages} {1600914} (\bibinfo {year} {2017})},\ \Eprint {http://arxiv.org/abs/https://advanced.onlinelibrary.wiley.com/doi/pdf/10.1002/adom.201600914} {https://advanced.onlinelibrary.wiley.com/doi/pdf/10.1002/adom.201600914} \BibitemShut {NoStop}%
\bibitem [{\citenamefont {Brown}\ \emph {et~al.}(2016)\citenamefont {Brown}, \citenamefont {Sundararaman}, \citenamefont {Narang}, \citenamefont {Goddard},\ and\ \citenamefont {Atwater}}]{PhysRevB.94.075120}%
  \BibitemOpen
  \bibfield  {author} {\bibinfo {author} {\bibfnamefont {A.~M.}\ \bibnamefont {Brown}}, \bibinfo {author} {\bibfnamefont {R.}~\bibnamefont {Sundararaman}}, \bibinfo {author} {\bibfnamefont {P.}~\bibnamefont {Narang}}, \bibinfo {author} {\bibfnamefont {W.~A.}\ \bibnamefont {Goddard}}, \ and\ \bibinfo {author} {\bibfnamefont {H.~A.}\ \bibnamefont {Atwater}},\ }\href {\doibase 10.1103/PhysRevB.94.075120} {\bibfield  {journal} {\bibinfo  {journal} {Phys. Rev. B}\ }\textbf {\bibinfo {volume} {94}},\ \bibinfo {pages} {075120} (\bibinfo {year} {2016})}\BibitemShut {NoStop}%
\bibitem [{\citenamefont {Sundararaman}\ \emph {et~al.}(2017)\citenamefont {Sundararaman}, \citenamefont {Letchworth-Weaver}, \citenamefont {Schwarz}, \citenamefont {Gunceler}, \citenamefont {Ozhabes},\ and\ \citenamefont {Arias}}]{SUNDARARAMAN2017278}%
  \BibitemOpen
  \bibfield  {author} {\bibinfo {author} {\bibfnamefont {R.}~\bibnamefont {Sundararaman}}, \bibinfo {author} {\bibfnamefont {K.}~\bibnamefont {Letchworth-Weaver}}, \bibinfo {author} {\bibfnamefont {K.~A.}\ \bibnamefont {Schwarz}}, \bibinfo {author} {\bibfnamefont {D.}~\bibnamefont {Gunceler}}, \bibinfo {author} {\bibfnamefont {Y.}~\bibnamefont {Ozhabes}}, \ and\ \bibinfo {author} {\bibfnamefont {T.}~\bibnamefont {Arias}},\ }\href {\doibase https://doi.org/10.1016/j.softx.2017.10.006} {\bibfield  {journal} {\bibinfo  {journal} {SoftwareX}\ }\textbf {\bibinfo {volume} {6}},\ \bibinfo {pages} {278} (\bibinfo {year} {2017})}\BibitemShut {NoStop}%
\bibitem [{Note1()}]{Note1}%
  \BibitemOpen
  \bibinfo {note} {The same considerations are valid for the incoherent electron-radiation scattering term}\BibitemShut {NoStop}%
\bibitem [{\citenamefont {Dr{\"o}geler}\ \emph {et~al.}(2016)\citenamefont {Dr{\"o}geler}, \citenamefont {Franzen}, \citenamefont {Volmer}, \citenamefont {Pohlmann}, \citenamefont {Banszerus}, \citenamefont {Wolter}, \citenamefont {Watanabe}, \citenamefont {Taniguchi}, \citenamefont {Stampfer},\ and\ \citenamefont {Beschoten}}]{Drogeler2016}%
  \BibitemOpen
  \bibfield  {author} {\bibinfo {author} {\bibfnamefont {M.}~\bibnamefont {Dr{\"o}geler}}, \bibinfo {author} {\bibfnamefont {C.}~\bibnamefont {Franzen}}, \bibinfo {author} {\bibfnamefont {F.}~\bibnamefont {Volmer}}, \bibinfo {author} {\bibfnamefont {T.}~\bibnamefont {Pohlmann}}, \bibinfo {author} {\bibfnamefont {L.}~\bibnamefont {Banszerus}}, \bibinfo {author} {\bibfnamefont {M.}~\bibnamefont {Wolter}}, \bibinfo {author} {\bibfnamefont {K.}~\bibnamefont {Watanabe}}, \bibinfo {author} {\bibfnamefont {T.}~\bibnamefont {Taniguchi}}, \bibinfo {author} {\bibfnamefont {C.}~\bibnamefont {Stampfer}}, \ and\ \bibinfo {author} {\bibfnamefont {B.}~\bibnamefont {Beschoten}},\ }\href {\doibase 10.1021/acs.nanolett.6b00497} {\bibfield  {journal} {\bibinfo  {journal} {Nano Lett.}\ }\textbf {\bibinfo {volume} {16}},\ \bibinfo {pages} {3533} (\bibinfo {year} {2016})}\BibitemShut {NoStop}%
\bibitem [{\citenamefont {Wu}\ \emph {et~al.}(2010)\citenamefont {Wu}, \citenamefont {Jiang},\ and\ \citenamefont {Weng}}]{WU201061}%
  \BibitemOpen
  \bibfield  {author} {\bibinfo {author} {\bibfnamefont {M.}~\bibnamefont {Wu}}, \bibinfo {author} {\bibfnamefont {J.}~\bibnamefont {Jiang}}, \ and\ \bibinfo {author} {\bibfnamefont {M.}~\bibnamefont {Weng}},\ }\href {\doibase https://doi.org/10.1016/j.physrep.2010.04.002} {\bibfield  {journal} {\bibinfo  {journal} {Physics Reports}\ }\textbf {\bibinfo {volume} {493}},\ \bibinfo {pages} {61} (\bibinfo {year} {2010})}\BibitemShut {NoStop}%
\bibitem [{\citenamefont {Elliott}(1954)}]{Elli1954}%
  \BibitemOpen
  \bibfield  {author} {\bibinfo {author} {\bibfnamefont {R.~J.}\ \bibnamefont {Elliott}},\ }\href {\doibase 10.1103/PhysRev.96.266} {\bibfield  {journal} {\bibinfo  {journal} {Phys. Rev.}\ }\textbf {\bibinfo {volume} {96}},\ \bibinfo {pages} {266} (\bibinfo {year} {1954})}\BibitemShut {NoStop}%
\bibitem [{\citenamefont {Yafet}(1963)}]{Yafet1963}%
  \BibitemOpen
  \bibfield  {author} {\bibinfo {author} {\bibfnamefont {Y.}~\bibnamefont {Yafet}},\ }in\ \href@noop {} {\emph {\bibinfo {booktitle} {Solid state physics}}},\ Vol.~\bibinfo {volume} {14}\ (\bibinfo  {publisher} {Elsevier},\ \bibinfo {year} {1963})\ pp.\ \bibinfo {pages} {1--98}\BibitemShut {NoStop}%
\bibitem [{\citenamefont {D{\'{y}}akonov}\ and\ \citenamefont {Pere{\'{l}}}(1971)}]{Dyakonov71}%
  \BibitemOpen
  \bibfield  {author} {\bibinfo {author} {\bibfnamefont {M.}~\bibnamefont {D{\'{y}}akonov}}\ and\ \bibinfo {author} {\bibfnamefont {V.}~\bibnamefont {Pere{\'{l}}}},\ }\href@noop {} {\bibfield  {journal} {\bibinfo  {journal} {Soviet Physics JETP}\ }\textbf {\bibinfo {volume} {33}},\ \bibinfo {pages} {1053} (\bibinfo {year} {1971})}\BibitemShut {NoStop}%
\bibitem [{\citenamefont {{Pikus}}\ and\ \citenamefont {{Bir}}(1971)}]{Pikus1971}%
  \BibitemOpen
  \bibfield  {author} {\bibinfo {author} {\bibfnamefont {G.~E.}\ \bibnamefont {{Pikus}}}\ and\ \bibinfo {author} {\bibfnamefont {G.~L.}\ \bibnamefont {{Bir}}},\ }\href@noop {} {\bibfield  {journal} {\bibinfo  {journal} {Soviet Journal of Experimental and Theoretical Physics}\ }\textbf {\bibinfo {volume} {33}},\ \bibinfo {pages} {108} (\bibinfo {year} {1971})}\BibitemShut {NoStop}%
\bibitem [{\citenamefont {Xu}\ \emph {et~al.}(2020)\citenamefont {Xu}, \citenamefont {Habib}, \citenamefont {Kumar}, \citenamefont {Wu}, \citenamefont {Sundararaman},\ and\ \citenamefont {Ping}}]{XuJ2020}%
  \BibitemOpen
  \bibfield  {author} {\bibinfo {author} {\bibfnamefont {J.}~\bibnamefont {Xu}}, \bibinfo {author} {\bibfnamefont {A.}~\bibnamefont {Habib}}, \bibinfo {author} {\bibfnamefont {S.}~\bibnamefont {Kumar}}, \bibinfo {author} {\bibfnamefont {F.}~\bibnamefont {Wu}}, \bibinfo {author} {\bibfnamefont {R.}~\bibnamefont {Sundararaman}}, \ and\ \bibinfo {author} {\bibfnamefont {Y.}~\bibnamefont {Ping}},\ }\href {\doibase 10.1038/s41467-020-16063-5} {\bibfield  {journal} {\bibinfo  {journal} {Nat Commun.}\ }\textbf {\bibinfo {volume} {11}},\ \bibinfo {pages} {2780} (\bibinfo {year} {2020})}\BibitemShut {NoStop}%
\bibitem [{\citenamefont {Xu}\ \emph {et~al.}(2021{\natexlab{b}})\citenamefont {Xu}, \citenamefont {Wang}, \citenamefont {Zhou},\ and\ \citenamefont {Li}}]{Xu2021}%
  \BibitemOpen
  \bibfield  {author} {\bibinfo {author} {\bibfnamefont {H.}~\bibnamefont {Xu}}, \bibinfo {author} {\bibfnamefont {H.}~\bibnamefont {Wang}}, \bibinfo {author} {\bibfnamefont {J.}~\bibnamefont {Zhou}}, \ and\ \bibinfo {author} {\bibfnamefont {J.}~\bibnamefont {Li}},\ }\href {\doibase https://doi.org/10.1038/s41467-021-24541-7} {\bibfield  {journal} {\bibinfo  {journal} {Nat. Commun.}\ }\textbf {\bibinfo {volume} {12}},\ \bibinfo {pages} {4330} (\bibinfo {year} {2021}{\natexlab{b}})}\BibitemShut {NoStop}%
\bibitem [{\citenamefont {Xu}\ \emph {et~al.}(2024{\natexlab{b}})\citenamefont {Xu}, \citenamefont {Li}, \citenamefont {Huynh}, \citenamefont {Fadel}, \citenamefont {Huang}, \citenamefont {Sundararaman}, \citenamefont {Vardeny},\ and\ \citenamefont {Ping}}]{JXu2024}%
  \BibitemOpen
  \bibfield  {author} {\bibinfo {author} {\bibfnamefont {J.}~\bibnamefont {Xu}}, \bibinfo {author} {\bibfnamefont {K.}~\bibnamefont {Li}}, \bibinfo {author} {\bibfnamefont {U.}~\bibnamefont {Huynh}}, \bibinfo {author} {\bibfnamefont {M.}~\bibnamefont {Fadel}}, \bibinfo {author} {\bibfnamefont {J.}~\bibnamefont {Huang}}, \bibinfo {author} {\bibfnamefont {R.}~\bibnamefont {Sundararaman}}, \bibinfo {author} {\bibfnamefont {V.}~\bibnamefont {Vardeny}}, \ and\ \bibinfo {author} {\bibfnamefont {Y.}~\bibnamefont {Ping}},\ }\href {\doibase https://doi.org/10.1038/s41467-023-42835-w} {\bibfield  {journal} {\bibinfo  {journal} {Nat. Commun.}\ }\textbf {\bibinfo {volume} {15}} (\bibinfo {year} {2024}{\natexlab{b}}),\ https://doi.org/10.1038/s41467-023-42835-w}\BibitemShut {NoStop}%
\bibitem [{\citenamefont {Li}\ \emph {et~al.}(2024)\citenamefont {Li}, \citenamefont {Xu}, \citenamefont {Huynh}, \citenamefont {Bodin}, \citenamefont {Gupta}, \citenamefont {Multunas}, \citenamefont {Simoni}, \citenamefont {Sundararaman}, \citenamefont {Verdany},\ and\ \citenamefont {Ping}}]{KLi2024}%
  \BibitemOpen
  \bibfield  {author} {\bibinfo {author} {\bibfnamefont {K.}~\bibnamefont {Li}}, \bibinfo {author} {\bibfnamefont {J.}~\bibnamefont {Xu}}, \bibinfo {author} {\bibfnamefont {U.~N.}\ \bibnamefont {Huynh}}, \bibinfo {author} {\bibfnamefont {R.}~\bibnamefont {Bodin}}, \bibinfo {author} {\bibfnamefont {M.}~\bibnamefont {Gupta}}, \bibinfo {author} {\bibfnamefont {C.}~\bibnamefont {Multunas}}, \bibinfo {author} {\bibfnamefont {J.}~\bibnamefont {Simoni}}, \bibinfo {author} {\bibfnamefont {R.}~\bibnamefont {Sundararaman}}, \bibinfo {author} {\bibfnamefont {Z.~V.}\ \bibnamefont {Verdany}}, \ and\ \bibinfo {author} {\bibfnamefont {Y.}~\bibnamefont {Ping}},\ }\href {\doibase 10.1021/acs.jpclett.4c02708} {\bibfield  {journal} {\bibinfo  {journal} {J. Phys. Chem. Lett.}\ }\textbf {\bibinfo {volume} {15}},\ \bibinfo {pages} {12156} (\bibinfo {year} {2024})}\BibitemShut {NoStop}%
\bibitem [{\citenamefont {Xu}\ \emph {et~al.}(2021{\natexlab{c}})\citenamefont {Xu}, \citenamefont {Takenaka}, \citenamefont {Habib}, \citenamefont {Sundararaman},\ and\ \citenamefont {Ping}}]{XuJ2021}%
  \BibitemOpen
  \bibfield  {author} {\bibinfo {author} {\bibfnamefont {J.}~\bibnamefont {Xu}}, \bibinfo {author} {\bibfnamefont {H.}~\bibnamefont {Takenaka}}, \bibinfo {author} {\bibfnamefont {A.}~\bibnamefont {Habib}}, \bibinfo {author} {\bibfnamefont {R.}~\bibnamefont {Sundararaman}}, \ and\ \bibinfo {author} {\bibfnamefont {Y.}~\bibnamefont {Ping}},\ }\href {\doibase 10.1021/acs.nanolett.1c03345} {\bibfield  {journal} {\bibinfo  {journal} {Nano Lett.}\ }\textbf {\bibinfo {volume} {21}},\ \bibinfo {pages} {9594} (\bibinfo {year} {2021}{\natexlab{c}})}\BibitemShut {NoStop}%
\bibitem [{\citenamefont {Quinton}\ \emph {et~al.}(2025)\citenamefont {Quinton}, \citenamefont {Fadel}, \citenamefont {Xu}, \citenamefont {Habib}, \citenamefont {Chandra}, \citenamefont {Ping},\ and\ \citenamefont {Sundararaman}}]{PhysRevB.111.115113}%
  \BibitemOpen
  \bibfield  {author} {\bibinfo {author} {\bibfnamefont {J.}~\bibnamefont {Quinton}}, \bibinfo {author} {\bibfnamefont {M.}~\bibnamefont {Fadel}}, \bibinfo {author} {\bibfnamefont {J.}~\bibnamefont {Xu}}, \bibinfo {author} {\bibfnamefont {A.}~\bibnamefont {Habib}}, \bibinfo {author} {\bibfnamefont {M.}~\bibnamefont {Chandra}}, \bibinfo {author} {\bibfnamefont {Y.}~\bibnamefont {Ping}}, \ and\ \bibinfo {author} {\bibfnamefont {R.}~\bibnamefont {Sundararaman}},\ }\href {\doibase 10.1103/PhysRevB.111.115113} {\bibfield  {journal} {\bibinfo  {journal} {Phys. Rev. B}\ }\textbf {\bibinfo {volume} {111}},\ \bibinfo {pages} {115113} (\bibinfo {year} {2025})}\BibitemShut {NoStop}%
\bibitem [{\citenamefont {Gu}\ \emph {et~al.}(2024)\citenamefont {Gu}, \citenamefont {Zheng}, \citenamefont {Jia}, \citenamefont {Shi}, \citenamefont {Zhao}, \citenamefont {Zeng},\ and\ \citenamefont {Zhang}}]{Gu2024}%
  \BibitemOpen
  \bibfield  {author} {\bibinfo {author} {\bibfnamefont {Y.}~\bibnamefont {Gu}}, \bibinfo {author} {\bibfnamefont {Z.}~\bibnamefont {Zheng}}, \bibinfo {author} {\bibfnamefont {L.}~\bibnamefont {Jia}}, \bibinfo {author} {\bibfnamefont {S.}~\bibnamefont {Shi}}, \bibinfo {author} {\bibfnamefont {T.}~\bibnamefont {Zhao}}, \bibinfo {author} {\bibfnamefont {T.}~\bibnamefont {Zeng}}, \ and\ \bibinfo {author} {\bibfnamefont {Q.~a.}\ \bibnamefont {Zhang}},\ }\href {\doibase https://doi.org/10.1002/adfm.202406444} {\bibfield  {journal} {\bibinfo  {journal} {Advanced Functional Materials}\ }\textbf {\bibinfo {volume} {34}},\ \bibinfo {pages} {2406444} (\bibinfo {year} {2024})}\BibitemShut {NoStop}%
\bibitem [{\citenamefont {Watanabe}\ and\ \citenamefont {Yanase}(2021)}]{PhysRevX.11.011001}%
  \BibitemOpen
  \bibfield  {author} {\bibinfo {author} {\bibfnamefont {H.}~\bibnamefont {Watanabe}}\ and\ \bibinfo {author} {\bibfnamefont {Y.}~\bibnamefont {Yanase}},\ }\href {\doibase 10.1103/PhysRevX.11.011001} {\bibfield  {journal} {\bibinfo  {journal} {Phys. Rev. X}\ }\textbf {\bibinfo {volume} {11}},\ \bibinfo {pages} {011001} (\bibinfo {year} {2021})}\BibitemShut {NoStop}%
\bibitem [{\citenamefont {de~Juan}\ \emph {et~al.}(2017)\citenamefont {de~Juan}, \citenamefont {Grushin}, \citenamefont {Morimoto},\ and\ \citenamefont {Moore}}]{deJuan2017}%
  \BibitemOpen
  \bibfield  {author} {\bibinfo {author} {\bibfnamefont {F.}~\bibnamefont {de~Juan}}, \bibinfo {author} {\bibfnamefont {A.}~\bibnamefont {Grushin}}, \bibinfo {author} {\bibfnamefont {T.}~\bibnamefont {Morimoto}}, \ and\ \bibinfo {author} {\bibfnamefont {J.}~\bibnamefont {Moore}},\ }\href {\doibase https://doi.org/10.1038/ncomms15995} {\bibfield  {journal} {\bibinfo  {journal} {Nat. Commun.}\ }\textbf {\bibinfo {volume} {8}},\ \bibinfo {pages} {15995} (\bibinfo {year} {2017})}\BibitemShut {NoStop}%
\bibitem [{\citenamefont {Tan}\ and\ \citenamefont {Rappe}(2016)}]{PhysRevLett.116.237402}%
  \BibitemOpen
  \bibfield  {author} {\bibinfo {author} {\bibfnamefont {L.~Z.}\ \bibnamefont {Tan}}\ and\ \bibinfo {author} {\bibfnamefont {A.~M.}\ \bibnamefont {Rappe}},\ }\href {\doibase 10.1103/PhysRevLett.116.237402} {\bibfield  {journal} {\bibinfo  {journal} {Phys. Rev. Lett.}\ }\textbf {\bibinfo {volume} {116}},\ \bibinfo {pages} {237402} (\bibinfo {year} {2016})}\BibitemShut {NoStop}%
\bibitem [{\citenamefont {Huang}\ \emph {et~al.}(2021)\citenamefont {Huang}, \citenamefont {Taniguchi}, \citenamefont {Shigefuji}, \citenamefont {Kobayashi}, \citenamefont {Matsubara}, \citenamefont {Sasagawa}, \citenamefont {Sato},\ and\ \citenamefont {Miyasaka}}]{Huang2021}%
  \BibitemOpen
  \bibfield  {author} {\bibinfo {author} {\bibfnamefont {P.-J.}\ \bibnamefont {Huang}}, \bibinfo {author} {\bibfnamefont {K.}~\bibnamefont {Taniguchi}}, \bibinfo {author} {\bibfnamefont {M.}~\bibnamefont {Shigefuji}}, \bibinfo {author} {\bibfnamefont {T.}~\bibnamefont {Kobayashi}}, \bibinfo {author} {\bibfnamefont {M.}~\bibnamefont {Matsubara}}, \bibinfo {author} {\bibfnamefont {T.}~\bibnamefont {Sasagawa}}, \bibinfo {author} {\bibfnamefont {H.}~\bibnamefont {Sato}}, \ and\ \bibinfo {author} {\bibfnamefont {H.}~\bibnamefont {Miyasaka}},\ }\href {\doibase https://doi.org/10.1002/adma.202008611} {\bibfield  {journal} {\bibinfo  {journal} {Adv. Mater.}\ }\textbf {\bibinfo {volume} {33}},\ \bibinfo {pages} {2008611} (\bibinfo {year} {2021})}\BibitemShut {NoStop}%
\bibitem [{\citenamefont {Mu}\ \emph {et~al.}(2021)\citenamefont {Mu}, \citenamefont {Pan},\ and\ \citenamefont {Zhou}}]{Mu2021}%
  \BibitemOpen
  \bibfield  {author} {\bibinfo {author} {\bibfnamefont {X.}~\bibnamefont {Mu}}, \bibinfo {author} {\bibfnamefont {Y.}~\bibnamefont {Pan}}, \ and\ \bibinfo {author} {\bibfnamefont {J.}~\bibnamefont {Zhou}},\ }\href {\doibase https://doi.org/10.1038/s41524-021-00531-7} {\bibfield  {journal} {\bibinfo  {journal} {npj Comput. Mater.}\ }\textbf {\bibinfo {volume} {7}},\ \bibinfo {pages} {21} (\bibinfo {year} {2021})}\BibitemShut {NoStop}%
\bibitem [{\citenamefont {Sturman}(2020)}]{Sturman2020-mu}%
  \BibitemOpen
  \bibfield  {author} {\bibinfo {author} {\bibfnamefont {B.~I.}\ \bibnamefont {Sturman}},\ }\href@noop {} {\bibfield  {journal} {\bibinfo  {journal} {Phys.--Usp.}\ }\textbf {\bibinfo {volume} {63}},\ \bibinfo {pages} {407} (\bibinfo {year} {2020})}\BibitemShut {NoStop}%
\bibitem [{\citenamefont {Zhu}\ and\ \citenamefont {Alexandradinata}(2024)}]{PhysRevB.110.115108}%
  \BibitemOpen
  \bibfield  {author} {\bibinfo {author} {\bibfnamefont {P.}~\bibnamefont {Zhu}}\ and\ \bibinfo {author} {\bibfnamefont {A.}~\bibnamefont {Alexandradinata}},\ }\href {\doibase 10.1103/PhysRevB.110.115108} {\bibfield  {journal} {\bibinfo  {journal} {Phys. Rev. B}\ }\textbf {\bibinfo {volume} {110}},\ \bibinfo {pages} {115108} (\bibinfo {year} {2024})}\BibitemShut {NoStop}%
\bibitem [{\citenamefont {Alexandradinata}(2024)}]{PhysRevB.110.075159}%
  \BibitemOpen
  \bibfield  {author} {\bibinfo {author} {\bibfnamefont {A.}~\bibnamefont {Alexandradinata}},\ }\href {\doibase 10.1103/PhysRevB.110.075159} {\bibfield  {journal} {\bibinfo  {journal} {Phys. Rev. B}\ }\textbf {\bibinfo {volume} {110}},\ \bibinfo {pages} {075159} (\bibinfo {year} {2024})}\BibitemShut {NoStop}%
\bibitem [{\citenamefont {Fecher}\ \emph {et~al.}(2022)\citenamefont {Fecher}, \citenamefont {K{\"u}bler},\ and\ \citenamefont {Felser}}]{ma15175812}%
  \BibitemOpen
  \bibfield  {author} {\bibinfo {author} {\bibfnamefont {G.~H.}\ \bibnamefont {Fecher}}, \bibinfo {author} {\bibfnamefont {J.}~\bibnamefont {K{\"u}bler}}, \ and\ \bibinfo {author} {\bibfnamefont {C.}~\bibnamefont {Felser}},\ }\href {\doibase 10.3390/ma15175812} {\bibfield  {journal} {\bibinfo  {journal} {Materials}\ }\textbf {\bibinfo {volume} {15}} (\bibinfo {year} {2022}),\ 10.3390/ma15175812}\BibitemShut {NoStop}%
\bibitem [{\citenamefont {Bloom}\ \emph {et~al.}(2024)\citenamefont {Bloom}, \citenamefont {Paltiel}, \citenamefont {Naaman},\ and\ \citenamefont {Waldeck}}]{CISS}%
  \BibitemOpen
  \bibfield  {author} {\bibinfo {author} {\bibfnamefont {B.}~\bibnamefont {Bloom}}, \bibinfo {author} {\bibfnamefont {Y.}~\bibnamefont {Paltiel}}, \bibinfo {author} {\bibfnamefont {R.}~\bibnamefont {Naaman}}, \ and\ \bibinfo {author} {\bibfnamefont {D.}~\bibnamefont {Waldeck}},\ }\href {\doibase 10.1021/acs.chemrev.3c00661} {\bibfield  {journal} {\bibinfo  {journal} {Chem. Rev.}\ }\textbf {\bibinfo {volume} {124}},\ \bibinfo {pages} {1950} (\bibinfo {year} {2024})}\BibitemShut {NoStop}%
\bibitem [{\citenamefont {Dalum}\ and\ \citenamefont {Hedegard}(2019)}]{Dalum2019}%
  \BibitemOpen
  \bibfield  {author} {\bibinfo {author} {\bibfnamefont {S.}~\bibnamefont {Dalum}}\ and\ \bibinfo {author} {\bibfnamefont {P.}~\bibnamefont {Hedegard}},\ }\href {\doibase 10.1021/acs.nanolett.9b01707} {\bibfield  {journal} {\bibinfo  {journal} {Nano Lett.}\ }\textbf {\bibinfo {volume} {19}},\ \bibinfo {pages} {5253} (\bibinfo {year} {2019})}\BibitemShut {NoStop}%
\bibitem [{\citenamefont {Yeganeh}\ \emph {et~al.}(2009)\citenamefont {Yeganeh}, \citenamefont {Ratner}, \citenamefont {Medina},\ and\ \citenamefont {Mujica}}]{10.1063/1.3167404}%
  \BibitemOpen
  \bibfield  {author} {\bibinfo {author} {\bibfnamefont {S.}~\bibnamefont {Yeganeh}}, \bibinfo {author} {\bibfnamefont {M.~A.}\ \bibnamefont {Ratner}}, \bibinfo {author} {\bibfnamefont {E.}~\bibnamefont {Medina}}, \ and\ \bibinfo {author} {\bibfnamefont {V.}~\bibnamefont {Mujica}},\ }\href {\doibase 10.1063/1.3167404} {\bibfield  {journal} {\bibinfo  {journal} {The Journal of Chemical Physics}\ }\textbf {\bibinfo {volume} {131}},\ \bibinfo {pages} {014707} (\bibinfo {year} {2009})},\ \Eprint {http://arxiv.org/abs/https://pubs.aip.org/aip/jcp/article-pdf/doi/10.1063/1.3167404/16105082/014707\_1\_online.pdf} {https://pubs.aip.org/aip/jcp/article-pdf/doi/10.1063/1.3167404/16105082/014707\_1\_online.pdf} \BibitemShut {NoStop}%
\bibitem [{\citenamefont {Boulougouris}(2013)}]{10.1063/1.4795319}%
  \BibitemOpen
  \bibfield  {author} {\bibinfo {author} {\bibfnamefont {G.~C.}\ \bibnamefont {Boulougouris}},\ }\href {\doibase 10.1063/1.4795319} {\bibfield  {journal} {\bibinfo  {journal} {The Journal of Chemical Physics}\ }\textbf {\bibinfo {volume} {138}},\ \bibinfo {pages} {114111} (\bibinfo {year} {2013})},\ \Eprint {http://arxiv.org/abs/https://pubs.aip.org/aip/jcp/article-pdf/doi/10.1063/1.4795319/13317773/114111\_1\_online.pdf} {https://pubs.aip.org/aip/jcp/article-pdf/doi/10.1063/1.4795319/13317773/114111\_1\_online.pdf} \BibitemShut {NoStop}%
\bibitem [{\citenamefont {Liu}\ \emph {et~al.}(2021)\citenamefont {Liu}, \citenamefont {Xiao}, \citenamefont {Koo},\ and\ \citenamefont {Yan}}]{Liu2021}%
  \BibitemOpen
  \bibfield  {author} {\bibinfo {author} {\bibfnamefont {Y.}~\bibnamefont {Liu}}, \bibinfo {author} {\bibfnamefont {J.}~\bibnamefont {Xiao}}, \bibinfo {author} {\bibfnamefont {J.}~\bibnamefont {Koo}}, \ and\ \bibinfo {author} {\bibfnamefont {B.}~\bibnamefont {Yan}},\ }\href {\doibase https://doi.org/10.1038/s41563-021-00924-5} {\bibfield  {journal} {\bibinfo  {journal} {Nature Materials}\ }\textbf {\bibinfo {volume} {20}},\ \bibinfo {pages} {638} (\bibinfo {year} {2021})}\BibitemShut {NoStop}%
\bibitem [{\citenamefont {Kim}\ \emph {et~al.}(2023)\citenamefont {Kim}, \citenamefont {Vetter}, \citenamefont {Yang}, \citenamefont {Yang}, \citenamefont {Wang}, \citenamefont {Sun}, \citenamefont {Yang}, \citenamefont {Comstock}, \citenamefont {Li}, \citenamefont {Zhou}, \citenamefont {Zhang}, \citenamefont {You}, \citenamefont {Sun},\ and\ \citenamefont {Liu}}]{Kim2023}%
  \BibitemOpen
  \bibfield  {author} {\bibinfo {author} {\bibfnamefont {K.}~\bibnamefont {Kim}}, \bibinfo {author} {\bibfnamefont {E.}~\bibnamefont {Vetter}}, \bibinfo {author} {\bibfnamefont {L.}~\bibnamefont {Yang}}, \bibinfo {author} {\bibfnamefont {C.}~\bibnamefont {Yang}}, \bibinfo {author} {\bibfnamefont {Z.}~\bibnamefont {Wang}}, \bibinfo {author} {\bibfnamefont {R.}~\bibnamefont {Sun}}, \bibinfo {author} {\bibfnamefont {Y.}~\bibnamefont {Yang}}, \bibinfo {author} {\bibfnamefont {A.}~\bibnamefont {Comstock}}, \bibinfo {author} {\bibfnamefont {X.}~\bibnamefont {Li}}, \bibinfo {author} {\bibfnamefont {J.}~\bibnamefont {Zhou}}, \bibinfo {author} {\bibfnamefont {L.}~\bibnamefont {Zhang}}, \bibinfo {author} {\bibfnamefont {W.}~\bibnamefont {You}}, \bibinfo {author} {\bibfnamefont {D.}~\bibnamefont {Sun}}, \ and\ \bibinfo {author} {\bibfnamefont {J.}~\bibnamefont {Liu}},\ }\href {\doibase 10.1038/s41563-023-01473-9} {\bibfield  {journal} {\bibinfo  {journal} {Nat. Mater.}\ }\textbf {\bibinfo {volume} {22}},\ \bibinfo
  {pages} {322} (\bibinfo {year} {2023})}\BibitemShut {NoStop}%
\bibitem [{\citenamefont {Tauchert}\ \emph {et~al.}(2022)\citenamefont {Tauchert}, \citenamefont {Volkov}, \citenamefont {Ehberger}, \citenamefont {Kazenwadel}, \citenamefont {Evers}, \citenamefont {Lange}, \citenamefont {Donges}, \citenamefont {Book}, \citenamefont {Kreuzpaintner}, \citenamefont {Nowak},\ and\ \citenamefont {Baum}}]{Tauchert2022}%
  \BibitemOpen
  \bibfield  {author} {\bibinfo {author} {\bibfnamefont {S.}~\bibnamefont {Tauchert}}, \bibinfo {author} {\bibfnamefont {M.}~\bibnamefont {Volkov}}, \bibinfo {author} {\bibfnamefont {D.}~\bibnamefont {Ehberger}}, \bibinfo {author} {\bibfnamefont {D.}~\bibnamefont {Kazenwadel}}, \bibinfo {author} {\bibfnamefont {M.}~\bibnamefont {Evers}}, \bibinfo {author} {\bibfnamefont {H.}~\bibnamefont {Lange}}, \bibinfo {author} {\bibfnamefont {A.}~\bibnamefont {Donges}}, \bibinfo {author} {\bibfnamefont {A.}~\bibnamefont {Book}}, \bibinfo {author} {\bibfnamefont {W.}~\bibnamefont {Kreuzpaintner}}, \bibinfo {author} {\bibfnamefont {U.}~\bibnamefont {Nowak}}, \ and\ \bibinfo {author} {\bibfnamefont {P.}~\bibnamefont {Baum}},\ }\href {\doibase https://doi.org/10.1038/s41586-021-04306-4} {\bibfield  {journal} {\bibinfo  {journal} {Nature}\ }\textbf {\bibinfo {volume} {602}},\ \bibinfo {pages} {73} (\bibinfo {year} {2022})}\BibitemShut {NoStop}%
\bibitem [{\citenamefont {F\"ahnle}\ and\ \citenamefont {Illg}(2011)}]{Illg2011}%
  \BibitemOpen
  \bibfield  {author} {\bibinfo {author} {\bibfnamefont {M.}~\bibnamefont {F\"ahnle}}\ and\ \bibinfo {author} {\bibfnamefont {C.}~\bibnamefont {Illg}},\ }\href {\doibase 10.1088/0953-8984/23/49/493201} {\bibfield  {journal} {\bibinfo  {journal} {J. Phys.: Condens. Matt.}\ }\textbf {\bibinfo {volume} {23}},\ \bibinfo {pages} {493201} (\bibinfo {year} {2011})}\BibitemShut {NoStop}%
\bibitem [{\citenamefont {Beaurepaire}\ \emph {et~al.}(1996)\citenamefont {Beaurepaire}, \citenamefont {Merle}, \citenamefont {Daunois},\ and\ \citenamefont {Bigot}}]{PhysRevLett.76.4250}%
  \BibitemOpen
  \bibfield  {author} {\bibinfo {author} {\bibfnamefont {E.}~\bibnamefont {Beaurepaire}}, \bibinfo {author} {\bibfnamefont {J.-C.}\ \bibnamefont {Merle}}, \bibinfo {author} {\bibfnamefont {A.}~\bibnamefont {Daunois}}, \ and\ \bibinfo {author} {\bibfnamefont {J.-Y.}\ \bibnamefont {Bigot}},\ }\href {\doibase 10.1103/PhysRevLett.76.4250} {\bibfield  {journal} {\bibinfo  {journal} {Phys. Rev. Lett.}\ }\textbf {\bibinfo {volume} {76}},\ \bibinfo {pages} {4250} (\bibinfo {year} {1996})}\BibitemShut {NoStop}%
\bibitem [{\citenamefont {Krau\ss{}}\ \emph {et~al.}(2009)\citenamefont {Krau\ss{}}, \citenamefont {Roth}, \citenamefont {Alebrand}, \citenamefont {Steil}, \citenamefont {Cinchetti}, \citenamefont {Aeschlimann},\ and\ \citenamefont {Schneider}}]{PhysRevB.80.180407}%
  \BibitemOpen
  \bibfield  {author} {\bibinfo {author} {\bibfnamefont {M.}~\bibnamefont {Krau\ss{}}}, \bibinfo {author} {\bibfnamefont {T.}~\bibnamefont {Roth}}, \bibinfo {author} {\bibfnamefont {S.}~\bibnamefont {Alebrand}}, \bibinfo {author} {\bibfnamefont {D.}~\bibnamefont {Steil}}, \bibinfo {author} {\bibfnamefont {M.}~\bibnamefont {Cinchetti}}, \bibinfo {author} {\bibfnamefont {M.}~\bibnamefont {Aeschlimann}}, \ and\ \bibinfo {author} {\bibfnamefont {H.~C.}\ \bibnamefont {Schneider}},\ }\href {\doibase 10.1103/PhysRevB.80.180407} {\bibfield  {journal} {\bibinfo  {journal} {Phys. Rev. B}\ }\textbf {\bibinfo {volume} {80}},\ \bibinfo {pages} {180407} (\bibinfo {year} {2009})}\BibitemShut {NoStop}%
\bibitem [{\citenamefont {Hinschberger}\ and\ \citenamefont {Hervieux}(2012)}]{HINSCHBERGER2012813}%
  \BibitemOpen
  \bibfield  {author} {\bibinfo {author} {\bibfnamefont {Y.}~\bibnamefont {Hinschberger}}\ and\ \bibinfo {author} {\bibfnamefont {P.-A.}\ \bibnamefont {Hervieux}},\ }\href {\doibase https://doi.org/10.1016/j.physleta.2012.01.023} {\bibfield  {journal} {\bibinfo  {journal} {Physics Letters A}\ }\textbf {\bibinfo {volume} {376}},\ \bibinfo {pages} {813} (\bibinfo {year} {2012})}\BibitemShut {NoStop}%
\bibitem [{\citenamefont {Simoni}\ \emph {et~al.}(2017)\citenamefont {Simoni}, \citenamefont {Stamenova},\ and\ \citenamefont {Sanvito}}]{PhysRevB.95.024412}%
  \BibitemOpen
  \bibfield  {author} {\bibinfo {author} {\bibfnamefont {J.}~\bibnamefont {Simoni}}, \bibinfo {author} {\bibfnamefont {M.}~\bibnamefont {Stamenova}}, \ and\ \bibinfo {author} {\bibfnamefont {S.}~\bibnamefont {Sanvito}},\ }\href {\doibase 10.1103/PhysRevB.95.024412} {\bibfield  {journal} {\bibinfo  {journal} {Phys. Rev. B}\ }\textbf {\bibinfo {volume} {95}},\ \bibinfo {pages} {024412} (\bibinfo {year} {2017})}\BibitemShut {NoStop}%
\bibitem [{\citenamefont {Carpene}\ \emph {et~al.}(2008)\citenamefont {Carpene}, \citenamefont {Mancini}, \citenamefont {Dallera}, \citenamefont {Brenna}, \citenamefont {Puppin},\ and\ \citenamefont {De~Silvestri}}]{PhysRevB.78.174422}%
  \BibitemOpen
  \bibfield  {author} {\bibinfo {author} {\bibfnamefont {E.}~\bibnamefont {Carpene}}, \bibinfo {author} {\bibfnamefont {E.}~\bibnamefont {Mancini}}, \bibinfo {author} {\bibfnamefont {C.}~\bibnamefont {Dallera}}, \bibinfo {author} {\bibfnamefont {M.}~\bibnamefont {Brenna}}, \bibinfo {author} {\bibfnamefont {E.}~\bibnamefont {Puppin}}, \ and\ \bibinfo {author} {\bibfnamefont {S.}~\bibnamefont {De~Silvestri}},\ }\href {\doibase 10.1103/PhysRevB.78.174422} {\bibfield  {journal} {\bibinfo  {journal} {Phys. Rev. B}\ }\textbf {\bibinfo {volume} {78}},\ \bibinfo {pages} {174422} (\bibinfo {year} {2008})}\BibitemShut {NoStop}%
\bibitem [{\citenamefont {Battiato}\ \emph {et~al.}(2010)\citenamefont {Battiato}, \citenamefont {Carva},\ and\ \citenamefont {Oppeneer}}]{PhysRevLett.105.027203}%
  \BibitemOpen
  \bibfield  {author} {\bibinfo {author} {\bibfnamefont {M.}~\bibnamefont {Battiato}}, \bibinfo {author} {\bibfnamefont {K.}~\bibnamefont {Carva}}, \ and\ \bibinfo {author} {\bibfnamefont {P.~M.}\ \bibnamefont {Oppeneer}},\ }\href {\doibase 10.1103/PhysRevLett.105.027203} {\bibfield  {journal} {\bibinfo  {journal} {Phys. Rev. Lett.}\ }\textbf {\bibinfo {volume} {105}},\ \bibinfo {pages} {027203} (\bibinfo {year} {2010})}\BibitemShut {NoStop}%
\bibitem [{\citenamefont {Imamura}\ \emph {et~al.}(2022)\citenamefont {Imamura}, \citenamefont {Arai}, \citenamefont {Matsumoto}, \citenamefont {Yamaji},\ and\ \citenamefont {Tsukahara}}]{IMAMURA2022169209}%
  \BibitemOpen
  \bibfield  {author} {\bibinfo {author} {\bibfnamefont {H.}~\bibnamefont {Imamura}}, \bibinfo {author} {\bibfnamefont {H.}~\bibnamefont {Arai}}, \bibinfo {author} {\bibfnamefont {R.}~\bibnamefont {Matsumoto}}, \bibinfo {author} {\bibfnamefont {T.}~\bibnamefont {Yamaji}}, \ and\ \bibinfo {author} {\bibfnamefont {H.}~\bibnamefont {Tsukahara}},\ }\href {\doibase https://doi.org/10.1016/j.jmmm.2022.169209} {\bibfield  {journal} {\bibinfo  {journal} {Journal of Magnetism and Magnetic Materials}\ }\textbf {\bibinfo {volume} {553}},\ \bibinfo {pages} {169209} (\bibinfo {year} {2022})}\BibitemShut {NoStop}%
\bibitem [{\citenamefont {Yuan}\ \emph {et~al.}(2022)\citenamefont {Yuan}, \citenamefont {Cao}, \citenamefont {Kamra}, \citenamefont {Duine},\ and\ \citenamefont {Yan}}]{YUAN20221}%
  \BibitemOpen
  \bibfield  {author} {\bibinfo {author} {\bibfnamefont {H.}~\bibnamefont {Yuan}}, \bibinfo {author} {\bibfnamefont {Y.}~\bibnamefont {Cao}}, \bibinfo {author} {\bibfnamefont {A.}~\bibnamefont {Kamra}}, \bibinfo {author} {\bibfnamefont {R.~A.}\ \bibnamefont {Duine}}, \ and\ \bibinfo {author} {\bibfnamefont {P.}~\bibnamefont {Yan}},\ }\href {\doibase https://doi.org/10.1016/j.physrep.2022.03.002} {\bibfield  {journal} {\bibinfo  {journal} {Physics Reports}\ }\textbf {\bibinfo {volume} {965}},\ \bibinfo {pages} {1} (\bibinfo {year} {2022})},\ \bibinfo {note} {quantum magnonics: When magnon spintronics meets quantum information science}\BibitemShut {NoStop}%
\bibitem [{\citenamefont {Soykal}\ and\ \citenamefont {Flatt\'e}(2010)}]{PhysRevLett.104.077202}%
  \BibitemOpen
  \bibfield  {author} {\bibinfo {author} {\bibfnamefont {O.~O.}\ \bibnamefont {Soykal}}\ and\ \bibinfo {author} {\bibfnamefont {M.~E.}\ \bibnamefont {Flatt\'e}},\ }\href {\doibase 10.1103/PhysRevLett.104.077202} {\bibfield  {journal} {\bibinfo  {journal} {Phys. Rev. Lett.}\ }\textbf {\bibinfo {volume} {104}},\ \bibinfo {pages} {077202} (\bibinfo {year} {2010})}\BibitemShut {NoStop}%
\bibitem [{\citenamefont {Lachance-Quirion}\ \emph {et~al.}(2020)\citenamefont {Lachance-Quirion}, \citenamefont {Wolski}, \citenamefont {Tabuchi}, \citenamefont {Kono}, \citenamefont {Usami},\ and\ \citenamefont {Nakamura}}]{https://doi.org/10.1126/science.aaz9236}%
  \BibitemOpen
  \bibfield  {author} {\bibinfo {author} {\bibfnamefont {D.}~\bibnamefont {Lachance-Quirion}}, \bibinfo {author} {\bibfnamefont {S.}~\bibnamefont {Wolski}}, \bibinfo {author} {\bibfnamefont {Y.}~\bibnamefont {Tabuchi}}, \bibinfo {author} {\bibfnamefont {S.}~\bibnamefont {Kono}}, \bibinfo {author} {\bibfnamefont {K.}~\bibnamefont {Usami}}, \ and\ \bibinfo {author} {\bibfnamefont {Y.}~\bibnamefont {Nakamura}},\ }\href {\doibase https://doi.org/10.1126/science.aaz9236} {\bibfield  {journal} {\bibinfo  {journal} {Science}\ }\textbf {\bibinfo {volume} {367}},\ \bibinfo {pages} {425} (\bibinfo {year} {2020})}\BibitemShut {NoStop}%
\bibitem [{\citenamefont {Fang}\ \emph {et~al.}(2025)\citenamefont {Fang}, \citenamefont {Simoni},\ and\ \citenamefont {Ping}}]{PhysRevB.111.104431}%
  \BibitemOpen
  \bibfield  {author} {\bibinfo {author} {\bibfnamefont {W.}~\bibnamefont {Fang}}, \bibinfo {author} {\bibfnamefont {J.}~\bibnamefont {Simoni}}, \ and\ \bibinfo {author} {\bibfnamefont {Y.}~\bibnamefont {Ping}},\ }\href {\doibase 10.1103/PhysRevB.111.104431} {\bibfield  {journal} {\bibinfo  {journal} {Phys. Rev. B}\ }\textbf {\bibinfo {volume} {111}},\ \bibinfo {pages} {104431} (\bibinfo {year} {2025})}\BibitemShut {NoStop}%
\bibitem [{\citenamefont {Burkard}\ \emph {et~al.}(2023)\citenamefont {Burkard}, \citenamefont {Ladd}, \citenamefont {Pan}, \citenamefont {Nichol},\ and\ \citenamefont {Petta}}]{RevModPhys.95.025003}%
  \BibitemOpen
  \bibfield  {author} {\bibinfo {author} {\bibfnamefont {G.}~\bibnamefont {Burkard}}, \bibinfo {author} {\bibfnamefont {T.~D.}\ \bibnamefont {Ladd}}, \bibinfo {author} {\bibfnamefont {A.}~\bibnamefont {Pan}}, \bibinfo {author} {\bibfnamefont {J.~M.}\ \bibnamefont {Nichol}}, \ and\ \bibinfo {author} {\bibfnamefont {J.~R.}\ \bibnamefont {Petta}},\ }\href {\doibase 10.1103/RevModPhys.95.025003} {\bibfield  {journal} {\bibinfo  {journal} {Rev. Mod. Phys.}\ }\textbf {\bibinfo {volume} {95}},\ \bibinfo {pages} {025003} (\bibinfo {year} {2023})}\BibitemShut {NoStop}%
\bibitem [{\citenamefont {Hu}\ \emph {et~al.}(1997)\citenamefont {Hu}, \citenamefont {De~Sousa},\ and\ \citenamefont {Das~Sarma}}]{Hu_foundQM}%
  \BibitemOpen
  \bibfield  {author} {\bibinfo {author} {\bibfnamefont {X.}~\bibnamefont {Hu}}, \bibinfo {author} {\bibfnamefont {R.}~\bibnamefont {De~Sousa}}, \ and\ \bibinfo {author} {\bibfnamefont {S.}~\bibnamefont {Das~Sarma}},\ }\enquote {\bibinfo {title} {Decoherence and dephasing in spin-based solid state quantum computers},}\ in\ \href {\doibase 10.1142/9789812776716_0001} {\emph {\bibinfo {booktitle} {Foundations of Quantum Mechanics in the Light of New Technology}}}\ (\bibinfo  {publisher} {World Scientific Publishing Company},\ \bibinfo {year} {1997})\ pp.\ \bibinfo {pages} {3--11}\BibitemShut {NoStop}%
\bibitem [{\citenamefont {Awschalom}\ \emph {et~al.}(2018)\citenamefont {Awschalom}, \citenamefont {Hanson}, \citenamefont {Wrachtrup},\ and\ \citenamefont {B.B.}}]{Awschalom2018}%
  \BibitemOpen
  \bibfield  {author} {\bibinfo {author} {\bibfnamefont {D.}~\bibnamefont {Awschalom}}, \bibinfo {author} {\bibfnamefont {R.}~\bibnamefont {Hanson}}, \bibinfo {author} {\bibfnamefont {J.}~\bibnamefont {Wrachtrup}}, \ and\ \bibinfo {author} {\bibfnamefont {Z.}~\bibnamefont {B.B.}},\ }\href {\doibase https://doi.org/10.1038/s41566-018-0232-2} {\bibfield  {journal} {\bibinfo  {journal} {Nature Photon.}\ }\textbf {\bibinfo {volume} {12}},\ \bibinfo {pages} {516} (\bibinfo {year} {2018})}\BibitemShut {NoStop}%
\bibitem [{\citenamefont {Ghosh}\ \emph {et~al.}(2021)\citenamefont {Ghosh}, \citenamefont {Ma}, \citenamefont {Onizhuk}, \citenamefont {Gavini},\ and\ \citenamefont {Galli}}]{Ghosh2021}%
  \BibitemOpen
  \bibfield  {author} {\bibinfo {author} {\bibfnamefont {K.}~\bibnamefont {Ghosh}}, \bibinfo {author} {\bibfnamefont {H.}~\bibnamefont {Ma}}, \bibinfo {author} {\bibfnamefont {M.}~\bibnamefont {Onizhuk}}, \bibinfo {author} {\bibfnamefont {V.}~\bibnamefont {Gavini}}, \ and\ \bibinfo {author} {\bibfnamefont {G.}~\bibnamefont {Galli}},\ }\href {\doibase 10.1038/s41524-021-00590-w} {\bibfield  {journal} {\bibinfo  {journal} {npj Computational Materials}\ }\textbf {\bibinfo {volume} {7}},\ \bibinfo {pages} {123} (\bibinfo {year} {2021})}\BibitemShut {NoStop}%
\bibitem [{\citenamefont {Hanson}\ \emph {et~al.}(2008)\citenamefont {Hanson}, \citenamefont {Dobrovitski}, \citenamefont {Feiguin}, \citenamefont {Gywat},\ and\ \citenamefont {Awschalom}}]{doi:10.1126/science.1155400}%
  \BibitemOpen
  \bibfield  {author} {\bibinfo {author} {\bibfnamefont {R.}~\bibnamefont {Hanson}}, \bibinfo {author} {\bibfnamefont {V.~V.}\ \bibnamefont {Dobrovitski}}, \bibinfo {author} {\bibfnamefont {A.~E.}\ \bibnamefont {Feiguin}}, \bibinfo {author} {\bibfnamefont {O.}~\bibnamefont {Gywat}}, \ and\ \bibinfo {author} {\bibfnamefont {D.~D.}\ \bibnamefont {Awschalom}},\ }\href {\doibase 10.1126/science.1155400} {\bibfield  {journal} {\bibinfo  {journal} {Science}\ }\textbf {\bibinfo {volume} {320}},\ \bibinfo {pages} {352} (\bibinfo {year} {2008})},\ \Eprint {http://arxiv.org/abs/https://www.science.org/doi/pdf/10.1126/science.1155400} {https://www.science.org/doi/pdf/10.1126/science.1155400} \BibitemShut {NoStop}%
\bibitem [{\citenamefont {Xiang}\ \emph {et~al.}(2013)\citenamefont {Xiang}, \citenamefont {Ashhab}, \citenamefont {You},\ and\ \citenamefont {Nori}}]{RevModPhys.85.623}%
  \BibitemOpen
  \bibfield  {author} {\bibinfo {author} {\bibfnamefont {Z.-L.}\ \bibnamefont {Xiang}}, \bibinfo {author} {\bibfnamefont {S.}~\bibnamefont {Ashhab}}, \bibinfo {author} {\bibfnamefont {J.~Q.}\ \bibnamefont {You}}, \ and\ \bibinfo {author} {\bibfnamefont {F.}~\bibnamefont {Nori}},\ }\href {\doibase 10.1103/RevModPhys.85.623} {\bibfield  {journal} {\bibinfo  {journal} {Rev. Mod. Phys.}\ }\textbf {\bibinfo {volume} {85}},\ \bibinfo {pages} {623} (\bibinfo {year} {2013})}\BibitemShut {NoStop}%
\bibitem [{\citenamefont {Hope}(1997)}]{PhysRevA.55.R2531}%
  \BibitemOpen
  \bibfield  {author} {\bibinfo {author} {\bibfnamefont {J.~J.}\ \bibnamefont {Hope}},\ }\href {\doibase 10.1103/PhysRevA.55.R2531} {\bibfield  {journal} {\bibinfo  {journal} {Phys. Rev. A}\ }\textbf {\bibinfo {volume} {55}},\ \bibinfo {pages} {R2531} (\bibinfo {year} {1997})}\BibitemShut {NoStop}%
\bibitem [{\citenamefont {Hope}\ \emph {et~al.}(2000)\citenamefont {Hope}, \citenamefont {Moy}, \citenamefont {Collett},\ and\ \citenamefont {Savage}}]{PhysRevA.61.023603}%
  \BibitemOpen
  \bibfield  {author} {\bibinfo {author} {\bibfnamefont {J.~J.}\ \bibnamefont {Hope}}, \bibinfo {author} {\bibfnamefont {G.~M.}\ \bibnamefont {Moy}}, \bibinfo {author} {\bibfnamefont {M.~J.}\ \bibnamefont {Collett}}, \ and\ \bibinfo {author} {\bibfnamefont {C.~M.}\ \bibnamefont {Savage}},\ }\href {\doibase 10.1103/PhysRevA.61.023603} {\bibfield  {journal} {\bibinfo  {journal} {Phys. Rev. A}\ }\textbf {\bibinfo {volume} {61}},\ \bibinfo {pages} {023603} (\bibinfo {year} {2000})}\BibitemShut {NoStop}%
\bibitem [{\citenamefont {Jaksch}\ and\ \citenamefont {Zoller}(2005)}]{JAKSCH200552}%
  \BibitemOpen
  \bibfield  {author} {\bibinfo {author} {\bibfnamefont {D.}~\bibnamefont {Jaksch}}\ and\ \bibinfo {author} {\bibfnamefont {P.}~\bibnamefont {Zoller}},\ }\href {\doibase https://doi.org/10.1016/j.aop.2004.09.010} {\bibfield  {journal} {\bibinfo  {journal} {Annals of Physics}\ }\textbf {\bibinfo {volume} {315}},\ \bibinfo {pages} {52} (\bibinfo {year} {2005})},\ \bibinfo {note} {special Issue}\BibitemShut {NoStop}%
\bibitem [{\citenamefont {Sokolovski}\ and\ \citenamefont {Gurvitz}(2009)}]{PhysRevA.79.032106}%
  \BibitemOpen
  \bibfield  {author} {\bibinfo {author} {\bibfnamefont {D.}~\bibnamefont {Sokolovski}}\ and\ \bibinfo {author} {\bibfnamefont {S.~A.}\ \bibnamefont {Gurvitz}},\ }\href {\doibase 10.1103/PhysRevA.79.032106} {\bibfield  {journal} {\bibinfo  {journal} {Phys. Rev. A}\ }\textbf {\bibinfo {volume} {79}},\ \bibinfo {pages} {032106} (\bibinfo {year} {2009})}\BibitemShut {NoStop}%
\bibitem [{\citenamefont {Navarrete-Benlloch}\ \emph {et~al.}(2011)\citenamefont {Navarrete-Benlloch}, \citenamefont {de~Vega}, \citenamefont {Porras},\ and\ \citenamefont {Cirac}}]{Navarrete2011}%
  \BibitemOpen
  \bibfield  {author} {\bibinfo {author} {\bibfnamefont {C.}~\bibnamefont {Navarrete-Benlloch}}, \bibinfo {author} {\bibfnamefont {I.}~\bibnamefont {de~Vega}}, \bibinfo {author} {\bibfnamefont {D.}~\bibnamefont {Porras}}, \ and\ \bibinfo {author} {\bibfnamefont {J.}~\bibnamefont {Cirac}},\ }\href {\doibase 10.1088/1367-2630/13/2/023024} {\bibfield  {journal} {\bibinfo  {journal} {New J. Phys.}\ }\textbf {\bibinfo {volume} {13}},\ \bibinfo {pages} {023024} (\bibinfo {year} {2011})}\BibitemShut {NoStop}%
\bibitem [{\citenamefont {Spataru}\ \emph {et~al.}(2005)\citenamefont {Spataru}, \citenamefont {Ismail-Beigi}, \citenamefont {Capaz},\ and\ \citenamefont {Louie}}]{Spat2005}%
  \BibitemOpen
  \bibfield  {author} {\bibinfo {author} {\bibfnamefont {C.~D.}\ \bibnamefont {Spataru}}, \bibinfo {author} {\bibfnamefont {S.}~\bibnamefont {Ismail-Beigi}}, \bibinfo {author} {\bibfnamefont {R.~B.}\ \bibnamefont {Capaz}}, \ and\ \bibinfo {author} {\bibfnamefont {S.~G.}\ \bibnamefont {Louie}},\ }\href {\doibase 10.1103/PhysRevLett.95.247402} {\bibfield  {journal} {\bibinfo  {journal} {Phys. Rev. Lett.}\ }\textbf {\bibinfo {volume} {95}},\ \bibinfo {pages} {247402} (\bibinfo {year} {2005})}\BibitemShut {NoStop}%
\bibitem [{\citenamefont {Jankovi\ifmmode~\acute{c}\else \'{c}\fi{}}\ and\ \citenamefont {Vukmirovi\ifmmode~\acute{c}\else \'{c}\fi{}}(2015)}]{PhysRevB.92.235208}%
  \BibitemOpen
  \bibfield  {author} {\bibinfo {author} {\bibfnamefont {V.}~\bibnamefont {Jankovi\ifmmode~\acute{c}\else \'{c}\fi{}}}\ and\ \bibinfo {author} {\bibfnamefont {N.}~\bibnamefont {Vukmirovi\ifmmode~\acute{c}\else \'{c}\fi{}}},\ }\href {\doibase 10.1103/PhysRevB.92.235208} {\bibfield  {journal} {\bibinfo  {journal} {Phys. Rev. B}\ }\textbf {\bibinfo {volume} {92}},\ \bibinfo {pages} {235208} (\bibinfo {year} {2015})}\BibitemShut {NoStop}%
\bibitem [{\citenamefont {Malakhov}\ \emph {et~al.}(2024)\citenamefont {Malakhov}, \citenamefont {Cistaro}, \citenamefont {Mart{\'i}n},\ and\ \citenamefont {Pic{\'o}n}}]{Malakhov2024}%
  \BibitemOpen
  \bibfield  {author} {\bibinfo {author} {\bibfnamefont {M.}~\bibnamefont {Malakhov}}, \bibinfo {author} {\bibfnamefont {G.}~\bibnamefont {Cistaro}}, \bibinfo {author} {\bibfnamefont {F.}~\bibnamefont {Mart{\'i}n}}, \ and\ \bibinfo {author} {\bibfnamefont {A.}~\bibnamefont {Pic{\'o}n}},\ }\href {https://doi.org/10.1038/s42005-024-01689-4} {\bibfield  {journal} {\bibinfo  {journal} {Commun. Phys.}\ }\textbf {\bibinfo {volume} {7}},\ \bibinfo {pages} {196} (\bibinfo {year} {2024})}\BibitemShut {NoStop}%
\bibitem [{\citenamefont {Antonius}\ and\ \citenamefont {Louie}(2022)}]{ant2022}%
  \BibitemOpen
  \bibfield  {author} {\bibinfo {author} {\bibfnamefont {G.}~\bibnamefont {Antonius}}\ and\ \bibinfo {author} {\bibfnamefont {S.~G.}\ \bibnamefont {Louie}},\ }\href {\doibase 10.1103/PhysRevB.105.085111} {\bibfield  {journal} {\bibinfo  {journal} {Phys. Rev. B}\ }\textbf {\bibinfo {volume} {105}},\ \bibinfo {pages} {085111} (\bibinfo {year} {2022})}\BibitemShut {NoStop}%
\bibitem [{\citenamefont {Cohen}\ \emph {et~al.}(2024)\citenamefont {Cohen}, \citenamefont {Haber}, \citenamefont {Neaton}, \citenamefont {Qiu},\ and\ \citenamefont {Refaely-Abramson}}]{PhysRevLett.132.126902}%
  \BibitemOpen
  \bibfield  {author} {\bibinfo {author} {\bibfnamefont {G.}~\bibnamefont {Cohen}}, \bibinfo {author} {\bibfnamefont {J.~B.}\ \bibnamefont {Haber}}, \bibinfo {author} {\bibfnamefont {J.~B.}\ \bibnamefont {Neaton}}, \bibinfo {author} {\bibfnamefont {D.~Y.}\ \bibnamefont {Qiu}}, \ and\ \bibinfo {author} {\bibfnamefont {S.}~\bibnamefont {Refaely-Abramson}},\ }\href {\doibase 10.1103/PhysRevLett.132.126902} {\bibfield  {journal} {\bibinfo  {journal} {Phys. Rev. Lett.}\ }\textbf {\bibinfo {volume} {132}},\ \bibinfo {pages} {126902} (\bibinfo {year} {2024})}\BibitemShut {NoStop}%
\bibitem [{\citenamefont {Amit}\ \emph {et~al.}(2025)\citenamefont {Amit}, \citenamefont {Vosco}, \citenamefont {Ben},\ and\ \citenamefont {Refaely-Abramson}}]{amit2025abinitiodensitymatrixapproachexciton}%
  \BibitemOpen
  \bibfield  {author} {\bibinfo {author} {\bibfnamefont {T.}~\bibnamefont {Amit}}, \bibinfo {author} {\bibfnamefont {G.}~\bibnamefont {Vosco}}, \bibinfo {author} {\bibfnamefont {M.~D.}\ \bibnamefont {Ben}}, \ and\ \bibinfo {author} {\bibfnamefont {S.}~\bibnamefont {Refaely-Abramson}},\ }\href {https://arxiv.org/abs/2505.07021} {\enquote {\bibinfo {title} {Ab-initio density-matrix approach to exciton coherence: phonon scattering, coulomb interactions and radiative recombination},}\ } (\bibinfo {year} {2025}),\ \Eprint {http://arxiv.org/abs/2505.07021} {arXiv:2505.07021 [cond-mat.mtrl-sci]} \BibitemShut {NoStop}%
\bibitem [{\citenamefont {Guo}\ \emph {et~al.}(2025)\citenamefont {Guo}, \citenamefont {Riva}, \citenamefont {Simoni}, \citenamefont {Xu},\ and\ \citenamefont {Ping}}]{guo2025phononassistedradiativelifetimesexciton}%
  \BibitemOpen
  \bibfield  {author} {\bibinfo {author} {\bibfnamefont {C.}~\bibnamefont {Guo}}, \bibinfo {author} {\bibfnamefont {G.}~\bibnamefont {Riva}}, \bibinfo {author} {\bibfnamefont {J.}~\bibnamefont {Simoni}}, \bibinfo {author} {\bibfnamefont {J.}~\bibnamefont {Xu}}, \ and\ \bibinfo {author} {\bibfnamefont {Y.}~\bibnamefont {Ping}},\ }\href {https://arxiv.org/abs/2504.18071} {\enquote {\bibinfo {title} {Phonon-assisted radiative lifetimes and exciton dynamics from first principles},}\ } (\bibinfo {year} {2025}),\ \Eprint {http://arxiv.org/abs/2504.18071} {arXiv:2504.18071 [cond-mat.mtrl-sci]} \BibitemShut {NoStop}%
\bibitem [{\citenamefont {Davis}\ \emph {et~al.}(2020)\citenamefont {Davis}, \citenamefont {Angermeier}, \citenamefont {Hermsmeier},\ and\ \citenamefont {White}}]{PhysRevResearch.2.043139}%
  \BibitemOpen
  \bibfield  {author} {\bibinfo {author} {\bibfnamefont {R.~A.}\ \bibnamefont {Davis}}, \bibinfo {author} {\bibfnamefont {W.~A.}\ \bibnamefont {Angermeier}}, \bibinfo {author} {\bibfnamefont {R.~K.~T.}\ \bibnamefont {Hermsmeier}}, \ and\ \bibinfo {author} {\bibfnamefont {T.~G.}\ \bibnamefont {White}},\ }\href {\doibase 10.1103/PhysRevResearch.2.043139} {\bibfield  {journal} {\bibinfo  {journal} {Phys. Rev. Res.}\ }\textbf {\bibinfo {volume} {2}},\ \bibinfo {pages} {043139} (\bibinfo {year} {2020})}\BibitemShut {NoStop}%
\bibitem [{\citenamefont {Dornheim}\ \emph {et~al.}(2023)\citenamefont {Dornheim}, \citenamefont {Moldabekov}, \citenamefont {Ramakrishna}, \citenamefont {Tolias}, \citenamefont {Baczewski}, \citenamefont {Kraus}, \citenamefont {Preston}, \citenamefont {Chapman}, \citenamefont {B{\"o}hme}, \citenamefont {D{\"o}ppner}, \citenamefont {Graziani}, \citenamefont {Bonitz}, \citenamefont {Cangi},\ and\ \citenamefont {Vorberger}}]{10.1063/5.0138955}%
  \BibitemOpen
  \bibfield  {author} {\bibinfo {author} {\bibfnamefont {T.}~\bibnamefont {Dornheim}}, \bibinfo {author} {\bibfnamefont {Z.~A.}\ \bibnamefont {Moldabekov}}, \bibinfo {author} {\bibfnamefont {K.}~\bibnamefont {Ramakrishna}}, \bibinfo {author} {\bibfnamefont {P.}~\bibnamefont {Tolias}}, \bibinfo {author} {\bibfnamefont {A.~D.}\ \bibnamefont {Baczewski}}, \bibinfo {author} {\bibfnamefont {D.}~\bibnamefont {Kraus}}, \bibinfo {author} {\bibfnamefont {T.~R.}\ \bibnamefont {Preston}}, \bibinfo {author} {\bibfnamefont {D.~A.}\ \bibnamefont {Chapman}}, \bibinfo {author} {\bibfnamefont {M.~P.}\ \bibnamefont {B{\"o}hme}}, \bibinfo {author} {\bibfnamefont {T.}~\bibnamefont {D{\"o}ppner}}, \bibinfo {author} {\bibfnamefont {F.}~\bibnamefont {Graziani}}, \bibinfo {author} {\bibfnamefont {M.}~\bibnamefont {Bonitz}}, \bibinfo {author} {\bibfnamefont {A.}~\bibnamefont {Cangi}}, \ and\ \bibinfo {author} {\bibfnamefont {J.}~\bibnamefont {Vorberger}},\ }\href {\doibase 10.1063/5.0138955} {\bibfield  {journal} {\bibinfo  {journal}
  {Physics of Plasmas}\ }\textbf {\bibinfo {volume} {30}},\ \bibinfo {pages} {032705} (\bibinfo {year} {2023})},\ \Eprint {http://arxiv.org/abs/https://pubs.aip.org/aip/pop/article-pdf/doi/10.1063/5.0138955/16795723/032705\_1\_online.pdf} {https://pubs.aip.org/aip/pop/article-pdf/doi/10.1063/5.0138955/16795723/032705\_1\_online.pdf} \BibitemShut {NoStop}%
\bibitem [{\citenamefont {Mercadier}\ \emph {et~al.}(2024)\citenamefont {Mercadier}, \citenamefont {Benediktovitch}, \citenamefont {Kru{\v s}i{\v c}}, \citenamefont {Kas}, \citenamefont {Schlappa}, \citenamefont {Ag{\aa}ker}, \citenamefont {Carley}, \citenamefont {Fazio}, \citenamefont {Gerasimova}, \citenamefont {Kim}, \citenamefont {{Le Guyader}}, \citenamefont {Mercurio}, \citenamefont {Parchenko}, \citenamefont {Rehr}, \citenamefont {Rubensson}, \citenamefont {Serkez}, \citenamefont {Stransky}, \citenamefont {Teichmann}, \citenamefont {Yin}, \citenamefont {{\v Z}itnik}, \citenamefont {Scherz}, \citenamefont {Ziaja},\ and\ \citenamefont {Rohringer}}]{a90936bc1d9c41fe8f767ace2bf4e491}%
  \BibitemOpen
  \bibfield  {author} {\bibinfo {author} {\bibfnamefont {L.}~\bibnamefont {Mercadier}}, \bibinfo {author} {\bibfnamefont {A.}~\bibnamefont {Benediktovitch}}, \bibinfo {author} {\bibfnamefont {{\v S}.}~\bibnamefont {Kru{\v s}i{\v c}}}, \bibinfo {author} {\bibfnamefont {J.}~\bibnamefont {Kas}}, \bibinfo {author} {\bibfnamefont {J.}~\bibnamefont {Schlappa}}, \bibinfo {author} {\bibfnamefont {M.}~\bibnamefont {Ag{\aa}ker}}, \bibinfo {author} {\bibfnamefont {R.}~\bibnamefont {Carley}}, \bibinfo {author} {\bibfnamefont {G.}~\bibnamefont {Fazio}}, \bibinfo {author} {\bibfnamefont {N.}~\bibnamefont {Gerasimova}}, \bibinfo {author} {\bibfnamefont {Y.}~\bibnamefont {Kim}}, \bibinfo {author} {\bibfnamefont {L.}~\bibnamefont {{Le Guyader}}}, \bibinfo {author} {\bibfnamefont {G.}~\bibnamefont {Mercurio}}, \bibinfo {author} {\bibfnamefont {S.}~\bibnamefont {Parchenko}}, \bibinfo {author} {\bibfnamefont {J.}~\bibnamefont {Rehr}}, \bibinfo {author} {\bibfnamefont {J.}~\bibnamefont {Rubensson}}, \bibinfo {author}
  {\bibfnamefont {S.}~\bibnamefont {Serkez}}, \bibinfo {author} {\bibfnamefont {M.}~\bibnamefont {Stransky}}, \bibinfo {author} {\bibfnamefont {M.}~\bibnamefont {Teichmann}}, \bibinfo {author} {\bibfnamefont {Z.}~\bibnamefont {Yin}}, \bibinfo {author} {\bibfnamefont {M.}~\bibnamefont {{\v Z}itnik}}, \bibinfo {author} {\bibfnamefont {A.}~\bibnamefont {Scherz}}, \bibinfo {author} {\bibfnamefont {B.}~\bibnamefont {Ziaja}}, \ and\ \bibinfo {author} {\bibfnamefont {N.}~\bibnamefont {Rohringer}},\ }\href {\doibase 10.1038/s41567-024-02587-w} {\bibfield  {journal} {\bibinfo  {journal} {Nature Physics}\ }\textbf {\bibinfo {volume} {20}},\ \bibinfo {pages} {1564} (\bibinfo {year} {2024})},\ \bibinfo {note} {publisher Copyright: {\textcopyright} The Author(s) 2024.}\BibitemShut {Stop}%
\bibitem [{\citenamefont {Moldabekov}\ \emph {et~al.}(2025)\citenamefont {Moldabekov}, \citenamefont {Vorberger},\ and\ \citenamefont {Dornheim}}]{MOLDABEKOV2025104144}%
  \BibitemOpen
  \bibfield  {author} {\bibinfo {author} {\bibfnamefont {Z.}~\bibnamefont {Moldabekov}}, \bibinfo {author} {\bibfnamefont {J.}~\bibnamefont {Vorberger}}, \ and\ \bibinfo {author} {\bibfnamefont {T.}~\bibnamefont {Dornheim}},\ }\href {\doibase https://doi.org/10.1016/j.ppnp.2024.104144} {\bibfield  {journal} {\bibinfo  {journal} {Progress in Particle and Nuclear Physics}\ }\textbf {\bibinfo {volume} {140}},\ \bibinfo {pages} {104144} (\bibinfo {year} {2025})}\BibitemShut {NoStop}%
\bibitem [{\citenamefont {Velsko}\ and\ \citenamefont {Oxtoby}(1980)}]{velsko80}%
  \BibitemOpen
  \bibfield  {author} {\bibinfo {author} {\bibfnamefont {S.}~\bibnamefont {Velsko}}\ and\ \bibinfo {author} {\bibfnamefont {D.~W.}\ \bibnamefont {Oxtoby}},\ }\href {\doibase 10.1063/1.439470} {\bibfield  {journal} {\bibinfo  {journal} {The Journal of Chemical Physics}\ }\textbf {\bibinfo {volume} {72}},\ \bibinfo {pages} {2260} (\bibinfo {year} {1980})},\ \Eprint {http://arxiv.org/abs/https://pubs.aip.org/aip/jcp/article-pdf/72/4/2260/18921174/2260\_1\_online.pdf} {https://pubs.aip.org/aip/jcp/article-pdf/72/4/2260/18921174/2260\_1\_online.pdf} \BibitemShut {NoStop}%
\bibitem [{\citenamefont {Berkowitz}\ and\ \citenamefont {Gerber}(1977)}]{BERKOWITZ1977260}%
  \BibitemOpen
  \bibfield  {author} {\bibinfo {author} {\bibfnamefont {M.}~\bibnamefont {Berkowitz}}\ and\ \bibinfo {author} {\bibfnamefont {R.}~\bibnamefont {Gerber}},\ }\href {\doibase https://doi.org/10.1016/0009-2614(77)80582-4} {\bibfield  {journal} {\bibinfo  {journal} {Chemical Physics Letters}\ }\textbf {\bibinfo {volume} {49}},\ \bibinfo {pages} {260} (\bibinfo {year} {1977})}\BibitemShut {NoStop}%
\bibitem [{\citenamefont {Head-Gordon}\ and\ \citenamefont {Tully}(1995)}]{Tully1995}%
  \BibitemOpen
  \bibfield  {author} {\bibinfo {author} {\bibfnamefont {M.}~\bibnamefont {Head-Gordon}}\ and\ \bibinfo {author} {\bibfnamefont {J.~C.}\ \bibnamefont {Tully}},\ }\href {https://doi.org/10.1063/1.469915} {\bibfield  {journal} {\bibinfo  {journal} {The Journal of Chemical Physics}\ }\textbf {\bibinfo {volume} {103}},\ \bibinfo {pages} {10137} (\bibinfo {year} {1995})},\ \Eprint {http://arxiv.org/abs/https://pubs.aip.org/aip/jcp/article-pdf/103/23/10137/19249503/10137\_1\_online.pdf} {https://pubs.aip.org/aip/jcp/article-pdf/103/23/10137/19249503/10137\_1\_online.pdf} \BibitemShut {NoStop}%
\bibitem [{\citenamefont {Rudge}\ \emph {et~al.}(2024)\citenamefont {Rudge}, \citenamefont {Kaspar}, \citenamefont {Grether}, \citenamefont {Wolf}, \citenamefont {Stock},\ and\ \citenamefont {Thoss}}]{Rudge2024}%
  \BibitemOpen
  \bibfield  {author} {\bibinfo {author} {\bibfnamefont {S.~L.}\ \bibnamefont {Rudge}}, \bibinfo {author} {\bibfnamefont {C.}~\bibnamefont {Kaspar}}, \bibinfo {author} {\bibfnamefont {R.~L.}\ \bibnamefont {Grether}}, \bibinfo {author} {\bibfnamefont {S.}~\bibnamefont {Wolf}}, \bibinfo {author} {\bibfnamefont {G.}~\bibnamefont {Stock}}, \ and\ \bibinfo {author} {\bibfnamefont {M.}~\bibnamefont {Thoss}},\ }\href {\doibase 10.1063/5.0204307} {\bibfield  {journal} {\bibinfo  {journal} {The Journal of Chemical Physics}\ }\textbf {\bibinfo {volume} {160}},\ \bibinfo {pages} {184106} (\bibinfo {year} {2024})},\ \Eprint {http://arxiv.org/abs/https://pubs.aip.org/aip/jcp/article-pdf/doi/10.1063/5.0204307/19932893/184106\_1\_5.0204307.pdf} {https://pubs.aip.org/aip/jcp/article-pdf/doi/10.1063/5.0204307/19932893/184106\_1\_5.0204307.pdf} \BibitemShut {NoStop}%
\bibitem [{Note2()}]{Note2}%
  \BibitemOpen
  \bibinfo {note} {These would not survive after we take the average and compute the energy}\BibitemShut {NoStop}%
\bibitem [{\citenamefont {Nielsen}\ and\ \citenamefont {Chuang}(2000)}]{nielsen00}%
  \BibitemOpen
  \bibfield  {author} {\bibinfo {author} {\bibfnamefont {M.~A.}\ \bibnamefont {Nielsen}}\ and\ \bibinfo {author} {\bibfnamefont {I.~L.}\ \bibnamefont {Chuang}},\ }\href@noop {} {\emph {\bibinfo {title} {Quantum Computation and Quantum Information}}}\ (\bibinfo  {publisher} {Cambridge University Press},\ \bibinfo {year} {2000})\BibitemShut {NoStop}%
\bibitem [{\citenamefont {McArdle}\ \emph {et~al.}(2020)\citenamefont {McArdle}, \citenamefont {Suguru}, \citenamefont {Aspuru-Guzik}, \citenamefont {Benjamin},\ and\ \citenamefont {Yuan}}]{McArdle_2020}%
  \BibitemOpen
  \bibfield  {author} {\bibinfo {author} {\bibfnamefont {S.}~\bibnamefont {McArdle}}, \bibinfo {author} {\bibfnamefont {E.}~\bibnamefont {Suguru}}, \bibinfo {author} {\bibfnamefont {A.}~\bibnamefont {Aspuru-Guzik}}, \bibinfo {author} {\bibfnamefont {S.~C.}\ \bibnamefont {Benjamin}}, \ and\ \bibinfo {author} {\bibfnamefont {X.}~\bibnamefont {Yuan}},\ }\href {\doibase https://doi.org/10.1103/RevModPhys.92.015003} {\bibfield  {journal} {\bibinfo  {journal} {Rev. Mod. Phys.}\ }\textbf {\bibinfo {volume} {92}},\ \bibinfo {pages} {015003} (\bibinfo {year} {2020})}\BibitemShut {NoStop}%
\bibitem [{\citenamefont {Ryabinkin}\ \emph {et~al.}(2018)\citenamefont {Ryabinkin}, \citenamefont {Yen}, \citenamefont {Genin},\ and\ \citenamefont {Izmaylov}}]{iQCC}%
  \BibitemOpen
  \bibfield  {author} {\bibinfo {author} {\bibfnamefont {I.~G.}\ \bibnamefont {Ryabinkin}}, \bibinfo {author} {\bibfnamefont {T.-C.}\ \bibnamefont {Yen}}, \bibinfo {author} {\bibfnamefont {S.~N.}\ \bibnamefont {Genin}}, \ and\ \bibinfo {author} {\bibfnamefont {A.~F.}\ \bibnamefont {Izmaylov}},\ }\href {\doibase 10.1021/acs.jctc.8b00932} {\bibfield  {journal} {\bibinfo  {journal} {Journal of Chemical Theory and Computation}\ }\textbf {\bibinfo {volume} {14}},\ \bibinfo {pages} {6317} (\bibinfo {year} {2018})}\BibitemShut {NoStop}%
\bibitem [{\citenamefont {Yamamoto}\ \emph {et~al.}(2022)\citenamefont {Yamamoto}, \citenamefont {Manrique}, \citenamefont {Kahn}, \citenamefont {Sawada},\ and\ \citenamefont {Ramo}}]{Yamamoto_2022}%
  \BibitemOpen
  \bibfield  {author} {\bibinfo {author} {\bibfnamefont {K.}~\bibnamefont {Yamamoto}}, \bibinfo {author} {\bibfnamefont {D.~Z.}\ \bibnamefont {Manrique}}, \bibinfo {author} {\bibfnamefont {I.~T.}\ \bibnamefont {Kahn}}, \bibinfo {author} {\bibfnamefont {H.}~\bibnamefont {Sawada}}, \ and\ \bibinfo {author} {\bibfnamefont {D.~M.}\ \bibnamefont {Ramo}},\ }\href {\doibase https://doi.org/10.1103/PhysRevResearch.4.033110} {\bibfield  {journal} {\bibinfo  {journal} {Phys. Rev. Research}\ }\textbf {\bibinfo {volume} {4}},\ \bibinfo {pages} {033110} (\bibinfo {year} {2022})}\BibitemShut {NoStop}%
\bibitem [{\citenamefont {Montanaro}(2016)}]{Montanaro_2016}%
  \BibitemOpen
  \bibfield  {author} {\bibinfo {author} {\bibfnamefont {A.}~\bibnamefont {Montanaro}},\ }\href {\doibase https://doi.org/10.1038/npjqi.2015.23} {\bibfield  {journal} {\bibinfo  {journal} {npj Quantum Inf.}\ }\textbf {\bibinfo {volume} {2}},\ \bibinfo {pages} {15023} (\bibinfo {year} {2016})}\BibitemShut {NoStop}%
\bibitem [{\citenamefont {Huang}\ \emph {et~al.}(2023)\citenamefont {Huang}, \citenamefont {Sheng}, \citenamefont {Govoni},\ and\ \citenamefont {Galli}}]{Huang_2023}%
  \BibitemOpen
  \bibfield  {author} {\bibinfo {author} {\bibfnamefont {B.}~\bibnamefont {Huang}}, \bibinfo {author} {\bibfnamefont {N.}~\bibnamefont {Sheng}}, \bibinfo {author} {\bibfnamefont {M.}~\bibnamefont {Govoni}}, \ and\ \bibinfo {author} {\bibfnamefont {G.}~\bibnamefont {Galli}},\ }\href {\doibase https://doi.org/10.1021/acs.jctc.2c01119} {\bibfield  {journal} {\bibinfo  {journal} {J. Chem. Theory Comput.}\ }\textbf {\bibinfo {volume} {19}},\ \bibinfo {pages} {1487} (\bibinfo {year} {2023})}\BibitemShut {NoStop}%
\bibitem [{\citenamefont {Vorwerk}\ \emph {et~al.}(2022)\citenamefont {Vorwerk}, \citenamefont {Sheng}, \citenamefont {Govoni}, \citenamefont {Huang},\ and\ \citenamefont {Galli}}]{Vorwerk_2022}%
  \BibitemOpen
  \bibfield  {author} {\bibinfo {author} {\bibfnamefont {C.}~\bibnamefont {Vorwerk}}, \bibinfo {author} {\bibfnamefont {N.}~\bibnamefont {Sheng}}, \bibinfo {author} {\bibfnamefont {M.}~\bibnamefont {Govoni}}, \bibinfo {author} {\bibfnamefont {B.}~\bibnamefont {Huang}}, \ and\ \bibinfo {author} {\bibfnamefont {G.}~\bibnamefont {Galli}},\ }\href {\doibase https://doi.org/10.1038/s43588-022-00279-0} {\bibfield  {journal} {\bibinfo  {journal} {Nat. Comput. Sci.}\ }\textbf {\bibinfo {volume} {2}},\ \bibinfo {pages} {424} (\bibinfo {year} {2022})}\BibitemShut {NoStop}%
\bibitem [{\citenamefont {Ullrich}\ and\ \citenamefont {Vignale}(2002)}]{Vignale_2002}%
  \BibitemOpen
  \bibfield  {author} {\bibinfo {author} {\bibfnamefont {C.~A.}\ \bibnamefont {Ullrich}}\ and\ \bibinfo {author} {\bibfnamefont {G.}~\bibnamefont {Vignale}},\ }\href {\doibase https://doi.org/10.1103/PhysRevB.65.245102} {\bibfield  {journal} {\bibinfo  {journal} {Phys. Rev. B}\ }\textbf {\bibinfo {volume} {65}},\ \bibinfo {pages} {245102} (\bibinfo {year} {2002})}\BibitemShut {NoStop}%
\bibitem [{\citenamefont {Shor}(1995)}]{Shor_1995}%
  \BibitemOpen
  \bibfield  {author} {\bibinfo {author} {\bibfnamefont {P.~W.}\ \bibnamefont {Shor}},\ }\href {\doibase https://doi.org/10.1103/PhysRevA.52.R2493} {\bibfield  {journal} {\bibinfo  {journal} {Phys. Rev. A}\ }\textbf {\bibinfo {volume} {52}},\ \bibinfo {pages} {R2493} (\bibinfo {year} {1995})}\BibitemShut {NoStop}%
\bibitem [{\citenamefont {Peres}(1985)}]{Peres_1985}%
  \BibitemOpen
  \bibfield  {author} {\bibinfo {author} {\bibfnamefont {A.}~\bibnamefont {Peres}},\ }\href {\doibase https://doi.org/10.1103/PhysRevA.32.3266} {\bibfield  {journal} {\bibinfo  {journal} {Phys. Rev. A}\ }\textbf {\bibinfo {volume} {32}},\ \bibinfo {pages} {3266} (\bibinfo {year} {1985})}\BibitemShut {NoStop}%
\bibitem [{\citenamefont {Takahashi}\ \emph {et~al.}(2011)\citenamefont {Takahashi}, \citenamefont {Tupitsyn}, \citenamefont {van Tol}, \citenamefont {Beedle}, \citenamefont {Hendrickson},\ and\ \citenamefont {Stamp}}]{Takahashi_2011}%
  \BibitemOpen
  \bibfield  {author} {\bibinfo {author} {\bibfnamefont {S.}~\bibnamefont {Takahashi}}, \bibinfo {author} {\bibfnamefont {I.}~\bibnamefont {Tupitsyn}}, \bibinfo {author} {\bibfnamefont {J.}~\bibnamefont {van Tol}}, \bibinfo {author} {\bibfnamefont {C.}~\bibnamefont {Beedle}}, \bibinfo {author} {\bibfnamefont {D.}~\bibnamefont {Hendrickson}}, \ and\ \bibinfo {author} {\bibfnamefont {P.}~\bibnamefont {Stamp}},\ }\href {\doibase https://doi.org/10.1038/nature10314} {\bibfield  {journal} {\bibinfo  {journal} {Nature}\ }\textbf {\bibinfo {volume} {476}},\ \bibinfo {pages} {76} (\bibinfo {year} {2011})}\BibitemShut {NoStop}%
\bibitem [{\citenamefont {Stamp}(2006)}]{Stamp_2006}%
  \BibitemOpen
  \bibfield  {author} {\bibinfo {author} {\bibfnamefont {P.}~\bibnamefont {Stamp}},\ }\href {\doibase 10.1016/j.shpsb.2006.04.003} {\bibfield  {journal} {\bibinfo  {journal} {Stud. Hist. Phil. Mod. Phys.}\ }\textbf {\bibinfo {volume} {37}},\ \bibinfo {pages} {467} (\bibinfo {year} {2006})}\BibitemShut {NoStop}%
\bibitem [{\citenamefont {Leggett}(2002)}]{Leggett_2002}%
  \BibitemOpen
  \bibfield  {author} {\bibinfo {author} {\bibfnamefont {A.~J.}\ \bibnamefont {Leggett}},\ }\href@noop {} {\bibfield  {journal} {\bibinfo  {journal} {J. Phys.: Condens. Matter}\ }\textbf {\bibinfo {volume} {14}},\ \bibinfo {pages} {R415} (\bibinfo {year} {2002})}\BibitemShut {NoStop}%
\bibitem [{\citenamefont {Penrose}(1996)}]{Penrose_1996}%
  \BibitemOpen
  \bibfield  {author} {\bibinfo {author} {\bibfnamefont {R.}~\bibnamefont {Penrose}},\ }\href@noop {} {\bibfield  {journal} {\bibinfo  {journal} {Gen. Relativ. Gravit.}\ }\textbf {\bibinfo {volume} {28}},\ \bibinfo {pages} {581} (\bibinfo {year} {1996})}\BibitemShut {NoStop}%
\bibitem [{\citenamefont {'t~Hooft}(1999)}]{Hooft99}%
  \BibitemOpen
  \bibfield  {author} {\bibinfo {author} {\bibfnamefont {G.}~\bibnamefont {'t~Hooft}},\ }\href@noop {} {\bibfield  {journal} {\bibinfo  {journal} {Class. Quantum Grav.}\ }\textbf {\bibinfo {volume} {16}},\ \bibinfo {pages} {3263} (\bibinfo {year} {1999})}\BibitemShut {NoStop}%
\bibitem [{\citenamefont {Xu}\ \emph {et~al.}(2019)\citenamefont {Xu}, \citenamefont {Garc\'ia-Pintos}, \citenamefont {Chenu},\ and\ \citenamefont {del Campo~A.}}]{Xu19}%
  \BibitemOpen
  \bibfield  {author} {\bibinfo {author} {\bibfnamefont {Z.}~\bibnamefont {Xu}}, \bibinfo {author} {\bibfnamefont {L.}~\bibnamefont {Garc\'ia-Pintos}}, \bibinfo {author} {\bibfnamefont {A.}~\bibnamefont {Chenu}}, \ and\ \bibinfo {author} {\bibnamefont {del Campo~A.}},\ }\href@noop {} {\bibfield  {journal} {\bibinfo  {journal} {Phys. Rev. Lett.}\ }\textbf {\bibinfo {volume} {122}},\ \bibinfo {pages} {014103} (\bibinfo {year} {2019})}\BibitemShut {NoStop}%
\bibitem [{\citenamefont {Chatterjee}\ \emph {et~al.}(2021)\citenamefont {Chatterjee}, \citenamefont {Stevenson}, \citenamefont {De~Franceschi}, \citenamefont {Morello}, \citenamefont {de~Leon},\ and\ \citenamefont {Kuemmeth}}]{Chatterjee21}%
  \BibitemOpen
  \bibfield  {author} {\bibinfo {author} {\bibfnamefont {A.}~\bibnamefont {Chatterjee}}, \bibinfo {author} {\bibfnamefont {P.}~\bibnamefont {Stevenson}}, \bibinfo {author} {\bibfnamefont {S.}~\bibnamefont {De~Franceschi}}, \bibinfo {author} {\bibfnamefont {A.}~\bibnamefont {Morello}}, \bibinfo {author} {\bibfnamefont {N.}~\bibnamefont {de~Leon}}, \ and\ \bibinfo {author} {\bibfnamefont {F.}~\bibnamefont {Kuemmeth}},\ }\href {\doibase https://doi.org/10.1038/s42254-021-00283-9} {\bibfield  {journal} {\bibinfo  {journal} {Nature Reviews Physics}\ }\textbf {\bibinfo {volume} {3}},\ \bibinfo {pages} {157} (\bibinfo {year} {2021})}\BibitemShut {NoStop}%
\bibitem [{\citenamefont {Kjaergaard}\ \emph {et~al.}(2020)\citenamefont {Kjaergaard}, \citenamefont {Schwartz}, \citenamefont {Braum\"uller}, \citenamefont {Krantz}, \citenamefont {Wang}, \citenamefont {Gustavsson},\ and\ \citenamefont {Oliver}}]{SCQ20}%
  \BibitemOpen
  \bibfield  {author} {\bibinfo {author} {\bibfnamefont {M.}~\bibnamefont {Kjaergaard}}, \bibinfo {author} {\bibfnamefont {M.~E.}\ \bibnamefont {Schwartz}}, \bibinfo {author} {\bibfnamefont {J.}~\bibnamefont {Braum\"uller}}, \bibinfo {author} {\bibfnamefont {P.}~\bibnamefont {Krantz}}, \bibinfo {author} {\bibfnamefont {J.~I.-J.}\ \bibnamefont {Wang}}, \bibinfo {author} {\bibfnamefont {S.}~\bibnamefont {Gustavsson}}, \ and\ \bibinfo {author} {\bibfnamefont {W.~D.}\ \bibnamefont {Oliver}},\ }\href {\doibase https://doi.org/10.1146/annurev-conmatphys-031119-050605} {\bibfield  {journal} {\bibinfo  {journal} {Annual Review of Condensed Matter Physics}\ }\textbf {\bibinfo {volume} {11}},\ \bibinfo {pages} {369} (\bibinfo {year} {2020})}\BibitemShut {NoStop}%
\bibitem [{\citenamefont {Nakajima}(1958)}]{Nakajima58}%
  \BibitemOpen
  \bibfield  {author} {\bibinfo {author} {\bibfnamefont {S.}~\bibnamefont {Nakajima}},\ }\href {\doibase https://doi.org/10.1143/PTP.20.948} {\bibfield  {journal} {\bibinfo  {journal} {Progr. Theor. Phys.}\ }\textbf {\bibinfo {volume} {20}},\ \bibinfo {pages} {948} (\bibinfo {year} {1958})}\BibitemShut {NoStop}%
\bibitem [{\citenamefont {Zwanzig}(1960)}]{Zwanzig60}%
  \BibitemOpen
  \bibfield  {author} {\bibinfo {author} {\bibfnamefont {R.}~\bibnamefont {Zwanzig}},\ }\href {\doibase https://doi.org/10.1063/1.1731409} {\bibfield  {journal} {\bibinfo  {journal} {J. Chem. Phys.}\ }\textbf {\bibinfo {volume} {33}},\ \bibinfo {pages} {1338} (\bibinfo {year} {1960})}\BibitemShut {NoStop}%
\bibitem [{\citenamefont {Schlosshauer}(2022)}]{Schlosshauer2022}%
  \BibitemOpen
  \bibfield  {author} {\bibinfo {author} {\bibfnamefont {M.}~\bibnamefont {Schlosshauer}},\ }\href {\doibase 10.1007/978-3-030-88781-0_3} {\emph {\bibinfo {title} {From Quantum to Classical: Essays in Honour of H.-Dieter Zeh}}},\ edited by\ \bibinfo {editor} {\bibfnamefont {C.}~\bibnamefont {Kiefer}}\ (\bibinfo  {publisher} {Springer International Publishing},\ \bibinfo {address} {Cham},\ \bibinfo {year} {2022})\ pp.\ \bibinfo {pages} {45--64}\BibitemShut {NoStop}%
\bibitem [{\citenamefont {Santos}(2022)}]{Santos22}%
  \BibitemOpen
  \bibfield  {author} {\bibinfo {author} {\bibfnamefont {E.}~\bibnamefont {Santos}},\ }in\ \href {\doibase 10.1093/oxfordhb/9780198844495.013.0052} {\emph {\bibinfo {booktitle} {{The Oxford Handbook of the History of Quantum Interpretations}}}}\ (\bibinfo  {publisher} {Oxford University Press},\ \bibinfo {year} {2022})\ \Eprint {http://arxiv.org/abs/https://academic.oup.com/book/0/chapter/364220643/chapter-ag-pdf/45614667/book\_43513\_section\_364220643.ag.pdf} {https://academic.oup.com/book/0/chapter/364220643/chapter-ag-pdf/45614667/book\_43513\_section\_364220643.ag.pdf} \BibitemShut {NoStop}%
\bibitem [{\citenamefont {Zurek}(2003)}]{Zurek2003}%
  \BibitemOpen
  \bibfield  {author} {\bibinfo {author} {\bibfnamefont {W.}~\bibnamefont {Zurek}},\ }\href {\doibase https://doi.org/10.1103/RevModPhys.75.715} {\bibfield  {journal} {\bibinfo  {journal} {Rev. Mod. Phys.}\ }\textbf {\bibinfo {volume} {75}},\ \bibinfo {pages} {715} (\bibinfo {year} {2003})}\BibitemShut {NoStop}%
\bibitem [{\citenamefont {Brune}\ \emph {et~al.}(1996)\citenamefont {Brune}, \citenamefont {Hagley}, \citenamefont {Dreyer}, \citenamefont {Ma\^{i}tre}, \citenamefont {Maali}, \citenamefont {Wunderlich}, \citenamefont {Raimond},\ and\ \citenamefont {Haroche}}]{Brune96}%
  \BibitemOpen
  \bibfield  {author} {\bibinfo {author} {\bibfnamefont {M.}~\bibnamefont {Brune}}, \bibinfo {author} {\bibfnamefont {E.}~\bibnamefont {Hagley}}, \bibinfo {author} {\bibfnamefont {J.}~\bibnamefont {Dreyer}}, \bibinfo {author} {\bibfnamefont {X.}~\bibnamefont {Ma\^{i}tre}}, \bibinfo {author} {\bibfnamefont {A.}~\bibnamefont {Maali}}, \bibinfo {author} {\bibfnamefont {C.}~\bibnamefont {Wunderlich}}, \bibinfo {author} {\bibfnamefont {J.}~\bibnamefont {Raimond}}, \ and\ \bibinfo {author} {\bibfnamefont {S.}~\bibnamefont {Haroche}},\ }\href {\doibase https://doi.org/10.1103/PhysRevLett.77.4887} {\bibfield  {journal} {\bibinfo  {journal} {Phys. Rev. Lett.}\ }\textbf {\bibinfo {volume} {77}},\ \bibinfo {pages} {4887} (\bibinfo {year} {1996})}\BibitemShut {NoStop}%
\bibitem [{\citenamefont {Raimond}\ \emph {et~al.}(2001)\citenamefont {Raimond}, \citenamefont {Brune},\ and\ \citenamefont {Haroche}}]{Raimond01}%
  \BibitemOpen
  \bibfield  {author} {\bibinfo {author} {\bibfnamefont {J.}~\bibnamefont {Raimond}}, \bibinfo {author} {\bibfnamefont {M.}~\bibnamefont {Brune}}, \ and\ \bibinfo {author} {\bibfnamefont {S.}~\bibnamefont {Haroche}},\ }\href {\doibase https://doi.org/10.1103/RevModPhys.73.565} {\bibfield  {journal} {\bibinfo  {journal} {Rev. Mod. Phys.}\ }\textbf {\bibinfo {volume} {73}},\ \bibinfo {pages} {565} (\bibinfo {year} {2001})}\BibitemShut {NoStop}%
\bibitem [{\citenamefont {Del\'eglise}\ \emph {et~al.}(2008)\citenamefont {Del\'eglise}, \citenamefont {Dotsenko}, \citenamefont {Sayrin}, \citenamefont {Bernu}, \citenamefont {Brune}, \citenamefont {Raimond},\ and\ \citenamefont {Haroche}}]{Deleglise08}%
  \BibitemOpen
  \bibfield  {author} {\bibinfo {author} {\bibfnamefont {S.}~\bibnamefont {Del\'eglise}}, \bibinfo {author} {\bibfnamefont {I.}~\bibnamefont {Dotsenko}}, \bibinfo {author} {\bibfnamefont {C.}~\bibnamefont {Sayrin}}, \bibinfo {author} {\bibfnamefont {J.}~\bibnamefont {Bernu}}, \bibinfo {author} {\bibfnamefont {M.}~\bibnamefont {Brune}}, \bibinfo {author} {\bibfnamefont {J.}~\bibnamefont {Raimond}}, \ and\ \bibinfo {author} {\bibfnamefont {S.}~\bibnamefont {Haroche}},\ }\href {\doibase https://doi.org/10.1038/nature07288} {\bibfield  {journal} {\bibinfo  {journal} {Nature}\ }\textbf {\bibinfo {volume} {455}},\ \bibinfo {pages} {510} (\bibinfo {year} {2008})}\BibitemShut {NoStop}%
\bibitem [{\citenamefont {Lidar}(2014)}]{Lidar14}%
  \BibitemOpen
  \bibfield  {author} {\bibinfo {author} {\bibfnamefont {D.}~\bibnamefont {Lidar}},\ }\href {\doibase https://doi.org/10.1002/9781118742631} {\bibfield  {journal} {\bibinfo  {journal} {Adv. Chem. Phys.}\ }\textbf {\bibinfo {volume} {154}},\ \bibinfo {pages} {295} (\bibinfo {year} {2014})}\BibitemShut {NoStop}%
\bibitem [{\citenamefont {Dalvit}\ \emph {et~al.}(2000)\citenamefont {Dalvit}, \citenamefont {Dziarmaga},\ and\ \citenamefont {Zurek}}]{Dalvit00}%
  \BibitemOpen
  \bibfield  {author} {\bibinfo {author} {\bibfnamefont {D.}~\bibnamefont {Dalvit}}, \bibinfo {author} {\bibfnamefont {J.}~\bibnamefont {Dziarmaga}}, \ and\ \bibinfo {author} {\bibfnamefont {W.}~\bibnamefont {Zurek}},\ }\href {\doibase https://doi.org/10.1103/PhysRevA.62.013607} {\bibfield  {journal} {\bibinfo  {journal} {Phys. Rev. A}\ }\textbf {\bibinfo {volume} {62}},\ \bibinfo {pages} {013607} (\bibinfo {year} {2000})}\BibitemShut {NoStop}%
\bibitem [{\citenamefont {Poyatos}\ \emph {et~al.}(1996)\citenamefont {Poyatos}, \citenamefont {Cirac},\ and\ \citenamefont {Zoller}}]{Poyatos96}%
  \BibitemOpen
  \bibfield  {author} {\bibinfo {author} {\bibfnamefont {J.}~\bibnamefont {Poyatos}}, \bibinfo {author} {\bibfnamefont {J.}~\bibnamefont {Cirac}}, \ and\ \bibinfo {author} {\bibfnamefont {P.}~\bibnamefont {Zoller}},\ }\href {\doibase https://doi.org/10.1103/PhysRevLett.77.4728} {\bibfield  {journal} {\bibinfo  {journal} {Phys. Rev. Lett.}\ }\textbf {\bibinfo {volume} {77}},\ \bibinfo {pages} {4728} (\bibinfo {year} {1996})}\BibitemShut {NoStop}%
\bibitem [{\citenamefont {Carvalho}\ \emph {et~al.}(2001)\citenamefont {Carvalho}, \citenamefont {Milman}, \citenamefont {de~Matos~Filho},\ and\ \citenamefont {Davidovich}}]{Carvalho01}%
  \BibitemOpen
  \bibfield  {author} {\bibinfo {author} {\bibfnamefont {A.}~\bibnamefont {Carvalho}}, \bibinfo {author} {\bibfnamefont {P.}~\bibnamefont {Milman}}, \bibinfo {author} {\bibfnamefont {R.}~\bibnamefont {de~Matos~Filho}}, \ and\ \bibinfo {author} {\bibfnamefont {L.}~\bibnamefont {Davidovich}},\ }\href {\doibase https://doi.org/10.1103/PhysRevLett.86.4988} {\bibfield  {journal} {\bibinfo  {journal} {Phys. Rev. Lett.}\ }\textbf {\bibinfo {volume} {86}},\ \bibinfo {pages} {4988} (\bibinfo {year} {2001})}\BibitemShut {NoStop}%
\bibitem [{\citenamefont {Wu}\ and\ \citenamefont {Lidar}(2002)}]{Wu02}%
  \BibitemOpen
  \bibfield  {author} {\bibinfo {author} {\bibfnamefont {L.-A.}\ \bibnamefont {Wu}}\ and\ \bibinfo {author} {\bibfnamefont {D.}~\bibnamefont {Lidar}},\ }\href {\doibase https://doi.org/10.1103/PhysRevLett.88.207902} {\bibfield  {journal} {\bibinfo  {journal} {Phys. Rev. Lett.}\ }\textbf {\bibinfo {volume} {88}},\ \bibinfo {pages} {207902} (\bibinfo {year} {2002})}\BibitemShut {NoStop}%
\bibitem [{\citenamefont {Viola}\ \emph {et~al.}(2000)\citenamefont {Viola}, \citenamefont {Knill},\ and\ \citenamefont {Lloyd}}]{Viola00}%
  \BibitemOpen
  \bibfield  {author} {\bibinfo {author} {\bibfnamefont {L.}~\bibnamefont {Viola}}, \bibinfo {author} {\bibfnamefont {E.}~\bibnamefont {Knill}}, \ and\ \bibinfo {author} {\bibfnamefont {S.}~\bibnamefont {Lloyd}},\ }\href {\doibase https://doi.org/10.1103/PhysRevLett.85.3520} {\bibfield  {journal} {\bibinfo  {journal} {Phys. Rev. Lett.}\ }\textbf {\bibinfo {volume} {85}},\ \bibinfo {pages} {3520} (\bibinfo {year} {2000})}\BibitemShut {NoStop}%
\bibitem [{\citenamefont {Viola}\ \emph {et~al.}(1999)\citenamefont {Viola}, \citenamefont {Knill},\ and\ \citenamefont {Lloyd}}]{Viola99}%
  \BibitemOpen
  \bibfield  {author} {\bibinfo {author} {\bibfnamefont {L.}~\bibnamefont {Viola}}, \bibinfo {author} {\bibfnamefont {E.}~\bibnamefont {Knill}}, \ and\ \bibinfo {author} {\bibfnamefont {S.}~\bibnamefont {Lloyd}},\ }\href {\doibase https://doi.org/10.1103/PhysRevLett.82.2417} {\bibfield  {journal} {\bibinfo  {journal} {Phys. Rev. Lett.}\ }\textbf {\bibinfo {volume} {82}},\ \bibinfo {pages} {2417} (\bibinfo {year} {1999})}\BibitemShut {NoStop}%
\bibitem [{\citenamefont {Viola}\ and\ \citenamefont {Lloyd}(1998)}]{Viola98}%
  \BibitemOpen
  \bibfield  {author} {\bibinfo {author} {\bibfnamefont {L.}~\bibnamefont {Viola}}\ and\ \bibinfo {author} {\bibfnamefont {S.}~\bibnamefont {Lloyd}},\ }\href {\doibase https://doi.org/10.1103/PhysRevA.58.2733} {\bibfield  {journal} {\bibinfo  {journal} {Phys. Rev. A}\ }\textbf {\bibinfo {volume} {58}},\ \bibinfo {pages} {2733} (\bibinfo {year} {1998})}\BibitemShut {NoStop}%
\bibitem [{\citenamefont {Arndt}\ \emph {et~al.}(1999)\citenamefont {Arndt}, \citenamefont {Nairz}, \citenamefont {Vos-Andreae}, \citenamefont {Keller}, \citenamefont {van~der Zouw},\ and\ \citenamefont {Zeilinger}}]{Arndt99}%
  \BibitemOpen
  \bibfield  {author} {\bibinfo {author} {\bibfnamefont {M.}~\bibnamefont {Arndt}}, \bibinfo {author} {\bibfnamefont {O.}~\bibnamefont {Nairz}}, \bibinfo {author} {\bibfnamefont {J.}~\bibnamefont {Vos-Andreae}}, \bibinfo {author} {\bibfnamefont {C.}~\bibnamefont {Keller}}, \bibinfo {author} {\bibfnamefont {G.}~\bibnamefont {van~der Zouw}}, \ and\ \bibinfo {author} {\bibfnamefont {A.}~\bibnamefont {Zeilinger}},\ }\href {\doibase https://doi.org/10.1038/44348} {\bibfield  {journal} {\bibinfo  {journal} {Nature}\ }\textbf {\bibinfo {volume} {401}},\ \bibinfo {pages} {680} (\bibinfo {year} {1999})}\BibitemShut {NoStop}%
\bibitem [{\citenamefont {Hornberger}\ \emph {et~al.}(2003)\citenamefont {Hornberger}, \citenamefont {Uttenthaler}, \citenamefont {Brezger}, \citenamefont {Hackerm\"uller}, \citenamefont {Arndt},\ and\ \citenamefont {Zeilinger}}]{Hornberger03}%
  \BibitemOpen
  \bibfield  {author} {\bibinfo {author} {\bibfnamefont {K.}~\bibnamefont {Hornberger}}, \bibinfo {author} {\bibfnamefont {S.}~\bibnamefont {Uttenthaler}}, \bibinfo {author} {\bibfnamefont {B.}~\bibnamefont {Brezger}}, \bibinfo {author} {\bibfnamefont {L.}~\bibnamefont {Hackerm\"uller}}, \bibinfo {author} {\bibfnamefont {M.}~\bibnamefont {Arndt}}, \ and\ \bibinfo {author} {\bibfnamefont {A.}~\bibnamefont {Zeilinger}},\ }\href {\doibase https://doi.org/10.1103/PhysRevLett.90.160401} {\bibfield  {journal} {\bibinfo  {journal} {Phys. Rev. Lett.}\ }\textbf {\bibinfo {volume} {90}},\ \bibinfo {pages} {160401} (\bibinfo {year} {2003})}\BibitemShut {NoStop}%
\bibitem [{\citenamefont {Fein}\ \emph {et~al.}(2019)\citenamefont {Fein}, \citenamefont {Geyer}, \citenamefont {Zwick}, \citenamefont {Kialka}, \citenamefont {Pedalino}, \citenamefont {Mayor}, \citenamefont {Gerlich},\ and\ \citenamefont {Arndt}}]{Fein19}%
  \BibitemOpen
  \bibfield  {author} {\bibinfo {author} {\bibfnamefont {Y.}~\bibnamefont {Fein}}, \bibinfo {author} {\bibfnamefont {P.}~\bibnamefont {Geyer}}, \bibinfo {author} {\bibfnamefont {P.}~\bibnamefont {Zwick}}, \bibinfo {author} {\bibfnamefont {F.}~\bibnamefont {Kialka}}, \bibinfo {author} {\bibfnamefont {S.}~\bibnamefont {Pedalino}}, \bibinfo {author} {\bibfnamefont {M.}~\bibnamefont {Mayor}}, \bibinfo {author} {\bibfnamefont {S.}~\bibnamefont {Gerlich}}, \ and\ \bibinfo {author} {\bibfnamefont {M.}~\bibnamefont {Arndt}},\ }\href {\doibase https://doi.org/10.1038/s41567-019-0663-9} {\bibfield  {journal} {\bibinfo  {journal} {Nat. Phys.}\ }\textbf {\bibinfo {volume} {15}},\ \bibinfo {pages} {1242} (\bibinfo {year} {2019})}\BibitemShut {NoStop}%
\bibitem [{\citenamefont {Ernst}\ \emph {et~al.}(1990)\citenamefont {Ernst}, \citenamefont {Bodenhausen},\ and\ \citenamefont {Wokaun}}]{ernst1990principles}%
  \BibitemOpen
  \bibfield  {author} {\bibinfo {author} {\bibfnamefont {R.}~\bibnamefont {Ernst}}, \bibinfo {author} {\bibfnamefont {G.}~\bibnamefont {Bodenhausen}}, \ and\ \bibinfo {author} {\bibfnamefont {A.}~\bibnamefont {Wokaun}},\ }\href {https://books.google.com/books?id=4bohtceju9kC} {\emph {\bibinfo {title} {Principles of Nuclear Magnetic Resonance in One and Two Dimensions}}},\ International series of monographs on chemistry\ (\bibinfo  {publisher} {Clarendon Press},\ \bibinfo {year} {1990})\BibitemShut {NoStop}%
\bibitem [{\citenamefont {Slichter}(1996)}]{slichter1996principles}%
  \BibitemOpen
  \bibfield  {author} {\bibinfo {author} {\bibfnamefont {C.}~\bibnamefont {Slichter}},\ }\href {https://books.google.com/books?id=zgnrRkaIhFoC} {\emph {\bibinfo {title} {Principles of Magnetic Resonance}}},\ Springer Series in Solid-State Sciences\ (\bibinfo  {publisher} {Springer Berlin Heidelberg},\ \bibinfo {year} {1996})\BibitemShut {NoStop}%
\bibitem [{\citenamefont {Shen}(2003)}]{shen2003principles}%
  \BibitemOpen
  \bibfield  {author} {\bibinfo {author} {\bibfnamefont {Y.}~\bibnamefont {Shen}},\ }\href {https://books.google.com/books?id=xeg9AQAAIAAJ} {\emph {\bibinfo {title} {The Principles of Nonlinear Optics}}},\ Wiley classics library\ (\bibinfo  {publisher} {Wiley},\ \bibinfo {year} {2003})\BibitemShut {NoStop}%
\bibitem [{\citenamefont {Schlosshauer}(2019)}]{Schlosshauer19}%
  \BibitemOpen
  \bibfield  {author} {\bibinfo {author} {\bibfnamefont {M.}~\bibnamefont {Schlosshauer}},\ }\href {\doibase https://doi.org/10.1016/j.physrep.2019.10.001} {\bibfield  {journal} {\bibinfo  {journal} {Phys. Rep.}\ }\textbf {\bibinfo {volume} {831}},\ \bibinfo {pages} {1} (\bibinfo {year} {2019})}\BibitemShut {NoStop}%
\bibitem [{\citenamefont {Schlosshauer}(2005)}]{Schlosshauer05}%
  \BibitemOpen
  \bibfield  {author} {\bibinfo {author} {\bibfnamefont {M.}~\bibnamefont {Schlosshauer}},\ }\href {\doibase https://doi.org/10.1103/RevModPhys.76.1267} {\bibfield  {journal} {\bibinfo  {journal} {Rev. Mod. Phys.}\ }\textbf {\bibinfo {volume} {76}},\ \bibinfo {pages} {1267} (\bibinfo {year} {2005})}\BibitemShut {NoStop}%
\bibitem [{\citenamefont {Dyakonov}(2008)}]{dyakonov2008spin}%
  \BibitemOpen
  \bibfield  {author} {\bibinfo {author} {\bibfnamefont {M.}~\bibnamefont {Dyakonov}},\ }\href {https://books.google.com/books?id=DoZu5QHoHOQC} {\emph {\bibinfo {title} {Spin Physics in Semiconductors}}},\ Springer Series in Solid-State Sciences\ (\bibinfo  {publisher} {Springer Berlin Heidelberg},\ \bibinfo {year} {2008})\BibitemShut {NoStop}%
\bibitem [{\citenamefont {Vignale}(2004)}]{Vignale2004}%
  \BibitemOpen
  \bibfield  {author} {\bibinfo {author} {\bibfnamefont {G.}~\bibnamefont {Vignale}},\ }\href {\doibase https://doi.org/10.1103/PhysRevB.70.201102} {\bibfield  {journal} {\bibinfo  {journal} {Phys. Rev. B}\ }\textbf {\bibinfo {volume} {70}},\ \bibinfo {pages} {201102(R)} (\bibinfo {year} {2004})}\BibitemShut {NoStop}%
\bibitem [{\citenamefont {Vignale}\ and\ \citenamefont {Kohn}(1996{\natexlab{a}})}]{Vignale1996}%
  \BibitemOpen
  \bibfield  {author} {\bibinfo {author} {\bibfnamefont {G.}~\bibnamefont {Vignale}}\ and\ \bibinfo {author} {\bibfnamefont {W.}~\bibnamefont {Kohn}},\ }\href {\doibase https://doi.org/10.1103/PhysRevLett.77.2037} {\bibfield  {journal} {\bibinfo  {journal} {Phys. Rev. Lett.}\ }\textbf {\bibinfo {volume} {77}},\ \bibinfo {pages} {2037} (\bibinfo {year} {1996}{\natexlab{a}})}\BibitemShut {NoStop}%
\bibitem [{\citenamefont {Rutckaia}\ and\ \citenamefont {Schilling}(2020)}]{Rutckaia2020}%
  \BibitemOpen
  \bibfield  {author} {\bibinfo {author} {\bibfnamefont {V.}~\bibnamefont {Rutckaia}}\ and\ \bibinfo {author} {\bibfnamefont {J.}~\bibnamefont {Schilling}},\ }\href {\doibase https://doi.org/10.1038/s41566-019-0571-7} {\bibfield  {journal} {\bibinfo  {journal} {Nat. Photonics}\ }\textbf {\bibinfo {volume} {14}},\ \bibinfo {pages} {4} (\bibinfo {year} {2020})}\BibitemShut {NoStop}%
\bibitem [{\citenamefont {Jiang}\ and\ \citenamefont {Wu}(2009)}]{PhysRevB.79.125206}%
  \BibitemOpen
  \bibfield  {author} {\bibinfo {author} {\bibfnamefont {J.~H.}\ \bibnamefont {Jiang}}\ and\ \bibinfo {author} {\bibfnamefont {M.~W.}\ \bibnamefont {Wu}},\ }\href {\doibase 10.1103/PhysRevB.79.125206} {\bibfield  {journal} {\bibinfo  {journal} {Phys. Rev. B}\ }\textbf {\bibinfo {volume} {79}},\ \bibinfo {pages} {125206} (\bibinfo {year} {2009})}\BibitemShut {NoStop}%
\bibitem [{\citenamefont {Chovan}\ and\ \citenamefont {Perakis}(2008)}]{PhysRevB.77.085321}%
  \BibitemOpen
  \bibfield  {author} {\bibinfo {author} {\bibfnamefont {J.}~\bibnamefont {Chovan}}\ and\ \bibinfo {author} {\bibfnamefont {I.~E.}\ \bibnamefont {Perakis}},\ }\href {\doibase 10.1103/PhysRevB.77.085321} {\bibfield  {journal} {\bibinfo  {journal} {Phys. Rev. B}\ }\textbf {\bibinfo {volume} {77}},\ \bibinfo {pages} {085321} (\bibinfo {year} {2008})}\BibitemShut {NoStop}%
\bibitem [{\citenamefont {Li}\ \emph {et~al.}(2013)\citenamefont {Li}, \citenamefont {Patz}, \citenamefont {Mouchliadis}, \citenamefont {Yan}, \citenamefont {Lograsso}, \citenamefont {Perakis},\ and\ \citenamefont {Wang}}]{LiNat2013}%
  \BibitemOpen
  \bibfield  {author} {\bibinfo {author} {\bibfnamefont {T.}~\bibnamefont {Li}}, \bibinfo {author} {\bibfnamefont {A.}~\bibnamefont {Patz}}, \bibinfo {author} {\bibfnamefont {L.}~\bibnamefont {Mouchliadis}}, \bibinfo {author} {\bibfnamefont {J.}~\bibnamefont {Yan}}, \bibinfo {author} {\bibfnamefont {T.}~\bibnamefont {Lograsso}}, \bibinfo {author} {\bibfnamefont {I.}~\bibnamefont {Perakis}}, \ and\ \bibinfo {author} {\bibfnamefont {J.}~\bibnamefont {Wang}},\ }\href {https://doi.org/10.1038/nature11934} {\bibfield  {journal} {\bibinfo  {journal} {Nature}\ }\textbf {\bibinfo {volume} {496}},\ \bibinfo {pages} {69} (\bibinfo {year} {2013})}\BibitemShut {NoStop}%
\bibitem [{\citenamefont {Zhang}\ \emph {et~al.}(2016)\citenamefont {Zhang}, \citenamefont {Bai},\ and\ \citenamefont {George}}]{Zhang2016}%
  \BibitemOpen
  \bibfield  {author} {\bibinfo {author} {\bibfnamefont {G.}~\bibnamefont {Zhang}}, \bibinfo {author} {\bibfnamefont {Y.}~\bibnamefont {Bai}}, \ and\ \bibinfo {author} {\bibfnamefont {T.}~\bibnamefont {George}},\ }\href {\doibase 10.1088/0953-8984/28/23/236004} {\bibfield  {journal} {\bibinfo  {journal} {J. Phys.: Condens. Matter}\ }\textbf {\bibinfo {volume} {28}},\ \bibinfo {pages} {236004} (\bibinfo {year} {2016})}\BibitemShut {NoStop}%
\bibitem [{\citenamefont {Hybertsen}\ and\ \citenamefont {Louie}(1986)}]{PhysRevB.34.5390}%
  \BibitemOpen
  \bibfield  {author} {\bibinfo {author} {\bibfnamefont {M.~S.}\ \bibnamefont {Hybertsen}}\ and\ \bibinfo {author} {\bibfnamefont {S.~G.}\ \bibnamefont {Louie}},\ }\href {\doibase 10.1103/PhysRevB.34.5390} {\bibfield  {journal} {\bibinfo  {journal} {Phys. Rev. B}\ }\textbf {\bibinfo {volume} {34}},\ \bibinfo {pages} {5390} (\bibinfo {year} {1986})}\BibitemShut {NoStop}%
\bibitem [{\citenamefont {Joost}\ \emph {et~al.}(2020{\natexlab{b}})\citenamefont {Joost}, \citenamefont {Schl\"unzen},\ and\ \citenamefont {Bonitz}}]{PhysRevB.101.245101}%
  \BibitemOpen
  \bibfield  {author} {\bibinfo {author} {\bibfnamefont {J.-P.}\ \bibnamefont {Joost}}, \bibinfo {author} {\bibfnamefont {N.}~\bibnamefont {Schl\"unzen}}, \ and\ \bibinfo {author} {\bibfnamefont {M.}~\bibnamefont {Bonitz}},\ }\href {\doibase 10.1103/PhysRevB.101.245101} {\bibfield  {journal} {\bibinfo  {journal} {Phys. Rev. B}\ }\textbf {\bibinfo {volume} {101}},\ \bibinfo {pages} {245101} (\bibinfo {year} {2020}{\natexlab{b}})}\BibitemShut {NoStop}%
\bibitem [{\citenamefont {Elliott}\ \emph {et~al.}(2012)\citenamefont {Elliott}, \citenamefont {Fuks}, \citenamefont {Rubio},\ and\ \citenamefont {Maitra}}]{PhysRevLett.109.266404}%
  \BibitemOpen
  \bibfield  {author} {\bibinfo {author} {\bibfnamefont {P.}~\bibnamefont {Elliott}}, \bibinfo {author} {\bibfnamefont {J.~I.}\ \bibnamefont {Fuks}}, \bibinfo {author} {\bibfnamefont {A.}~\bibnamefont {Rubio}}, \ and\ \bibinfo {author} {\bibfnamefont {N.~T.}\ \bibnamefont {Maitra}},\ }\href {\doibase 10.1103/PhysRevLett.109.266404} {\bibfield  {journal} {\bibinfo  {journal} {Phys. Rev. Lett.}\ }\textbf {\bibinfo {volume} {109}},\ \bibinfo {pages} {266404} (\bibinfo {year} {2012})}\BibitemShut {NoStop}%
\bibitem [{\citenamefont {Helbig}\ \emph {et~al.}(2011)\citenamefont {Helbig}, \citenamefont {Fuks}, \citenamefont {Tokatly}, \citenamefont {Appel}, \citenamefont {Gross},\ and\ \citenamefont {Rubio}}]{HELBIG20111}%
  \BibitemOpen
  \bibfield  {author} {\bibinfo {author} {\bibfnamefont {N.}~\bibnamefont {Helbig}}, \bibinfo {author} {\bibfnamefont {J.}~\bibnamefont {Fuks}}, \bibinfo {author} {\bibfnamefont {I.}~\bibnamefont {Tokatly}}, \bibinfo {author} {\bibfnamefont {H.}~\bibnamefont {Appel}}, \bibinfo {author} {\bibfnamefont {E.}~\bibnamefont {Gross}}, \ and\ \bibinfo {author} {\bibfnamefont {A.}~\bibnamefont {Rubio}},\ }\href {\doibase https://doi.org/10.1016/j.chemphys.2011.06.010} {\bibfield  {journal} {\bibinfo  {journal} {Chemical Physics}\ }\textbf {\bibinfo {volume} {391}},\ \bibinfo {pages} {1} (\bibinfo {year} {2011})},\ \bibinfo {note} {open problems and new solutions in time dependent density functional theory}\BibitemShut {NoStop}%
\bibitem [{\citenamefont {Vignale}\ and\ \citenamefont {Kohn}(1996{\natexlab{b}})}]{PhysRevLett.77.2037}%
  \BibitemOpen
  \bibfield  {author} {\bibinfo {author} {\bibfnamefont {G.}~\bibnamefont {Vignale}}\ and\ \bibinfo {author} {\bibfnamefont {W.}~\bibnamefont {Kohn}},\ }\href {\doibase 10.1103/PhysRevLett.77.2037} {\bibfield  {journal} {\bibinfo  {journal} {Phys. Rev. Lett.}\ }\textbf {\bibinfo {volume} {77}},\ \bibinfo {pages} {2037} (\bibinfo {year} {1996}{\natexlab{b}})}\BibitemShut {NoStop}%
\bibitem [{\citenamefont {Vignale}\ \emph {et~al.}(1997)\citenamefont {Vignale}, \citenamefont {Ullrich},\ and\ \citenamefont {Conti}}]{PhysRevLett.79.4878}%
  \BibitemOpen
  \bibfield  {author} {\bibinfo {author} {\bibfnamefont {G.}~\bibnamefont {Vignale}}, \bibinfo {author} {\bibfnamefont {C.~A.}\ \bibnamefont {Ullrich}}, \ and\ \bibinfo {author} {\bibfnamefont {S.}~\bibnamefont {Conti}},\ }\href {\doibase 10.1103/PhysRevLett.79.4878} {\bibfield  {journal} {\bibinfo  {journal} {Phys. Rev. Lett.}\ }\textbf {\bibinfo {volume} {79}},\ \bibinfo {pages} {4878} (\bibinfo {year} {1997})}\BibitemShut {NoStop}%
\bibitem [{\citenamefont {Ullrich}\ and\ \citenamefont {Tokatly}(2006)}]{PhysRevB.73.235102}%
  \BibitemOpen
  \bibfield  {author} {\bibinfo {author} {\bibfnamefont {C.~A.}\ \bibnamefont {Ullrich}}\ and\ \bibinfo {author} {\bibfnamefont {I.~V.}\ \bibnamefont {Tokatly}},\ }\href {\doibase 10.1103/PhysRevB.73.235102} {\bibfield  {journal} {\bibinfo  {journal} {Phys. Rev. B}\ }\textbf {\bibinfo {volume} {73}},\ \bibinfo {pages} {235102} (\bibinfo {year} {2006})}\BibitemShut {NoStop}%
\bibitem [{\citenamefont {Simoni}\ and\ \citenamefont {Sanvito}(2022)}]{PhysRevB.105.104437}%
  \BibitemOpen
  \bibfield  {author} {\bibinfo {author} {\bibfnamefont {J.}~\bibnamefont {Simoni}}\ and\ \bibinfo {author} {\bibfnamefont {S.}~\bibnamefont {Sanvito}},\ }\href {\doibase 10.1103/PhysRevB.105.104437} {\bibfield  {journal} {\bibinfo  {journal} {Phys. Rev. B}\ }\textbf {\bibinfo {volume} {105}},\ \bibinfo {pages} {104437} (\bibinfo {year} {2022})}\BibitemShut {NoStop}%
\bibitem [{\citenamefont {Ghosh}\ and\ \citenamefont {Dhara}(1988)}]{PhysRevA.38.1149}%
  \BibitemOpen
  \bibfield  {author} {\bibinfo {author} {\bibfnamefont {S.~K.}\ \bibnamefont {Ghosh}}\ and\ \bibinfo {author} {\bibfnamefont {A.~K.}\ \bibnamefont {Dhara}},\ }\href {\doibase 10.1103/PhysRevA.38.1149} {\bibfield  {journal} {\bibinfo  {journal} {Phys. Rev. A}\ }\textbf {\bibinfo {volume} {38}},\ \bibinfo {pages} {1149} (\bibinfo {year} {1988})}\BibitemShut {NoStop}%
\bibitem [{\citenamefont {van Leeuwen}(1999)}]{PhysRevLett.82.3863}%
  \BibitemOpen
  \bibfield  {author} {\bibinfo {author} {\bibfnamefont {R.}~\bibnamefont {van Leeuwen}},\ }\href {\doibase 10.1103/PhysRevLett.82.3863} {\bibfield  {journal} {\bibinfo  {journal} {Phys. Rev. Lett.}\ }\textbf {\bibinfo {volume} {82}},\ \bibinfo {pages} {3863} (\bibinfo {year} {1999})}\BibitemShut {NoStop}%
\bibitem [{\citenamefont {Spohn}(1980)}]{RevModPhys.52.569}%
  \BibitemOpen
  \bibfield  {author} {\bibinfo {author} {\bibfnamefont {H.}~\bibnamefont {Spohn}},\ }\href {\doibase 10.1103/RevModPhys.52.569} {\bibfield  {journal} {\bibinfo  {journal} {Rev. Mod. Phys.}\ }\textbf {\bibinfo {volume} {52}},\ \bibinfo {pages} {569} (\bibinfo {year} {1980})}\BibitemShut {NoStop}%
\bibitem [{\citenamefont {Berkelbach}\ and\ \citenamefont {Thoss}(2020)}]{10.1063/1.5142731}%
  \BibitemOpen
  \bibfield  {author} {\bibinfo {author} {\bibfnamefont {T.~C.}\ \bibnamefont {Berkelbach}}\ and\ \bibinfo {author} {\bibfnamefont {M.}~\bibnamefont {Thoss}},\ }\href {\doibase 10.1063/1.5142731} {\bibfield  {journal} {\bibinfo  {journal} {The Journal of Chemical Physics}\ }\textbf {\bibinfo {volume} {152}},\ \bibinfo {pages} {020401} (\bibinfo {year} {2020})},\ \Eprint {http://arxiv.org/abs/https://pubs.aip.org/aip/jcp/article-pdf/doi/10.1063/1.5142731/15568495/020401\_1\_online.pdf} {https://pubs.aip.org/aip/jcp/article-pdf/doi/10.1063/1.5142731/15568495/020401\_1\_online.pdf} \BibitemShut {NoStop}%
\bibitem [{\citenamefont {Cai}\ \emph {et~al.}(2010)\citenamefont {Cai}, \citenamefont {Popescu},\ and\ \citenamefont {Briegel}}]{PhysRevE.82.021921}%
  \BibitemOpen
  \bibfield  {author} {\bibinfo {author} {\bibfnamefont {J.}~\bibnamefont {Cai}}, \bibinfo {author} {\bibfnamefont {S.}~\bibnamefont {Popescu}}, \ and\ \bibinfo {author} {\bibfnamefont {H.~J.}\ \bibnamefont {Briegel}},\ }\href {\doibase 10.1103/PhysRevE.82.021921} {\bibfield  {journal} {\bibinfo  {journal} {Phys. Rev. E}\ }\textbf {\bibinfo {volume} {82}},\ \bibinfo {pages} {021921} (\bibinfo {year} {2010})}\BibitemShut {NoStop}%
\bibitem [{\citenamefont {Collini}\ \emph {et~al.}(2010)\citenamefont {Collini}, \citenamefont {Wong}, \citenamefont {Wilk}, \citenamefont {Curmi}, \citenamefont {Brumer},\ and\ \citenamefont {Scholes}}]{Qbiol1}%
  \BibitemOpen
  \bibfield  {author} {\bibinfo {author} {\bibfnamefont {E.}~\bibnamefont {Collini}}, \bibinfo {author} {\bibfnamefont {C.}~\bibnamefont {Wong}}, \bibinfo {author} {\bibfnamefont {K.}~\bibnamefont {Wilk}}, \bibinfo {author} {\bibfnamefont {P.}~\bibnamefont {Curmi}}, \bibinfo {author} {\bibfnamefont {P.}~\bibnamefont {Brumer}}, \ and\ \bibinfo {author} {\bibfnamefont {G.}~\bibnamefont {Scholes}},\ }\href {\doibase https://doi.org/10.1038/nature08811} {\bibfield  {journal} {\bibinfo  {journal} {Nature}\ }\textbf {\bibinfo {volume} {463}},\ \bibinfo {pages} {644} (\bibinfo {year} {2010})}\BibitemShut {NoStop}%
\bibitem [{\citenamefont {Engel}\ \emph {et~al.}(2007)\citenamefont {Engel}, \citenamefont {Calhoun}, \citenamefont {Read}, \citenamefont {Ahn}, \citenamefont {Mancal}, \citenamefont {Cheng}, \citenamefont {Blankenship},\ and\ \citenamefont {Fleming}}]{Qbiol2}%
  \BibitemOpen
  \bibfield  {author} {\bibinfo {author} {\bibfnamefont {G.}~\bibnamefont {Engel}}, \bibinfo {author} {\bibfnamefont {T.}~\bibnamefont {Calhoun}}, \bibinfo {author} {\bibfnamefont {E.}~\bibnamefont {Read}}, \bibinfo {author} {\bibfnamefont {T.-K.}\ \bibnamefont {Ahn}}, \bibinfo {author} {\bibfnamefont {T.}~\bibnamefont {Mancal}}, \bibinfo {author} {\bibfnamefont {Y.-C.}\ \bibnamefont {Cheng}}, \bibinfo {author} {\bibfnamefont {R.}~\bibnamefont {Blankenship}}, \ and\ \bibinfo {author} {\bibfnamefont {G.}~\bibnamefont {Fleming}},\ }\href {\doibase https://doi.org/10.1038/nature05678} {\bibfield  {journal} {\bibinfo  {journal} {Nature}\ }\textbf {\bibinfo {volume} {446}},\ \bibinfo {pages} {782} (\bibinfo {year} {2007})}\BibitemShut {NoStop}%
\bibitem [{\citenamefont {Baek}\ \emph {et~al.}(2024)\citenamefont {Baek}, \citenamefont {Han}, \citenamefont {Cheon},\ and\ \citenamefont {Lee}}]{Baek2024}%
  \BibitemOpen
  \bibfield  {author} {\bibinfo {author} {\bibfnamefont {I.}~\bibnamefont {Baek}}, \bibinfo {author} {\bibfnamefont {S.}~\bibnamefont {Han}}, \bibinfo {author} {\bibfnamefont {S.}~\bibnamefont {Cheon}}, \ and\ \bibinfo {author} {\bibfnamefont {H.-W.}\ \bibnamefont {Lee}},\ }\href {\doibase https://doi.org/10.1038/s44306-024-00041-4} {\bibfield  {journal} {\bibinfo  {journal} {npj Spintronics}\ }\textbf {\bibinfo {volume} {2}} (\bibinfo {year} {2024}),\ https://doi.org/10.1038/s44306-024-00041-4}\BibitemShut {NoStop}%
\end{thebibliography}%
\bibliographystyle{apsrev4-1}

\end{document}